\documentclass[twocolumn]{article}
\pdfoutput=1

\usepackage{authblk}
\usepackage{amsfonts}
\usepackage{amsmath}
\usepackage{amssymb}
\usepackage{xfrac}
\usepackage{graphicx}
\usepackage{booktabs,array}
\usepackage[backend=biber,style=phys]{biblatex}
\usepackage{comment}
\usepackage[skip=2pt,font=footnotesize]{caption}
\usepackage{titlesec}
\usepackage[flushleft]{threeparttable}
\usepackage{color}
\usepackage[normalem]{ulem}
\usepackage{tablefootnote}
\usepackage{hyperref}

\newcolumntype{L}[1]{>{\raggedright\arraybackslash}m{#1}}
\newcolumntype{C}[1]{>{\centering\arraybackslash}m{#1}}
\newcolumntype{R}[1]{>{\raggedleft\arraybackslash}m{#1}}

\titleformat*{\section}{\normalsize\bfseries\rmfamily}
\titleformat*{\subsection}{\normalsize\bfseries\rmfamily}
\titleformat*{\subsubsection}{\normalsize\bfseries\rmfamily}

\graphicspath{{./Figures/},{./Workflows/},{./OrderedFigures/}}
\DeclareGraphicsExtensions{.jpg,.png,.pdf}
\makeatletter
\def\@fnsymbol#1{\ensuremath{\ifcase#1\or \dagger\or \ddagger\or
	\mathsection\or \mathparagraph\or \|\or **\or \dagger\dagger
	\or \ddagger\ddagger \else\@ctrerr\fi}}
\makeatother

\title{Neural network surrogate of QuaLiKiz using JET experimental data to populate training space}
\author[1$\dagger$]{A. Ho}
\author[1]{J. Citrin}
\author[2]{C. Bourdelle}
\author[3]{Y. Camenen}
\author[4]{F. J. Casson}
\author[1]{K. L. van de Plassche}
\author[5]{\\H. Weisen}
\author[*]{JET Contributors}
\affil[1]{DIFFER -- Dutch Institute for Fundamental Energy Research, De Zaale 20, 5612 AJ Eindhoven, the Netherlands}
\affil[2]{CEA, IRFM, F-13108 Saint Paul Lez Durance, France}
\affil[3]{Aix-Marseille University, CNRS, PIIM, UMR 7345, 13013 Marseille, France}
\affil[4]{CCFE, Culham Science Centre, Abingdon, Oxon, OX14 3DB, United Kingdom of Great Britain and Northern Ireland}
\affil[5]{Ecole Polytechnique F{\'e}d{\'e}rale de Lausanne (EPFL), Swiss Plasma Center (SPC), CH-1015 Lausanne, Switzerland}
\affil[*]{See the author list of E. Joffrin et al. 2019 Nucl. Fusion 59 112021}
\affil[$\dagger$]{Corresponding author: a.ho@differ.nl}

\date{}

\begin{document}

\maketitle

\begin{abstract}
Within integrated tokamak plasma modelling, turbulent transport codes are typically the computational bottleneck limiting their routine use outside of post-discharge analysis. Neural network (NN) surrogates have been used to accelerate these calculations while retaining the desired accuracy of the physics-based models. This paper extends a previous NN model, known as QLKNN-hyper-10D, by incorporating the impact of impurities, plasma rotation and magnetic equilibrium effects. This is achieved by adding a light impurity fractional density ($n_{\text{imp,light}} / n_e$) and its normalized gradient, the normalized pressure gradient ($\alpha$), the toroidal Mach number ($M_{\text{tor}}$) and the normalized toroidal flow velocity gradient. The input space was sampled based on experimental data from the JET tokamak to avoid the curse of dimensionality. The resulting networks, named QLKNN-jetexp-15D, show good agreement with the original QuaLiKiz model, both by comparing individual transport quantity predictions as well as comparing its impact within the integrated model, JINTRAC. The profile-averaged RMS of the integrated modelling simulations is $<$10\% for each of the 5 scenarios tested. This is non-trivial given the potential numerical instabilities present within the highly nonlinear system of equations governing plasma transport, especially considering the novel addition of momentum flux predictions to the model proposed here. An evaluation of all 25 NN output quantities at one radial location takes $\sim$0.1~ms, $10^4$ times faster than the original QuaLiKiz model. Within the JINTRAC integrated modelling tests performed in this study, using QLKNN-jetexp-15D resulted in a speed increase of only 60--100 as other physics modules outside of turbulent transport become the bottleneck.

\end{abstract}

\section{Introduction}
\label{sec:Introduction}

With the development of increasingly powerful high performance computing resources, machine learning (ML) techniques are becoming practical tools in model construction and large data exploitation. These tools open new exploratory options to address current issues in fusion plasma analysis and operation~\cite{aNNDisruption-KatesHarbeck,aDisruptionRandomForest-Rea,aNNMagneticRecon-Bockenhoff,aBayesianPlasmaBoundary-Skvara,aNNRunaways-Hesslow}. Specifically, the usage of neural networks (NNs)~\cite{aDeepLearningNN-Schmidhuber} as fast surrogate models has already been applied to plasma microturbulent transport calculations~\cite{aProof-Citrin,aTGLFNN-Meneghini}. These NN implementations improve the applicability of nuanced physics-based calculations within integrated tokamak plasma transport modelling due to their fast evaluation times. In turn, this improves the capability of these complex models to assist in exploratory and optimization studies within a meaningful timeframe, such as those involved with the deuterium-tritium extrapolation at JET~\cite{aDTChallenges-Garcia}.

\begin{figure*}[tbp]
	\centering
	\includegraphics[scale=1.25]{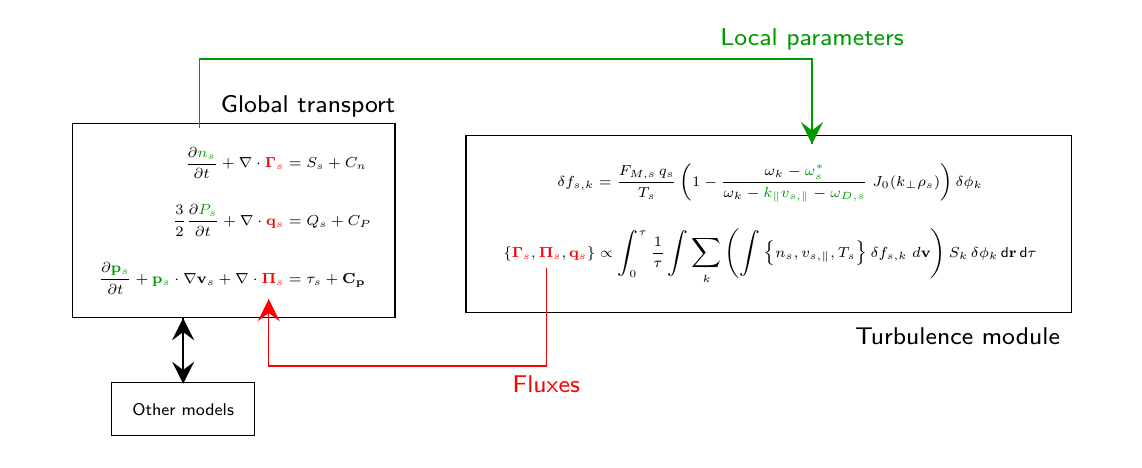}
	\caption{A schematic representation of modular integration coupling the global transport code, JINTRAC, and the turbulence module, QuaLiKiz. Under this architecture, the turbulence module can be replaced with any component which accepts the local plasma parameters and returns the transport fluxes.}
	\label{fig:IntegratedModelSchematic}
\end{figure*}

Integrated plasma transport models interconnect multiple independent physics models to consistently evaluate the plasma state under different concurrent phenomena. This consistent calculation allows for a deeper understanding of the interactions between various plasma physics phenomena and can be used to determine key macroscopic parameters for identifying various plasma regimes, as exampled by previous integrated modelling efforts~\cite{aICRHOptimization-Casson,aTungstenModelling-Breton,aTiTe-Linder}. Each of these component models are responsible for computing a specific plasma phenomenon, e.g. transport fluxes, sources or sinks. The JINTRAC integrated model~\cite{aJINTRAC-Romanelli} adopts a modular integration approach. This means that each model, known as a \emph{module}, is evaluated independently and its outputs are explicitly converted into appropriate inputs for the others, depicted schematically in Figure~\ref{fig:IntegratedModelSchematic}. This approach allows any module to be replaced with any other which has the same inputs and outputs, a property which supports the development of specialized reduced physics models.

The current bottleneck for these integrated transport models is the evaluation of microturbulent transport properties. High-fidelity nonlinear models, e.g. GENE~\cite{aETG-Jenko} or GKW~\cite{aGKW-Peeters}, require $\gtrsim$10000~CPUh per evaluation at a single radial location, prohibiting their use in routine integrated modelling analysis. Validated reduced turbulent transport models, such as QuaLiKiz~\cite{aQLK-Bourdelle}, speed up these calculations to $\sim$10~CPUs per radial location by applying careful approximations suitable for tokamak plasma scenarios. This currently allows integrated models to simulate 1~s of plasma in a couple of days.

However, the replacement modules in this integrated approach are not limited to first-principles models. QuaLiKiz NN regressions, trained on data from QuaLiKiz evaluations, provide further reduction of the computation time to $\sim$1~ms per radial location while still providing an accuracy comparable to the original model~\cite{aQLKNN-vdPlassche,aProof-Citrin}. Within integrated modelling applications, this reduces the turbulent transport prediction time to the point that it is no longer the observed bottleneck. These NNs allow the use of more complete physics models within iterative applications provided that the NN training dataset encapsulates the required plasma scenarios.

This paper extends the previous NN approach~\cite{aQLKNN-vdPlassche} by incorporating the known impact of fuel dilution~\cite{aDilutionITG-Ennever}, plasma rotation~\cite{aRotationNonLinear-Camenen} and magnetic equilibrium effects~\cite{aQLK-Citrin}. While several key aspects of these dependencies are already included in the previous work, referred to as the \emph{QLKNN-hyper-10D} in this paper, their treatment is made more explicit by allowing the NN to learn from a more complete set of input variables within the context of the underlying model, QuaLiKiz. This is achieved by including an additional 5 input parameters: a light impurity fractional density ($n_{\text{imp,light}} / n_e$) and its normalized gradient ($R/L_{n_\text{imp,light}}$), the normalized pressure gradient ($\alpha_{\text{MHD}}$), the toroidal Mach number ($M_{\text{tor}}$), and the normalized toroidal flow velocity gradient ($R/L_{u_{\text{tor}}}$).

QLKNN-hyper-10D used a lattice sampling method, referred to as a \emph{hyper-rectangular grid} in Ref.~\cite{aQLKNN-vdPlassche}, over 9 of its 10 input dimensions to populate its training dataset, with all of the input values carefully selected from domain expertise. The remaining parameter was applied via a reduced-physics model after NN evaluation. This lattice sampling approach ensures both adequate coverage of the input parameter space while simultaneously providing a convenient framework for visualizing the data for validation purposes. However, it suffers from the curse of dimensionality, which increases the dataset size exponentially with the number of input parameters.

Due to this limitation, a different dataset generation methodology was needed to include the 5 extra dimensions. Applying the proposed 5D extension to that dataset via the lattice sampling method would approach the limit of a tractable dataset size, even for modern supercomputing resources, as shown in Table~\ref{tbl:TrainingDatasetSizes}. Instead, this study populated the training dataset by extracting experimental data from the JET tokamak data repository and deriving the necessary input parameters from the processed data. This avoids the curse of dimensionality by emulating a multivariate Monte Carlo sampling method while simultaneously limiting the dataset to a relevant input space. However, the primary drawback is the loss of clear dataset boundaries, which are necessary in defining the NN applicability region as they are known to perform poorly when extrapolating. The QuaLiKiz code was executed using these experimentally-derived inputs to generate the dataset used in this study to train the proposed 15D NNs, referred to in this paper as \emph{QLKNN-jetexp-15D}.

\begin{table}
	\centering
	\caption{A summary of NN training set sizes. The projected 15D size assumes a minimal resolution of 3 points for each new input dimension.}
	\begin{tabular}{c|cc}
		Dataset & Points & Time [CPUh] \\
		\midrule
		10D & $\sim 3 \times 10^8$ & $\sim10^6$ \\
		Projected 15D & $\gtrsim 7.2 \times 10^{10}$ & $\gtrsim 3 \times 10^8$ \\
		Actual 15D & $3.38 \times 10^7$ & $3.5 \times 10^5$ \\
	\end{tabular}
	\label{tbl:TrainingDatasetSizes}
\end{table}

In order to address the dataset boundary issue, the concept of committee NNs was adopted to identify NN extrapolation regions via an increasing committee prediction variance, similar to other NN regression studies~\cite{aTGLFNN-Meneghini,aTGLFNN2-Meneghini}. The performance of these networks were then compared to the original model, both via individual local transport predictions and via its integration into JINTRAC. The latter approach demonstrates a typical use case of the original model and verifies that the contours of the output space are effectively approximated by the surrogate, due to the time evolution aspect of the integrated model. The applicability of using the committee NN variance as indicators of the training dataset boundary was also evaluated. Although the variance, $\sigma^2$, is the term typically used in NN literature, the actual metric evaluated in this paper is the standard deviation, $\sigma$.

Section~\ref{sec:DatasetGeneration} outlines the steps taken in dataset generation in order to ensure the trained NN gives decent predictions within experimentally-relevant parameter space and Section~\ref{sec:NNTraining} describes the NN training procedure itself. Section~\ref{sec:ComparisonStudies} discusses the comparison studies performed and their implications. Finally, a summary is provided in Section~\ref{sec:Conclusions} and comments are made on any potential future work.

\section{Dataset generation}
\label{sec:DatasetGeneration}

\begin{figure*}[t]
	\centering
	\includegraphics[scale=0.5]{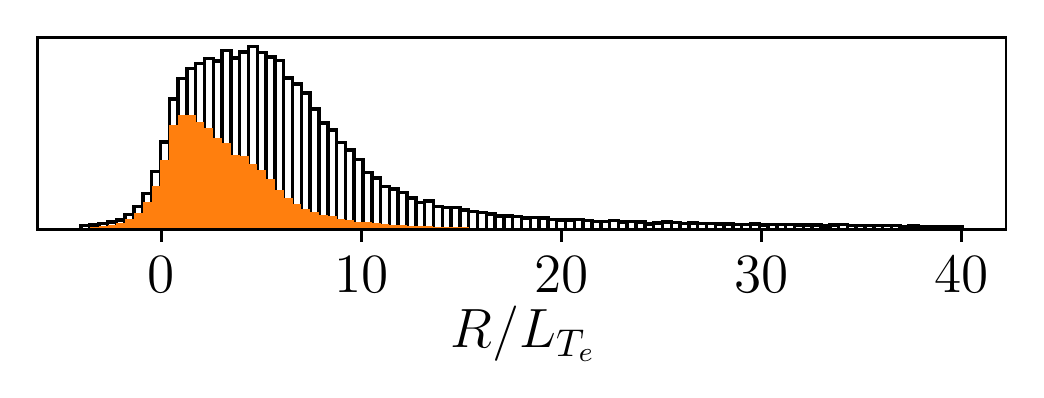}%
	\includegraphics[scale=0.5]{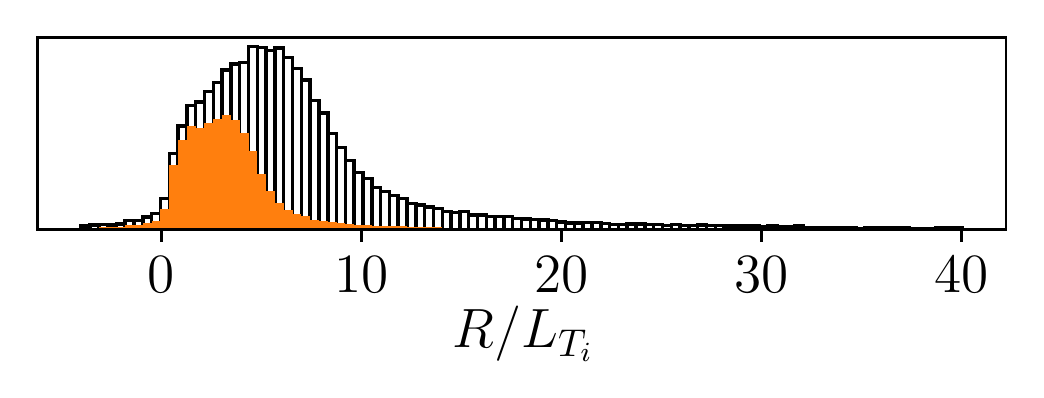}%
	\includegraphics[scale=0.5]{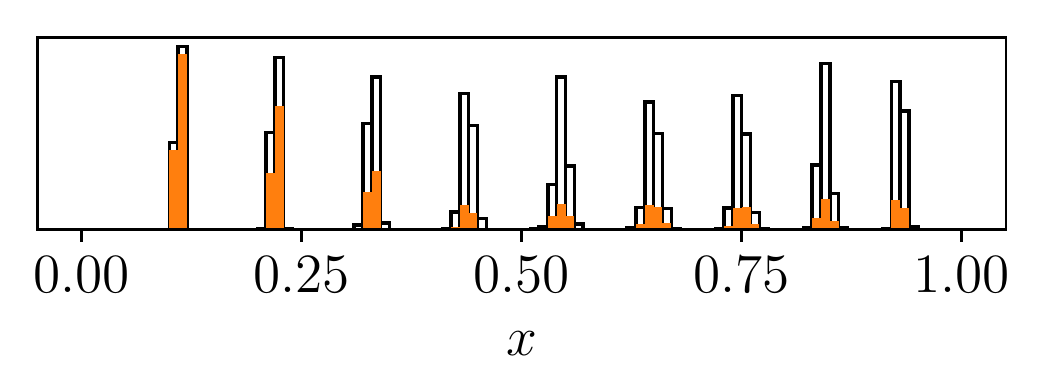}\\
	\vspace{-1mm}\includegraphics[scale=0.5]{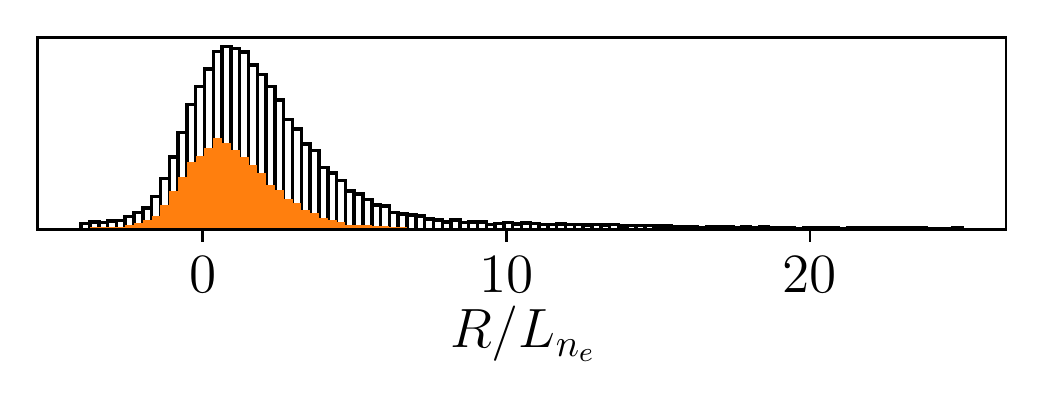}%
	\includegraphics[scale=0.5]{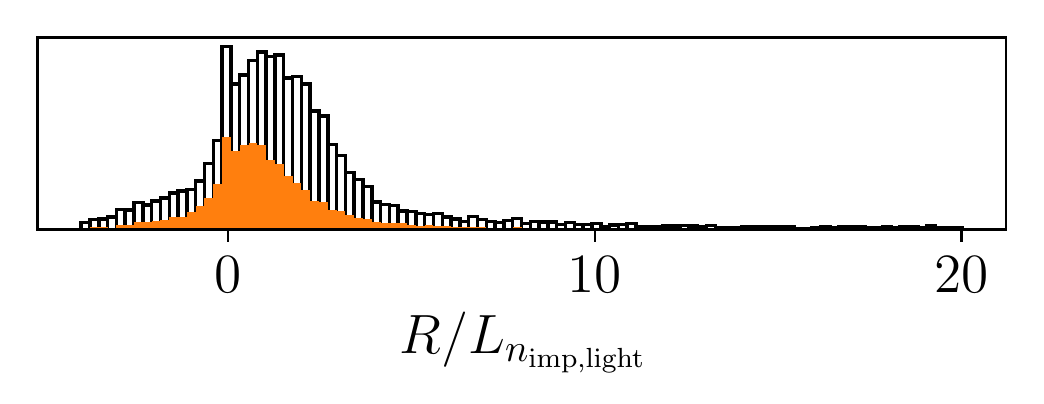}%
	\includegraphics[scale=0.5]{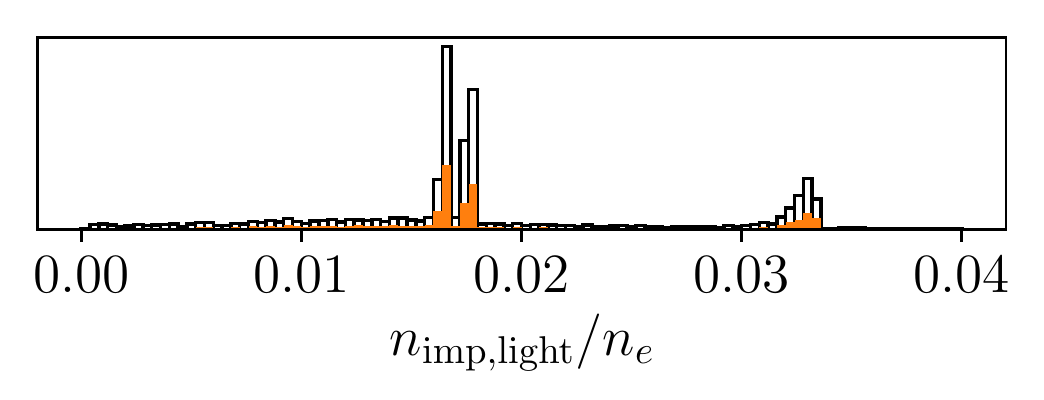}\\
	\vspace{-1mm}\includegraphics[scale=0.5]{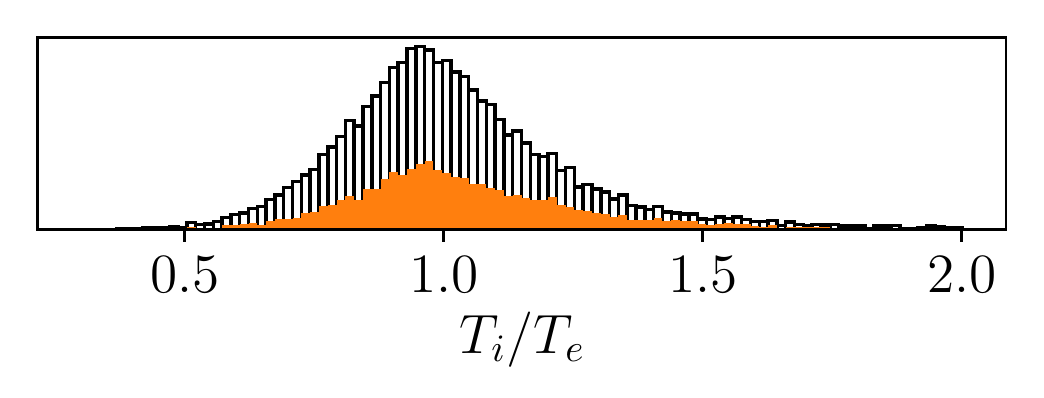}%
	\includegraphics[scale=0.5]{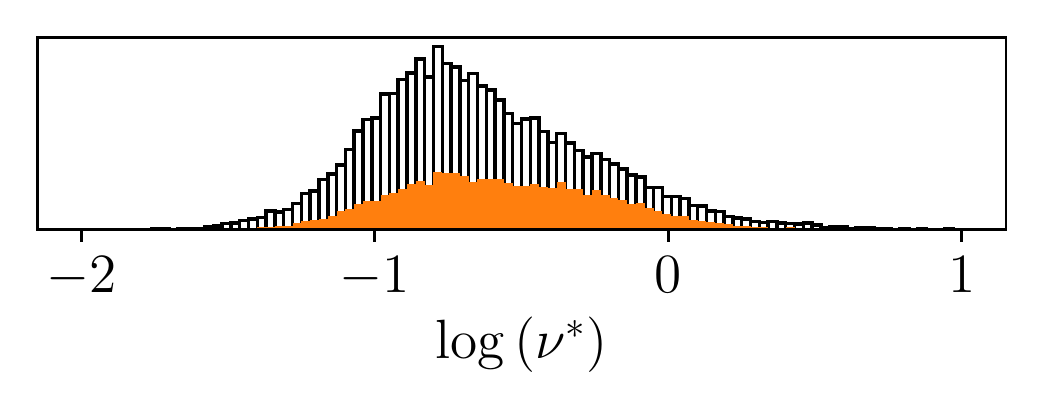}%
	\includegraphics[scale=0.5]{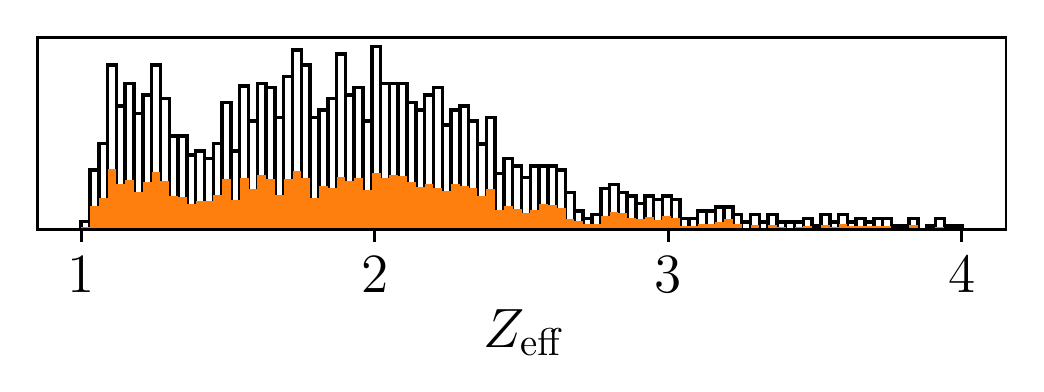}\\
	\vspace{-1mm}\includegraphics[scale=0.5]{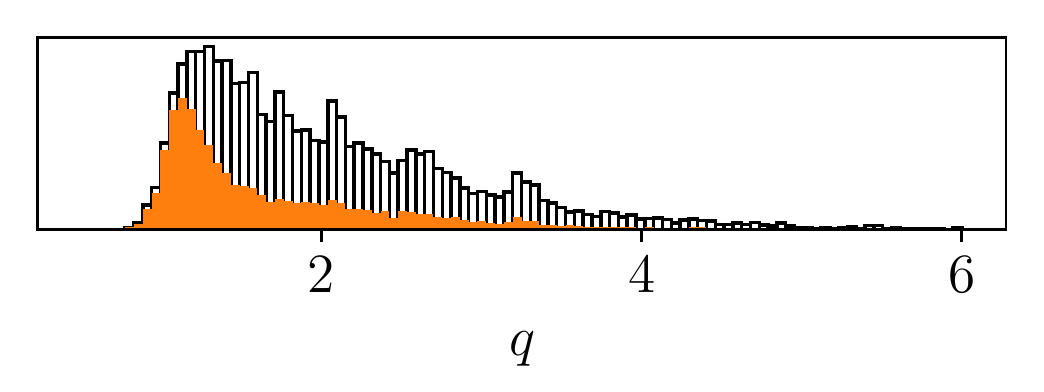}%
	\includegraphics[scale=0.5]{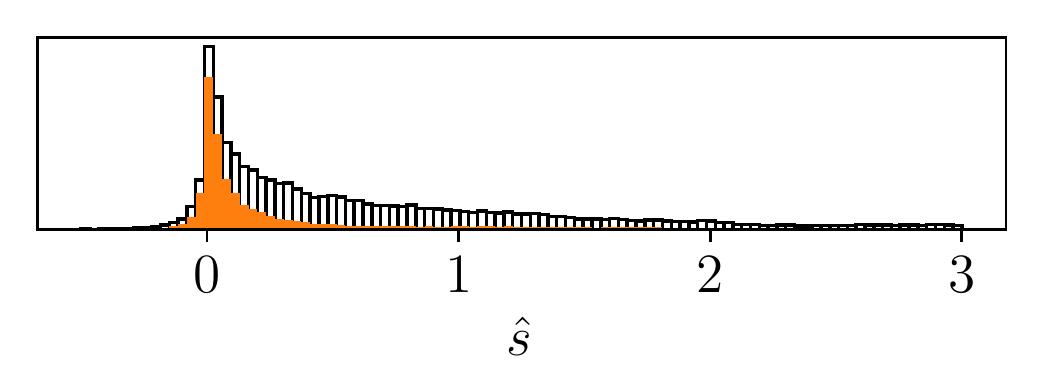}%
	\includegraphics[scale=0.5]{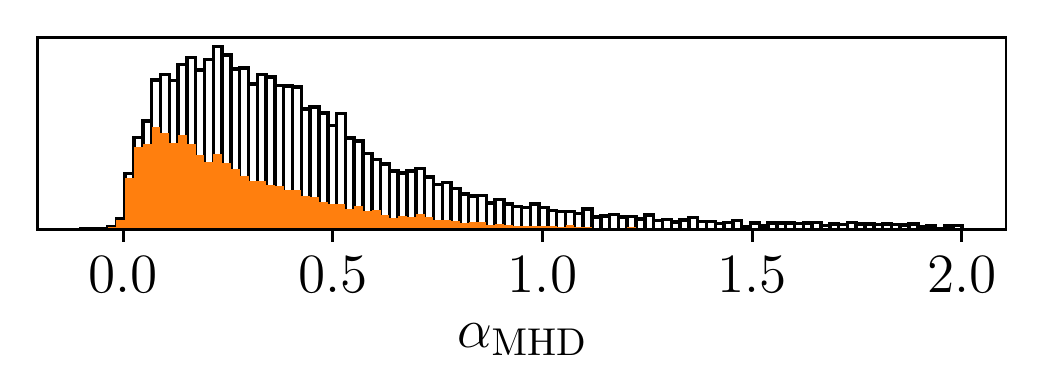}\\
	\vspace{-1mm}\includegraphics[scale=0.5]{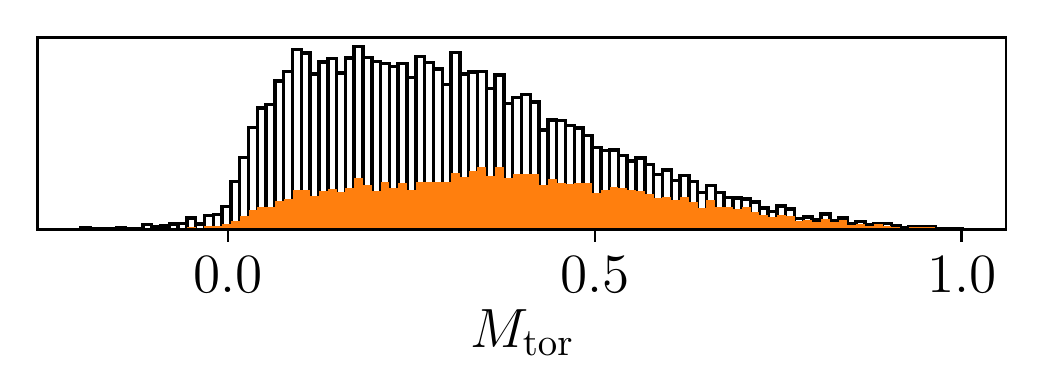}%
	\includegraphics[scale=0.5]{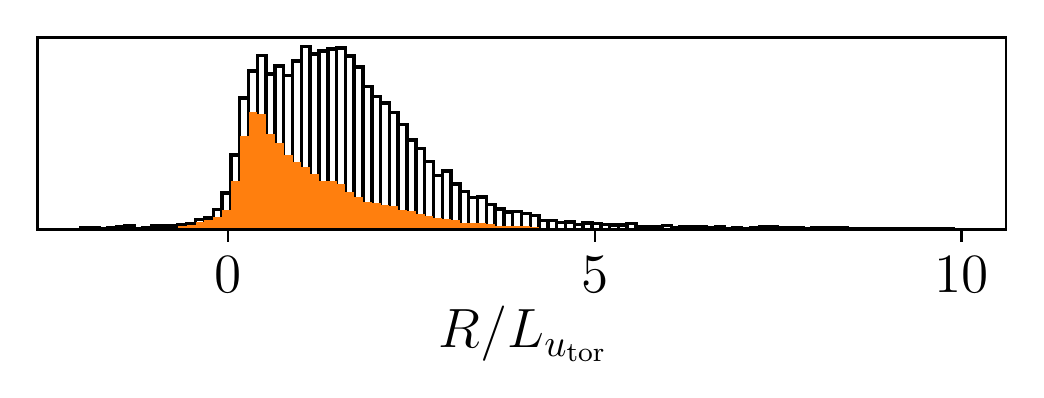}%
	\includegraphics[scale=0.5]{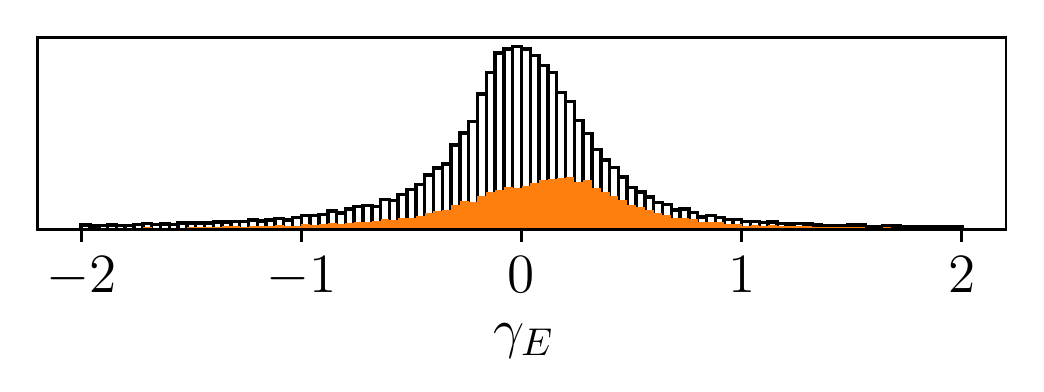}
	\caption{QuaLiKiz input parameter distributions (black unfilled bars) for the data set completely characterised by available experimental data, i.e. only time windows in which rotation measurements were available. $T_i = T_{\text{imp}}$ was assumed for the construction of this data set. The main ion densities were estimated via quasineutrality, using the measured light impurity ion density, an assumed heavy impurity species (either tungsten or iron) and a flat $Z_{\text{eff}}$ profile. For comparison, the distribution of points resulting in a completely stable scenario (orange filled bars), i.e. no unstable ITG, TEM or ETG modes, from the QuaLiKiz calculation. These distributions are only intended for gaining an intuition on JET parameter space, as the one-dimensional projection obscures any correlations between parameters.}
	\label{fig:FullJETQuaLiKizStats}
\end{figure*}

The generation and correct labelling of the training dataset is a crucial aspect of the success of ML applications. A dataset which contains hidden systematic biases or large portions of misrepresented data can lead to unexpected results. The general opaqueness of the salient features learned by ML algorithms make these errors difficult to identify without exhaustive testing. A large emphasis has been placed into the components required for effective dataset generation to reduce the risk of propagating these errors. To provide an idea of the parameter space involved, Figure~\ref{fig:FullJETQuaLiKizStats} shows the distributions of the chosen 15 input parameters within the NN training set used in this study. This section describes the considerations taken in generating this dataset from both the simulation code and the experimental data aspects of the problem.

\subsection{Input parameter selection}
\label{subsec:InputParameterSelection}

Plasma transport resulting from microturbulent behaviour can be calculated from local plasma parameters using nonlinear codes, e.g. GENE or GKW. This behaviour starts as a set of linear instabilities which grow until mode coupling mechanisms balance the instability drive. This causes the effective microturbulent transport properties to reach a macroscopic steady-state called saturation. The quasilinear tokamak turbulent transport code, QuaLiKiz, uses these same local plasma parameters to calculate the linear electrostatic drift wave and interchange instabilities. It then  determines these saturated fluid transport quantities directly from the computed linear growth rate spectra using quasilinear saturation rules~\cite{aSaturationVnV-Casati,aLowShear-Citrin}.

Similar to previous NN studies, this study focuses on predicting these saturated transport quantites given specific plasma conditions as inputs. Most of these inputs are represented as normalized or dimensionless values, due to the size scalability of turbulent phenomena. Table~\ref{tbl:DimensionlessInputs} provides the definition of these dimensionless parameters, specific to the QuaLiKiz code. Within the collisionality parameter, QuaLiKiz uses the following expression for the Coulomb logarithm, $\Lambda$:
\begin{equation}
\label{eq:CoulombLogarithm}
	\Lambda = 15.2 - 0.5 \ln\!\left(\frac{n_e}{10^{20} \, \text{m}^{-3}}\right) + \ln\!\left(\frac{T_e}{10^3 \, \text{eV}}\right)
\end{equation}
and the following expression for the bounce period, $\tau_b$:
\begin{equation}
\label{eq:BouncePeriod}
	\tau_b = q R_0 \left(\frac{r}{R_0}\right)^{-3/2} \left(\frac{-q_e T_e}{m_e}\right)^{-1/2}
\end{equation}
where $q_e$ is the electron charge in C, $m_e$ is the electron mass in kg. $r$ is the midplane-averaged minor radius of the flux surface, and $R_0$ is the midplane-averaged major radius at the last-closed-flux-surface.

\begin{table*}[tb]
	\centering
	\caption{List of dimensionless parameters used as inputs within QuaLiKiz code and the formulae for calculating them. Additional details concerning the formulae and other code inputs are available online on \url{https://gitlab.com/qualikiz-group/QuaLiKiz/-/wikis/Input-and-output-variables}~\cite{mInputOutput-QuaLiKiz}.}
	\begin{tabular}{lcc}
		Name & Variable & Conversion from physical units \\
		\midrule
		\addlinespace[2.5mm]
		Species charge number & $Z_s$ & $\frac{q_s}{q_e}$ \\
		\addlinespace[1.5mm]
		Species mass number & $A_s$ & $\frac{m_s}{m_p}$ \\
		\addlinespace[1.5mm]
		Fractional species density & $N_s$ & $\frac{n_s}{n_e}$ \\
		\addlinespace[1.5mm]
		Logarithmic density gradient & $R/L_{n_s}$ & $- \frac{R_0 \nabla n_s}{n_s}$ \\
		\addlinespace[1.5mm]
		Species temperature ratio & $T_s/T_e$ & $\frac{T_s}{T_e}$ \\
		\addlinespace[1.5mm]
		Logarithmic temperature gradient & $R/L_{T_s}$ & $- \frac{R_0 \nabla T_s}{T_s}$ \\
		\addlinespace[1.5mm]
		Rotation Mach number & $M_{\text{tor}}$ & $\frac{u_{\text{tor}}}{c_s}$ \\
		\addlinespace[1.5mm]
		Normalized rotation gradient & $R/L_{u_{\text{tor}}}$ & $- \frac{R_0 \nabla u_{\text{tor}}}{c_s}$ \\
		\addlinespace[1.5mm]
		Radial coordinate & $x$ & $\frac{r}{a}$ \\
		\addlinespace[1.5mm]
		Inverse aspect ratio at last-closed-flux-surface & $\epsilon$ & $\frac{a}{R_0}$ \\
		\addlinespace[1.5mm]
		Safety factor & $q$ & $\frac{\partial \psi_{\text{tor}}}{\partial \psi_{\text{pol}}}$ \\
		\addlinespace[1.5mm]
		Magnetic shear & $\hat{s}$ & $\frac{r \nabla q}{q}$ \\
		\addlinespace[1.5mm]
		Normalized pressure gradient & $\alpha_{\text{MHD}}$ & $-2 \mu_0 \frac{q^2 R}{B^2} \sum_s \left(T_s \nabla n_s + n_s \nabla T_s\right)$ \\
		\addlinespace[1.5mm]
		Collisionality (implicit) & $\nu^*$ & $9174 \frac{n_e}{10^{20} \text{m}^{-3}} \left(\frac{T_e}{10^3 \text{eV}}\right)^{-3/2} Z_{\text{eff}} \, \Lambda \, \tau_b$ \\
		\addlinespace[1.5mm]
		$E \times B$ shearing rate~\cite{aExBShearSuppression-Hahm} & $\gamma_E$ & $\frac{r}{q} \nabla \left[\frac{q}{r} \left(u_{\text{tor}} \frac{B_{\text{pol}}}{B} - u_{\text{pol}} \frac{B_{\text{tor}}}{B} + \frac{1}{B} \frac{\sum_i \nabla \left(n_i T_i\right)}{\sum_i n_i q_i}\right)\right]$ \\
		\addlinespace[1mm]
		\bottomrule
	\end{tabular}
	\label{tbl:DimensionlessInputs}
\end{table*}

The outputs are likewise computed in dimensionless values. Within gyrokinetics, this is known as gyro-Bohm scaling, represented by the following multiplication factor:
\begin{equation}
\label{eq:GyroBohmNormalization}
	\chi_{\text{GB}} \equiv \frac{\sqrt{m_i} T_e^{1.5}}{q_e^2 B_0^2 a}
\end{equation}
where $m_i$ is the main ion mass, $T_e$ is the electron temperature, $q_e$ is the electron charge, $B_0$ is the magnetic field at the magnetic axis, and $a$ is the mid-plane averaged minor radius of the plasma. This particular form of the gyro-Bohm scaling is the definition used by the QuaLiKiz code. Further details about the output parameters relevant for this work can be found in Table~\ref{tbl:DimensionlessOutputs}.

\begin{table*}[tb]
	\centering
	\caption{List of dimensionless transport fluxes provided as output from the QuaLiKiz code. The code outputs these values both in dimensionless form and converted back to physical units, but the formulae for converting them back to physical units is provided nonetheless, for reference purposes. Additional details concerning the formulae and other code inputs are available on \url{https://gitlab.com/qualikiz-group/QuaLiKiz/-/wikis/Input-and-output-variables}~\cite{mInputOutput-QuaLiKiz}.}
	\begin{tabular}{lcr}
		Name & Variable & Conversion to physical units \\
		\midrule
		\addlinespace[1.5mm]
		Heat flux & $\mathbf{q}_{s,\text{GB}}$ & $\frac{n_s T_s}{a} \chi_{\text{GB}} \, \mathbf{q}_{s,\text{GB}}$ \\
		\addlinespace[1.5mm]
		Particle flux & $\mathbf{\Gamma}_{s,\text{GB}}$ & $\frac{n_s}{a} \chi_{\text{GB}} \, \mathbf{\Gamma}_{s,\text{GB}}$ \\
		\addlinespace[1.5mm]
		Momentum flux & $\mathbf{\Pi}_{s,\text{GB}}$ & $\sqrt{\frac{2 T_s}{m_s}} \frac{n_s m_s R_0}{a} \chi_{\text{GB}} \, \mathbf{\Pi}_{s,\text{GB}}$ \\
		\addlinespace[1.5mm]
		Particle diffusivity & $D_{s,\text{GB}}$ & $\chi_{\text{GB}} \, D_{s,\text{GB}}$ \\
		\addlinespace[1.5mm]
		Particle pinch & $V_{s,\text{GB}}$ & $\frac{1}{a} \chi_{\text{GB}} \, V_{s,\text{GB}}$ \\
		\addlinespace[1mm]
		\bottomrule
	\end{tabular}
	\label{tbl:DimensionlessOutputs}
\end{table*}

To separate the impact of effective charge, $Z_{\text{eff}}$, and main ion dilution, represented by $N_i$, two impurity species were specified in the QuaLiKiz simulations. This results in a total of 4 species per simulation, 3 ions and the electrons, and leads to 33 individual dimensionless input parameters which must be specified for a complete QuaLiKiz simulation. While NN techniques are capable of handling significantly more inputs, it is still advantageous to reduce this number as much as possible for ease of interpretability. The physical constraints and data availability of JET data allow this to be done in a physically justifiable manner.

The physical requirement of plasma quasineutrality provides a constraint on the densities and density gradients allowed in the simulation. These constraints are expressed as follows:
\begin{equation}
\label{eq:QuasineutralityConditions}
	\begin{gathered}
	\sum_{i} N_i Z_i = 1 \\
	\sum_{i} \left(R/L_{n_i}\right) N_i Z_i = R/L_{n_e}
	\end{gathered}
\end{equation}
where $Z_i$ is the charge of the ion species, $i$, and $N$ and $R/L_n$ are dimensionless parameters defined in Table~\ref{tbl:DimensionlessInputs}.

The impurities were categorized into a generic light impurity species, defined as any ion with charge $1 < Z_s \le 10$, and a generic heavy impurity species, defined as any ion with charge $Z_s > 10$. The specific light impurities used in the dataset cover multiple species (He, Be, C, N, Ne), whereas the heavy impurity was defined to be Ni. This removes the need to uniquely specify $Z_{\text{imp,light}}$, $Z_{\text{imp,heavy}}$, $A_{\text{imp,light}}$, and $A_{\text{imp,heavy}}$. The loss of explicit uniqueness brought by this categorization is considered acceptable as the discrepancies associated with this choice are assumed to be negligible in this work. Further studies are required to determine whether the transport characteristics of species within each group are similar enough to be categorized as such. The effective charge, $Z_{\text{eff}}$, is used in place of one the species charge numbers and the other is left floating due to its approximate invariance. This parameter is expressed as follows:
\begin{equation}
\label{eq:EffectiveCharge}
	Z_{\text{eff}} = \sum_i N_i Z_i^2
\end{equation}

Based on the general availability of processed diagnostic data within the JET experimental repository, the following assumptions were made in order to complete the specification of a QuaLiKiz run:
\begin{itemize}
	\itemsep -1mm
	\item $Z_{\text{eff}}$ is radially constant throughout the plasma, i.e. $\nabla Z_{\text{eff}} \equiv 0$
	\item $T_i = T_{\text{imp}}$, as the widely available diagnostics measure the temperature of impurity ion species, implying $R/L_{T_i} = R/L_{T_{\text{imp}}}$
	\item and the main fuel ion is deuterium, with $Z_i = 1$ and $A_i = 2$
\end{itemize}
The effective charge gradient was calculated assuming that the impurity charge numbers are radially constant. This assumption is generally not true, especially for impurities with $Z > 10$. However, due to the relatively low densities of heavy impurity species in real plasmas, it is expected that the discrpenacies introduced with this assumption are negligible. This results in the following constraint:
\begin{equation}
\label{eq:EffectiveChargeGradientConstraint}
	\sum_i \left(R/L_{n_i}\right) N_i Z_i^2 = \left(R/L_{n_e}\right) Z_{\text{eff}}
\end{equation}
In addition, the geometry of JET generally restricts the inverse aspect ratio to $\epsilon \simeq 0.33$. Applying these constraints leaves a total of 15 parameters, detailed in Table~\ref{tbl:NeuralNetworkInputs}. For reference, $\Omega_{\text{tor}}$ is the angular frequency of the toroidal rotation and the associated rotation velocity, $u_{\text{tor}}$, was calculated in this study using the following expression:
\begin{equation}
\label{eq:AngularRotationCalculation}
	u_{\text{tor}} = \frac{R_{\text{outer}} + R_{\text{inner}}}{2} \, \Omega_{\text{tor}}
\end{equation}
where $R_{\text{inner}}$ and $R_{\text{outer}}$ are the innermost and outermost major radius of the flux surface, respectively.

\begin{table}[t]
	\centering
	\caption{The 15 dimensionless input parameters chosen for the NNs trained in this study. Although each dimensionless parameter is a function of many physical values, computing them in order from top to bottom in this list allow each dimensionless value to be defined solely on its associated physical value. The ion pressure, $n_i T_i$, is estimated from the density and temperature measurements of the other species combined with the $Z_{\text{eff}}$ measurement. The gyro-Bohm scaling factor accounts for the last remaining experimental measurement, $T_e$, which should be placed at the top of this list.}
	\begin{tabular}{cc}
		Dimensionless & Associated Physical\\
		Parameter & Parameter (Measured) \\
		\midrule
		$x$ & $r$ \\
		$q$ & $q$ \\
		$\hat{s}$ & $\nabla q$ \\
		$R/L_{T_e}$ & $\nabla T_e$ \\
		$Z_{\text{eff}}$ & $Z_{\text{eff}}$ \\
		$\log_{10}\left(\nu^*\right)$ & $n_e$ \\
		$R/L_{n_e}$ & $\nabla n_e$ \\
		$T_i/T_e$ & $T_{\text{imp}}$ \\
		$R/L_{T_i}$ & $\nabla T_{\text{imp}}$ \\
		$N_{\text{imp,light}}$ & $n_{\text{imp,light}}$ \\
		$R/L_{n_{\text{imp,light}}}$ & $\nabla n_{\text{imp,light}}$ \\
		$\alpha$ & $B_0$ \\
		$M_{\text{tor}}$ & $\Omega_{\text{tor}}$ \\
		$R/L_{u_{\text{tor}}}$ & $\nabla \Omega_{\text{tor}}$ \\
		$\gamma_E$ & $\nabla^2 \left(n_i T_i\right)$ \\
	\end{tabular}
	\label{tbl:NeuralNetworkInputs}
\end{table}

\subsection{Sampling methodology}
\label{subsec:DatasetSamplingMethodology}

This study uses the data acquired from JET tokamak plasma experiments to populate the input space. This approach mimics a multivariate random sampling method applied to the required input parameters, due to the various experimental conditions available and the presence of measurement uncertainty. This method is expected to converge to an equivalent representation of the physically relevant input parameter space with much fewer samples than the lattice sampling approach~\cite{tMonteCarlo-Weinzierl,aMonteCarlo-James}. Additionally, the physical and technological limitations of the experimental device provide natural restrictions to the sampled region which exclude deeply non-physical conditions. The presence of experimental noise allows some limited excursion into non-physical space, which is beneficial for NN prediction robustness. The primary drawbacks of this approach include difficulties in:
\begin{itemize}
	\itemsep -1mm
	\item identifying gaps in the sampled data;
	\item identifying the input boundaries to determine when the NN begins to extrapolate;
	\item and visualizing the high-dimensional data to develop an intuition about the problem.
\end{itemize}

While Latin hypercube sampling is a more efficient solution for multivariate random sampling~\cite{aLatinHypercube-McKay}, it was not used in this study as the distributions of the output transport coefficients are unavailable prior to generating the dataset. As the final predictive capability is more dependent on whether this output space is sufficiently sampled, this method introduces the risk of undersampling when performed while only accounting for the distributions of the input parameters. The iterative process of adding more samples post-analysis renders it counterproductive in reducing the overall time required to generate the dataset. To avoid this issue, the dataset was generated using as many of the experimentally-derived input points as possible, accepting that the parameter space may still be overpopulated. Additional data clustering and reduction techniques, e.g. DBSCAN~\cite{aDBSCAN-Ester} or k-means~\cite{aKMeans-MacQueen} algorithms, can be used to minimize the degree of overpopulation, but this is not within the scope of this study and is left as future work.

However, tokamak experimental measurement data have irregular spatial and temporal structure and are subject to multiple sources of noise, described by its reported measurement uncertainties. This means it must first be screened for validity and fitted in order to provide a data format suitable for sampling. The workflow performing this task must be automatable and robust enough to handle both the volume and variety of data being processed.

This study uses a previously created and verified data pipeline~\cite{aQLKVnV-Ho} to satisfy these criteria, featuring the use of Gaussian process regression (GPR)~\cite{bGP-Rasmussen} for one-dimensional plasma profile fitting. Specifically, the fits were performed on the quantities, $q$, $n_e$, $T_e$, $T_i$, $\Omega_{\text{tor}}$, and $n_{\text{imp,light}}$, when measurements were available. The primary advantage of the GPR algorithm is its capability to rigourously propagate measurement uncertainties to the resulting fits and fit derivatives. The availability of meaningful derivative uncertainties is especially useful in this application, due to the sensitivity of plasma microinstabilities on local plasma gradients~\cite{aITGThreshold-Guo,aETG-Jenko}.

Experimental data was extracted from 2135 discharges within the JET data repository, where the selected discharges were inspired by those recorded within the JETPEAK database~\cite{aJETPEAK-Siren}. The experimental data was extracted over a pre-determined time window and averaged in time. For steady-state scenarios, the window was chosen to have a width of 500~ms. Steady-state scenarios are defined in this study as those which have:
\begin{itemize}
	\itemsep -1mm
	\item approximately constant plasma current;
	\item approximately constant toroidal magnetic field;
	\item approximately constant line-integrated $Z_{\text{eff}}$;
	\item no power transitions in the external heating systems
\end{itemize}
throughout the entire time window. For non-steady-state or transient scenarios, this window was chosen to have a width of 200~ms.

The time-averaged uncertainty, $\sigma$, of a given measurement was updated from Reference~\cite{aQLKVnV-Ho} to the following expression:
\begin{equation}
\label{eq:MeasurementUncertainty}
	\sigma^2 = \frac{1}{N} \sum_n^N \sigma_n^2 + \left[\frac{1}{c_4\!\left(N\right)}\right]^2 \frac{1}{N} \sum_n^N \left(y_n - \bar{y}_n\right)^2
\end{equation}
where $y_n$ is the measured value, $\sigma_n$ is its reported uncertainty, $N$ is the total number of measurement points in the time window, and $c_4\!\left(N\right)$ is a correction factor to obtain the unbiased estimate of the sample standard deviation~\cite{aUnbiasedSD-Cureton}. This correction factor is preferred over the commonly used $N-1$ correction for $N < 10$. Equation~\eqref{eq:MeasurementUncertainty} incorporates the expected diagnostic noise reduction from repeated measurements while still retaining information about the spread of values obtained from those same repeated measurements.

A minimum of 1 time window in each of the following phases were selected from each discharge:
\begin{itemize}
	\itemsep -1mm
	\item current ramp-up (3742 transient windows);
	\item current flat-top (5419 steady-state windows);
	\item current ramp-down (3167 transient windows).
\end{itemize}
These selection criteria ensure that a wide variety of plasma scenarios and parameters were sampled. A total of 12328 time windows were extracted from the 2135 discharges selected.

Certain processed data fields were only available for a subset of the sampled time windows. In these cases, the following assumptions were made in order to fill in any missing data within the 15 required parameters, standardizing the data for automated execution of the QuaLiKiz code:
\begin{itemize}
	\itemsep -1mm
	\item The $Z_{\text{eff}}$ contribution of the light impurity did not exceed 0.2 if insufficient impurity information is provided;
	\item $M_{\text{tor}} = R/L_{u_{\text{tor}}} = \gamma_E = 0$ if no plasma rotation measurements are available;
	\item $T_i = T_{\text{imp}} = T_e$ if no ion temperature measurements are available;
	\item $Z_{\text{eff}} = 1.25$ if no line-integrated effective charge measurements are available.
\end{itemize}

Also, the availability of magnetic information within the experimental repository generally restricts the extracted database to using the equilibrium and $q$ profile provided by EFIT executed with only magnetic measurement data, as it was automatically produced after each discharge. Although the accuracy of the EFIT algorithm under these conditions is generally insufficient for detailed transport analysis, it is considered less problematic for the purposes of sampling experimental space. This is because any such inaccuracies should be covered by the statistical distribution over all the sampled discharges, converting any individual bias into sampling noise. This same principle applies to many of the broad assumptions made in order to standardize the input parameters, provided that assumptions are not applied to an overwhelming majority of the sampled data points.

With the experimental data extracted and standardized, the resulting profile database can then be sampled to populate the input part of the training dataset. The specific values of the inputs were taken by radially sampling the fitted one-dimensional profiles obtained via GPR techniques along the radial coordinate, $\rho_{\text{tor}}$, defined as follows:
\begin{equation}
\label{eq:ToroidalRhoCoordinate}
	\rho_{\text{tor}}\!\left(r\right) = \sqrt{\frac{\psi_{\text{tor}}\!\left(r\right)}{\psi_{\text{tor}}|_{r = a}}}
\end{equation}
where $\psi_{\text{tor}}$ is the toroidal magnetic flux passing through the poloidal cross-section of the tokamak plasma. This radial sampling was done via a fixed grid on $\rho_{\text{tor}}$ specifically for testing the interpolative power of the NN regression method, as any large discrepancies in between radial points would also be easily observable in the integrated modelling application. Table~\ref{tbl:LatticeSamplingVariations} provides the sampled values for variables in which a fixed user-defined grid was used. The radial, $\rho_{\text{tor}}$, sampling range was determined from the historical region of applicability of the QuaLiKiz model within global transport simulations, while the interval was chosen to avoid generating an excessively large dataset. The sampled wavenumbers, $k_\theta \rho_s$, selected within QuaLiKiz are identical to the set chosen for the validation of QuaLiKiz within integrated modelling applications~\cite{aQLK-Citrin}.

\begin{table}[tbp]
	\centering
	\caption{A summary of the explicitly chosen sample values for NN training dataset generation. $\mu$ and $\sigma$ represent the mean and standard deviation of the variable in question, respectively, and both are given by the GPR fit routine. As indicated by *, the variation of the rotation shear parameter, $\gamma_E$, was only applied for data points which had non-zero rotation.}
	\begin{tabular}{c|l}
		Variable & Values \\
		\midrule
		$\rho_{\text{tor}}$ & 0.1, 0.2, 0.3, 0.4, 0.5, 0.6, 0.7, 0.8, 0.9 \\
		$k_\theta \rho_s$ & 0.1, 0.175, 0.25, 0.325, 0.4, 0.5, 0.7, \\
		& 1, 1.8, 3, 9, 15, 21, 27, 36, 45 \\
		$R/L_{T_e}$ & $\mu - 1\sigma$, $\mu - 0.5\sigma$, $\mu$, $\mu + 0.5\sigma$, $\mu + 1\sigma$ \\
		$R/L_{T_i}$ & $\mu - 1\sigma$, $\mu - 0.5\sigma$, $\mu$, $\mu + 0.5\sigma$, $\mu + 1\sigma$ \\
		$R/L_{n_e}$ & $\mu - 1\sigma$, $\mu$, $\mu + 1\sigma$ \\
		$\hat{s}$ & $\mu - 1\sigma$, $\mu$, $\mu + 1\sigma$ \\
		$\gamma_E$\textsuperscript{*} & 0, $\mu - 1\sigma$, $\mu$, $\mu + 1\sigma$ \\
	\end{tabular}
	\label{tbl:LatticeSamplingVariations}
\end{table}

\subsection{Dataset refinement}
\label{subsec:DatasetRefinement}

Due to the sensitivity of the turbulent behaviour on 5 of the 15 input parameters, namely $\left\lbrace R/L_{n_e},\,R/L_{T_e},\,R/L_{T_i},\,\hat{s},\,\gamma_E \right\rbrace$~\cite{aETG-Jenko}, the sampling statistics of these parameters were artificially enhanced, as detailed in Table~\ref{tbl:LatticeSamplingVariations}. This ensures that the linear critical thresholds, a crucial feature of the output space, are properly resolved. Since the GPR routine rigourously propagates the experimental uncertainties through to the fits, performing a lattice expansion of each data point within the normally-distributed $\pm\,1\sigma$ uncertainties of these quantities achieved this goal while remaining true to the underlying experimental data. A 5-point expansion was taken in $\left\lbrace R/L_{T_e},\,R/L_{T_i} \right\rbrace$ and a 3-point expansion in $\left\lbrace R/L_{n_e},\,\hat{s} \right\rbrace$, leading to a minimum sample multiplication factor of 225. All samples which contained rotation data were duplicated assuming zero rotation to enhance the ability of the NN to interpolate in the rotation variables. They were also subjected to an additional 3-point expansion in $\gamma_E$. This gives a sample multiplication factor of 900 for these samples with rotation measurements. The final sampled dataset size after this expansion is $\sim$37.65 million data points, where $\sim$16.91 million or $\sim$45\% of them come from time windows with rotation measurements and the remaining $\sim$20.74 million or $\sim$55\% are from the remaining time windows.

Next, the output half of the NN training dataset was populated by executing QuaLiKiz on all of the collected samples. However, not every sample yielded a valid set of QuaLiKiz inputs, due to unforeseen abnormalities in the experimental data repository and/or failures in the automated data extraction routine. This left a total of $\sim$33.8 million data points which took $\sim$350,000 CPUh to compute, notably less than the pure lattice approach. This data was trimmed into a single $\sim$12~GiB HDF5 data file (uncompressed) and is available upon request. One data point in the NN training set consists of a set of these 15 input parameters and the output transport quantities produced from the corresponding QuaLiKiz run. At this point, $\sim$41.6\% of the dataset are stable, i.e. they exhibit no ITG, TEM, or ETG linear instabilities. This statistic supports the viability of using experimental data for sampling as it reflects the notion that tokamak experimental conditions lie near the threshold of these linear instabilities.

Additional data filtering was performed on the QuaLiKiz outputs via a number of sanity checks computed in post-processing. This prevents the introduction of excessive noise to the NN training set. These check the consistency of the output with respect to:
\begin{itemize}
	\itemsep -1mm
	\item abnormally small fluxes caused by rounding errors, $\left|\mathbf{\Gamma}_{e,i,\text{GB}}\right| \ge 10^{-4}$ if $\mathbf{\Gamma}_{e,i,\text{GB}} \ne 0$
	\item non-negative heat fluxes, $\mathbf{q}_{e,i,\text{GB}} \ge 0$
	\item ambipolar particle flux, $\sum_i \mathbf{\Gamma}_{i,\text{GB}} \, Z_i = \mathbf{\Gamma}_{e,\text{GB}}$
	\item calculation of electron particle flux from diffusive and convective terms, $\mathbf{\Gamma}_{e,\text{GB}} = \epsilon D_{e,\text{GB}} \left(R/L_{n_e}\right) + V_{n,e,\text{GB}}$
	\item and calculation of electron heat flux from diffusive and convective terms, $\mathbf{q}_{e,\text{GB}} = \epsilon \chi_{e,GB} \left(R/L_{T_e}\right) + V_{T,e,\text{GB}}$
\end{itemize}
Any points which do not pass these checks are excluded from the training dataset. Since QuaLiKiz computes the transport flux separately from the corresponding diffusive and convective transport coefficients, the last two checks ensure that these numerical solutions are consistent with each other. While not all of the output variables validated by these checks are used in the NN training, passing all of the tests provides some confidence on the convergence of the numerical algorithms as well as the sufficiency of resolved wavenumbers. Overall, $\sim$57.9\% of the original expanded dataset were considered suitable for training the NNs. This degree of data loss is expected due to the relatively loose sampling criteria and brute-force lattice expansion of the gradient quantities. Additional details about the filters and their associated statistics are provided in Table~\ref{tbl:WorkflowDataRetention}.

\begin{table*}[t]
	\centering
	\caption{Number of data points in the training dataset at each stage of its generation, along with percentages of data retention. The consistency filters were applied cumulatively for performance reasons, meaning that their quoted statistics do not double-count data points which would be screened out by multiple filters.}
	\begin{tabular}{lcccc}
		Extraction step & Without rotation & With rotation & Data lost & Total data lost \\
		\midrule
		Sampled fitted profiles & 92160 & 18792 & -- & -- \\
		Expanded gradients within $\pm1\sigma$ & 24964200 & 12684600 & 0\% & 0\% \\
		Completed QuaLiKiz runs & 21916575 & 11478375 & 11.3\% & 11.3\% \\
		Bounded input ranges & 21744487 & 11232300 & 1.3\% & 12.4\% \\
		Applied consistency filters & 16187891 & 7000646 & 29.7\% & 38.4\% \\
		\hspace{5mm}Electron particle flux filter @ $\pm$5\% & & & 11.3\% & \\
		\hspace{5mm}Electron heat flux filter @ $\pm$5\% & & & 4.0\% & \\
		\hspace{5mm}Negative heat flux filter & & & 6.2\% & \\
		\hspace{5mm}Ambipolar filter @ $\pm$10\% & & & 6.8\% & \\
		\hspace{5mm}Rounding error filter & & & 1.3\% & \\
		Capped heat flux output, $\mathbf{q}_e, \mathbf{q}_i \le 100$ & 14476613 & 6582532 & 9.2\% & 44.1\% \\
	\end{tabular}
	\label{tbl:WorkflowDataRetention}
\end{table*}

As provided at the beginning of this section, Figure~\ref{fig:FullJETQuaLiKizStats} shows the single parameter distributions of the 15 dimensionless inputs chosen for the final NN training set. This particular visual representation of the data obscures any correlations present in the dataset, e.g $\hat{s}$ generally increasing with $q$. However, it provides valuable insight into the input parameter ranges present in the dataset. This is especially useful when for determining extrapolation regions. The filled bars in the figure represent the distribution of data points which result in a prediction of no transport resulting from microturbulent behaviour from the QuaLiKiz model, i.e. no unstable ITG, TEM or ETG modes. As expected from known literature, the large majority of these points occur at lower logarithmic gradients.

Within the NN training set, 65.0\% of the input-output pairs are completely stable; 13.5\% exhibit dominant ITG modes; 4.6\% exhibit dominant TEM modes; and 5.3\% exhibit only ETG modes. The remaining input-output pairs yielded a combination of these three instabilities, with 8.1\% being a combination of ITG and ETG modes and the remaining split ($\sim$1\% each) across the other possible combinations, including exhibiting all three instabilities.

\section{Neural network training}
\label{sec:NNTraining}

This section provides a basic description of NNs and detail the NN architecture and training hyperparameters used for this study. In general, the NNs trained within this study used the same TensorFlow-1.6 training pipeline used by the previous work~\cite{aQLKNN-vdPlassche} and further details can be found in the referenced paper. The weights and biases of the NNs are publicly available in a GitLab repository, Ref.~\cite{mQLKNNjetexp-Ho}, in which version 1.0.1 is discussed in this document.

\subsection{Fundamental concepts}
\label{subsec:NNFundamentalConcepts}

Neural networks are a type of ML regression model in which an arbitrary function is represented by a large number of simple non-linear components called neurons~\cite{bNN-Haykin}. Each neuron contains a small number of free parameters which are adjusted to fit the overall function to a given dataset. This process is called \emph{supervised learning}. The necessary free parameter adjustments for each update step in the learning process is determined by using gradient-based methods~\cite{aCNN-LeCun}. In general, all networks contain an input layer and an output layer of neurons, which can be used for data normalization or criteria enforcement. Any layers between these two interfaces are called \emph{hidden layers}, whose specific architecture can depend on the application.

The NNs used in this study are fully-connected feedforward NNs (FFNN), a network architecture in which the outputs of each neuron are:
\begin{itemize}
	\itemsep -1mm
	\item connected to the input of every neuron in the next layer;
	\item not connected to the input of any neurons in the current layer;
	\item not connected to the input of any neurons in any previous layers.
\end{itemize}
Information is passed forward from a neuron $i$ within a layer to a neuron $j$ within the next layer using the following expression:
\begin{equation}
\label{eq:NNInformationTransfer}
	x_j = \sum_i w_{ij} \, y_i + b_j \quad , \qquad y_j = f\!\left(x_j\right)
\end{equation}
where $x$ and $y$ represent the input and output of a given neuron, $w$ and $b$ represent the neuron weights and biases, and $f\!\left(x\right)$ is the activation function of the neuron. The non-linear regression capabilities of the NN model require that some portion of the activation functions within the network are themselves non-linear~\cite{aDeepLearning-Goodfellow,aReLU-Jarrett}.

The NNs in this study consist of 3 hidden layers with 150, 70, and 30 neurons going from the input layer to the output layer. This tapered multi-layer structure was proposed heuristically to minimize overfitting~\cite{bNNPractices-Masters} and later shown to be generally optimal for FFNN regression using adaptive network pruning techniques~\cite{aNNPruning-Stathakis}. All neurons within the hidden layers have the same activation function, $f\!\left(x\right) = \tanh\!\left(x\right)$. Both input and output layers do not have an activation function, i.e. $f\!\left(x\right) = x$. In practice, the input layer is not treated as a true NN layer as these neurons simply serve to accept the values given by a user and distribute it to the first hidden layer, i.e. they only have 1 input connection with $w=1$ and $b=0$. This architecture results in a total of 15131 free parameters per NN, significantly less than the number of data points multiplied by input parameters. This check ensures that the NN cannot simply memorize the entire input dataset exactly through tuning its free parameters.

The training algorithm computes the square difference, or a related measure, between the prediction of the network at a given input point and its corresponding known output and adjusts the weights and biases accordingly to reduce that difference. This measure is generally known as the \emph{cost function} and the training algorithm attempts to minimize it over all of the given inputs, turning it into an optimization problem. The backpropagation method~\cite{aBackprop-Rumelhart} allows this update to be done on all the free parameters of the NN simultaneously, making the optimization of such large numbers of free parameters feasible. Although there are generally less free parameters than input data points, there is still a large risk of overfitting as not every point in the dataset introduces new features that must be captured. Additionally, the presence of data noise and redundancy can bias the optimization scheme towards undesired solution spaces. Constraints based on regression complexity can be provided to the cost function to reduce this risk, a technique known as \emph{regularization}. The L2-norm was chosen to fulfill this function in this study, defined as a term in the cost function as follows:
\begin{equation}
\label{eq:L2NormRegularization}
	C_{\text{L2}} = \frac{\lambda}{2} \sum_l \sum_j \sum_i w_{ij,l}^2
\end{equation}
where $l$ represents the neuron layer and $\lambda$ is the hyperparameter to adjust the degree of regularization applied.

Although this paper used the same training pipeline as the previous NNs, slight differences in the training dataset required modifications to the hyperparameters to achieve the desired network performance. The hyperparameter values used for all the networks in this study are detailed in Appendix~\ref{app:NNHyperparameters}. It is noted that the NN training algorithm uses a built-in outlier filter, which trimmed the upper and lower 0.1\% of the values present in each input and output parameter. Although the exact number of filtered points depends on the specific output parameter that each NN is being trained on, it can be assumed that this process removes $\lesssim$1\% of the values in the NN training set.

\subsection{Committee neural networks}
\label{subsec:CommitteeeNN}

Similar to previous approaches~\cite{aTGLFNN2-Meneghini,aNUBEAMNN-Boyer}, this study employs a committee of NNs to improve the prediction quality. A \emph{committee NN} evaluates multiple separate NNs, known as \emph{members}, with identical input and output parameters and combines their predictions into a collective prediction via a weighted average. This technique also provides information on the spread or standard deviation, $\sigma$, of the predictions of each member. It is proposed to use this standard deviation to address the issue of identifying the extrapolation boundary of the training dataset.

Each member within the committee NNs used in this study have exactly the same architecture and were trained on the same dataset, although this is not strictly necessary. The random initialization of the NN training pipeline ensures that each member converges to a different local minimum, in terms of the optimized free parameter configuration, but with a similar overall solution. As a result, each member can be weighted equally and the mean prediction of all the members taken to be the prediction of the committee. Additionally, the standard deviation of the member predictions can then be used to identify when the networks are extrapolating. As the NN training hyperparameters were tuned specifically to avoid overfitting, large disagreements between the members could either be the result of a lack of data in that region or an extreme volatility in the output space in that region within the training dataset. The careful design of the input and output variables, combined with domain knowledge about the smoothness of the output space, rules out volatility due to an extremely high model sensitivity or a non-unique solution space.

A committee NN with 10 members was trained for each of the 25 output quantities, for a total of 250 individual NNs within the QLKNN-jetexp-15D model. The NNs predicting the main ion momentum flux is a novel addition within this study. This paper also employed the leading flux networks developed previously, in order to ensure the simultaneous threshold behaviour for each predicted quantity corresponding to the same microturbulent mode~\cite{aQLKNN-vdPlassche}. The leading fluxes for the three modes present in QuaLiKiz are:
\begin{center}
	ITG: $\mathbf{q}_i$, \hspace{3mm} ETG: $\mathbf{q}_e$, \hspace{3mm} TEM: $\mathbf{q}_e$
\end{center}
and the NNs representing the remaining transport quantities are fitted on the ratio of that quantity to the leading flux of the turbulent mode. For example, the ITG ion particle flux within the leading flux methodology is then calculated as follows:
\begin{equation}
\label{eq:ComboNNMeanExample}
	\mathbb{E}\!\left[\mathbf{\Gamma}_{i,\text{ITG}}\right] = \mathbb{E}\!\left[\left(\frac{\mathbf{\Gamma}_i}{\mathbf{q}_i}\right)_{\text{ITG}}\right] \mathbb{E}\!\left[\mathbf{q}_{i,\text{ITG}}\right]
\end{equation}
where $\mathbb{E}$ represents the mean prediction of the committee NN. While this operation is trivial for the mean, the impact on the committee standard deviation is less straightforward. For ease of implementation, the non-leading flux variances were calculated assuming each NN can be treated as an independent random variable, leading to the expression:
\begin{multline}
\label{eq:IndependentComboNNVarianceExample}
	\mathbb{V}\!\left[\mathbf{\Gamma}_{i,\text{ITG}}\right] = \mathbb{V}\!\left[\left(\frac{\mathbf{\Gamma}_i}{\mathbf{q}_i}\right)_{\text{ITG}}\right] \mathbb{E}\!\left[\mathbf{q}_{i,\text{ITG}}\right]^2 \\
	+ \mathbb{E}\!\left[\left(\frac{\mathbf{\Gamma}_i}{\mathbf{q}_i}\right)_{\text{ITG}}\right]^2 \mathbb{V}\!\left[\mathbf{q}_{i,\text{ITG}}\right]
\end{multline}
where $\mathbb{V}$ represents the prediction variance, i.e. $\sigma^2$, of the committee NN and the other symbols are described in Equation~\eqref{eq:ComboNNMeanExample}.

\section{Comparison studies}
\label{sec:ComparisonStudies}

This section attempts to validate the QLKNN-jetexp-15D model by comparing its predictions to the original QuaLiKiz model within a wide variety of conditions. The numerical accuracy of the NN and its capability to capture known trends were validated by directly comparing the predicted transport quantities using parameter scans, detailed in Section~\ref{subsec:QuaLiKizComparisons}. The general output topology and suitability of the NN training dataset were validated by comparing the results from a time-evolved plasma simulation within the integrated model, JINTRAC. This is detailed in Section~\ref{subsec:IntegratedModelComparisons}. These methods are chosen since the typical NN goodness metrics, e.g. root-mean-squared (RMS) error or descaled validation loss, only have relative meaning within the NN training process. This limits their uses to those of model optimization.

No comparisons are made between the QLKNN-jetexp-15D and the QLKNN-hyper-10D models within this study, due to fundamental differences in the NN training datasets. After considering the assumptions made during their respective dataset generation steps, the two training datasets had no overlapping regions in which a meaningful direct comparison could be made. Also, the QLKNN-jetexp-15D training dataset was generated using QuaLiKiz-v2.6.1, whereas the QLKNN-hyper-10D training dataset was generated using QuaLiKiz-v2.5.1. This study found that an adjustment to the ion heat flux, $\mathbf{q}_i$, saturation rule between these two versions had a non-negligible impact on the JINTRAC results, meaning no meaningful integrated modelling comparison could be made as well. For the purposes of validating the NN model, comparisons solely against the original model will suffice.

\subsection{Model prediction comparisons}
\label{subsec:QuaLiKizComparisons}

The irregularity of the input values in the training dataset make it difficult to visually compare the NN prediction directly against the dataset itself. This visualization would require the data to be sorted into bins around the desired input values. Since the ranges required to produce useful bins are partly dependent on the data density, it cannot be guaranteed that the corresponding output does not vary significantly within these ranges. This introduces an uncertainty over whether any observed discrepancies are from the inaccuracy of the surrogate model or the blurring of output values over the chosen input ranges.

To avoid this issue, the original QuaLiKiz model was independently executed with a set of input parameters chosen to emulate a one-dimensional scan in the dataset generation workflow. This means the QuaLiKiz inputs in the scan were chosen to keep the other 14 dimensionless parameters constant while adhering to the same assumptions discussed in Section~\ref{subsec:InputParameterSelection}. The resulting QuaLiKiz transport coefficients are then compared to the NN predictions.

\begin{figure}[tb]
	\centering
	\includegraphics[scale=0.25]{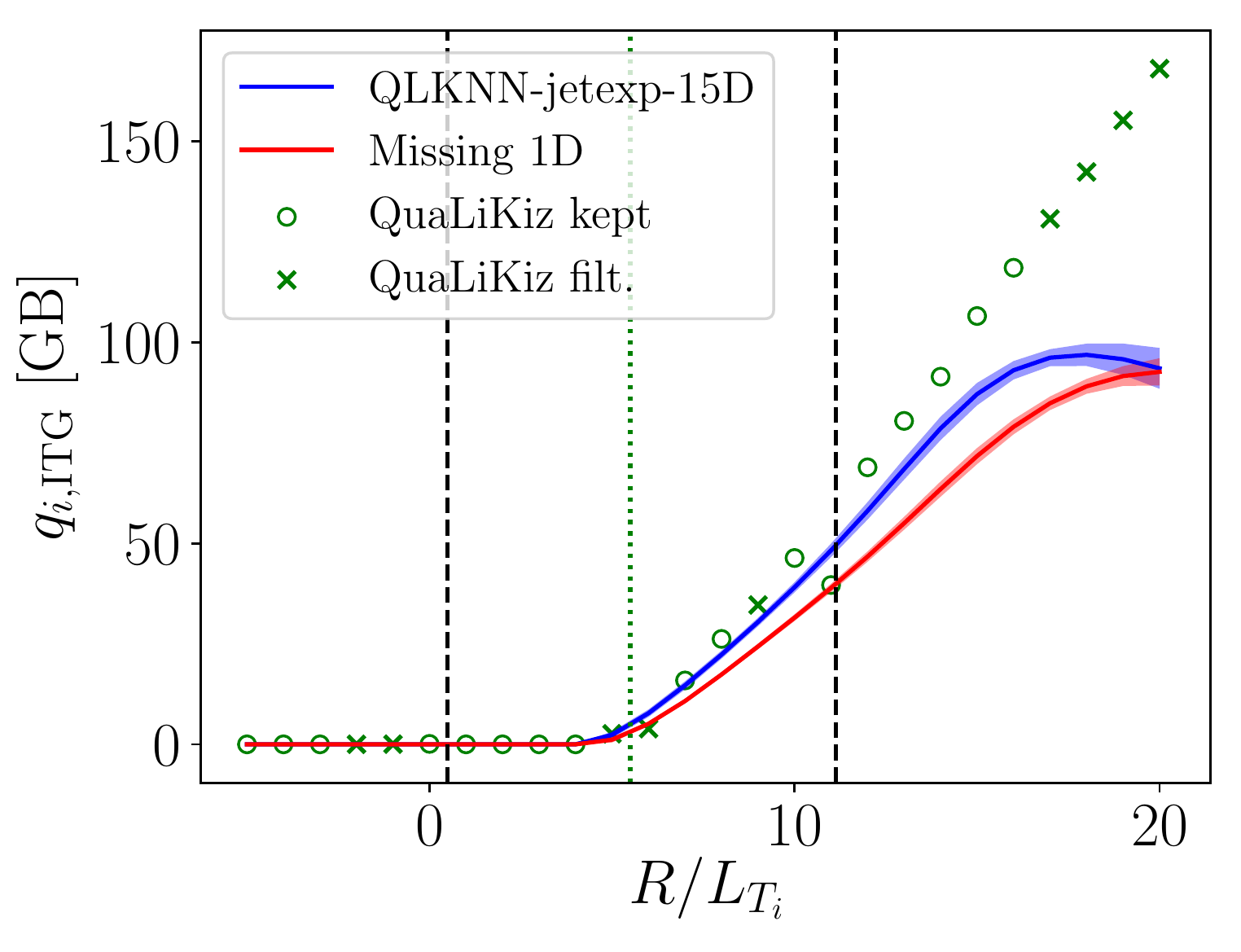}%
	\hspace{2mm}\includegraphics[scale=0.25]{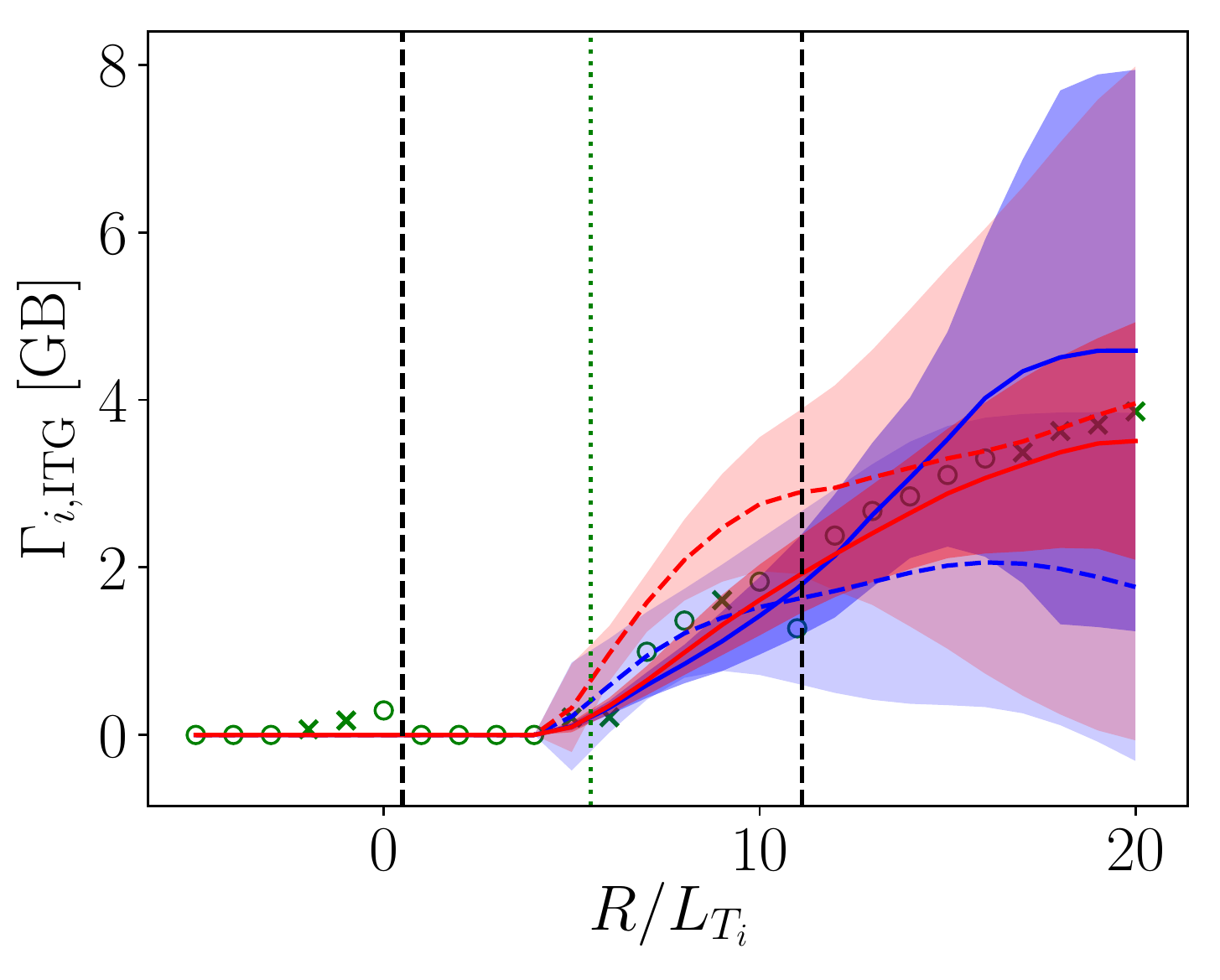} \\
	\hspace{0.25mm}\includegraphics[scale=0.25]{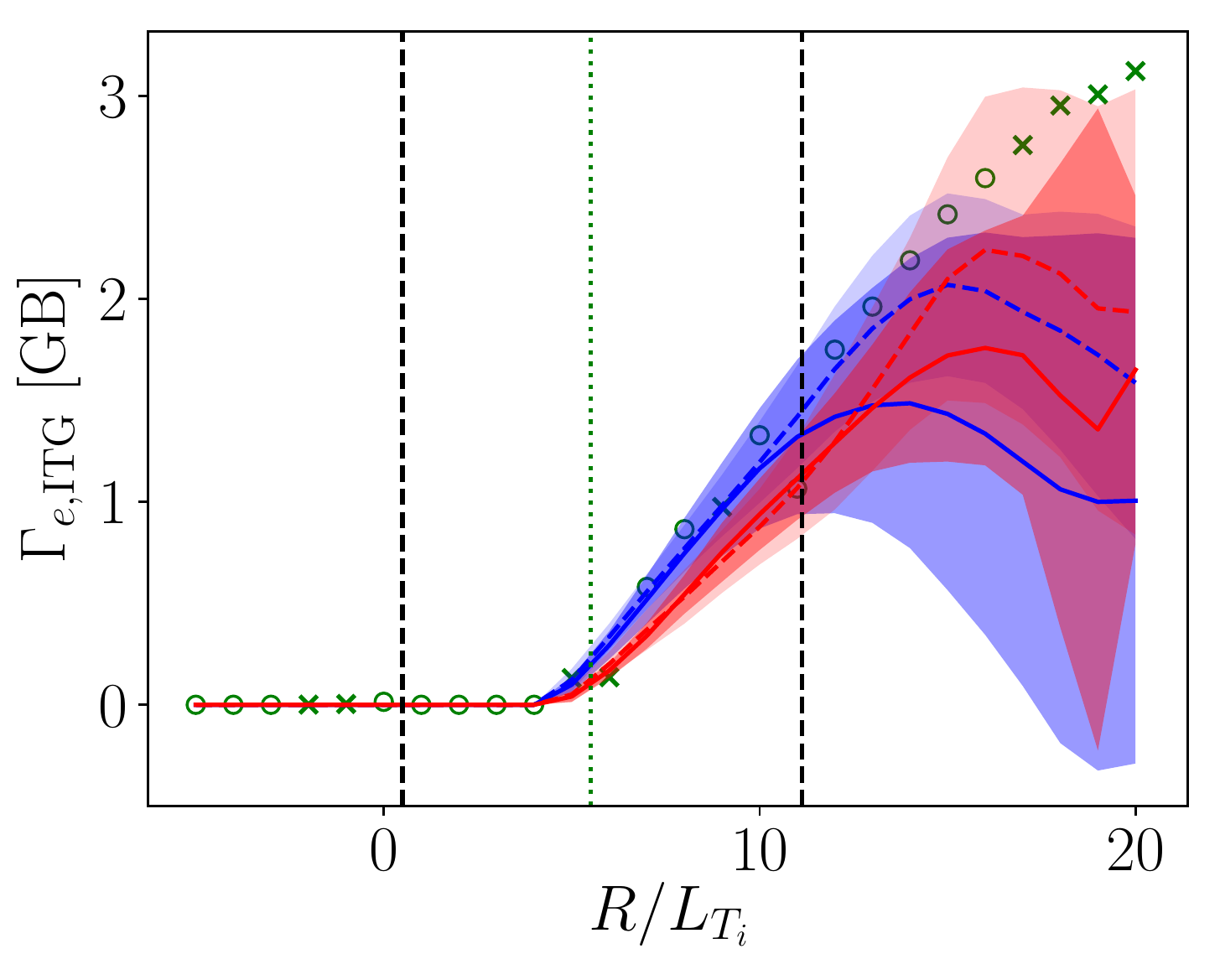}%
	\hspace{2mm}\includegraphics[scale=0.25]{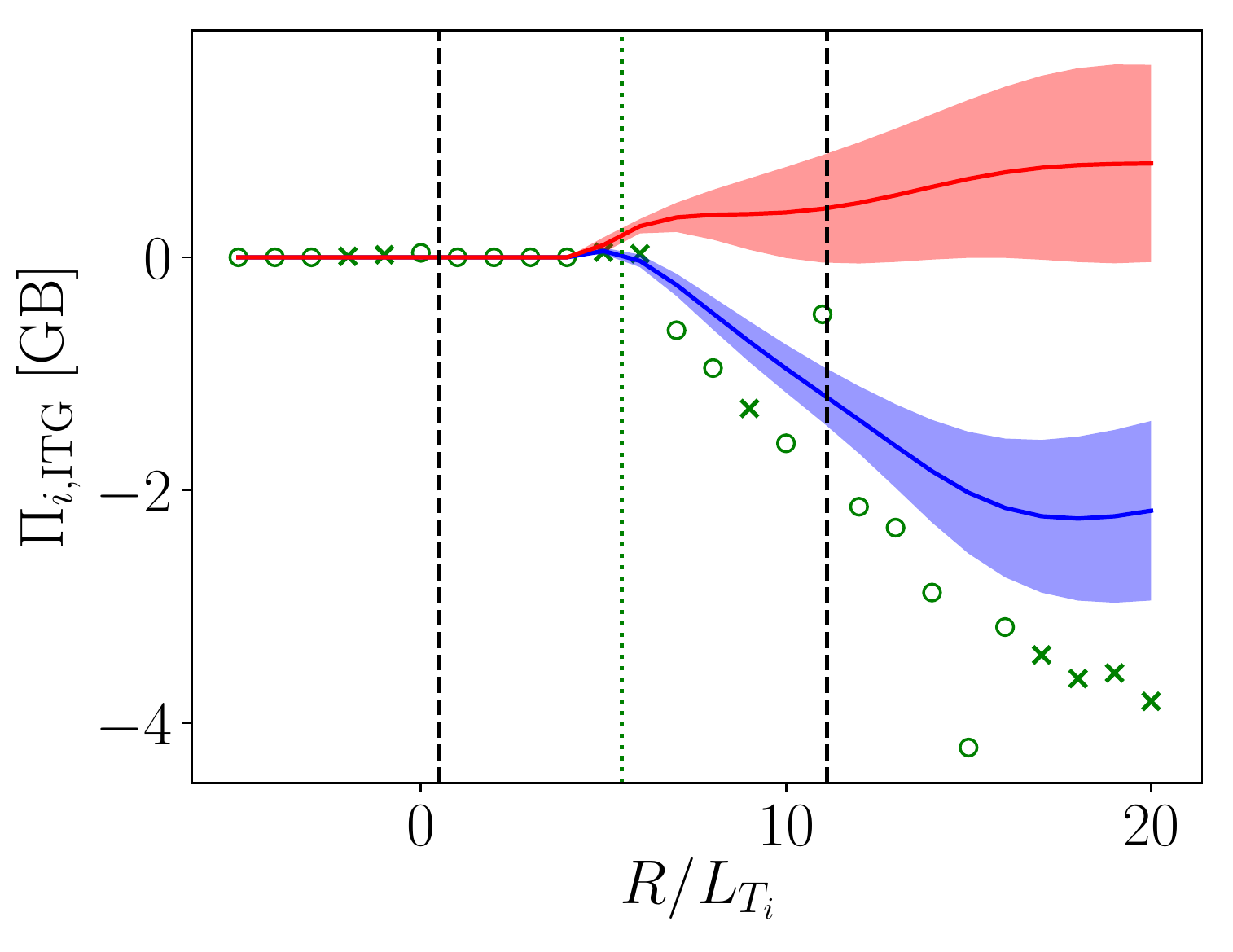}
	\caption{Comparison of main ITG-driven transport fluxes as a function of the logarithmic ion temperature gradient, $R/L_{T_i}$, predicted by QLKNN-jetexp-15D (blue lines) and QuaLiKiz (green points), showing points which would pass the data pipeline filters (circles) and those which would be screened out (crosses). The standard deviation of the committee NN (shaded regions) and the equivalent transport flux (dashed lines) reconstructed by combining simulation plasma parameters and NN predicted diffusion, $D$, and pinch coefficients, $V$, are also shown. The necessity of careful dataset generation is demonstrated via the regression quality reduction (red lines) by artificially removing one input parameter, $R/L_{u_{\text{tor}}}$. The base value (dotted green vertical line) and 2.5\%, 97.5\% quantiles (dashed black vertical lines) are shown to highlight the growing standard deviation as the NN leaves the training dataset boundaries.}
	\label{fig:FluxComparisonsIonTemperatureGradient}
\end{figure}

\begin{figure}[tb]
	\centering
	\includegraphics[scale=0.25]{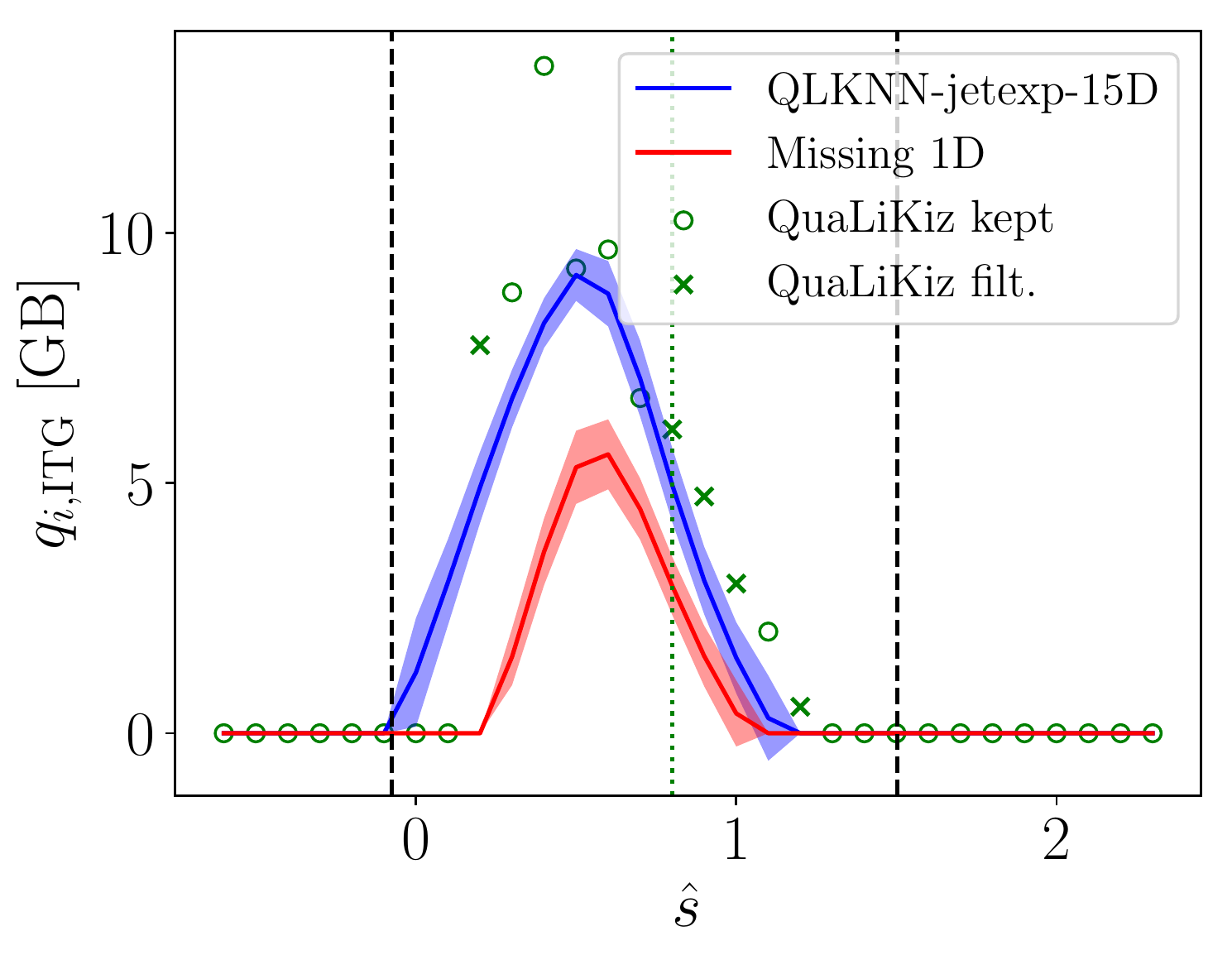}%
	\hspace{2mm}\includegraphics[scale=0.25]{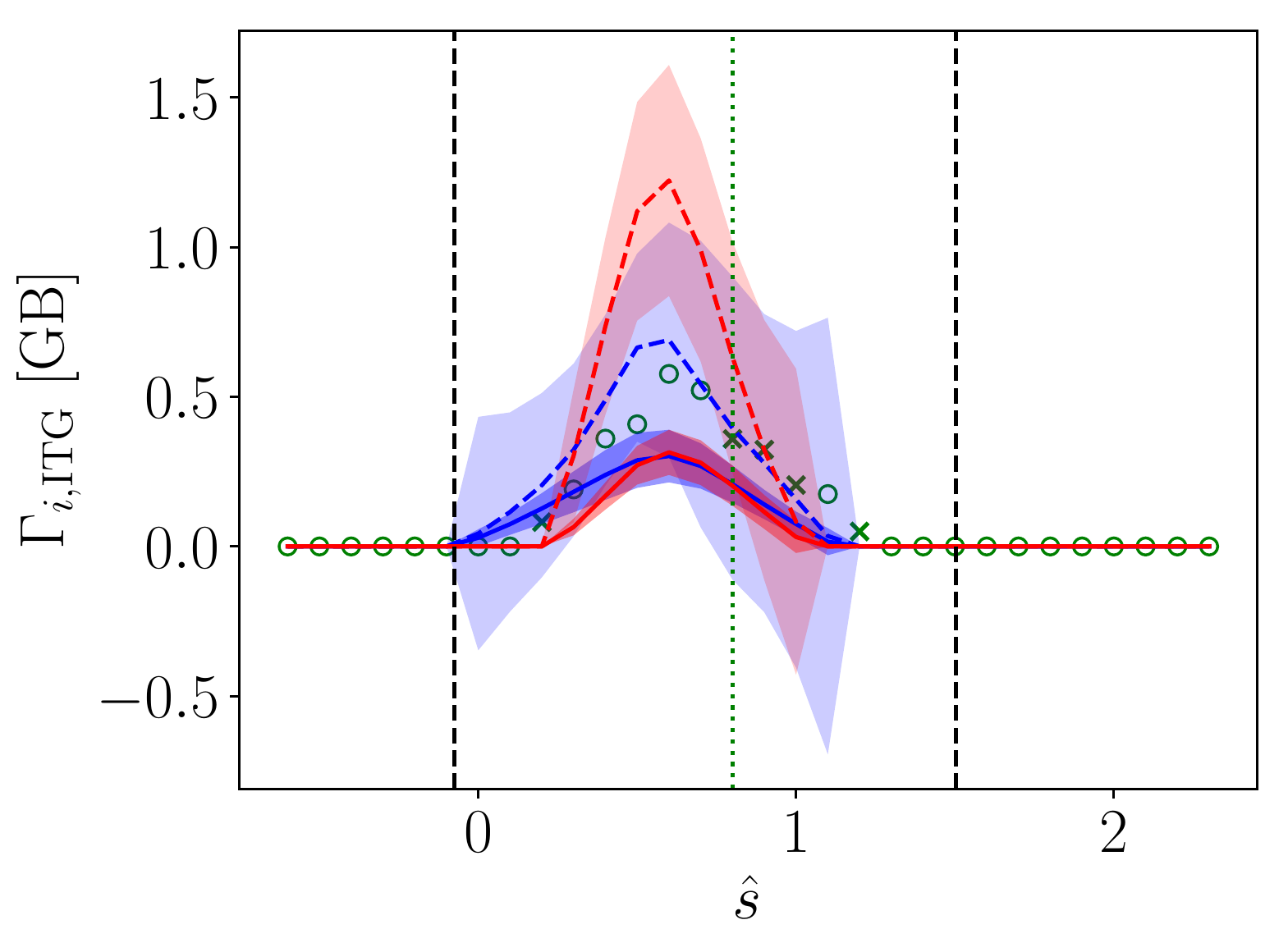} \\
	\hspace{0.25mm}\includegraphics[scale=0.25]{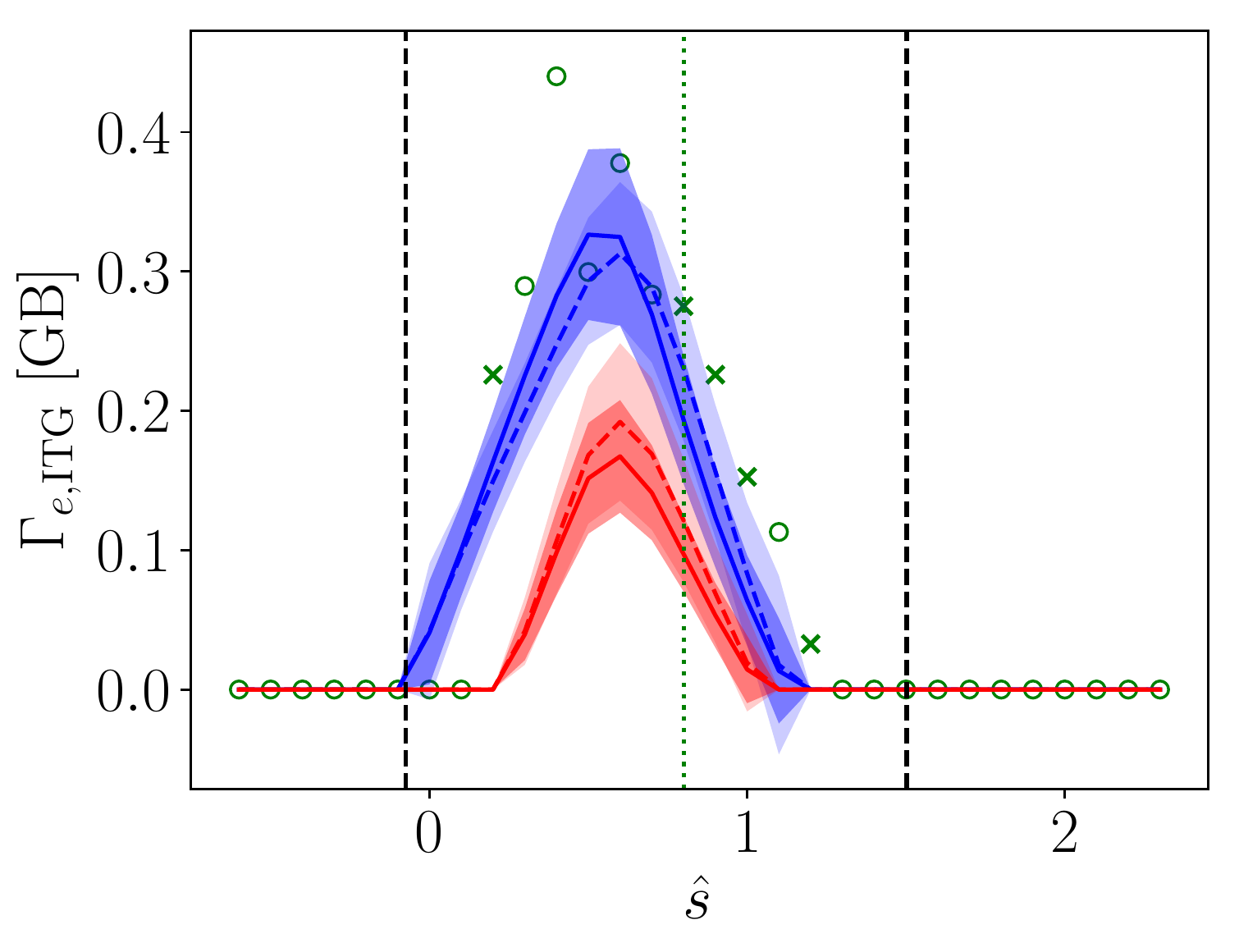}%
	\hspace{2mm}\includegraphics[scale=0.25]{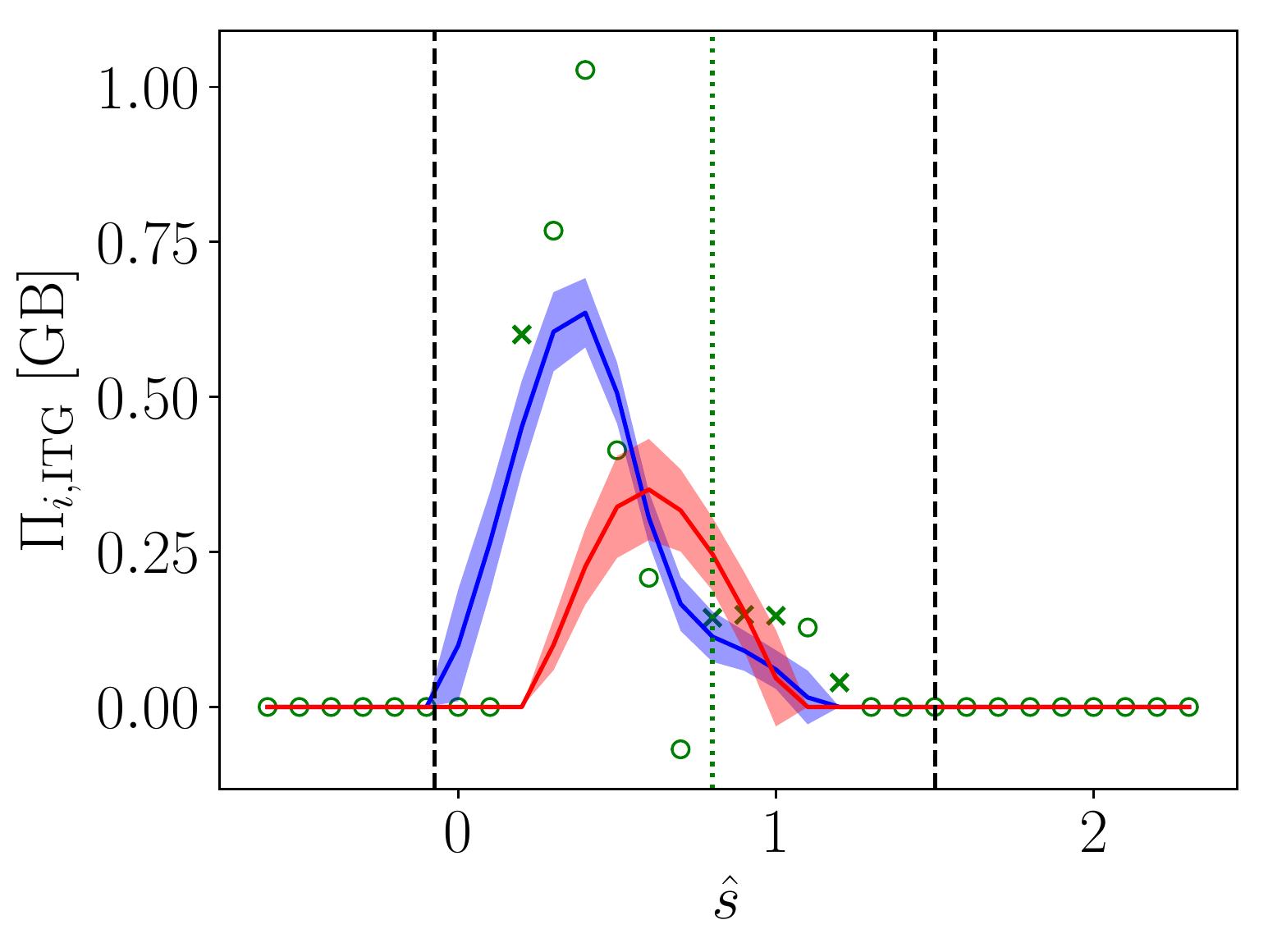}
	\caption{Comparison of main ITG-driven transport fluxes as a function of the magnetic shear, $\hat{s}$, predicted by QLKNN-jetexp-15D (lines) and QuaLiKiz (green points), showing points which would pass the data pipeline filters (circles) and those which would be screened out (crosses). The standard deviation of the committee NN (shaded regions) and the equivalent transport flux (dashed lines) reconstructed by combining simulation plasma parameters and NN predicted diffusion, $D$, and pinch coefficients, $V$, are also shown. The necessity of careful dataset generation is demonstrated via the regression quality reduction (red lines) by artificially removing one input parameter, $R/L_{u_{\text{tor}}}$. The base value (dotted green vertical line) and 2.5\%, 97.5\% quantiles (dashed black vertical lines) are shown to highlight the growing standard deviation as the NN leaves the training dataset boundaries.}
	\label{fig:FluxComparisonsMagneticShear}
\end{figure}

\begin{figure}[tb]
	\centering
	\includegraphics[scale=0.24]{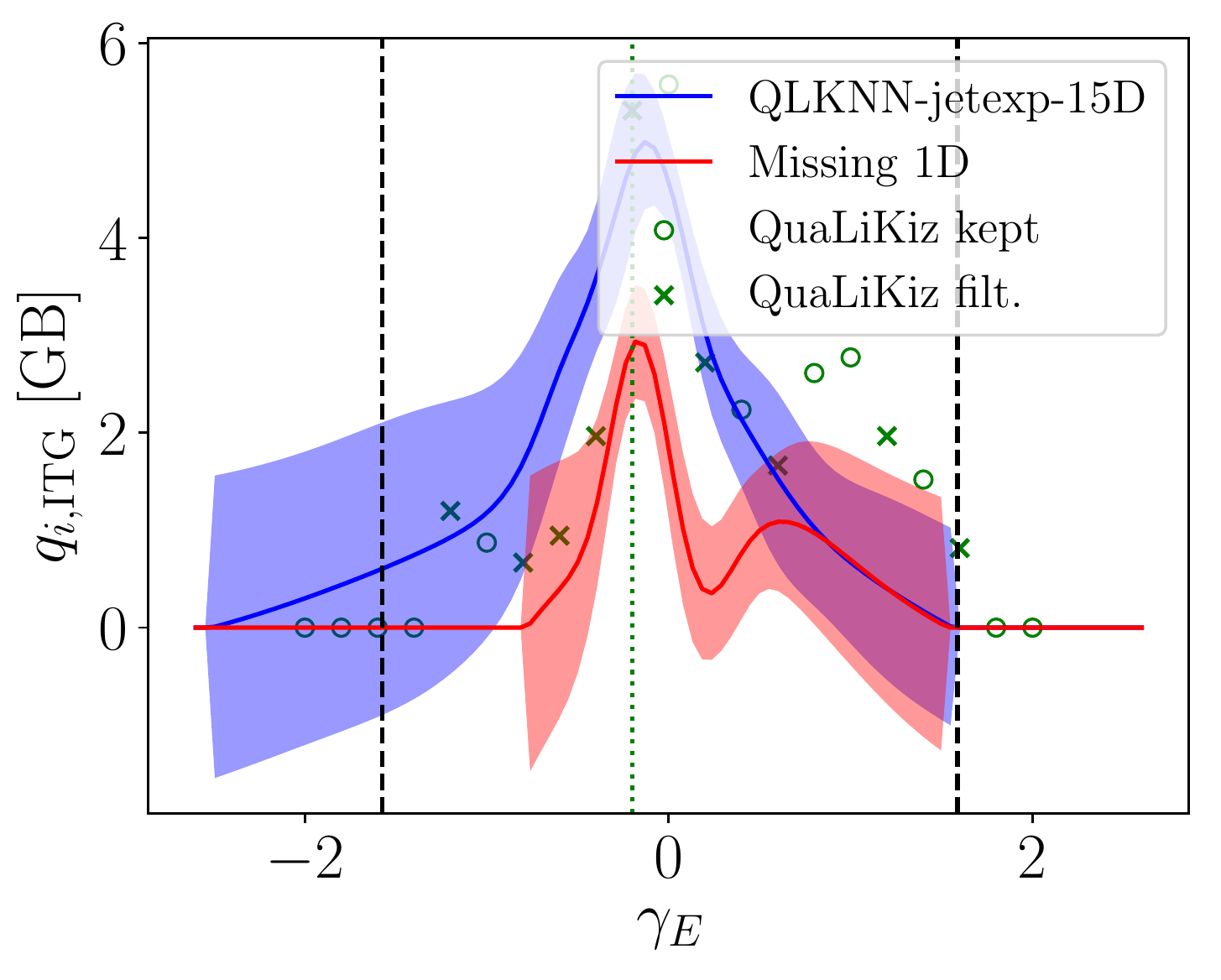}%
	\hspace{2mm}\includegraphics[scale=0.24]{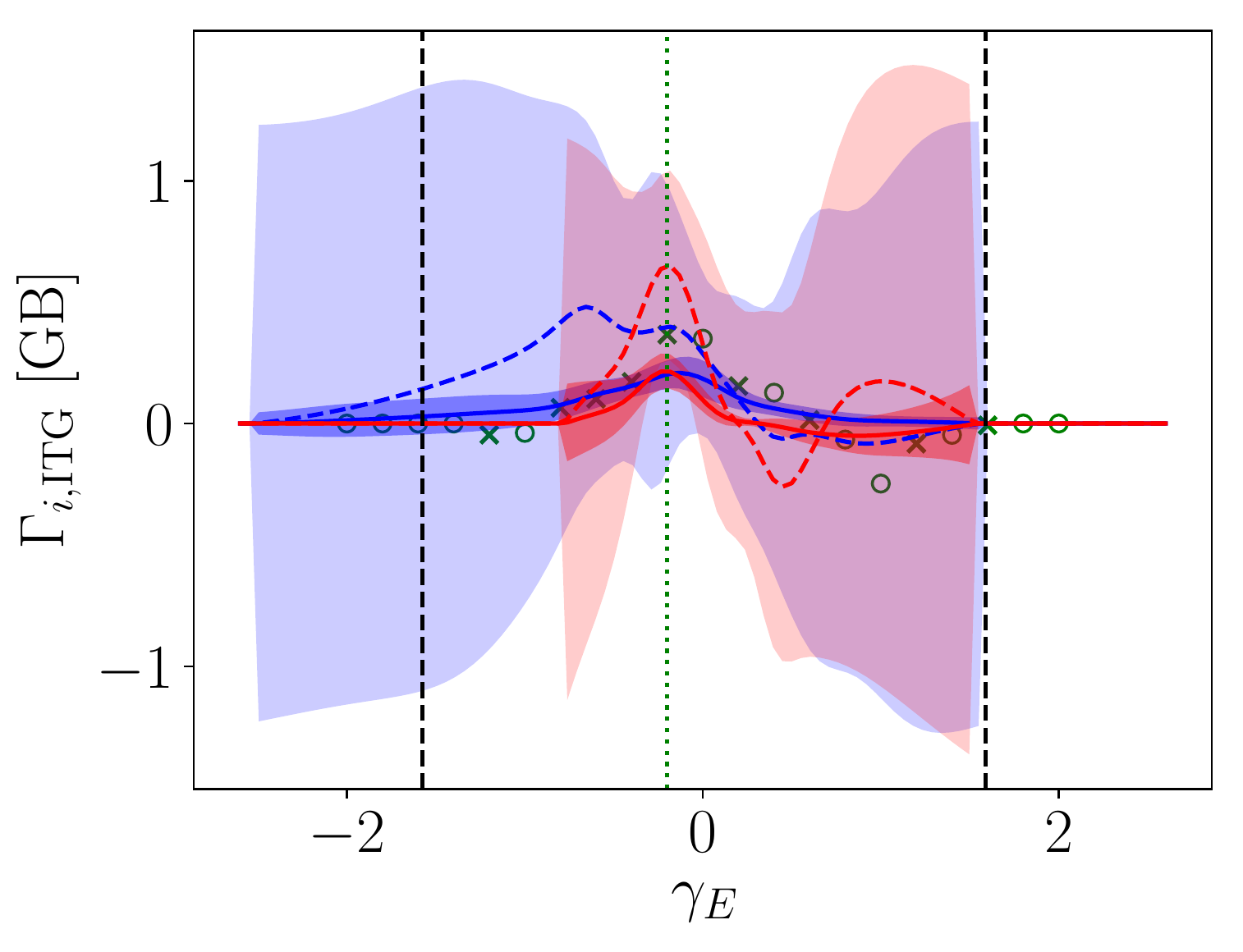} \\
	\hspace{0.25mm}\includegraphics[scale=0.24]{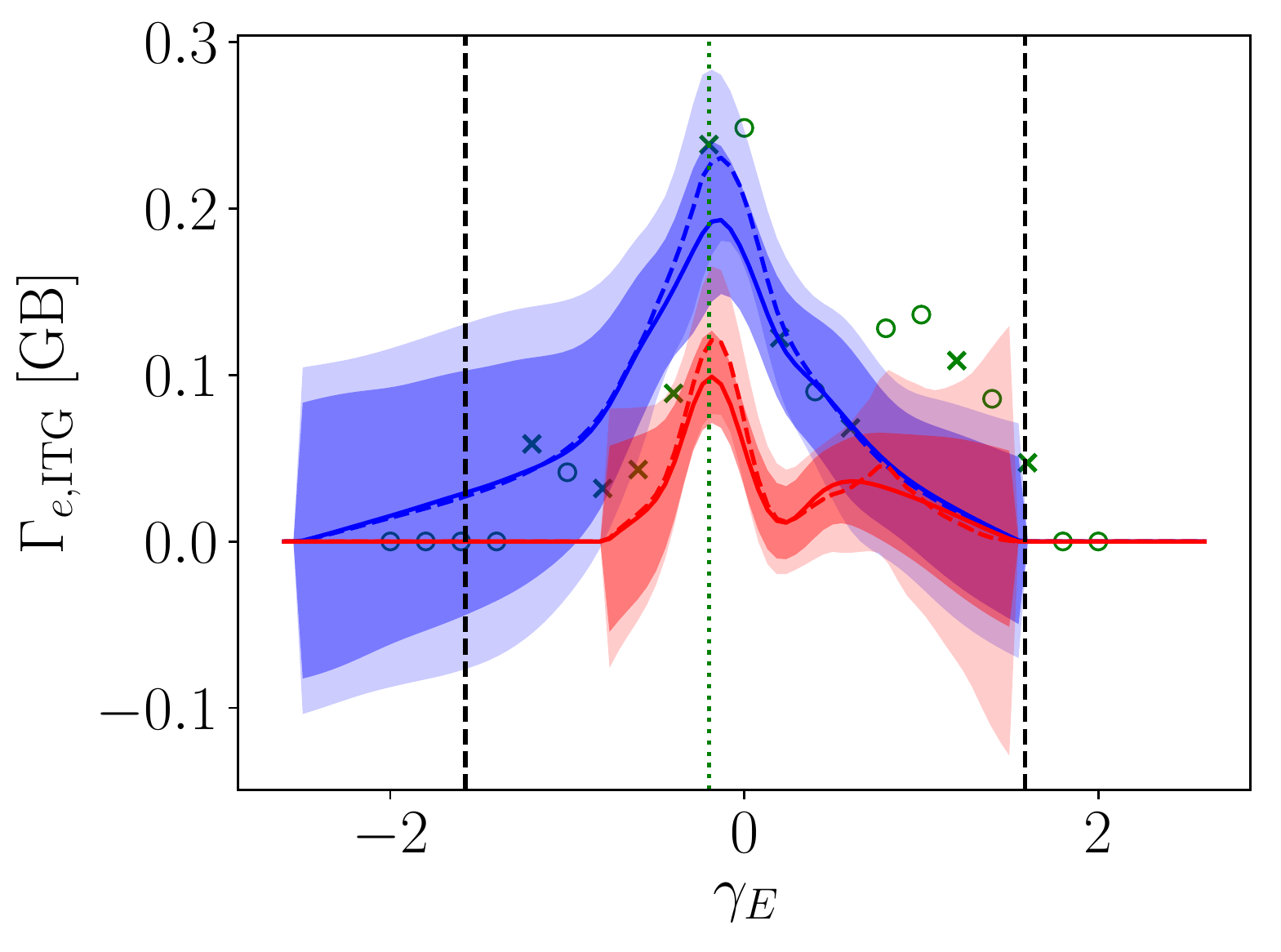}%
	\hspace{1.5mm}\includegraphics[scale=0.24]{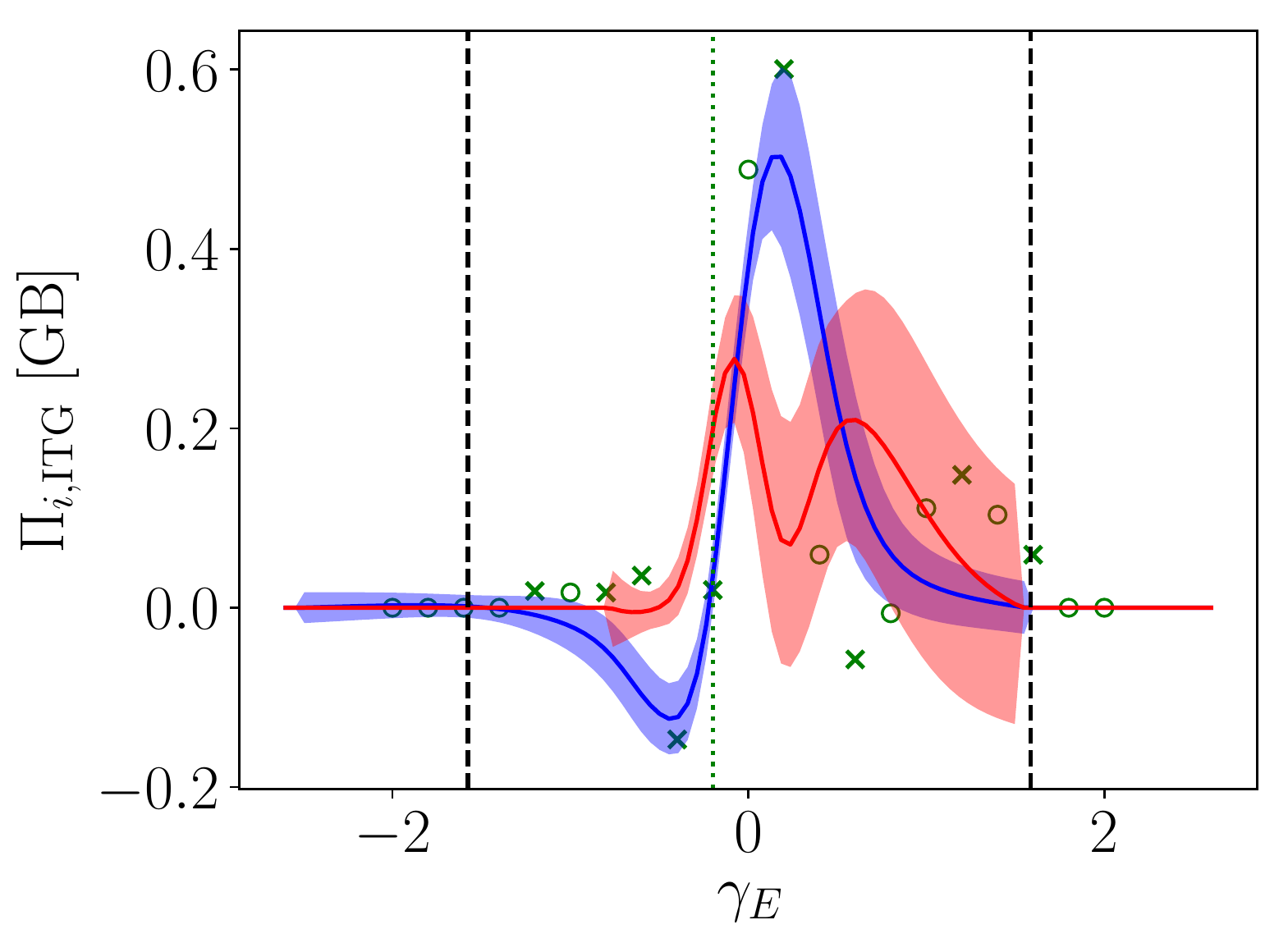}
	\caption{Comparison of main ITG-driven transport fluxes as a function of the normalized $E \times B$ shearing rate, $\gamma_E$, predicted by QLKNN-jetexp-15D (lines) and QuaLiKiz (green points), showing points which would pass the data pipeline filters (circles) and those which would be screened out (crosses). The standard deviation of the committee NN (shaded regions) and the equivalent transport flux (dashed lines) reconstructed by combining simulation plasma parameters and NN predicted diffusion, $D$, and pinch coefficients, $V$, are also shown. The necessity of careful dataset generation is demonstrated via the regression quality reduction (red lines) by artificially removing one input parameter, $R/L_{u_{\text{tor}}}$. The base value (dotted green vertical line) and 2.5\%, 97.5\% quantiles (dashed black vertical lines) are shown to highlight the growing standard deviation as the NN leaves the training dataset boundaries.}
	\label{fig:FluxComparisonsExBShearingRate}
\end{figure}

Figures~\ref{fig:FluxComparisonsIonTemperatureGradient}, \ref{fig:FluxComparisonsMagneticShear}, and \ref{fig:FluxComparisonsExBShearingRate} compare the NN surrogate models to the original QuaLiKiz evaluations over parameter scans of the logarithmic ion temperature gradient, $R/L_{T_i}$, magnetic shear, $\hat{s}$, and normalized $E \times B$ shearing rate, $\gamma_E$, respectively. These scans were performed with the other 14 parameters retaining the following base values: $R/L_{T_e} = 5.5$, $R/L_{T_i} = 5.5$, $x = 0.55$, $R/L_{n_e} = 2$, $R/L_{n_{\text{imp,light}}} = 2$, $N_{\text{imp,light}} = 0.017$, $T_i/T_e = 1$, $\log_{10}(\nu^*) = -0.9$, $Z_{\text{eff}} = 1.7$, $q = 1.8$, $\hat{s} = 0.8$, $\alpha = 0.3$, $M_{\text{tor}} = 0.3$, $R/L_{u_{\text{tor}}} = 1.5$, $\gamma_E = -0.2$. These base values were chosen to be near the highest data point density in the training set for the radial grid point, $\rho_{\text{tor}} = 0.5$. The minimum and maximum values of the scans were chosen to extend into the low density regions of the training set, whenever allowed by the constraints of the physical system. Since the leading flux NN output is clipped to zero when the prediction is negative, the leading flux NN standard deviation is also clipped to zero in these cases.

Although the parameter scans were performed for all 15 input parameters, these 3 input parameters were chosen to emphasize the capability of the NNs to resolve specific features in output space. The remaining 12 parameter scan plots can be found in Appendix~\ref{app:ExtraParameterScans}. Additionally, only the main transport fluxes, i.e. main ion heat flux, $\mathbf{q}_i$, electron particle flux, $\mathbf{\Gamma}_e$, main ion particle flux, $\mathbf{\Gamma}_i$, and momentum flux, $\mathbf{\Pi}$, for ITG turbulence are shown as it is the dominant turbulent regime found in JET. However, the agreement between QuaLiKiz and QLKNN-jetexp-15D is similar for the TEM and ETG turbulence as well.

From these figures, it can be concluded that the QLKNN-jetexp-15D model successfully replicates the original QuaLiKiz model in regions where the training set is sufficiently dense. As expected, the discrepancy between the committee NN mean predictions and the original QuaLiKiz model increases as the data density in the training set, shown in Figure~\ref{fig:FullJETQuaLiKizStats}, begins to decrease. More importanty, this increased discrepancy is accompanied by an increase in the committee NN standard deviation, demonstrating the proposed relation between the committee standard deviation and the underlying training set data density. With appropriately selected threshold, this standard deviation becomes a useful metric to identify when the committee NN prediction is within an extrapolation region, i.e. a region in input space for which there is minimal representation within the training set.

However, this metric is only meaningful provided that the possibility of excessive training set noise and NN overfitting are sufficiently reduced during the generation of the trained NN model, as was done in this study. This is shown by the red lines and shaded regions in Figures~\ref{fig:FluxComparisonsIonTemperatureGradient}, \ref{fig:FluxComparisonsMagneticShear} and \ref{fig:FluxComparisonsExBShearingRate}, where input noise was artificially added by removing the $R/L_{u_{\text{tor}}}$ parameter from the training set. This reduced the problem to a 14D description and broke the uniqueness criteria established in Section~\ref{sec:DatasetGeneration}. The resulting committee NNs not only make worse mean predictions in general but the standard deviation is no longer strongly tied to the input data density, best shown in Figure~\ref{fig:FluxComparisonsExBShearingRate}.

The effective flux reconstructed from the NN-predicted electron particle diffusion, $D_e$, and pinch, $V_e$, coefficients, shown in the figures with the dashed lines, generally agree with the NN-predicted electron particle flux itself, $\mathbf{\Gamma}_e$. This result is attributed to the presence of the strict input data filter on the electron particle flux consistency with the respective $D$ and $V$ from the same simulation, as described in Section~\ref{subsec:DatasetRefinement}. Conversely, this agreement is not generally observed with the ion particle flux, likely due to the absence of a similar consistency filter on the ion transport quantities.

Although the metric required to apply the filter is computed by QuaLiKiz, it was purposefully not done in the dataset generation step in this study. A brief examination showed that the ion particle transport consistency filter by itself would discard $\sim$50\% of the unstable points in the collected data if set with a tolerance of $\pm$5\%, and discard $\sim$25\% with a tolerance of $\pm$20\%. This amount of data loss was considered too much to be reasonable from a single filter. Since the NN applications do not rely solely on the $D_i$ and $V_i$ predictions, this was seen as an acceptable loss in NN performance for the purposes of this study. As a result, only the ion particle flux prediction, $\mathbf{\Gamma}_i$, is recommended for use inside an integrated model, although the detailed trials are provided in Section~\ref{subsec:IntegratedModelComparisons}.

Recent investigations into the internal consistency check revealed that the ion rotodiffusion pinch term of QuaLiKiz, which is set to zero from previous verification exercises, was distorting the calculation. This has been fixed for QuaLiKiz-v2.8.1 and higher, which is expected to improve the statistics of the ion particle transport consistency filter. However, this version was not available during the data generation phase of this study. To this effect, a further investigation into the impact of a strict ion transport quantity consistency filter, and/or other physics-informed consistency filters, on the overall NN performance are suggested as future work.

\subsection{Integrated modelling comparisons}
\label{subsec:IntegratedModelComparisons}

The QLKNN-jetexp-15D model was further tested within the integrated model, JINTRAC, based on their reasonable regression performance shown in the previous section. This paper applies these new NNs to the same 3 integrated modelling cases examined in the previous work~\cite{aQLKNN-vdPlassche}: a carbon wall baseline discharge (JET\#73342)~\cite{aQLK-Citrin}, an ITER-like wall hybrid discharge (JET\#92398)~\cite{aJETHybrid-Casson}, and an ITER-like wall high-performance baseline discharge (JET\#92436)~\cite{aQLKVnV-Ho}. In addition, 2 extra cases are included in this comparison in order to test the NN applicability within other common plasma scenarios: an ITER-like wall ohmic L-mode discharge (JET\#91637) and an hydrogenic isotope-mixing discharge (JET\#91227)~\cite{aIsotopeMixing-Marin}.

The integrated modelling simulations use the NNs to predict the electron and ion heat fluxes, $\mathbf{q}_e$ and $\mathbf{q}_i$, as well as the ion particle flux, $\mathbf{\Gamma}_i$. These are then given to JETTO~\cite{tJETTO-Cenacchi}, the global transport module in JINTRAC, which primarily solves the following equations for all species, labelled with $s$, within a one-dimensional plasma:
\begin{equation}
\label{eq:JETTOGlobalTransport}
	\begin{aligned}
	\frac{\partial n_s}{\partial t} + \mathbf{\nabla} \cdot \mathbf{\Gamma}_s &= S_s \\
	\frac{3}{2}\frac{\partial n_s T_s}{\partial t} + \mathbf{\nabla} \cdot \mathbf{q}_s &= Q_s
	\end{aligned}
\end{equation}
where $n$ is the number density, $T$ is the temperature, and $S$ and $Q$ represents the particle and heat sources and sinks, respectively. These equations then provide the evolution of the electron temperature, $T_e$, main ion temperature, $T_i$, and electron density, $n_e$, as a function of radius in time. Other physics models or experimental measurements were used to provide the values for the various sources and sinks, whose performance has already been validated for the test cases shown.

In some select cases, the momentum flux, $\mathbf{\Pi}_i$, is also provided to simultaneously solve the momentum transport equation, given as:
\begin{equation}
\label{eq:JETTOMomentumTransport}
	\frac{\partial \mathbf{p}_s}{\partial t} + \mathbf{p}_s \cdot \mathbf{\nabla} \mathbf{v}_s  + \mathbf{\nabla} \cdot \mathbf{\Pi}_s = \mathbf{R}_s + \mathbf{\tau}_s
\end{equation}
where $\mathbf{p}$ is the bulk plasma momentum, $\mathbf{v}$ is the bulk plasma velocity, $\mathbf{R}$ is the resistivity term, and $\mathbf{\tau}$ represents the remaining momentum sources and sinks. It is important to note that self-consistent momentum evolution, via the inclusion of Equation~\eqref{eq:JETTOMomentumTransport}, is not routinely performed within current integrated modelling workflows. However, a predict-first approach would necessarily include momentum transport. This motivated the novel inclusion of momentum flux predictions within QLKNN-jetexp-15D.

Unfortunately, the wide variety of possible impurity species in the plasma means that a complete and consistent description of the transport properties is not always guaranteed by the QLKNN-jetexp-15D output. This is due to the impurity categories implemented to standardize the input portion of the NN training dataset. For this reason, the NN ion transport coefficients are strictly for deuterium, where it is the main fuel ion. Due to the stiffness of heat transport and strength of the equipartition coupling, the differences in the heat transport properties of the different ion species are typically neglected. However, this cannot generally be assumed for particle transport~\cite{aIsotopeMixing-Marin}. In simulations where the density profiles of these additional species are not fixed, further assumptions are required to provide the transport coefficients for these species.

The plasma ambipolarity constraint can be used to specify one these extra coefficients and it is expressed as follows:
\begin{equation}
\label{eq:AmbipolarityConstraint}
	\mathbf{\Gamma}_e = \sum_i \mathbf{\Gamma}_i Z_i
\end{equation}
However, this is insufficient to fully specify the transport properties of simulations with 2 or more additional species, which is becoming more routine in tokamak plasma modelling. 6 different options were developed to address the remaining parameters, taking advantage of the fact that the committee NNs were trained to predict the total fluxes, the diffusion coefficients and the convection coefficients separately. Based on the test results, only 3 of these options are shown and discussed within this section. The remaining options are provided in Appendix~\ref{app:ParticleTransportOptions}.

The first option, labelled \emph{P1} in this document, assumes that the all ion particle fluxes are directly proportional to the electron particle flux, as follows:
\begin{equation}
\label{eq:ParticleTransportOption1}
	\mathbf{\Gamma}_i = \mathbf{\Gamma}_e \frac{n_i}{n_e}
\end{equation}
where $e$ represents the electrons and $i$ represents a generic ion species. Although this assumption is not generally valid, this is the current default option within the QLKNN implementation in JINTRAC and the closest option to the one used to benchmark the QLKNN-hyper-10D model~\cite{aQLKNN-vdPlassche}. It is kept here as a comparison to the previous implementation.

The second option, labelled \emph{P6} in this document, assumes that the ion particle diffusive and convective transport coefficients are equal to those of the deuterium ion, expressed as follows:
\begin{equation}
\label{eq:ParticleTransportOption6}
	\mathbf{\Gamma}_i = -D_{i0} \nabla n_i + V_{i0} n_i
\end{equation}
where $i0$ represents the deuterium ion and $i$ represents a generic ion species. This is the ideal option to transfer deuterium transport quantities to other hydrogenic species~\cite{aIsotopeMixing-Marin}. The performance of this option is questionable with QLKNN-jetexp-15D due to absence of ion particle transport consistency filters. However, it is shown here for comparison purposes.

The third option, labelled \emph{P5} in this document, assumes that the ion particle diffusive and convective transport coefficients are equal to those of the electrons, expressed as follows:
\begin{equation}
\label{eq:ParticleTransportOption5}
	\mathbf{\Gamma}_i = -D_e \nabla n_i + V_e n_i
\end{equation}
This is the current recommended option to transfer transport quantities to non-deuterium ion species, based solely on its performance in reproducing the total particle flux within the parameter scan comparisons in Section~\ref{subsec:QuaLiKizComparisons}.

\begin{figure}[tb]
	\centering
	\includegraphics[scale=0.27]{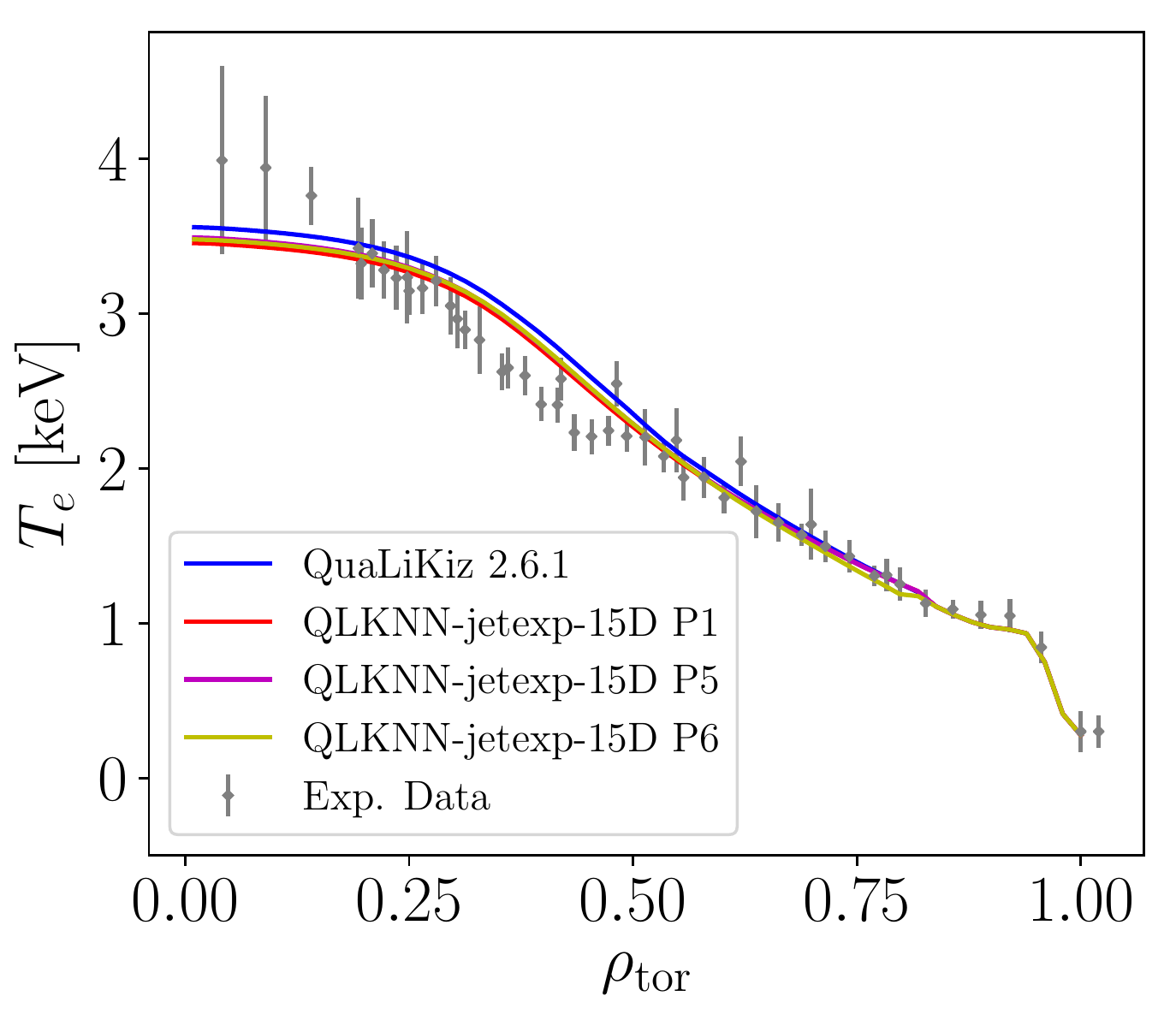}%
	\hspace{2mm}\includegraphics[scale=0.27]{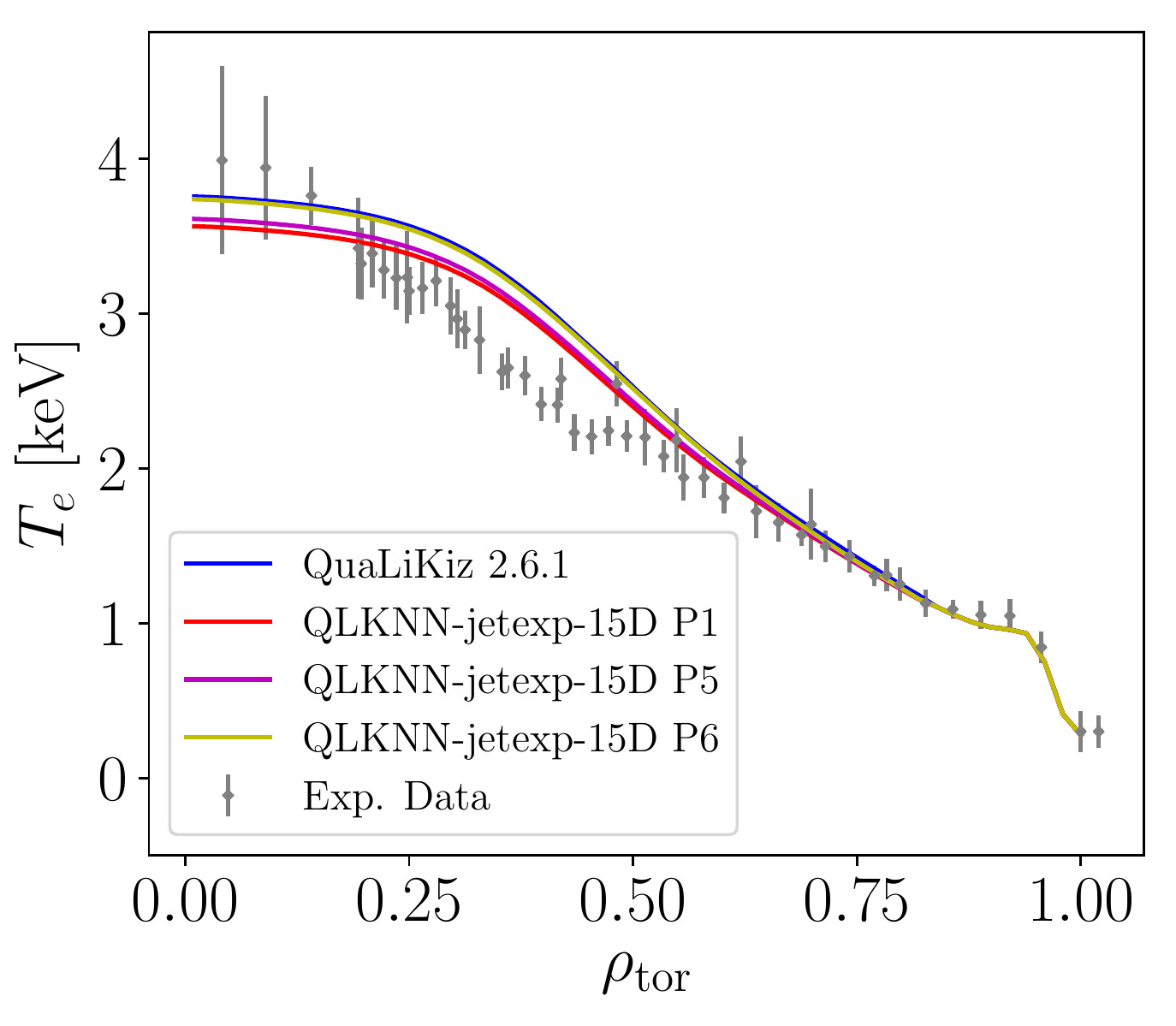}\\
	\includegraphics[scale=0.27]{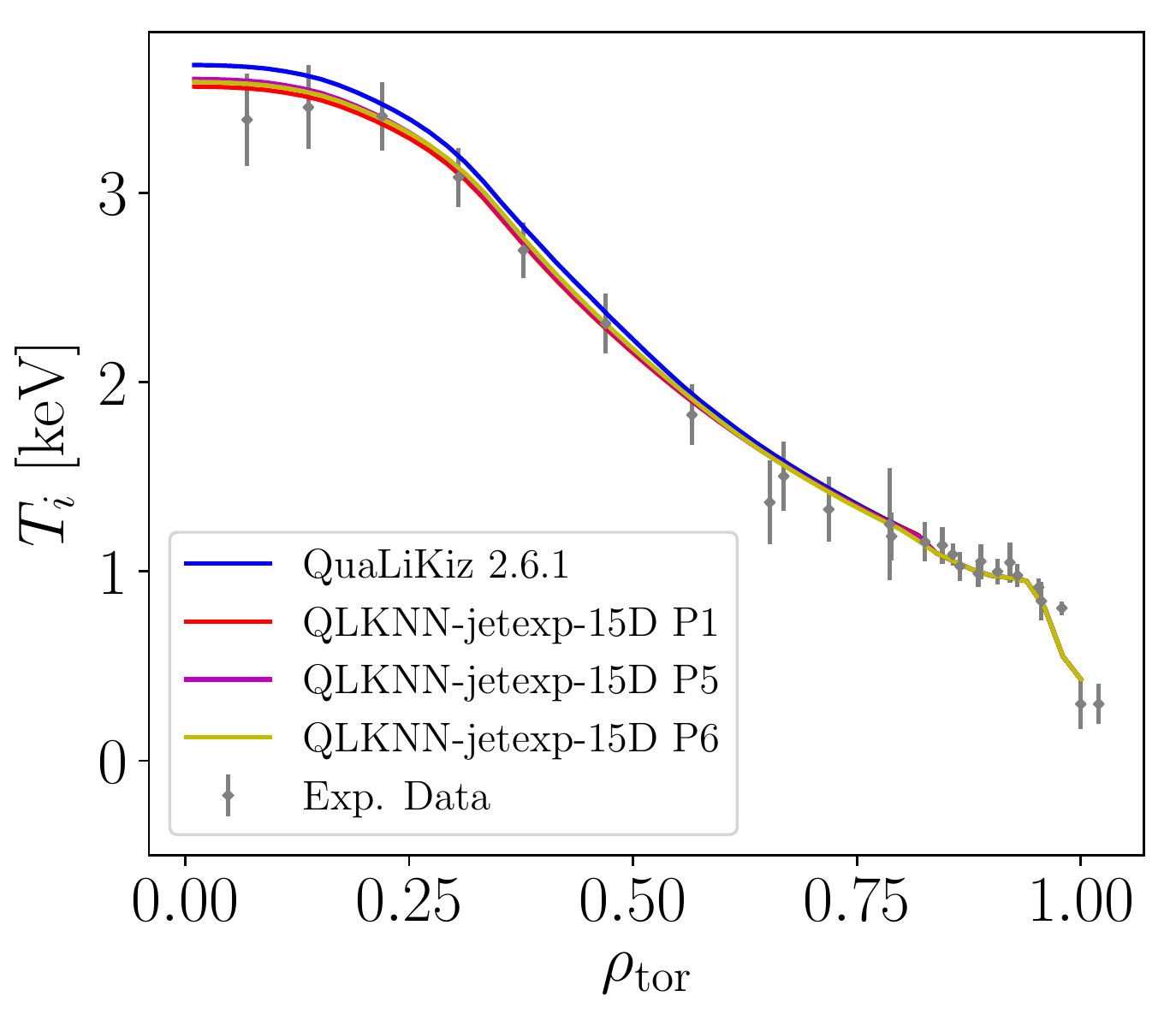}%
	\hspace{2mm}\includegraphics[scale=0.27]{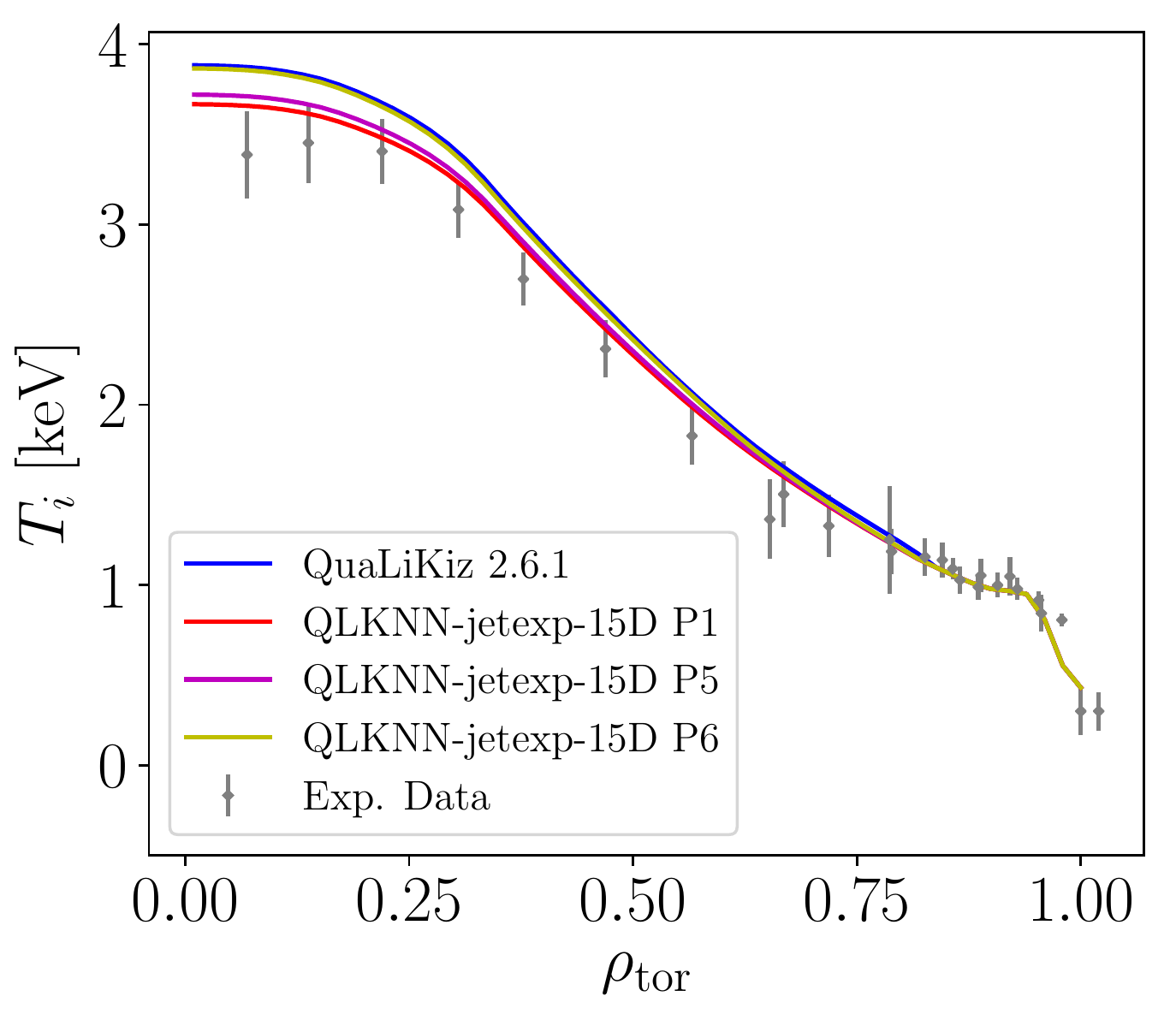}\\
	\hspace{-0.5mm}\includegraphics[scale=0.27]{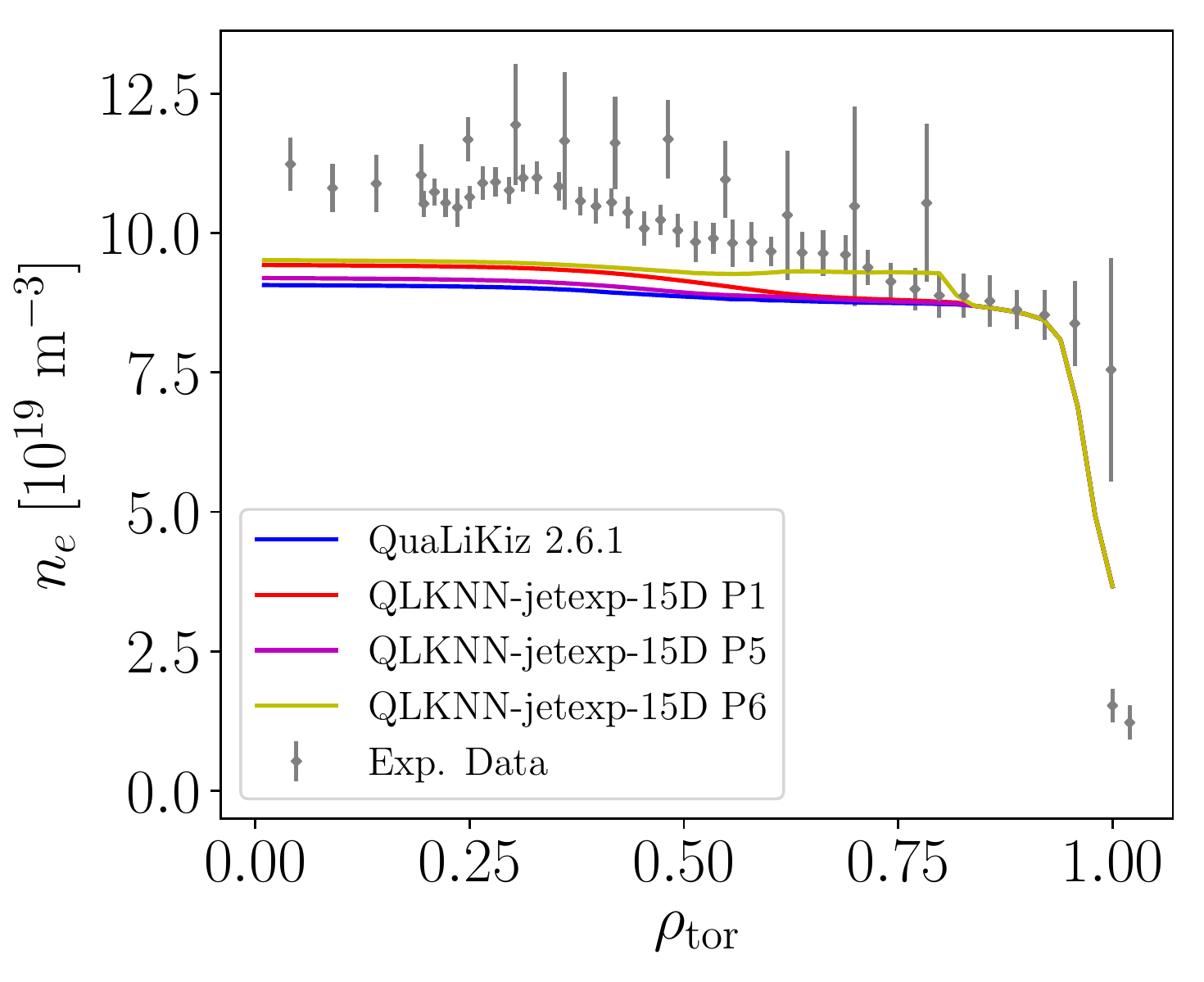}%
	\includegraphics[scale=0.27]{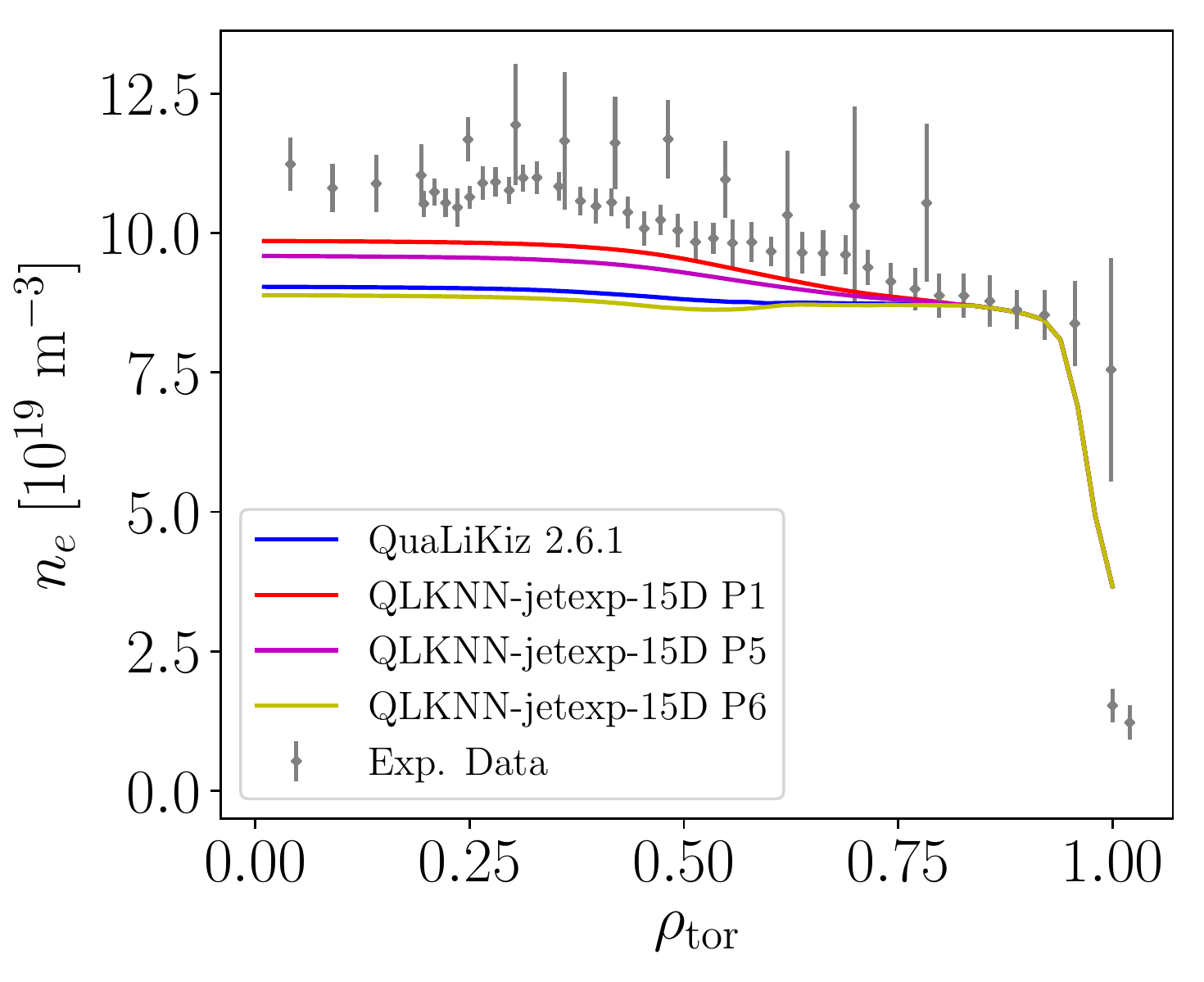}
	\caption{Comparison of integrated modelling results, $T_e,\,T_i,\,n_e$, for JET\#73342 using predicted transport fluxes from 15D NN against those of the original QuaLiKiz model, without the impact of plasma rotation (left) and with the impact of plasma rotation only applied to $\rho_{\text{tor}} > 0.4$ (right). The experimental data (gray points) used to determine the initial conditions of the simulation are also shown. The turbulent transport predictions are applied between $0.15 \le \rho_{\text{tor}} \le 0.85$.}
	\label{fig:IntegratedModelComparisonJET73342}
\end{figure}

\begin{figure}[tb]
	\centering
	\includegraphics[scale=0.27]{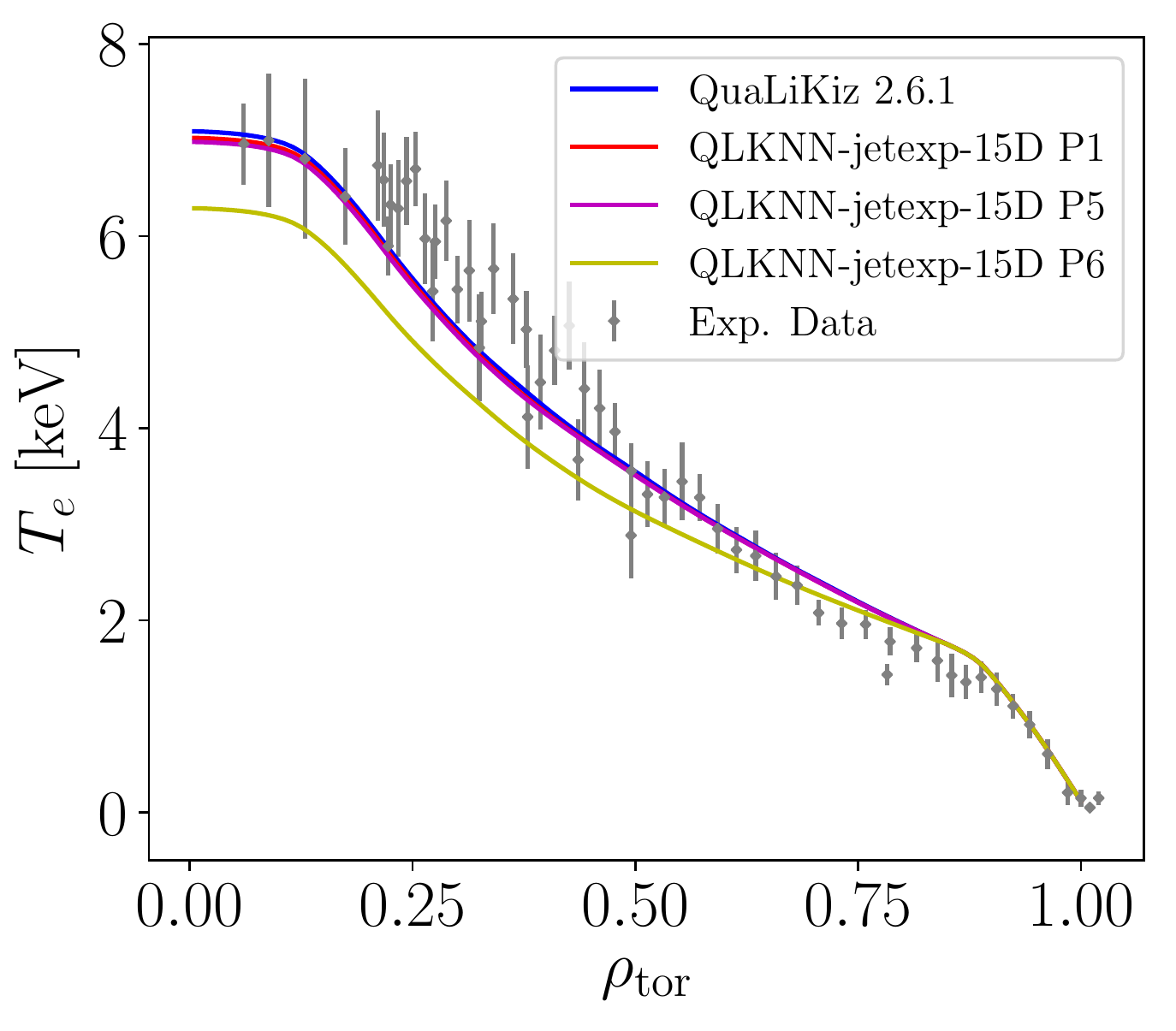}%
	\includegraphics[scale=0.27]{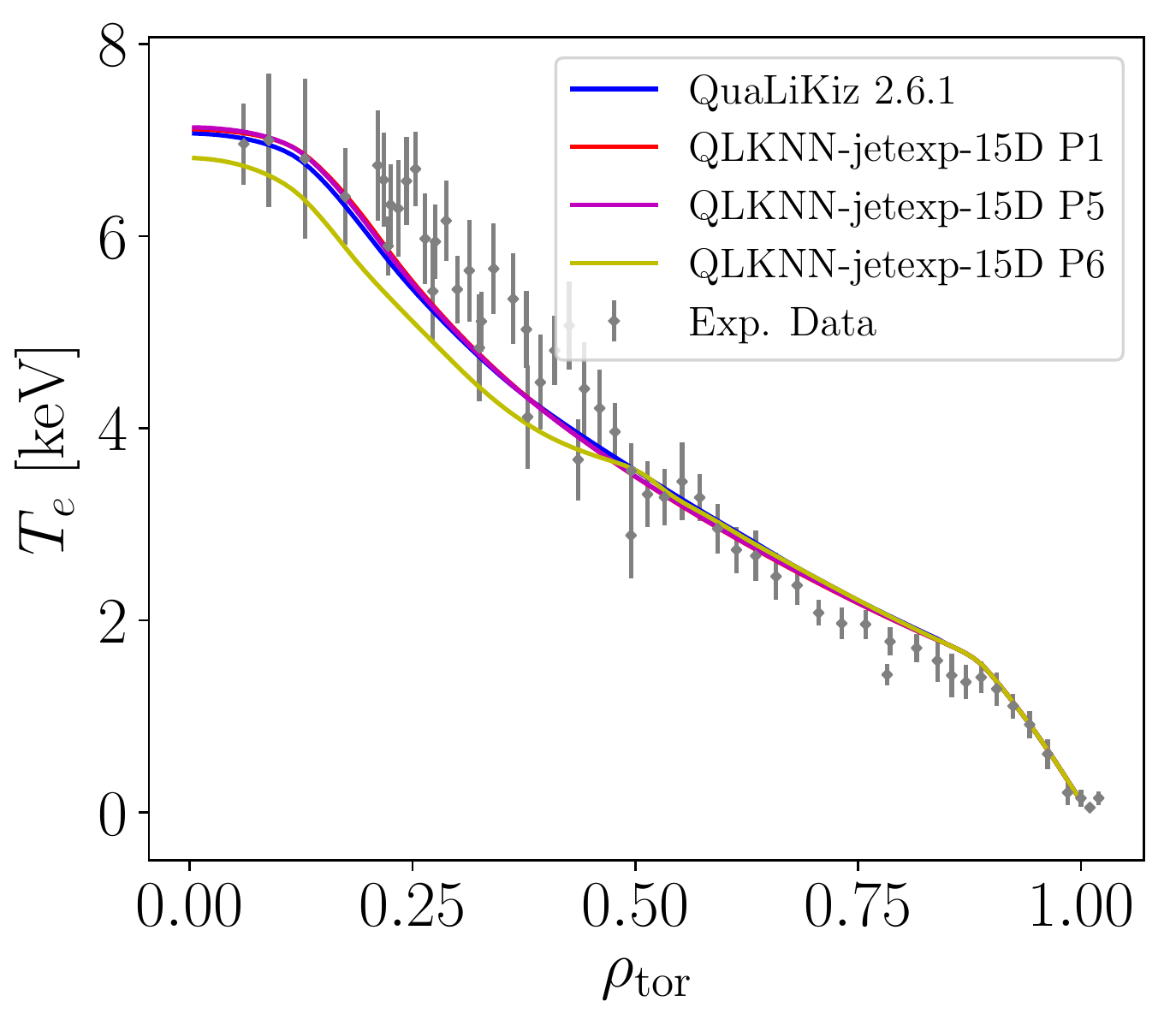}\\
	\includegraphics[scale=0.27]{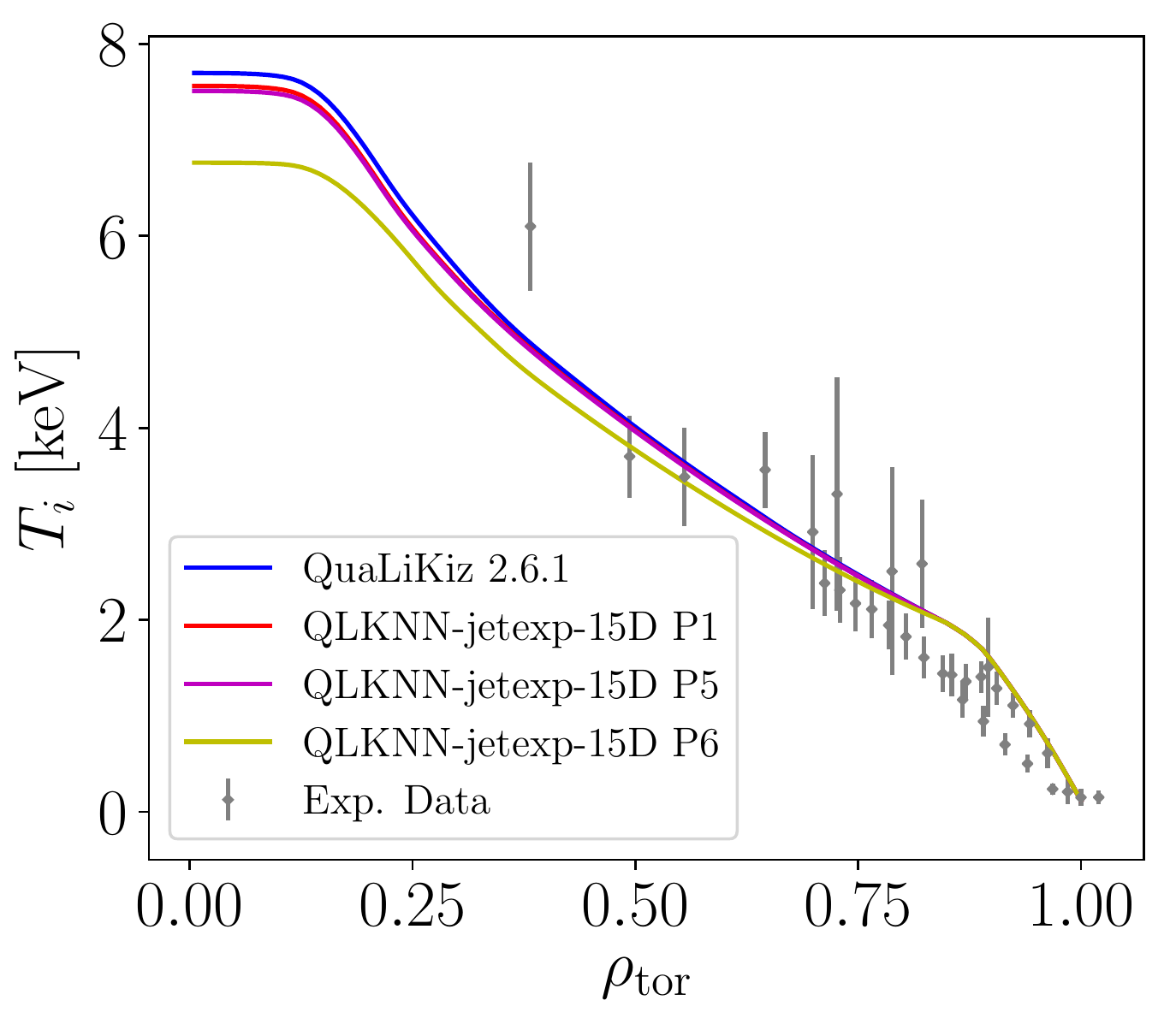}%
	\includegraphics[scale=0.27]{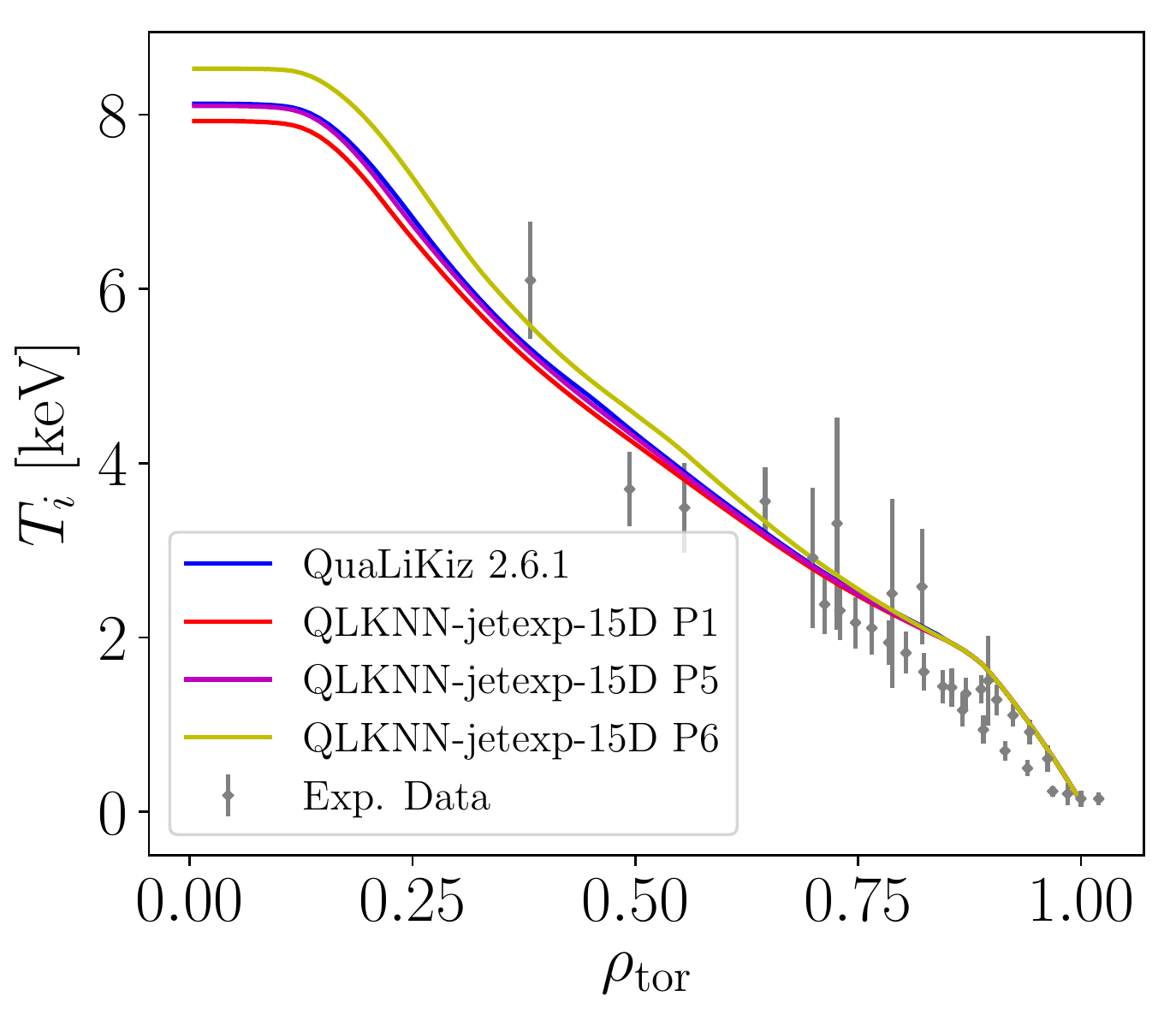}\\
	\includegraphics[scale=0.27]{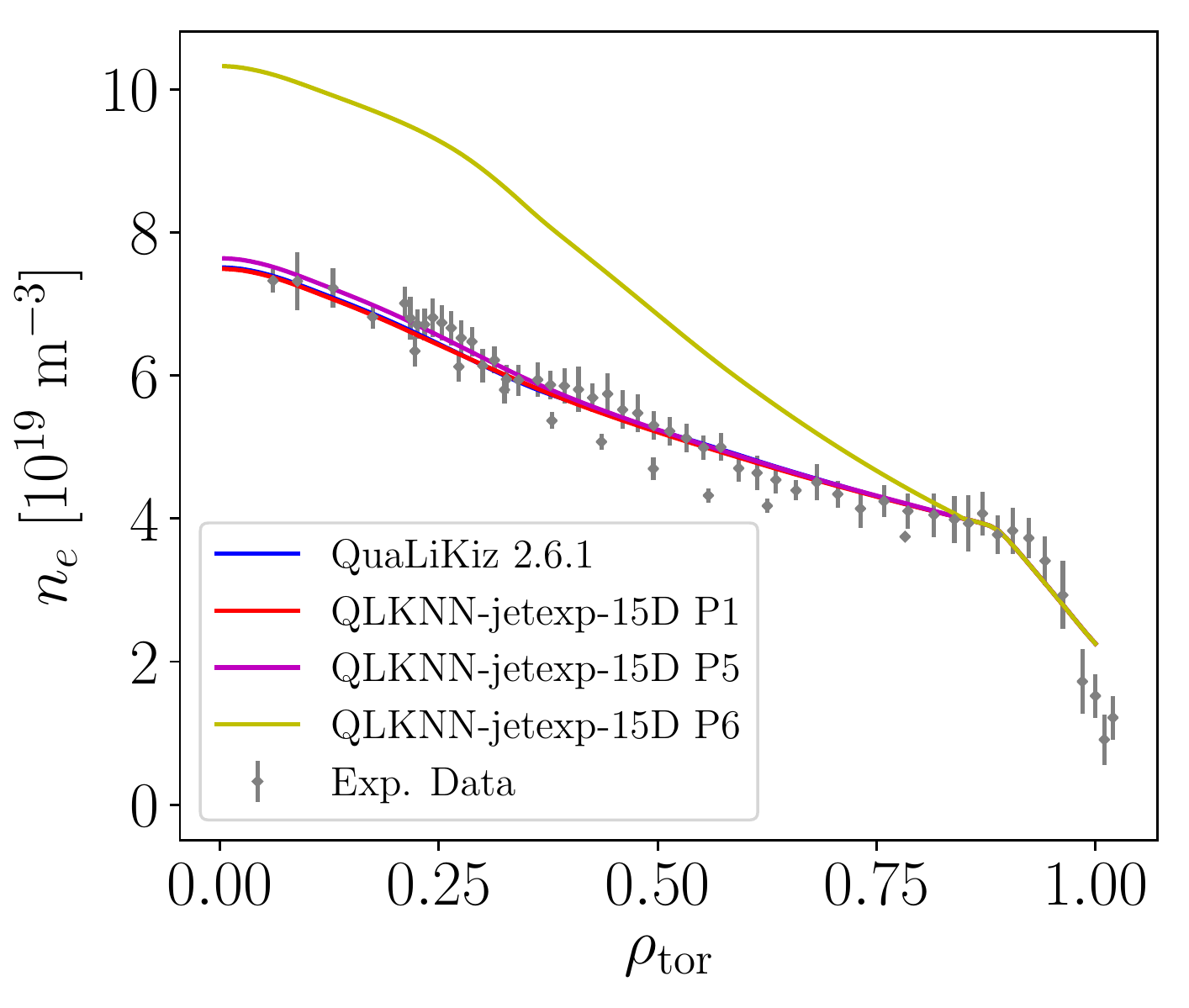}%
	\includegraphics[scale=0.27]{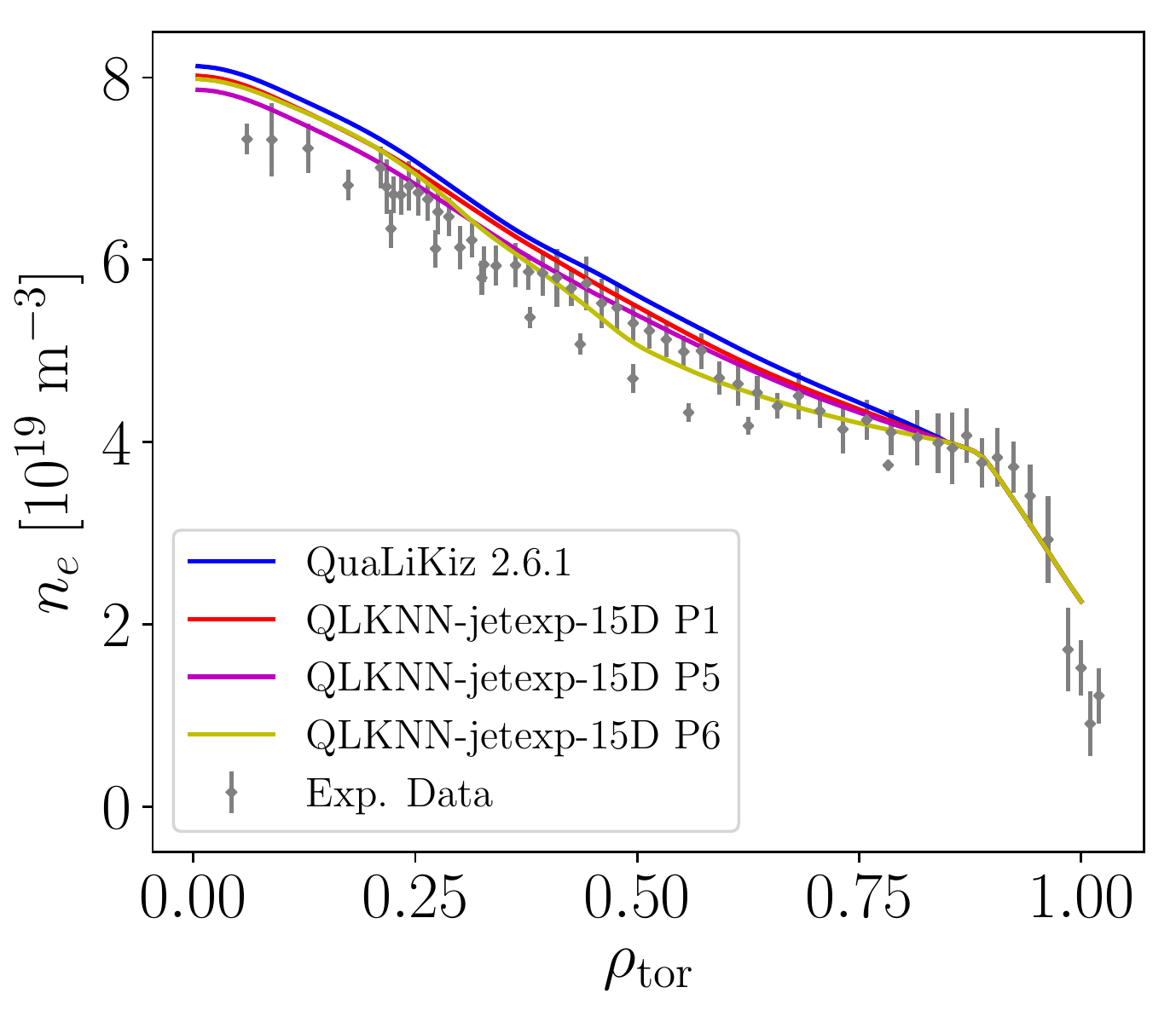}
	\caption{Comparison of integrated modelling results, $T_e,\,T_i,\,n_e$, for JET\#92398 using predicted transport fluxes from QLKNN-jetexp-15D model against those of the original QuaLiKiz model, without the impact of plasma rotation (left) and with the impact of plasma rotation only applied to $\rho_{\text{tor}} > 0.4$ (right). The experimental data (gray points) used to determine the initial conditions of the simulation are also shown. The turbulent transport predictions are applied between $0.03 \le \rho_{\text{tor}} \le 0.85$.}
	\label{fig:IntegratedModelComparisonJET92398}
\end{figure}

\begin{figure}[tb]
	\centering
	\includegraphics[scale=0.27]{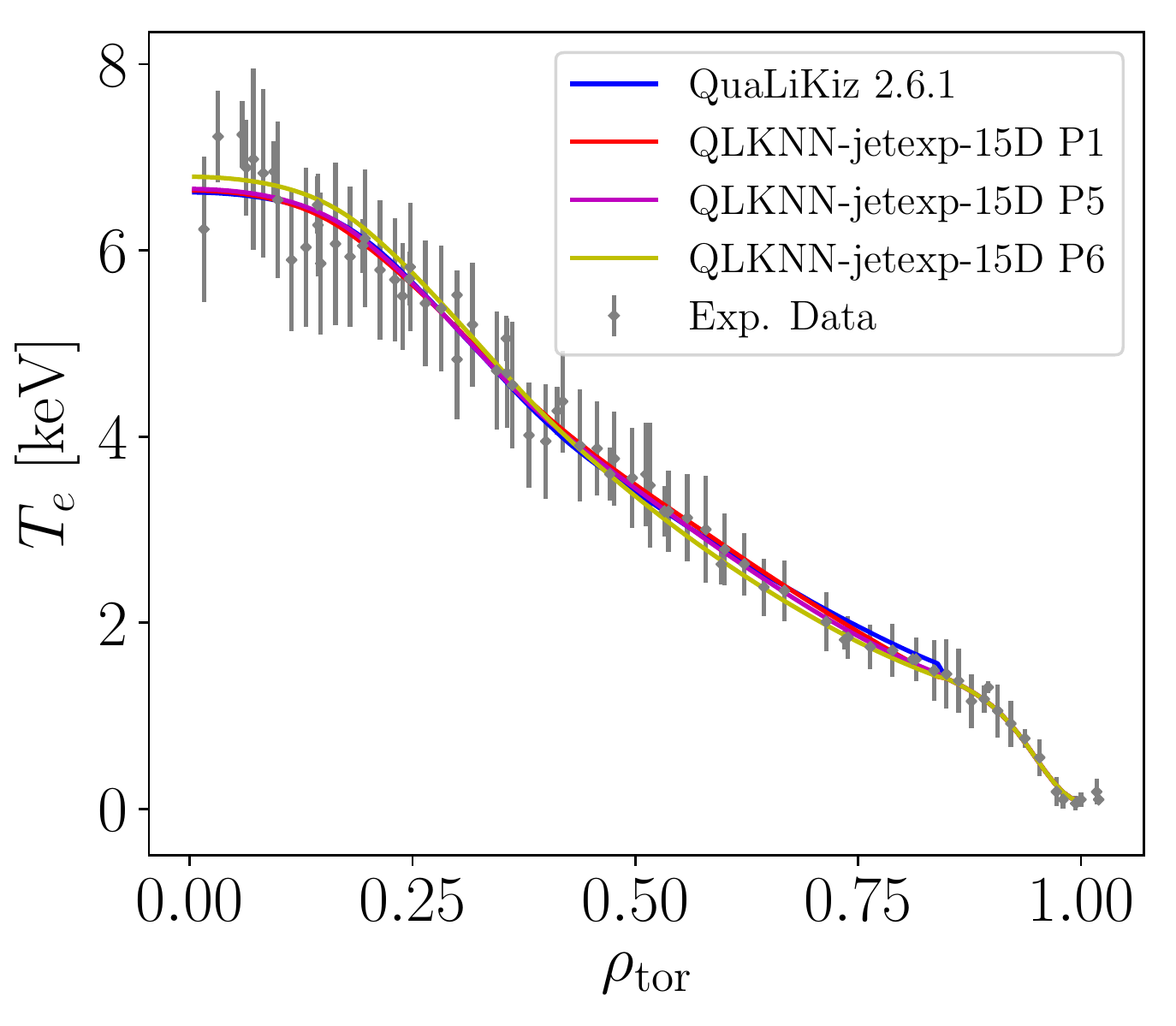}%
	\includegraphics[scale=0.27]{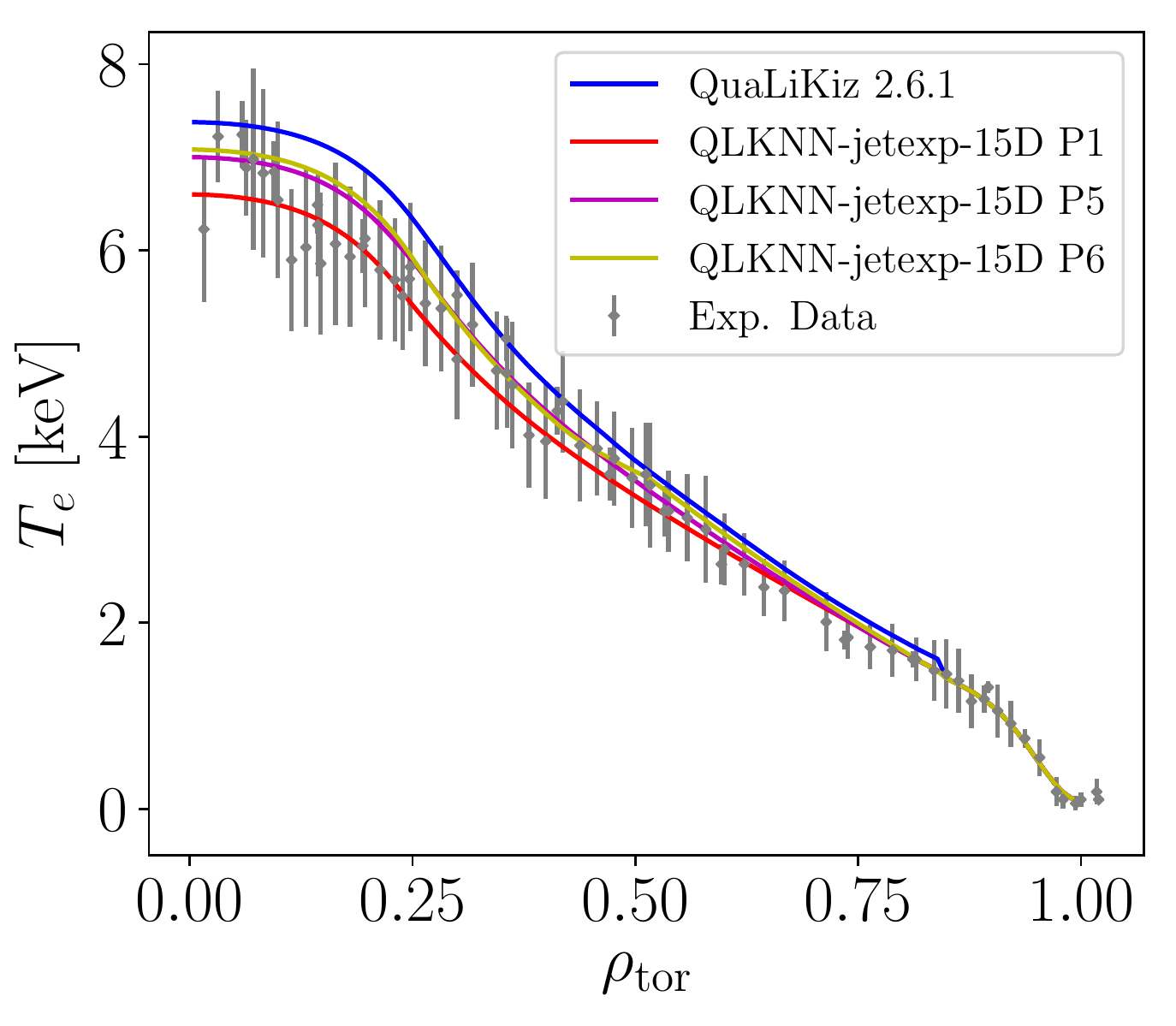}\\
	\includegraphics[scale=0.27]{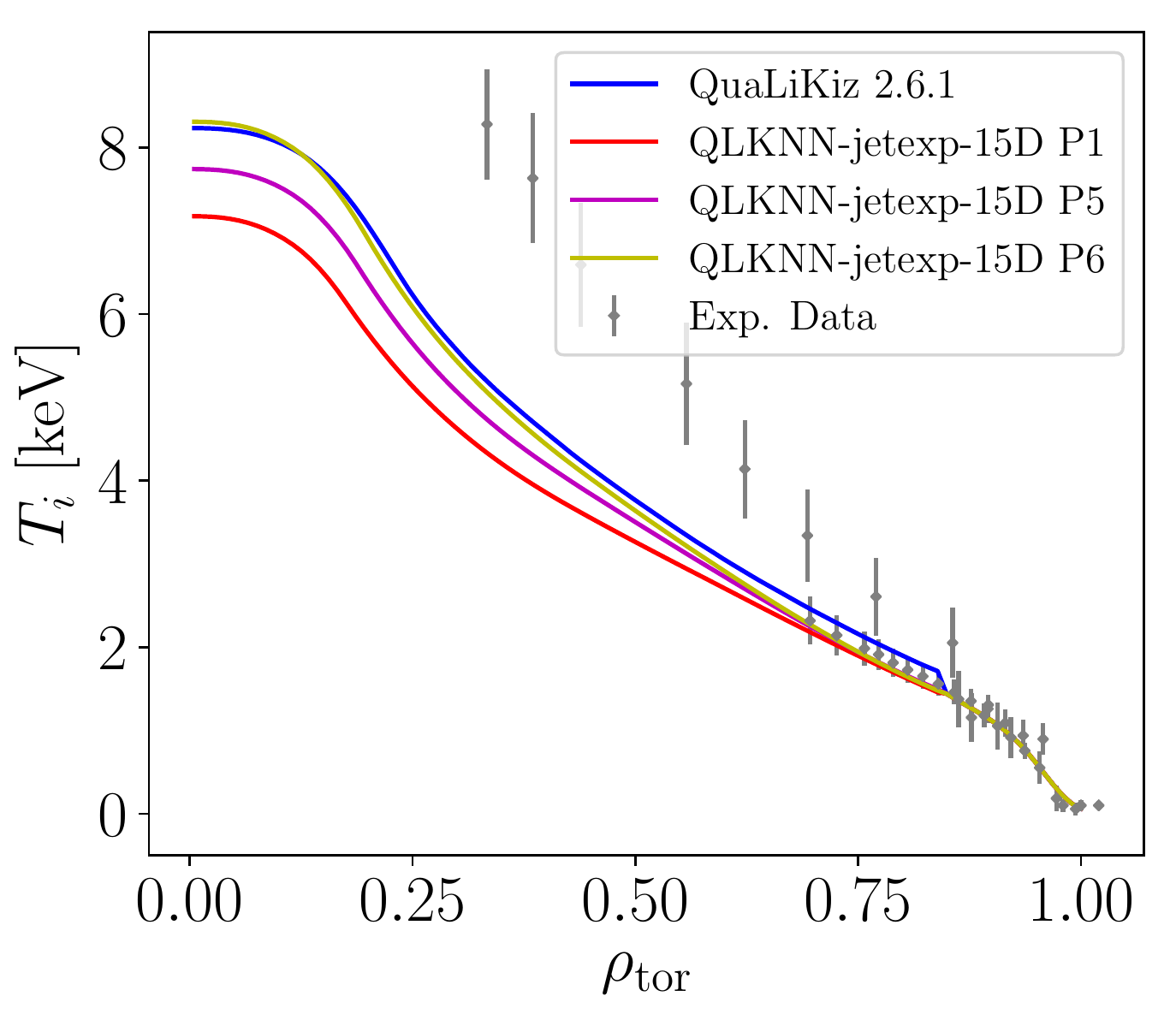}%
	\includegraphics[scale=0.27]{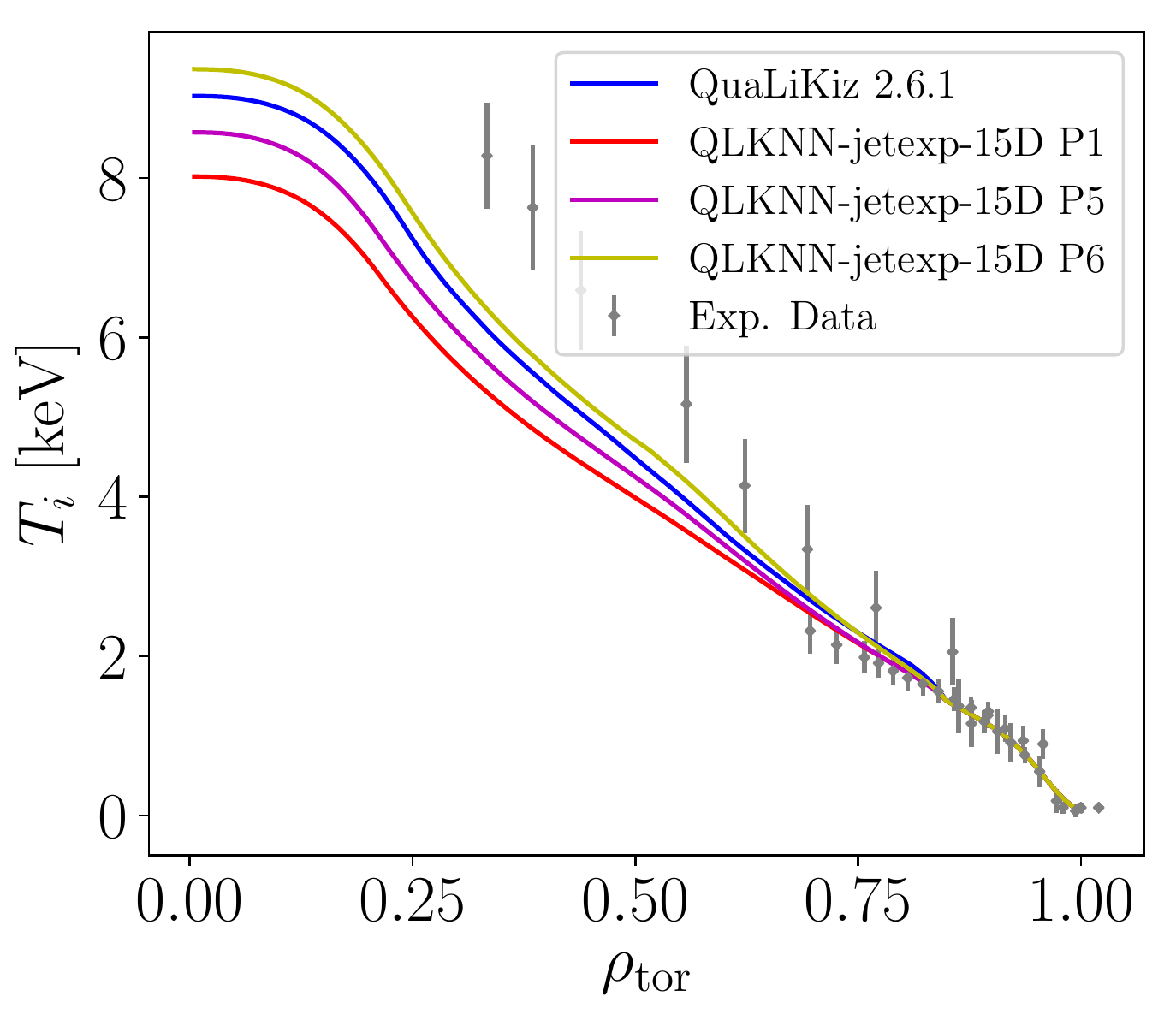}\\
	\includegraphics[scale=0.27]{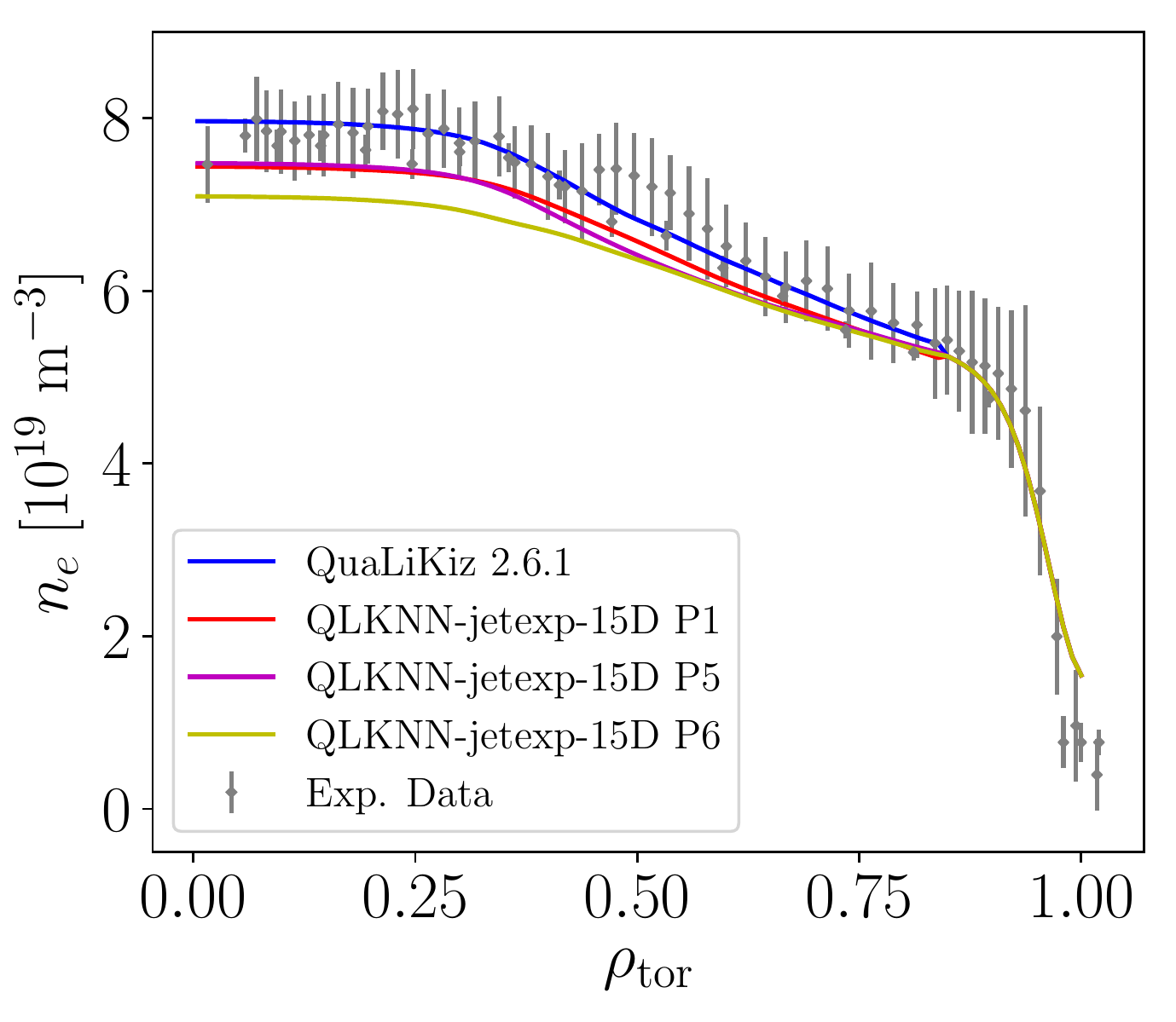}%
	\includegraphics[scale=0.27]{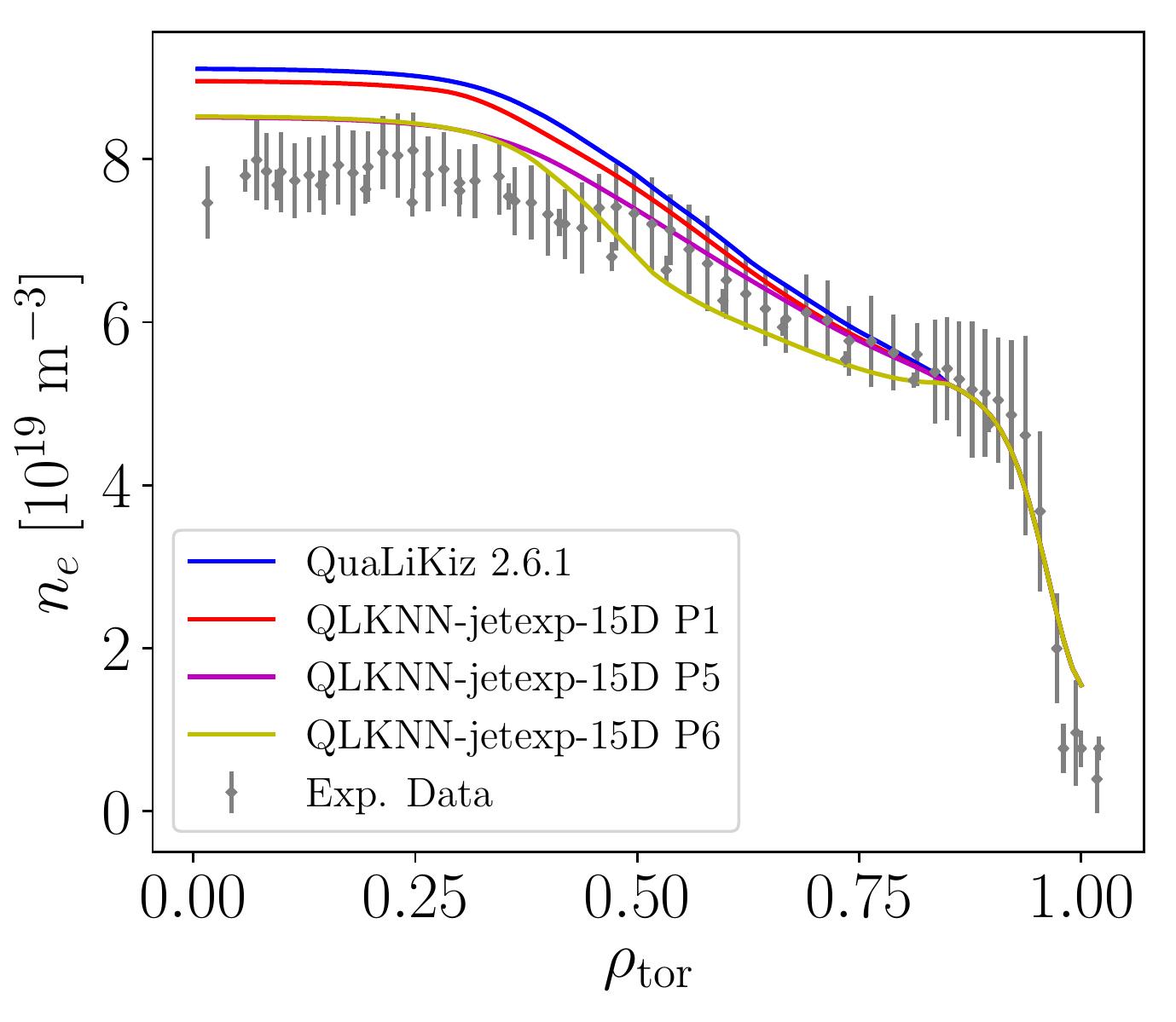}
	\caption{Comparison of integrated modelling results, $T_e,\,T_i,\,n_e$, for JET\#92436 using predicted transport fluxes from 15D NN against those of the original QuaLiKiz model, without the impact of plasma rotation (left) and with the impact of plasma rotation only applied to $\rho_{\text{tor}} > 0.4$ (right). The experimental data (gray points) used to determine the initial conditions of the simulation are also shown. The turbulent transport predictions are applied between $0.15 \le \rho_{\text{tor}} \le 0.85$.}
	\label{fig:IntegratedModelComparisonJET92436}
\end{figure}

\begin{figure}[tb]
	\centering
	\includegraphics[scale=0.27]{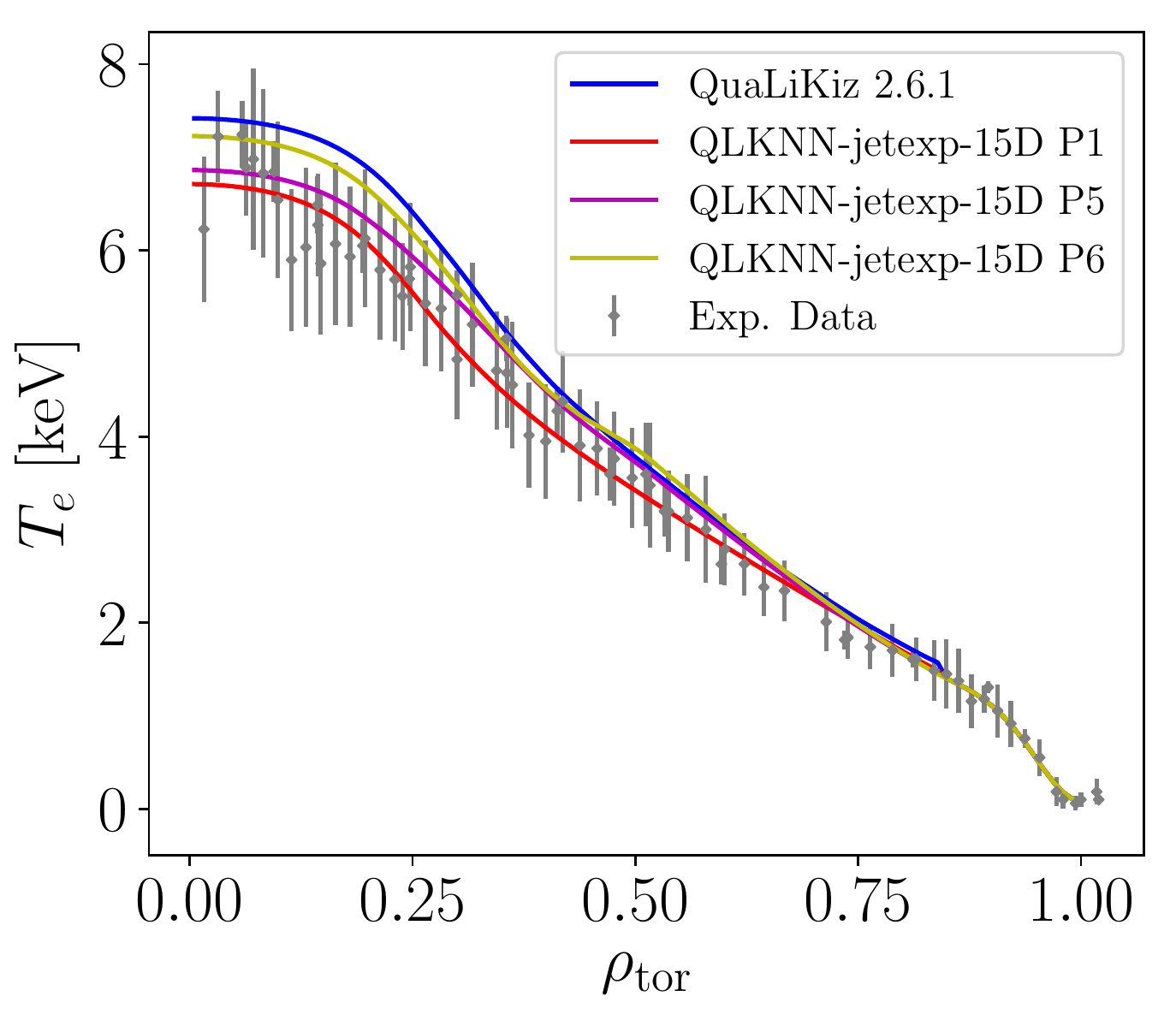}%
	\hspace{2.5mm}\includegraphics[scale=0.27]{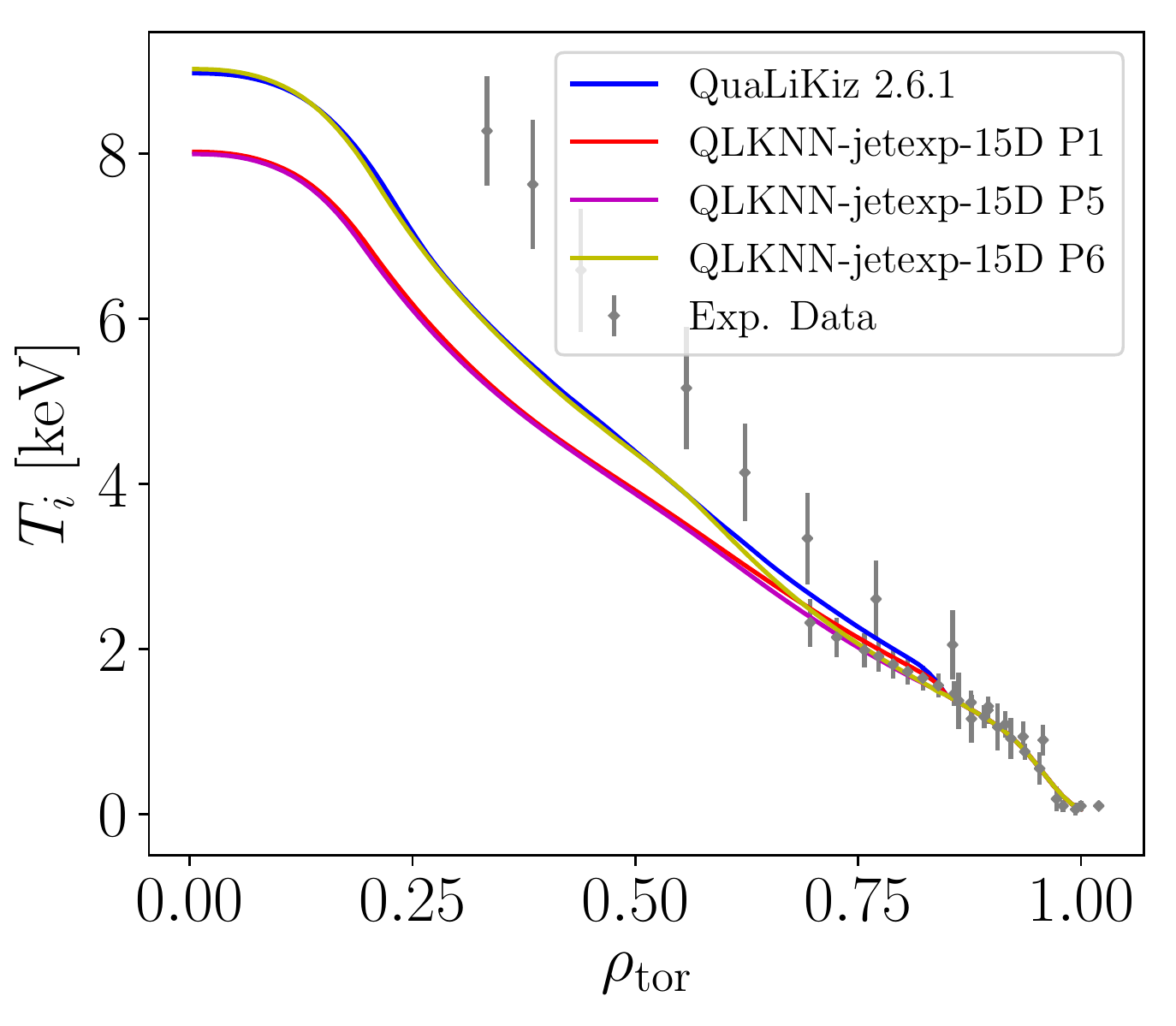}\\	\includegraphics[scale=0.27]{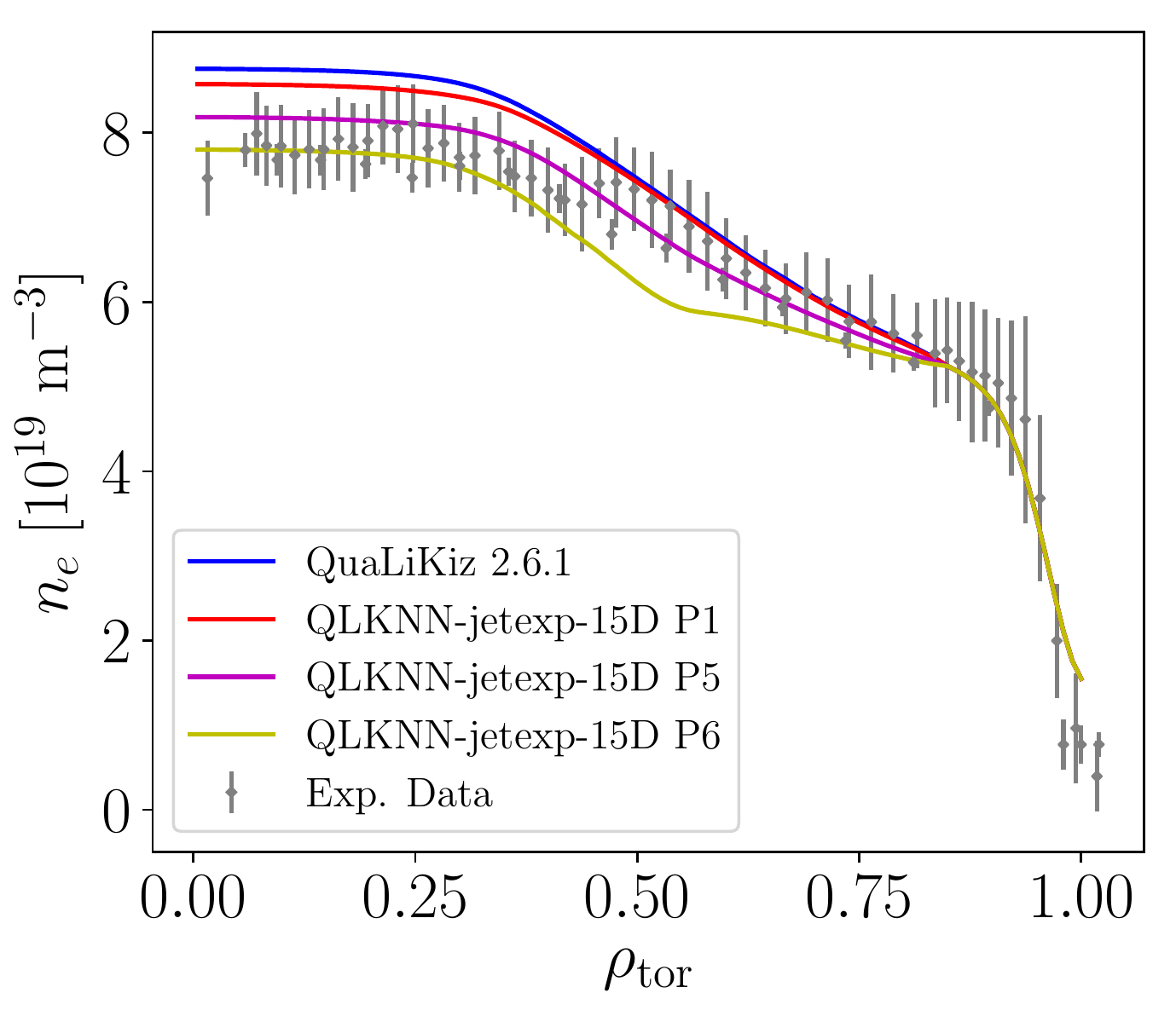}%
	\includegraphics[scale=0.27]{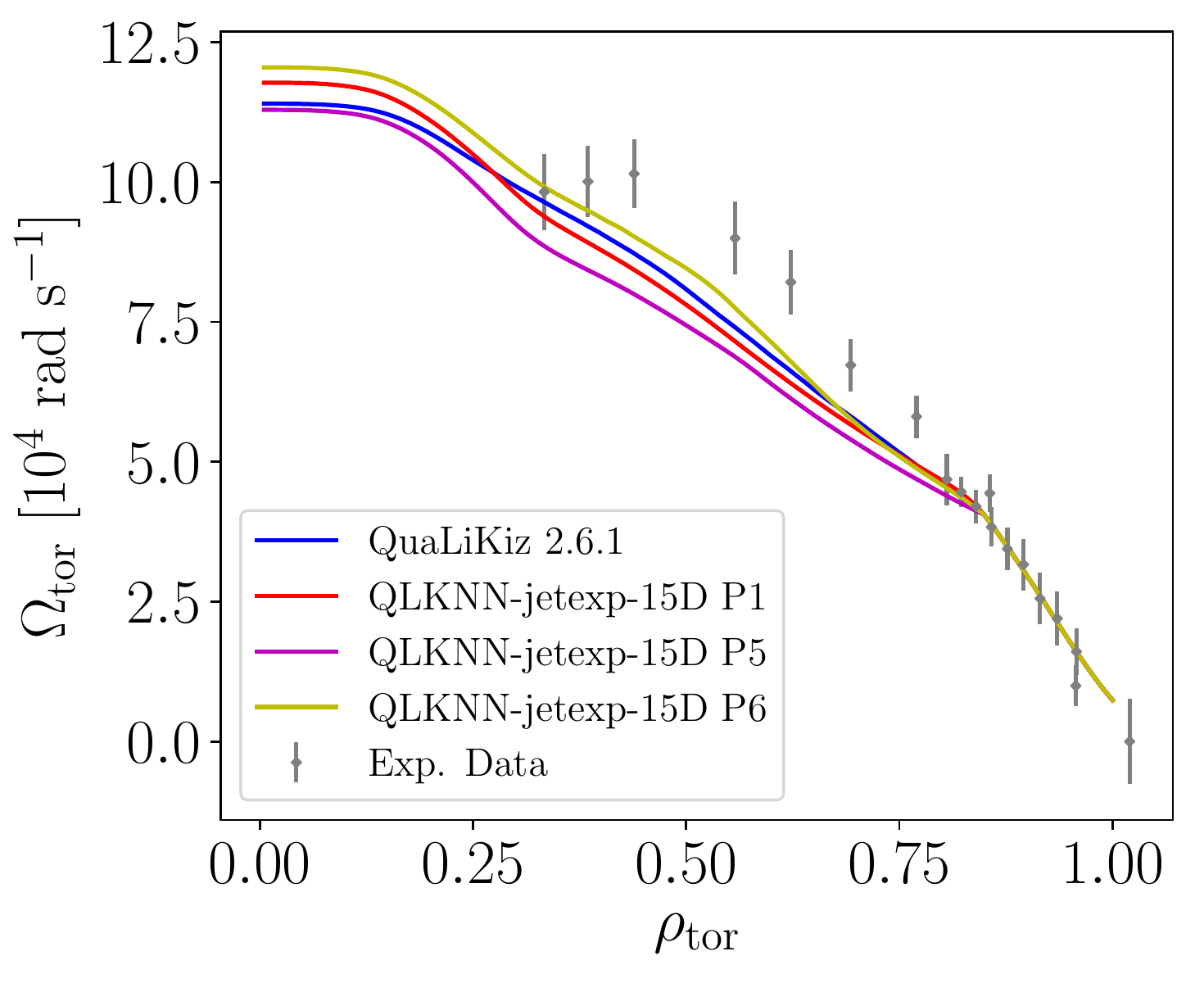}
	\caption{Comparison of integrated modelling results, $T_e,\,T_i,\,n_e,\,\Omega_{\text{tor}}$, for JET\#92436 using predicted transport fluxes from 15D NN against those of the original QuaLiKiz model, with the impact of plasma rotation only applied to $\rho_{\text{tor}} > 0.4$. The experimental data (gray points) used to determine the initial conditions of the simulation are also shown. The turbulent transport predictions are applied between $0.15 \le \rho_{\text{tor}} \le 0.85$.}
	\label{fig:IntegratedModelComparisonJET92436wRotation}
\end{figure}

\begin{figure}[tb]
	\centering
	\includegraphics[scale=0.26]{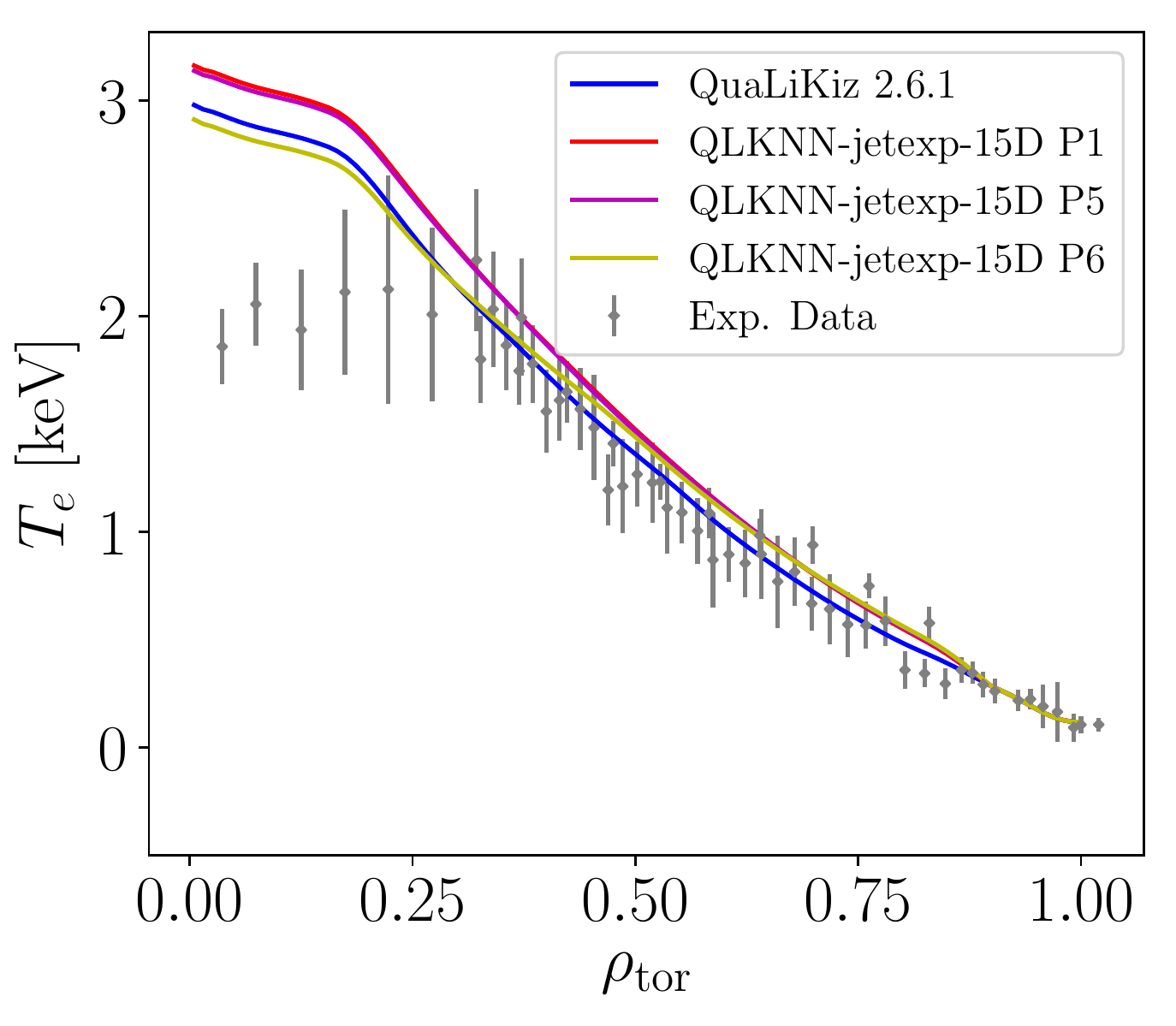}%
	\hspace{2.75mm}\includegraphics[scale=0.26]{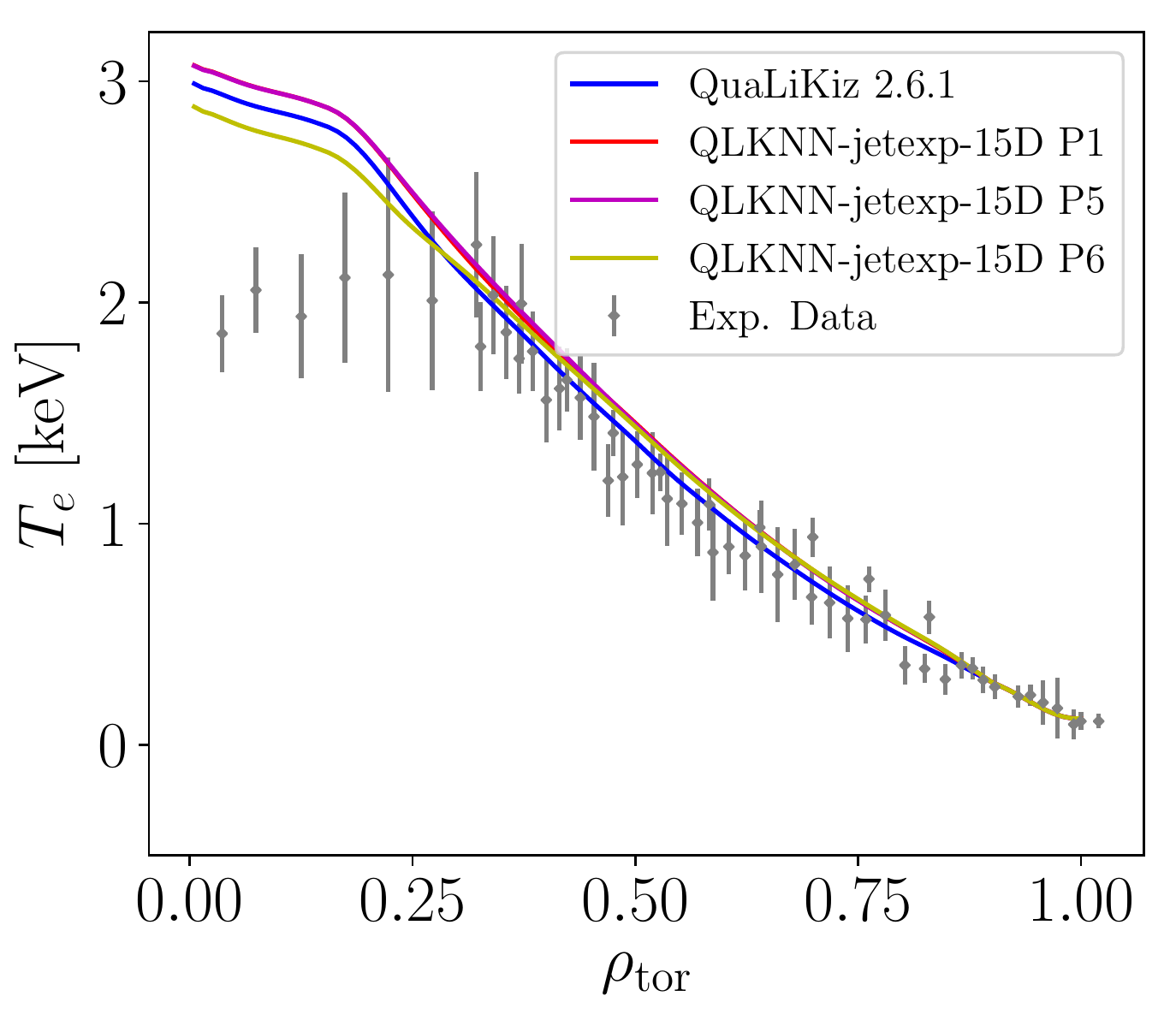}\\
	\hspace{-2.5mm}\includegraphics[scale=0.26]{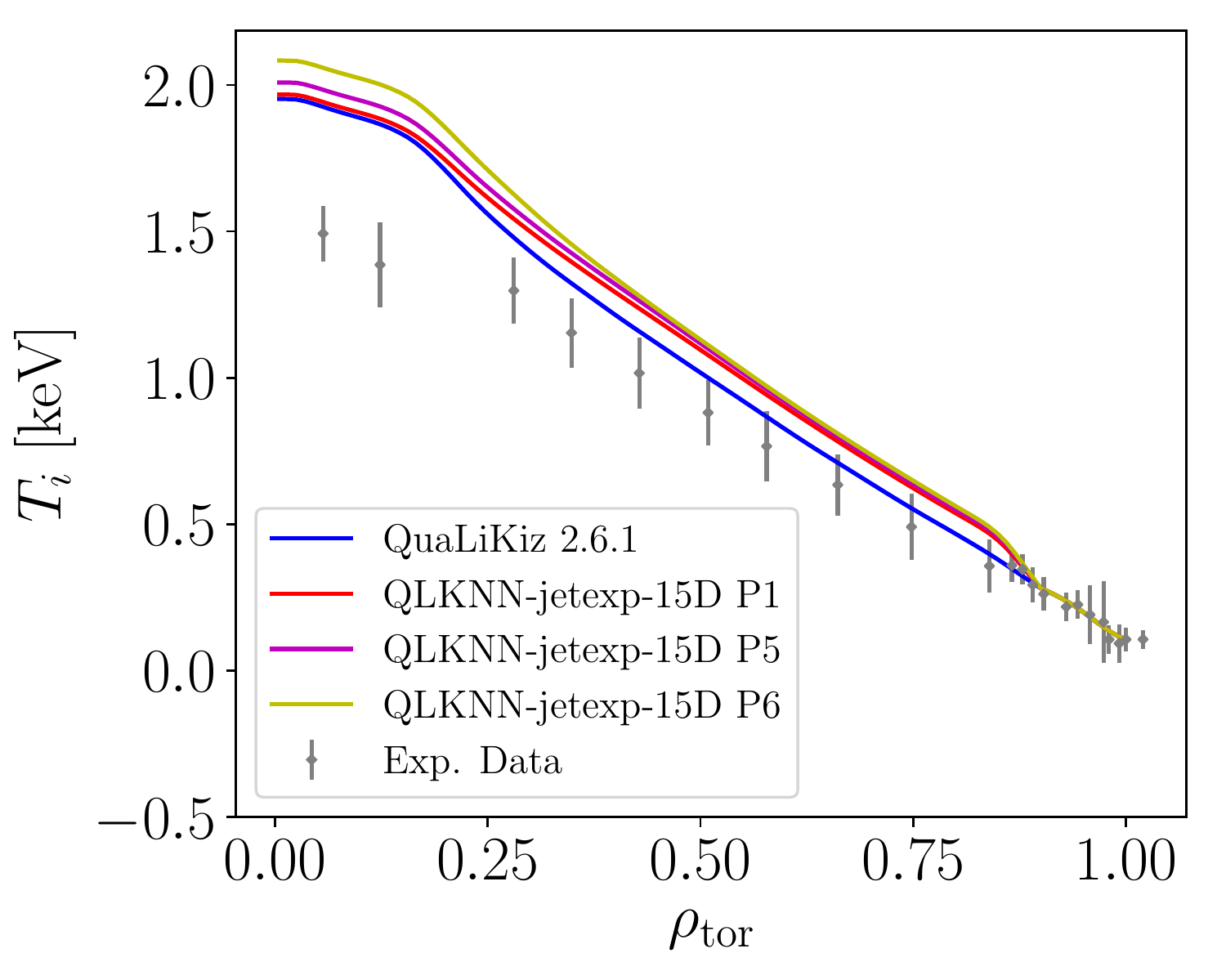}%
	\includegraphics[scale=0.26]{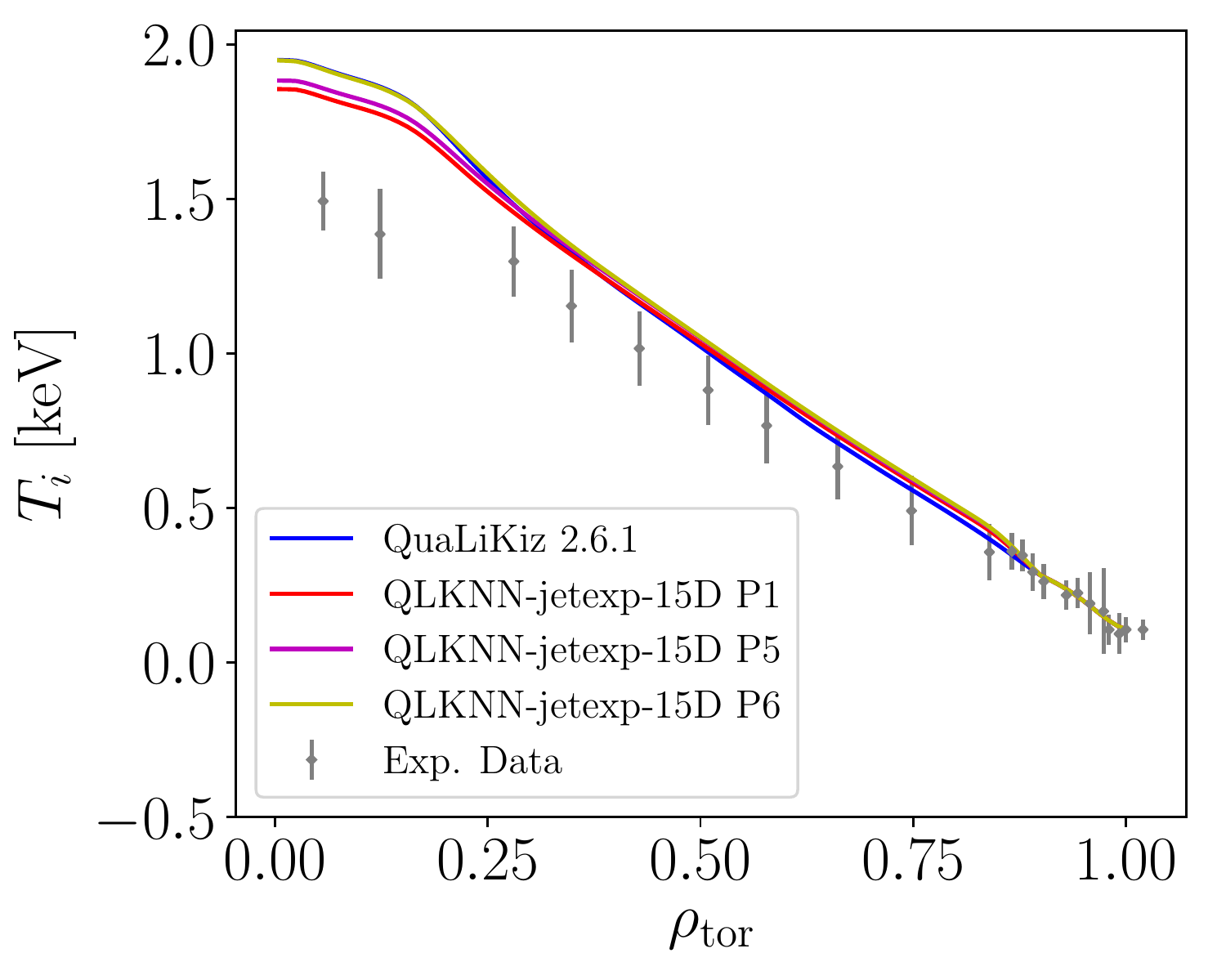}\\
	\hspace{-2.5mm}\includegraphics[scale=0.26]{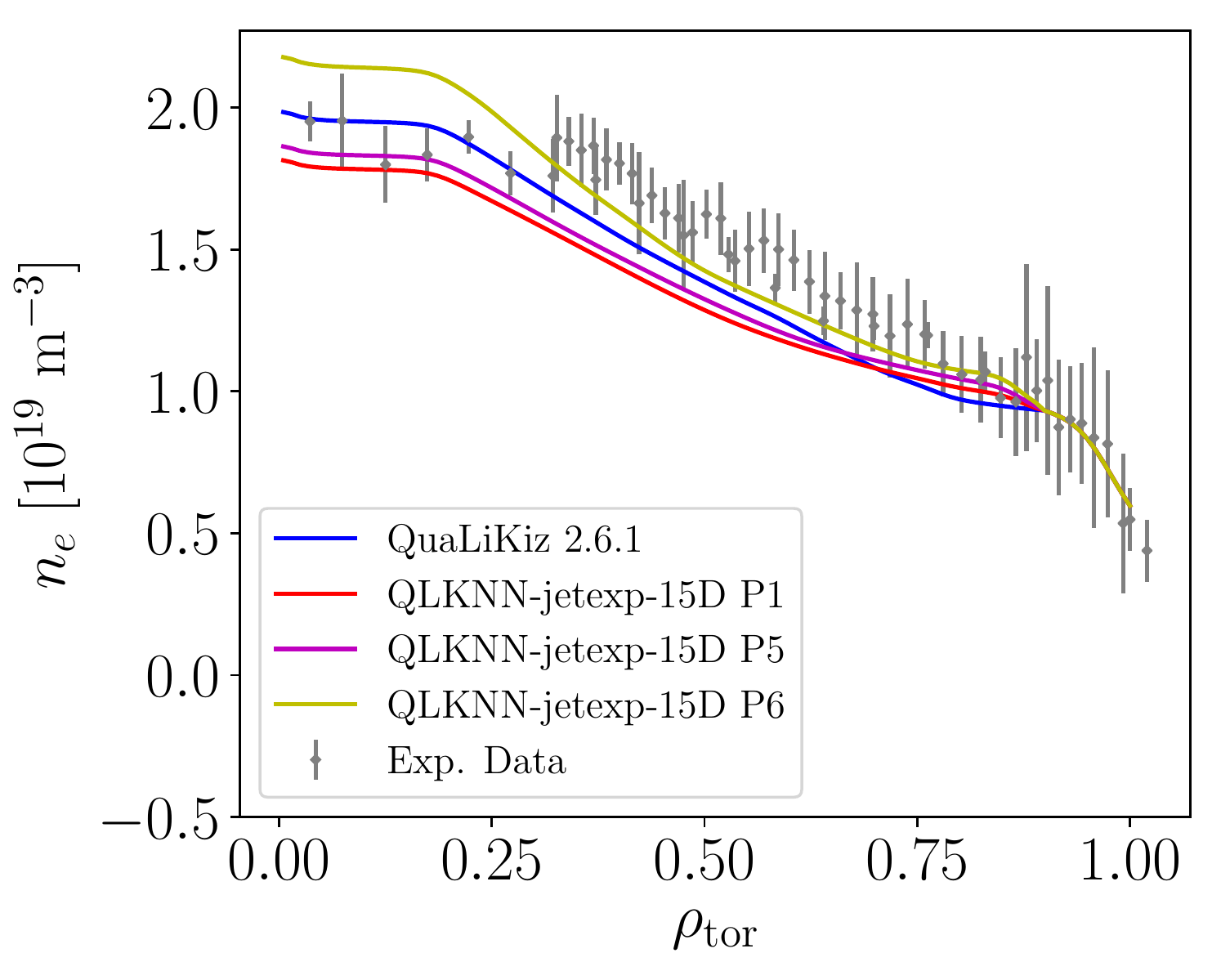}%
	\includegraphics[scale=0.26]{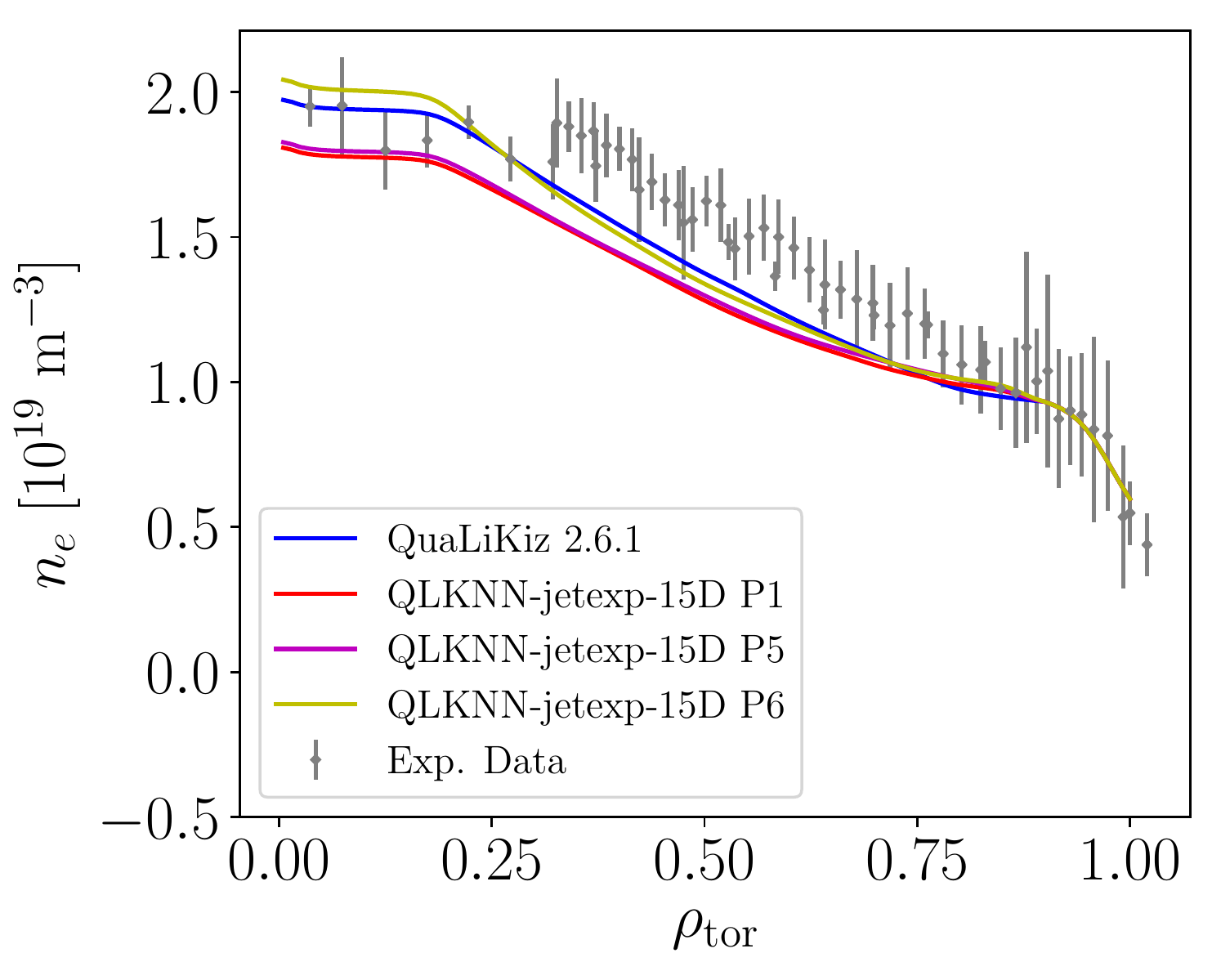}
	\caption{Comparison of integrated modelling results, $T_e,\,T_i$, for JET\#91637 using predicted transport fluxes from 15D NN against those of the original QuaLiKiz model, without the impact of plasma rotation (left) and with the impact of plasma rotation only applied to $\rho_{\text{tor}} > 0.4$ (right). The experimental data (gray points) used to determine the initial conditions of the simulation are also shown. The turbulent transport predictions are applied between $0.15 \le \rho_{\text{tor}} \le 0.90$.}
	\label{fig:IntegratedModelComparisonJET91637}
\end{figure}

\begin{figure}[tb]
	\centering
	\includegraphics[scale=0.27]{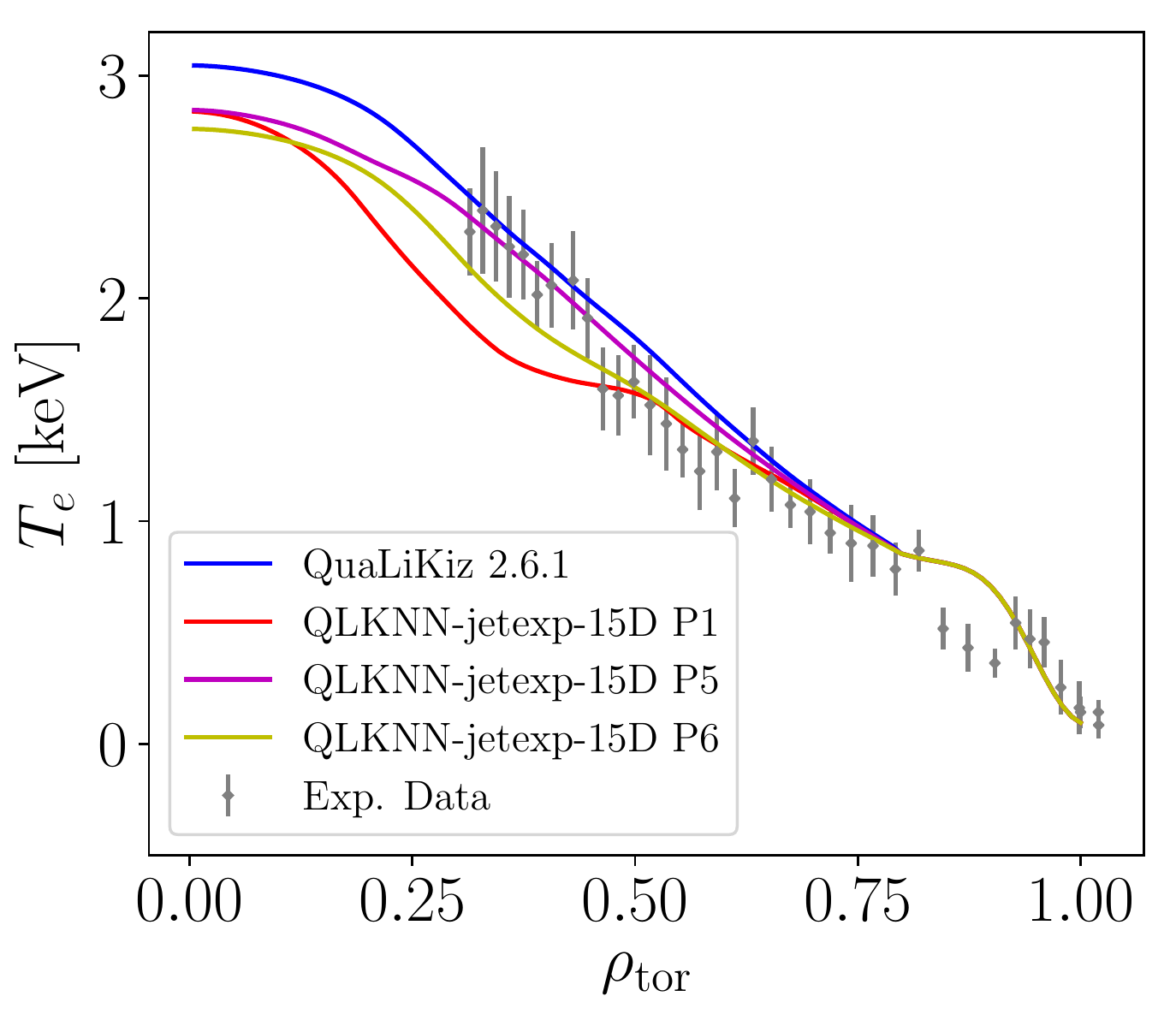}%
	\includegraphics[scale=0.27]{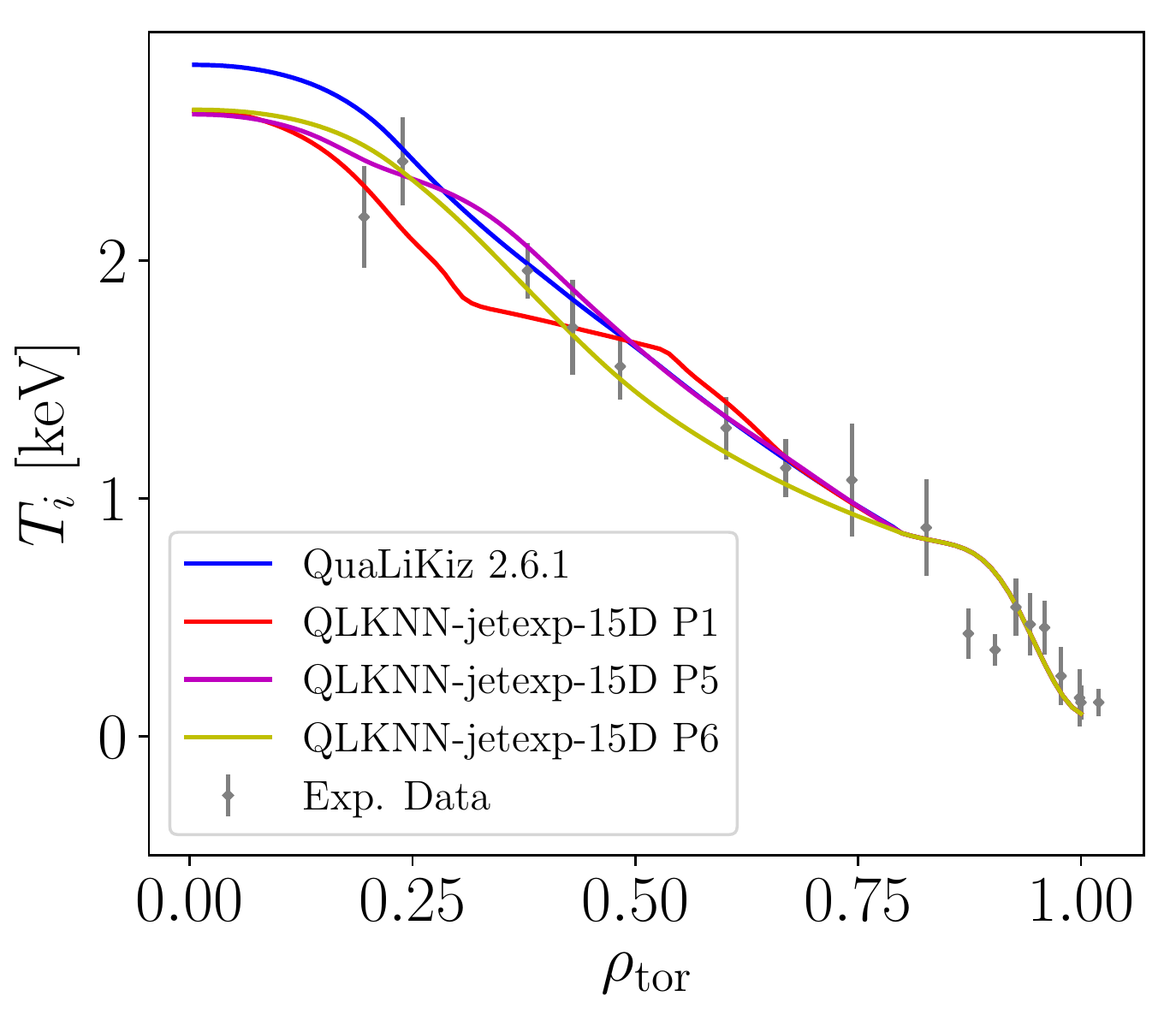}\\	\includegraphics[scale=0.27]{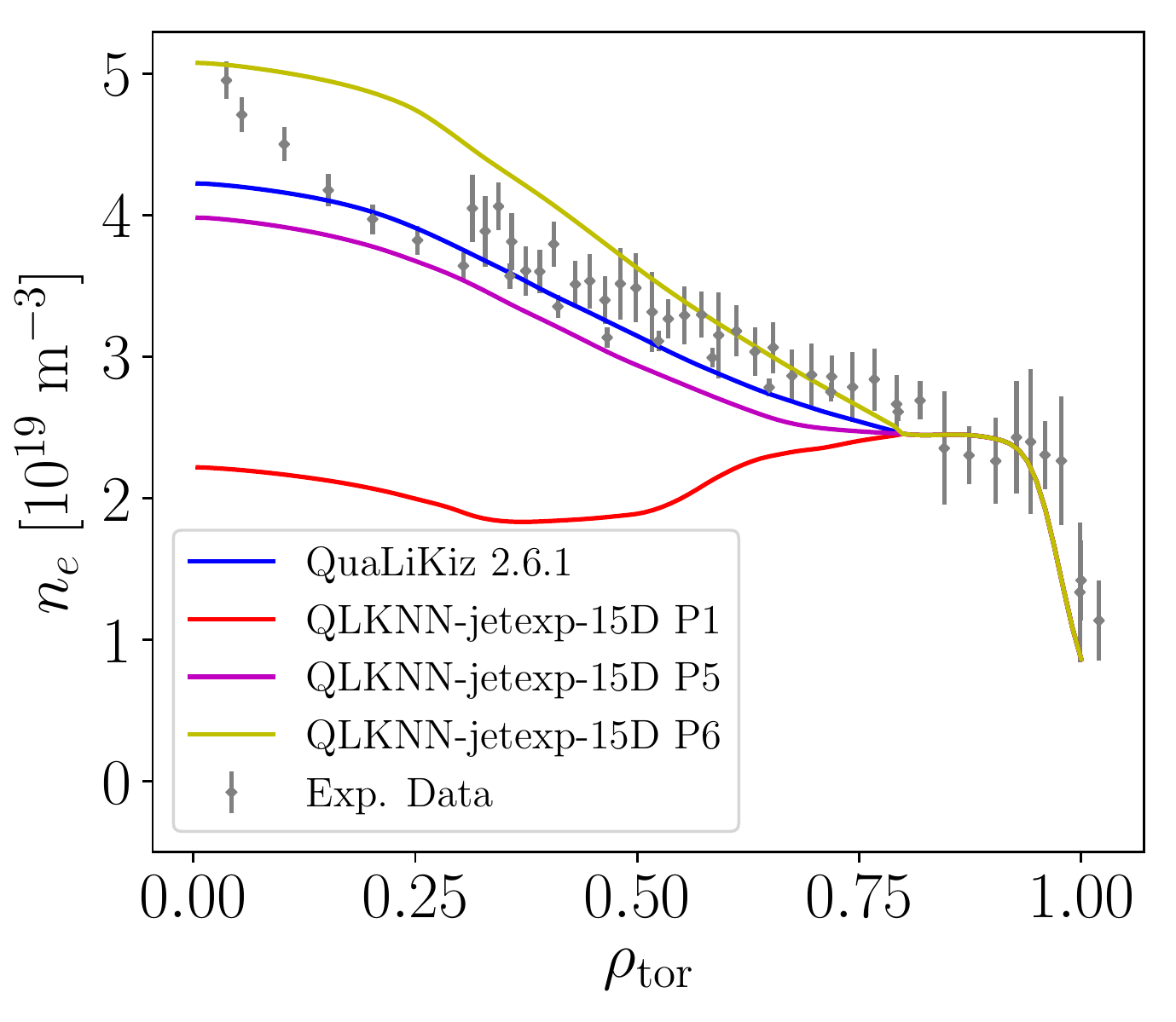}%
	\includegraphics[scale=0.27]{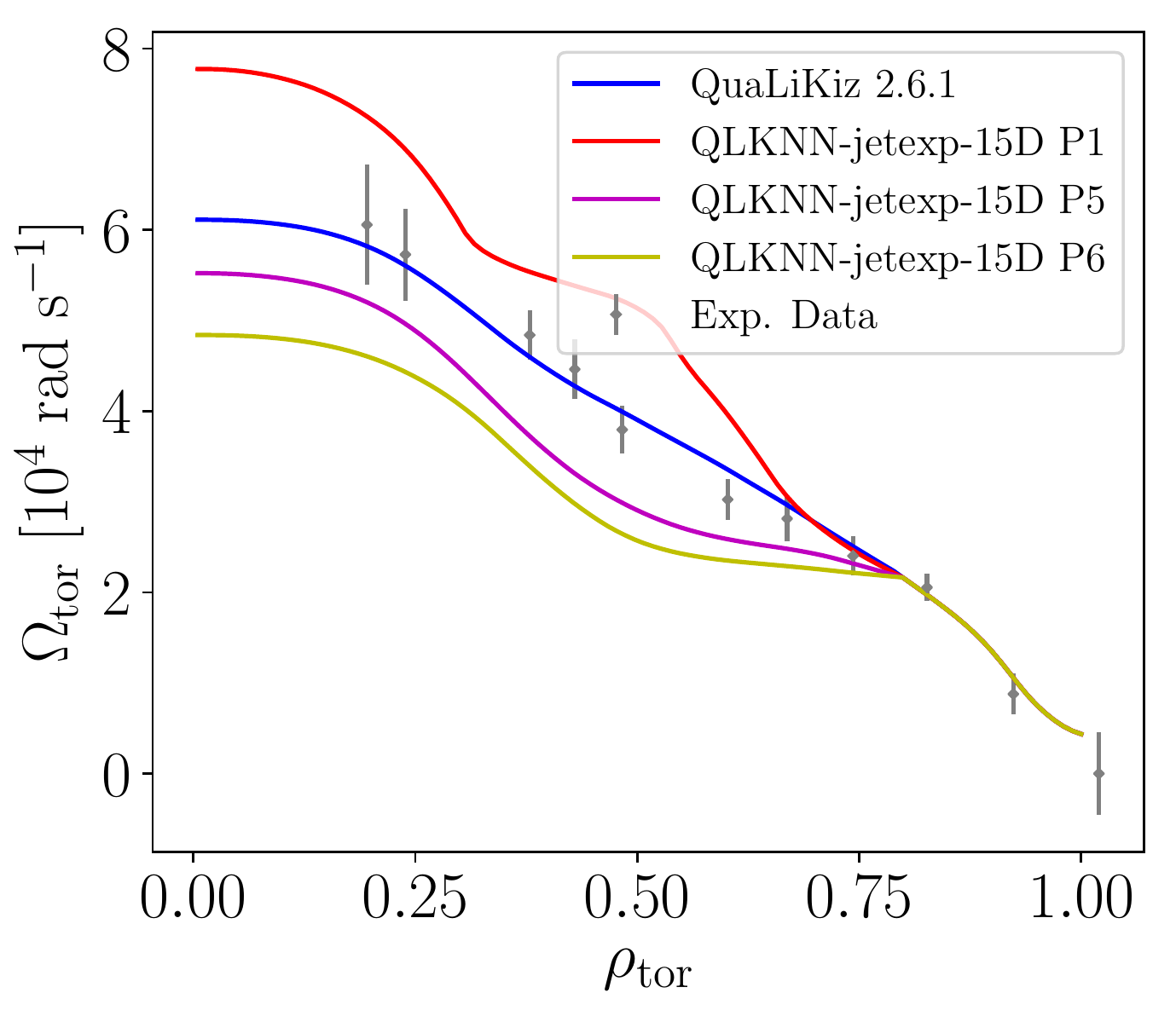}
	\caption{Comparison of integrated modelling results, $T_e,\,T_i,\,n_e,\,\Omega_{\text{tor}}$, for JET\#91227 using predicted transport fluxes from 15D NN against those of the original QuaLiKiz model, with the impact of plasma rotation only applied to $\rho_{\text{tor}} > 0.4$. The experimental data (gray points) used to determine the initial conditions of the simulation are also shown. The turbulent transport predictions are applied between $0.03 \le \rho_{\text{tor}} \le 0.80$.}
	\label{fig:IntegratedModelComparisonJET91227wRotation}
\end{figure}

Table~\ref{tbl:JETTOSettingsSummary} provides a summary of the most relevant simulation specifications and results. Figures~\ref{fig:IntegratedModelComparisonJET73342}~--~\ref{fig:IntegratedModelComparisonJET91227wRotation} show a comparison of the plasma profiles using QLKNN-jetexp-15D and the original QuaLiKiz model, along with the time-averaged measurements and uncertainties from the experimental discharge, for use as reference. Additional information regarding other integrated modelling runs performed within this study are available in Appendix~\ref{app:DetailedJETTOResults}.

To take advantage of the committee NN standard deviation predictions, the QLKNN implementation within JINTRAC was updated to evaluate and check the NN prediction standard deviation for each time step in the integrated model simulation. If the standard deviation of any given quantity exceeds a pre-defined threshold for a given time step, the integrated model switches back to the original QuaLiKiz model for all radial points on that time step. The exact threshold values used for the committee NN standard deviation checks and a detailed description of the method used to determine them can be found in Appendix~\ref{app:NNVarianceThresholds}. Even with relaxed thresholds, it was found that the NN standard deviation nearly always exceeds them near the edge of the simulation boundary. There are multiple potential reasons for this, such as:
\begin{itemize}
	\itemsep -1mm
	\item insufficient data density in this region due to the extreme plasma parameters there;
	\item insufficient data density in this region due to its proximity to the edge of the sampled parameter space;
	\item frequent deviations from the sampled parameter space due to the time evolution computation scheme;
	\item numerical abnormalities at the simulation boundary are causing  deviations from the expected parameter space;
	\item or any combination of the above.
\end{itemize}
This region only spans 2 or 3 radial grid points next to the simulation boundary and was considered to be a negligible discrepancy due to the fine grids used within the simulations in this study. To avoid future spurious tripping of this nature, the standard deviation check was set to trigger only when half of the points or more inside the turbulence prediction region exceeds the pre-defined threshold. Under these conditions, the extrapolation switch was never triggered within any of the simulations shown in this section. Nonetheless, the model switching scheme could be improved by running the original QuaLiKiz model only on the radial points where the standard deviation threshold was exceeded and/or only when a given radial point exceeds the threshold multiple times successively. Although this study demonstrates the viability of the proposed implementation, further studies into more optimal standard deviation threshold values and threshold triggering schemes are strongly recommended. 

\begin{table*}[tbp]
	\centering
	\begin{threeparttable}
		\caption{Summary table of most pertinent JINTRAC settings of the base case simulation. The values in the square brackets are taken from the benchmarking of QLKNN-hyper-10D to its corresponding version of the original QuaLiKiz model~\cite{aQLKNN-vdPlassche}.}
		\begin{tabular}{l|ccccc}
			\toprule[1.5pt]
			& JET\#73342 & JET\#92398 & JET\#92436 & JET\#91637 & JET\#91227 \\
			\midrule
			Description & High density & High perf.\tnote{1} & High perf.\tnote{1} & Ohmic & Mixed-isotope \\
			& H-mode & hybrid\tnote{2} & H-mode & L-mode & H-mode \\
			Simulation type & Stationary & Dynamic & Stationary & Stationary & Stationary \\
			\midrule
			\# of grid points & 51 & 101 & 101 & 101 & 101 \\
			Plasma time\tnote{3} & 60.75 -- 62.75~s & 46.4 -- 48.6~s & 50.0 -- 52.0~s & 58.75 -- 59.75~s & 45.2 -- 47.2~s \\
			Sim. boundary ($\rho_{\text{tor}}$) & 0.85 & 0.85 & 0.85 & 0.9 & 0.8 \\
			QuaLiKiz region & 0.15 -- 0.85 & 0.03 -- 0.85 & 0.15 -- 0.85 & 0.15 -- 0.90 & 0.03 -- 0.80 \\
			Impurity species & C & Ni & Be, Ni, W & Be & Be \\
			Part. trans. option\tnote{4} & 1 & 1 & 1 & 1 & 5 \\
			Impurity profile & Scaled & Scaled & Predicted & Scaled & Predicted \\
			QuaLiKiz rot. option & 2 & 2 & 2 & 2 & 2 \\ 
			Momentum profile & Fixed & Fixed & Predicted & Fixed & Predicted \\
			\midrule
			\multicolumn{6}{c}{\textbf{Using original QuaLiKiz model}} \\
			\midrule
			\# of cores & 16 & 16 & 16 & 16 & 16 \\
			Wall time & 26~h & 9~h & 217~h & 4~h & 79~h \\
			\midrule
			\multicolumn{6}{c}{\textbf{Using QLKNN-jetexp-15D model}} \\
			\midrule
			\# of cores & 1 [2] & 1 [2] & 1 [2] & 1 & 1 \\
			Wall time & 27~m [1~m] & 20~m [8~m] & 2~h [33~m] & 5~m & 1~h \\
			\midrule
			\multicolumn{6}{c}{\textbf{Predicted profile RRMS within QuaLiKiz region}} \\
			\midrule
			$T_e$ & 4.4\% [4.1\%] & 1.5\% [13.0\%] & 9.7\% [2.8\%] & 5.9\% & 6.7\% \\
			$T_i$ & 4.2\% [3.4\%] & 2.5\% [10.0\%] & 9.9\% [15.0\%] & 3.3\% & 8.3\% \\
			$n_e$ & 7.1\% [2.8\%] & 1.7\% [9.9\%] & 1.2\% [14.0\%] & 6.0\% & 4.3\% \\
			$\Omega_{\text{tor}}$ & -- [--] & -- [--] & 2.6\% [--] & -- & 2.9\% \\
			\bottomrule
		\end{tabular}
		\begin{tablenotes}
			\item[1] The term ``high performance" refers to an H-mode plasma scenario in which $T_i > T_e$ by a substantial amount within the central core~\cite{aJETHighPerformance-Nave}.
			\item[2] The term ``hybrid" refers to a plasma scenario in which strong $q$ profile shaping is applied~\cite{aJETHybrid-Becoulet} during the current ramp-up phase to achieve higher core confinement through a favourable magnetic shear, $\hat{s}$, profile.
			\item[3] The reference time, $t=0$, in the JET data system is when the magnetic coils start ramping up, instead of the usual time of plasma breakdown. These two events are typically $40$~s apart at JET.
			\item[4] The particle transport options are only applicable when using the QLKNN model. Further details about the different options available within QLKNN are given in Appendix~\ref{app:ParticleTransportOptions}.
		\end{tablenotes}
		\label{tbl:JETTOSettingsSummary}
	\end{threeparttable}
\end{table*}

From these figures, there is a generally good qualitative agreement between the predicted profiles resulting from using the QLKNN-jetexp-15D model and the original QuaLiKiz model across a wide variety of plasma simulation scenarios. The general profile shape and gradient are comparable in all cases except the mixed-isotope case, JET\#91227, where odd bends occur in all the predicted profiles. From Table~\ref{tbl:JETTOSettingsSummary}, it can be seen that the profile-averaged RMS between QLKNN-jetexp-15D and the original QuaLiKiz model for all the predicted profiles is $<$10\% with the recommended settings. As expected, option 6 provides results similar to option 1 in the majority of the test cases but is more prone to erratic behaviour due to the lack of a consistency filter on the ion diffusion and pinch coefficients in the dataset.

The particle transport option 5 yields comparable results to option 1, except for a non-negligible discrepancy in JET\#92436 with predictive momentum transport (right side of Figure~\ref{fig:IntegratedModelComparisonJET92436}) and a significant discrepancy in JET\#91227 (Figure~\ref{fig:IntegratedModelComparisonJET91227wRotation}). Both simulations include predictive momentum transport and predictive impurity transport. However, since the JET\#92436 simulation with predictive momentum does not affect this discrepancy significantly, it is more likely that the inclusion of predictive impurity transport is the cause. By using QLKNN with predictive impurity transport, the NN-predicted particle transport quantities are scaled according to the impurity species density and inserted as impurity transport quantities. For heavy impurities, this assumption may be reasonable due to the dominance of neoclassical transport over turbulent transport phenomena at high particle mass. However, this assumption may not be suitable for light impurities, which could have a large impact on electron density profile within JINTRAC depending on the $Z_{\text{eff}}$.

For the case of JET\#91227, the presence of both hydrogen and deuterium species in the plasma adds an additional complication to its interpretation. In spite of the improved agreement, it is actually surprising that option 5 gives reasonable results in this scenario. This is because it conflicts with the expectation that the electron $D$ and $V$ can be very different from the ion $D$ and $V$ in mixed-isotope plasmas~\cite{aIsotopeMixing-Marin}. Since this simulation was designed for analyzing long-timescale transport behaviour, it is suspected that this difference between electron and ion transport properties becomes overshadowed by the ambipolarity constraint in the plasma. This allows the substitution of electron coefficients over these long timescales to provide an equivalent plasma profile. Nonetheless, additional work in developing option 6 is required to extend the application of the NN into transient analysis, as it is expected to provide the most accurate description of the dynamic ion transport properties~\cite{aFastIsotopeMixing-Bourdelle}.

Finally, the generally good agreement between the two models within the ohmic L-mode discharge, JET\#91637, indicate that the model can also be suitable for analyzing transport behaviour in plasma ramp-up and ramp-down phases. The discrepancies of the temperature profiles in the inner core ($\rho_{\text{tor}} \le 0.2$) are expected to be reduced with the inclusion of predictive impurity transport and a heavy impurity species in the simulation. This allows some central accumulation of radiating ion species, which effectively lowers the core temperature by increasing the heat sink term. As this heavy impurity accumulation is determined mostly by neoclassical transport, this discrepancy does not invalidate the turbulent transport properties predicted by the NN.

Overall, it is important to highlight that the success of the QLKNN-jetexp-15D model is both unprecedented and not trivial. The highly nonlinear interaction between plasma microturbulent transport coefficients and global plasma transport can create undesired effects within the integrated model, typically in the form of numerical instabilities. From a plasma transport perspective, the creation of a surrogate model in which only simultaneous heat and particle transport predictions reproduce experimental results is already not guaranteed. This becomes considerably less trivial once momentum transport predictions are also included self-consistently. This demonstrates not only the robustness of the underlying model, QuaLiKiz, but also that of the NN training workflow used and extended in this study.

\section{Conclusions}
\label{sec:Conclusions}

The extension of the QLKNN-hyper-10D model to explicitly include dilution, plasma rotation, and magnetic equilibrium effects was successfully implemented. The newly-developed set of neural networks, named QLKNN-jetexp-15D, achieved this by including an additional 5 input parameters. The NN transport quantity predictions of this model, including the novel momentum flux predictions, showed good agreement with the original QuaLiKiz model over independent single parameter scans. It was also demonstrated to give comparable results to the original QuaLiKiz model, i.e. within $\sim$10\%, when integrated with plasma transport models for a variety of different plasma scenarios.

These integrated modelling applications using the QLKNN-jetexp-15D model were shown to be 60--100 times faster than the original QuaLiKiz model, although not as fast as 1000 times speedup reported by the QLKNN-hyper-10D model. This difference is largely attributed to the usage of committee NNs, which have 10 members each within this study. The novel aspects of the QLKNN-jetexp-15D model over the QLKNN-hyper-10D model is the addition of fast momentum transport predictions, which are consistent with the original QuaLiKiz model, and the explicit inclusion of the $E \times B$ turbulence suppression, instead of employing an approximated model in post-processing. In addition, this study also successfully shows that the NN model applicability can be extended to ohmic regimes and multiple isotope simulations. The combination of improved speed and wider applicability allows JINTRAC with QLKNN-jetexp-15D to be used as a preliminary tool for sensitivity and interconnectivity studies of the various plasma processes. This provides an ideal simulation test environment for future work in D-T extrapolation and scenario development at JET.

From a more technical standpoint, this paper also demonstrates the feasibility of sampling from post-processed experimental data to populate the input half of the NN training set. This allowed the inclusion of more input dimensions while avoiding the prohibitively large datasets caused by lattice sampling and the curse of dimensionality. The primary drawback of this sampling method, being the loss of clearly defined dataset boundaries for the identification of NN extrapolation regions, was successfully addressed by using committee NN. Specifically, the standard deviation of the committee predictions was shown to increase as the data density in the training set decreases. This property is only present if the regression problem definition and data filtering steps were performed sufficiently rigourously, such that other sources of input noise are removed in the training dataset.

Future work is foreseen in extending the NNs to include the transport fluxes for impurity species, specifically for the light impurity transport quantities as they are expected to be different than the main ion quantities. As the light impurities effectively determine the fuel dilution, accurately predicting its particle transport could be crucial for fusion power calculations in future modelling exercises. This may also require including the light impurity charge, $Z_{\text{imp,light}}$, as an NN input parameter. Additional work can also be done to improve the input data filters in the dataset generation step. It is expected that improvements in newer versions of QuaLiKiz will allow the ion consistency check filters to be included, as well as more strict filter tolerances.

A longer term goal would be to apply this procedure to similar data from other tokamak devices, such as ASDEX-Upgrade, Alcator C-Mod, and WEST. The dimensionless parameter distributions of the training datasets from each machine can be compared to gain insight into missing parameter regimes for experimental exploration. Ultimately, such a multi-machine database could improve the predictive capabilities of the NN by converting current extrapolation regions into interpolation regions. Furthermore, a certain risk of data bias was accepted along with the experimental-based sampling method. Data clustering and reduction techniques could be used to minimize this bias while simultaneously keeping enough data density to take advantage of the committee standard deviation metric for identifying regions of extrapolation.

Overall, the combined speed and accuracy of the QLKNN-jetexp-15D model enable its use in scenario optimization and tokamak controller design. These results can then be ratified using higher-fidelity models for both for accuracy and for deeper fundamental physics analysis.

\section{Acknowledgements}
\label{sec:Acknowledgements}

This work has been carried out within the framework of the EUROfusion Consortium and has received funding from the Euratom research and training programme 2014-2018 and 2019-2020 under grant agreement No 633053. The views and opinions expressed herein do not necessarily reflect those of the European Commission.

\section{Data availability}
\label{sec:DataAvailability}

Raw experimental data were generated at the Joint European Torus (JET) facility.  Derived data and simulation results supporting the findings of this study are available from the corresponding author upon reasonable request.

\printbibliography

\vfill
\pagebreak
\appendix

\section{NN training hyperparameters}
\label{app:NNHyperparameters}

\setcounter{equation}{0}
\renewcommand{\theequation}{\Alph{section}\arabic{equation}}

Table~\ref{tbl:NNTrainingHyperparameters} provides a list of the important hyperparameters within the NN training algorithm used in this study, along with their values. A coarse scan was performed on some of these hyperparameters, but there is no guarantee that the listed values are the precise optimum for this given problem. However, due to the acceptable agreement of the results with the original QuaLiKiz model, it was decided that the values listed here are sufficient enough to produce an accurate result.

\begin{table}[h]
	\centering
	\caption{Hyperparameter settings used within the NN training algorithm, implemented within TensorFlow-1.6. The settings marked with \textsuperscript{*} are custom hyperparameters specific to the training procedure developed for QuaLiKiz neural network regressions~\cite{aQLKNN-vdPlassche}, and are only applied to the leading flux networks. The values in square brackets are those used by the QLKNN-hyper networks, only provided when they are different than those used by the QLKNN-jetexp networks.}
	\begin{tabular}{cc}
		\toprule
		Hyperparameter name & Value \\
		\midrule
		Number of hidden layers & 3 \\
		Neurons in hidden layer & 150, 70, 30 \\
		& [128, 128, 128] \\
		L2 regularization factor & $5 \times 10^{-5}$ \\
		& [$1 \times 10^{-5}$] \\
		Positive penalty when stable\textsuperscript{*} & $10^{-3}$ \\
		Positive penalty offset\textsuperscript{*} & $-1$ [$-5$] \\
		Early stopping patience & 30 [15] \\
		Optimizer & Adam \\
		Learning rate & $10^{-3}$ \\
		Gradient memory factor & 0.9 \\
		Squared gradient memory factor & 0.999 \\
		Validation fraction & 5\% \\
		Test fraction & 5\% \\
		\bottomrule
	\end{tabular}
	\label{tbl:NNTrainingHyperparameters}
\end{table}

The training pipeline randomly splits the dataset into training, testing, and validation sets according to the ratios 90\%/5\%/5\%, respectively. This random split is done independently for each member of the committee NN, helping to reduce the potential bias in the final committee NN. In addition, the inputs and outputs in the training dataset were normalized to have a mean of 0 and a standard deviation of 1 before training. The NN evaluation implementation then reverses this transformation when making a prediction with a trained NN.

\section{Integrated modelling particle transport options}
\label{app:ParticleTransportOptions}

\setcounter{equation}{0}
\renewcommand{\theequation}{\Alph{section}\arabic{equation}}

This section details the 6 different options available in JINTRAC for defining the extra degrees of freedom present in multiple ion species simulations with QLKNN-jetexp-15D. JINTRAC evolves the ion density profiles and assigns the appropriate electron density profile via quasineutrality. For this reason, the options are all expressed in terms of calculating the ion transport coefficients. These options are as follows:
\begin{equation}
\label{eq:ParticleTransportOption1Appendix}
	1: \quad \mathbf{\Gamma}_i = \mathbf{\Gamma}_e \frac{n_i}{n_e}
\end{equation}
\begin{equation}
\label{eq:ParticleTransportOption2Appendix}
	2: \quad
	\begin{gathered}
	\tilde{V}_e = \frac{1}{n_e} \left(\mathbf{\Gamma}_e + D_e \nabla n_e\right) \\
	\mathbf{\Gamma}_i = -D_e \nabla n_i + \tilde{V}_e n_i
	\end{gathered}
\end{equation}
\begin{equation}
\label{eq:ParticleTransportOption3Appendix}
	3: \quad
	\begin{gathered}
	\tilde{V}_{i0} = \frac{1}{n_{i0}} \left(\mathbf{\Gamma}_{i0} + D_{i0} \nabla n_{i0}\right) \\
	\mathbf{\Gamma}_i = -D_{i0} \nabla n_i + \tilde{V}_{i0} n_i
	\end{gathered}
\end{equation}
\begin{equation}
\label{eq:ParticleTransportOption4Appendix}
	4: \quad \mathbf{\Gamma}_i = \mathbf{\Gamma}_{i0} \frac{n_i}{n_{i0}}
\end{equation}
\begin{equation}
\label{eq:ParticleTransportOption5Appendix}
	5: \quad \mathbf{\Gamma}_i = -D_e \nabla n_i + V_e n_i
\end{equation}
\begin{equation}
\label{eq:ParticleTransportOption6Appendix}
	6: \quad \mathbf{\Gamma}_i = -D_{i0} \nabla n_i + V_{i0} n_i
\end{equation}
where $e$ represents the electrons, $i0$ represents the deuterium ion, and $i$ represents a generic ion species.

All of these options have been implemented within JINTRAC, through the \texttt{NN\_part\_trans\_switch} setting. While the default option within the JINTRAC interface is option 1, this study did not investigate the applicability of these options in sufficient detail to make a concrete recommendation. This study notes that options 1, 2, and 5 have the greatest accuracy compared to the original QuaLiKiz model, due to the presence of the electron consistency filters, but options 5 and 6 are the most representative of the underlying physics. Due to ongoing QuaLiKiz improvements, it is expected that the discrepancy between the various options will be reduced significantly with the new version of QuaLiKiz, and consequently the new version of QLKNN.

\section{Detailed JINTRAC results}
\label{app:DetailedJETTOResults}

\setcounter{equation}{0}
\renewcommand{\theequation}{\Alph{section}\arabic{equation}}

Within the tables of this appendix, the following shorthands are used: M $\rightarrow$ momentum option; P $\rightarrow$ particle option; int. $\rightarrow$ interpretive; pred. $\rightarrow$ predictive.
In this context, ``interpretive" means that the profile is fixed at its initial condition and ``predictive" means that the profile is evolved in time according to its transport equation. In both of these cases, the profile values are used within the turbulence calculations whereas ``off" means that the values are treated as zero within the calculations.

The profile-averaged relative root-mean-square (RRMS) is calculated as follows:
\begin{equation}
\label{eq:}
	\text{RRMS} = \sqrt{\frac{1}{N} \sum_j^N \frac{\left(Y_{\text{QLKNN}} - Y_{\text{QuaLiKiz}}\right)^2}{Y_{\text{QuaLiKiz}}}}
\end{equation}
where $\rho_{\text{tor,lb}} \le \rho_{\text{tor},j} \le \rho_{\text{tor,ub}}$ and $Y$ represents a generic profile quantity.

\begin{table}[h]
	\centering
	\caption{RRMS results for JET\#73342}
	\begin{tabular}{cc|cccc}
		\toprule
		M & P & $T_e$ & $T_i$ & $n_e$ & $\Omega_{\text{tor}}$ \\
		\midrule
		off & 1 & 2.6\% & 2.5\% & 2.9\% & -- \\
		& 5 & 2.0\% & 2.0\% & 1.0\% & -- \\
		& 6 & 2.8\% & 1.9\% & 5.2\% & -- \\
		int. & 1 & 4.4\% & 4.2\% & 7.1\% & -- \\
		& 5 & 3.4\% & 3.3\% & 4.7\% & -- \\
		& 6 & 1.0\% & 1.3\% & 1.4\% & -- \\
		pred. & 1 & 4.9\% & 5.0\% & 7.3\% & 6.2\% \\
		& 5 & 3.7\% & 4.0\% & 5.1\% & 4.4\% \\
		& 6 & 1.0\% & 1.4\% & 1.1\% & 1.8\% \\
		\bottomrule
	\end{tabular}
	\label{tbl:DetailedJETTOResults73342}
\end{table}

\begin{table}[h]
	\centering
	\caption{RRMS results for JET\#92398}
	\begin{tabular}{cc|cccc}
		\toprule
		M & P & $T_e$ & $T_i$ & $n_e$ & $\Omega_{\text{tor}}$ \\
		\midrule
		off & 1 & 0.8\% & 1.3\% & 0.4\% & -- \\
		& 5 & 1.2\% & 1.7\% & 1.1\% & -- \\
		& 6 & 10.0\% & 7.2\% & 32.5\% & -- \\
		int. & P1 & 1.5\% & 2.5\% & 1.7\% & -- \\
		& 5 & 1.4\% & 0.9\% & 3.3\% & -- \\
		& 6 & 4.2\% & 4.8\% & 5.7\% & -- \\
		\bottomrule
	\end{tabular}
	\label{tbl:DetailedJETTOResults92398}
\end{table}

\begin{table}[h]
	\centering
	\caption{RRMS results for JET\#92436}
	\begin{tabular}{cc|cccc}
		\toprule
		M & P & $T_e$ & $T_i$ & $n_e$ & $\Omega_{\text{tor}}$ \\
		\midrule
		off & 1 & 2.6\% & 14.3\% & 4.6\% & -- \\
		& 5 & 2.8\% & 8.7\% & 5.3\% & -- \\
		& 6 & 4.8\% & 5.9\% & 7.8\% & -- \\
		int. & 1 & 10.5\% & 10.4\% & 1.8\% & -- \\
		& 5 & 6.3\% & 5.8\% & 5.0\% & -- \\
		& 6 & 5.4\% & 4.1\% & 9.1\% & -- \\
		pred. & 1 & 9.7\% & 9.9\% & 1.2\% & 2.6\% \\
		& 5 & 4.7\% & 11.7\% & 5.6\% & 6.5\% \\
		& 6 & 3.1\% & 4.6\% & 11.7\% & 3.5\% \\
		\bottomrule
	\end{tabular}
	\label{tbl:DetailedJETTOResults92436}
\end{table}

\begin{table}[h]
	\centering
	\caption{RRMS results for JET\#91227}
	\begin{tabular}{cc|cccc}
		\toprule
		M & P & $T_e$ & $T_i$ & $n_e$ & $\Omega_{\text{tor}}$ \\
		\midrule
		off & 1 & 6.5\% & 3.7\% & 23.4\% & -- \\
		& 5 & 8.4\% & 16.6\% & 5.7\% & -- \\
		& 6 & 16.0\% & 9.1\% & 23.5\% & -- \\
		int. & 1 & 10.0\% & 11.7\% & 33.4\% & -- \\
		& 5 & 6.8\% & 5.8\% & 7.3\% & -- \\
		& 6 & 10.2\% & 5.7\% & 11.2\% & -- \\
		pred. & 1 & 14.8\% & 8.6\% & 39.4\% & 21.5\% \\
		& 5 & 4.9\% & 4.1\% & 5.7\% & 16.8\% \\
		& 6 & 10.4\% & 7.5\% & 16.6\% & 24.4\% \\
		\bottomrule
	\end{tabular}
	\label{tbl:DetailedJETTOResults91227}
\end{table}

\begin{table}[h]
	\centering
	\caption{RRMS results for JET\#91637}
	\begin{tabular}{cc|cccc}
		\toprule
		M & P & $T_e$ & $T_i$ & $n_e$ & $\Omega_{\text{tor}}$ \\
		\midrule
		off & 1 & 9.5\% & 9.3\% & 6.2\% & -- \\
		& 5 & 9.6\% & 11.4\% & 4.9\% & -- \\
		& 6 & 8.5\% & 13.4\% & 7.0\% & -- \\
		int. & 1 & 5.9\% & 3.3\% & 6.0\% & -- \\
		& 5 & 6.0\% & 4.3\% & 5.3\% & -- \\
		& 6 & 5.6\% & 4.7\% & 2.2\% & -- \\
		pred. & 1 & 6.4\% & 6.0\% & 5.0\% & 2.2\% \\
		& 5 & 6.7\% & 8.3\% & 4.3\% & 2.9\% \\
		& 6 & 5.5\% & 8.9\% & 5.0\% & 5.0\% \\
		\bottomrule
	\end{tabular}
	\label{tbl:DetailedJETTOResults91637}
\end{table}

\section{Committee NN standard deviation acceptance thresholds}
\label{app:NNVarianceThresholds}

\setcounter{equation}{0}
\renewcommand{\theequation}{\Alph{section}\arabic{equation}}

This study shows that the committee NN prediction standard deviation can be strongly tied to the NN training set data density, provided that specific prerequisites are met in the problem definition and data collection phase. Setting acceptance thresholds for the various committee NN prediction standard deviations allows this correlation to be used within integrated models to flag simulation parameters where the NNs begin to extrapolate. This section describes how this information was translated into a practical logic switch for the integrated model implementation used in this study.

\begin{figure}[tb]
	\centering
	\includegraphics[scale=0.35]{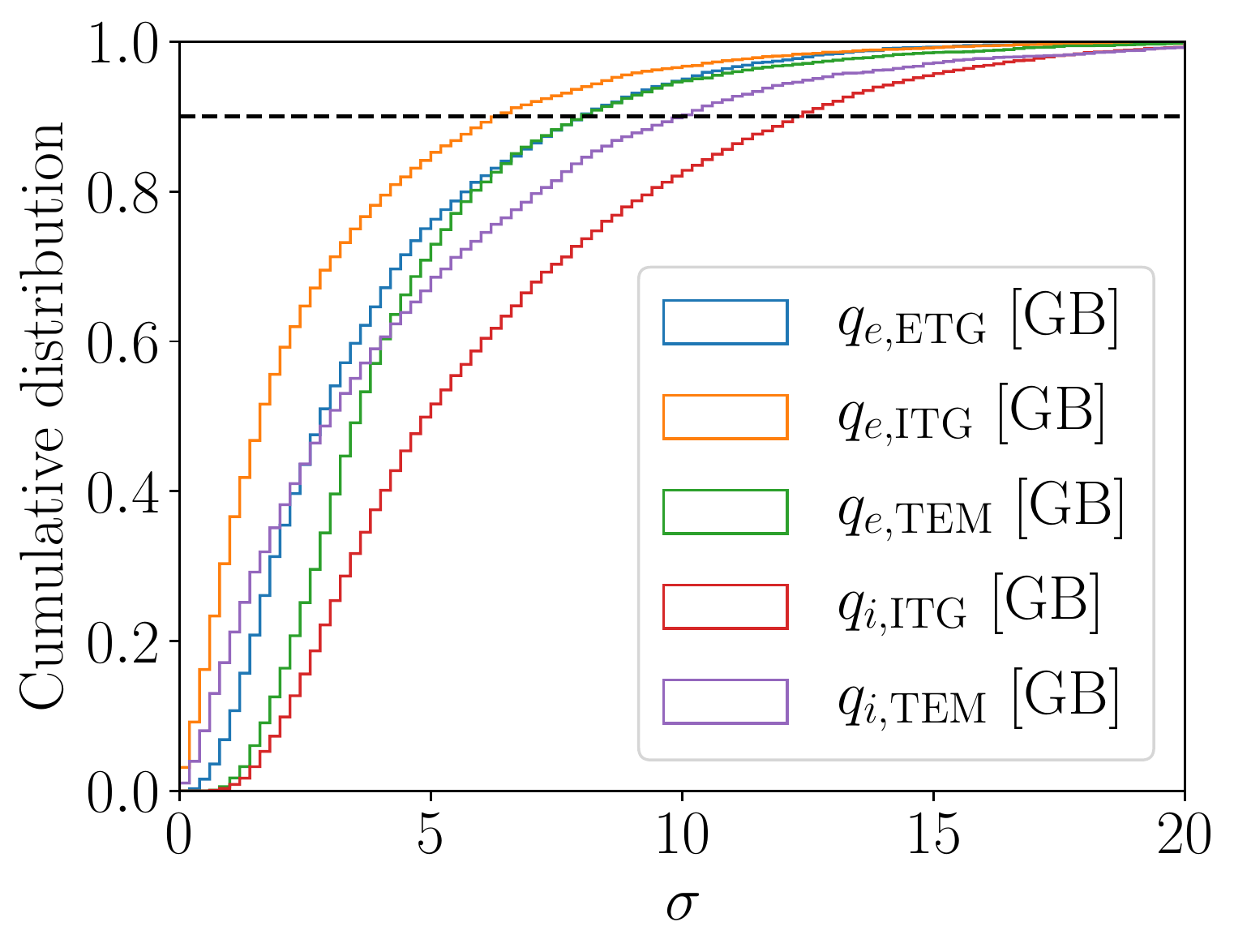}
	\includegraphics[scale=0.35]{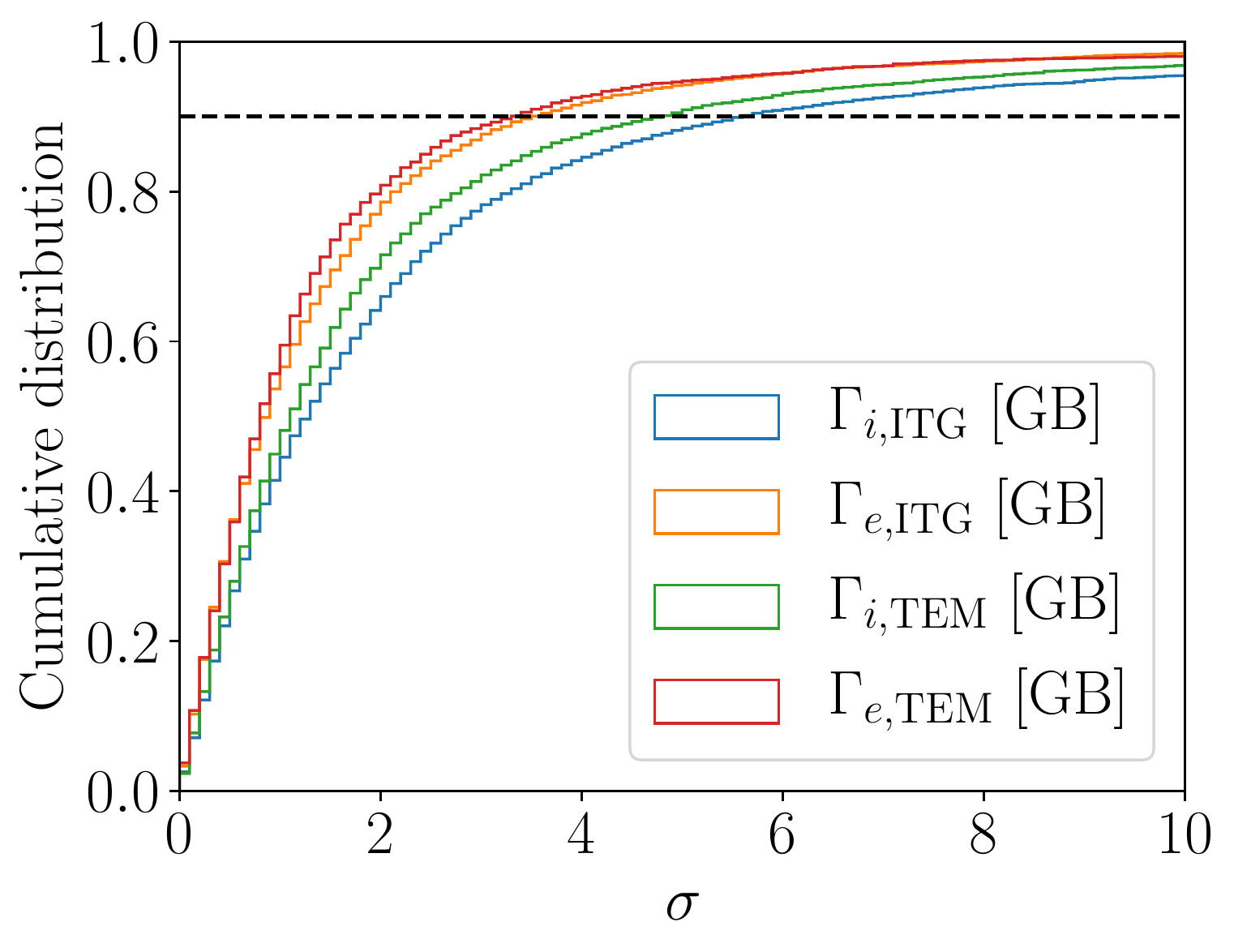}
	\includegraphics[scale=0.35]{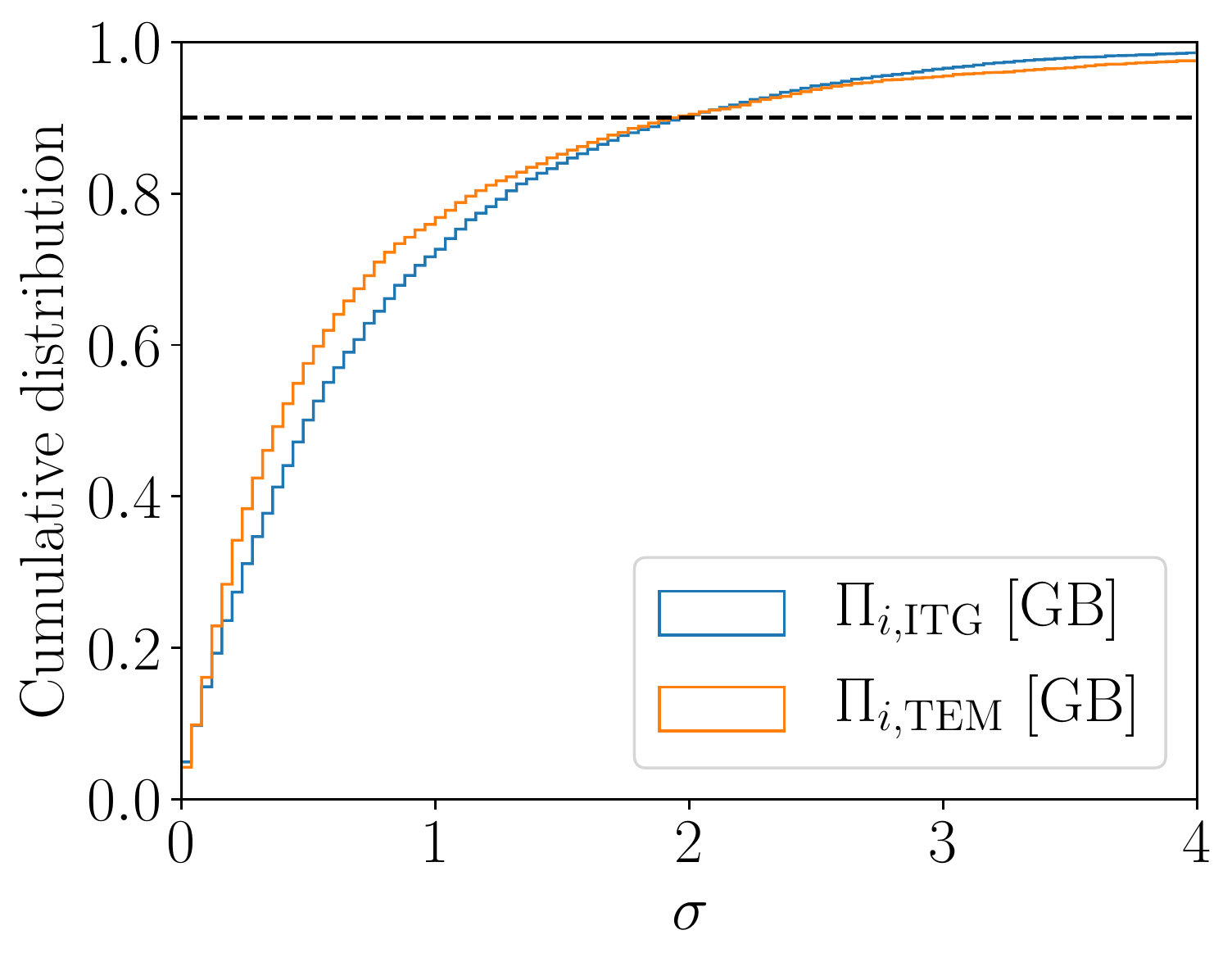}
	\caption{Cumulative distribution of the absolute standard deviation values for the heat flux (top), particle flux (center) and momentum flux (bottom) committee NN predictions within a random sample of 10000 points. The samples were drawn using a uniform distribution within the central 90\% of the parameter ranges inside the extracted JET database. The dashed horizontal line shows the 90\textsuperscript{th} percentile used to define the absolute standard deviation threshold.}
	\label{fig:CumulativeDistributionTransportFluxThreshold}
\end{figure}

Firstly, an NN prediction database was generated by uniformly sampling 10000 points within the central 90\% of the training dataset input space, as defined by the distributions shown in Figure~\ref{fig:FullJETQuaLiKizStats}. The prediction mean, $\mu$, and standard deviation, $\sigma$, of each committee NN was then evaluated at these randomly sampled points. A brief examination of the database revealed that absolute thresholds, based directly on $\sigma$, were better at flagging extrapolation regions where transport coefficients were large. Relative thresholds, based on $\sigma/\mu$, were better at flagging extrapolation regions where transport coefficients were low. A combination of both were implemented to adequately cover both cases.

The absolute acceptance threshold, $\sigma_{\text{lim}}$, was defined at the 90\textsuperscript{th} percentile of the absolute standard deviation, $\sigma$, distribution of a random sample of 10000 evaluations. Figure~\ref{fig:CumulativeDistributionTransportFluxThreshold} shows the cumulative distribution of absolute NN standard deviations within the sample. While only the transport fluxes are shown, the same procedure was repeated with the diffusive and convective coefficients as well. The relative acceptance threshold, $\left(\sigma/\mu\right)_{\text{lim}}$, was calculated using the remaining 10\% of the data, and was defined at the 50\textsuperscript{th} percentile of the relative standard deviation, $\sigma/\mu$, following a similar procedure as the previous step. This staged process ensures that the points with small transport coefficients and reasonable standard deviations do not bias the relative threshold result to be larger than necessary.

Table~\ref{tbl:NNStandardDeviationThresholds} provides the standard devation thresholds determined by this procedure for the QLKNN-jetexp-15D committee networks developed in this study. If the committee NN standard devation of any of the predicted transport quantities exceeds either of these two thresholds, that point is flagged as potentially within an extrapolation region. Further studies into more optimal schemes for determining these thresholds are strongly recommended, as the implementation presented here was only designed to demonstrate the feasibility of the threshold flagging concept.

\begin{table}[tb]
	\centering
	\begin{threeparttable}
		\caption{Absolute and relative standard deviation thresholds per transport quantity within the QLKNN-jetexp-15D model.}
		\begin{tabular}{C{2.5cm}|C{2cm}C{2cm}}
			\toprule
			Quantity & $\sigma_{\text{lim}}$ [GB] & $\left(\sigma/\mu\right)_{\text{lim}}$ \\
			\midrule
			$\mathbf{q}_{e,\text{ETG}}$ & 8.4 & 0.5 \\
			$\mathbf{q}_{e,\text{ITG}}$ & 5.6 & 1.0 \\
			$\mathbf{q}_{i,\text{ITG}}$ & 13.0 & 0.4 \\
			$\mathbf{q}_{e,\text{TEM}}$ & 9.0 & 0.9 \\
			$\mathbf{q}_{i,\text{TEM}}$ & 8.6 & 1.4 \\
			\midrule
			$\mathbf{\Gamma}_{e,\text{ITG}}$ & 3.4 & 2.5 \\
			$\mathbf{\Gamma}_{i,\text{ITG}}$ & 5.6 & 3.2 \\
			$\mathbf{\Gamma}_{e,\text{TEM}}$ & 3.6 & 2.1 \\
			$\mathbf{\Gamma}_{i,\text{TEM}}$ & 5.0 & 2.5 \\
			\midrule
			$\mathbf{\Pi}_{i,\text{ITG}}$ & 2.3 & 1.9 \\
			$\mathbf{\Pi}_{i,\text{TEM}}$ & 2.3 & 2.5 \\
			\midrule
			$D_{e,\text{ITG}}$ & 4.2 & 2.5 \\
			$D_{i,\text{ITG}}$ & 8.0 & 3.0 \\
			$D_{e,\text{TEM}}$ & 3.2 & 2.4 \\
			$D_{i,\text{TEM}}$ & 9.5 & 2.8 \\
			\midrule
			$V_{c,e,\text{ITG}}$ & 1.6 & 2.6 \\
			$V_{c,i,\text{ITG}}$ & 3.0 & 2.3 \\
			$V_{c,e,\text{TEM}}$ & 1.3 & 1.2 \\
			$V_{c,i,\text{TEM}}$ & 3.2 & 2.6 \\
			\midrule
			$V_{t,e,\text{ITG}}$ & 4.6 & 2.4 \\
			$V_{t,i,\text{ITG}}$ & 8.7 & 2.0 \\
			$V_{t,e,\text{TEM}}$ & 2.0 & 1.5 \\
			$V_{t,i,\text{TEM}}$ & 9.0 & 2.5 \\
			\midrule
			$V_{r,i,\text{ITG}}$\tnote{1} & 1.0 & 0.1 \\
			$V_{r,i,\text{TEM}}$\tnote{1} & 1.0 & 0.1 \\
		\end{tabular}
		\begin{tablenotes}
			\item[1] The rotodiffusion pinch coefficients, $V_r$, are currently defined as zero internally within QuaLiKiz itself. Thus, the NNs for these quantities are also hardcoded to return zeros. The thresholds provided here are simply placeholders for future expansion of the QuaLiKiz and QLKNN models.
		\end{tablenotes}
		\label{tbl:NNStandardDeviationThresholds}
	\end{threeparttable}
\end{table}

\clearpage

\section{Additional parameter scans comparisons}
\label{app:ExtraParameterScans}

\setcounter{equation}{0}
\renewcommand{\theequation}{\Alph{section}\arabic{equation}}

For all the plots in this appendix, QLKNN-jetexp-15D is represented by solid lines and QuaLiKiz by green points. Note that QuaLiKiz-v2.6.2 was used for data reproduction purposes, as it is identical to v2.6.1 except for changes to stabilize the code with more modern compilers. The points which would pass the data pipeline filters are denoted with circles and those which would be screened out are denoted with crosses. The standard deviation of the committee NN is represented by the shaded regions, with the darker regions belonging to the solid line and the lighter region belonging to the dashed line. The solid lines represent the NN predicted transport flux directly, while the dashed lines represent the equivalent transport flux reconstructed by combining simulation plasma parameters and NN predicted diffusive, $D$, and convective coefficients, $V$.

Additionally, the dotted green vertical line indicate the base value from which the parameter scan extends from, while the dashed black vertical lines indicated the 2.5\% and 97.5\% quantiles. These provide a rough estimate of the useful boundary of the training dataset, beyond which the variances are expected to grow. This behaviour is well demonstrated by the plots in this section, other than conditions where the leading flux is clipped to zero and the committee NN standard deviation is consequently also set to zero.

The primary discrepancies within this section occur in the scans along the normalized pressure gradient parameter, $\alpha_{\text{MHD}}$, and the effective charge, $Z_{\text{eff}}$. The normalized pressure gradient, $\alpha_{\text{MHD}}$, is internally modified in QuaLiKiz to avoid the parameter regime where slab modes are excited. This modification is such that $\alpha_{\text{MHD}} = \max\!\left(\hat{s} - 0.2, 0.0\right)$ when the input specifies $\left(\hat{s} - \alpha_{\text{MHD}}\right) < 0.2$. This breaks the uniqueness criteria established in Section~\ref{subsec:InputParameterSelection}, resulting in a non-negligible effect on the NN regression due to the effective noise added to the dataset. However, since the impact of $\alpha_{\text{MHD}}$ is relatively minor in most JET discharges, this does not significantly impact the integrated modelling results.

Regarding the momentum transport of $Z_{\text{eff}}$, it is suspected that the combined sizes of the stable and rotationless subsets of the dataset dominated over the unstable rotation cases at higher $Z_{\text{eff}}$, as indicated by the prediction remaining close to zero. This value of zero also alters the standard deviation prediction of the committee NN, according to Equation~\eqref{eq:IndependentComboNNVarianceExample}. However, it is important to note that most validated JET experimental data remains in the region of $Z_{\text{eff}} < 2.5$, where the NN prediction remains close the original QuaLiKiz prediction. Regardless, a more extensive sampling of rotation cases with higher $Z_{\text{eff}}$ is planned for future expansion of the NN model, to improve its range of applicability in momentum transport predictions.

\begin{figure}[h]
	\centering
	\includegraphics[scale=0.23]{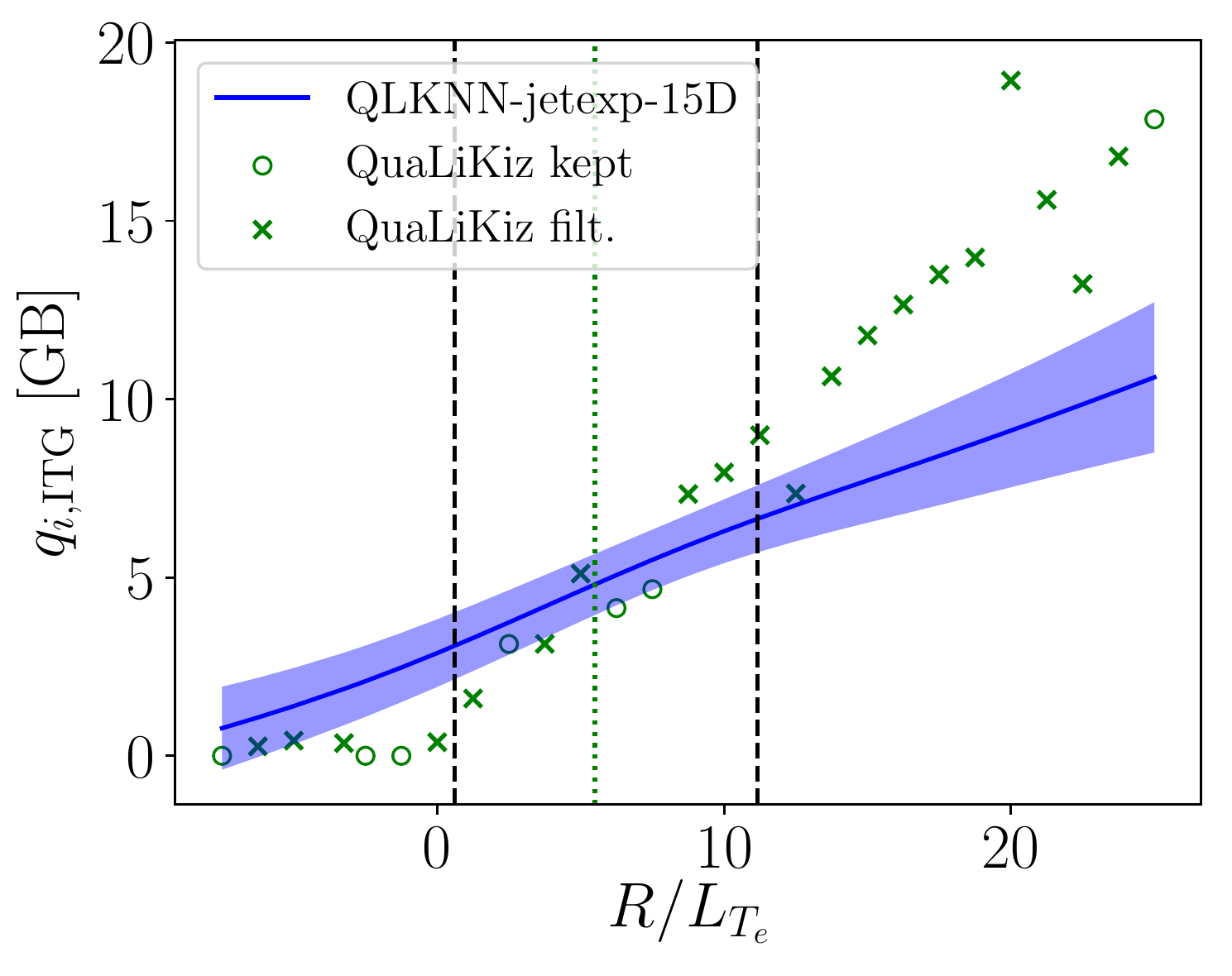}%
	\hspace{2mm}\includegraphics[scale=0.23]{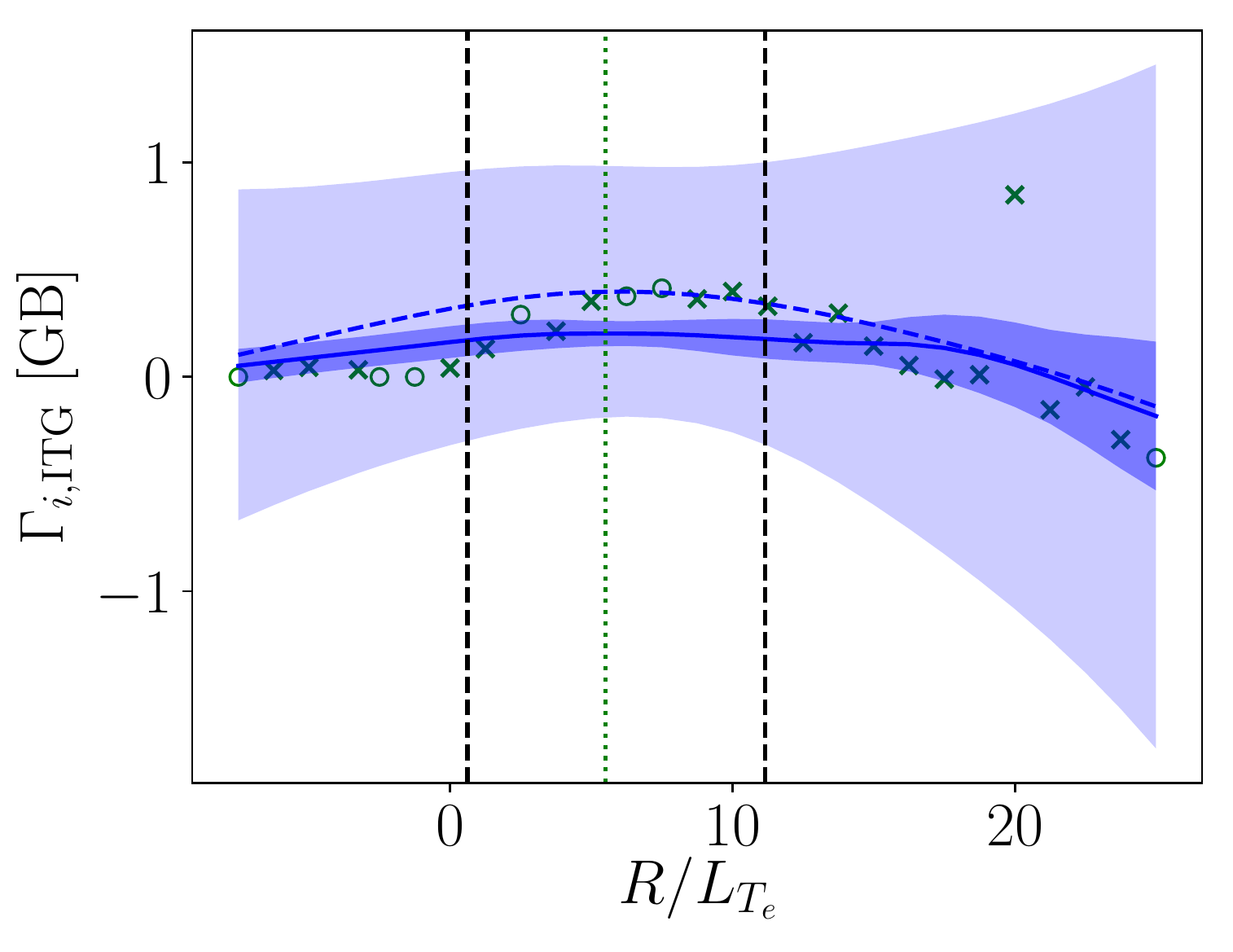} \\
	\hspace{0.25mm}\includegraphics[scale=0.23]{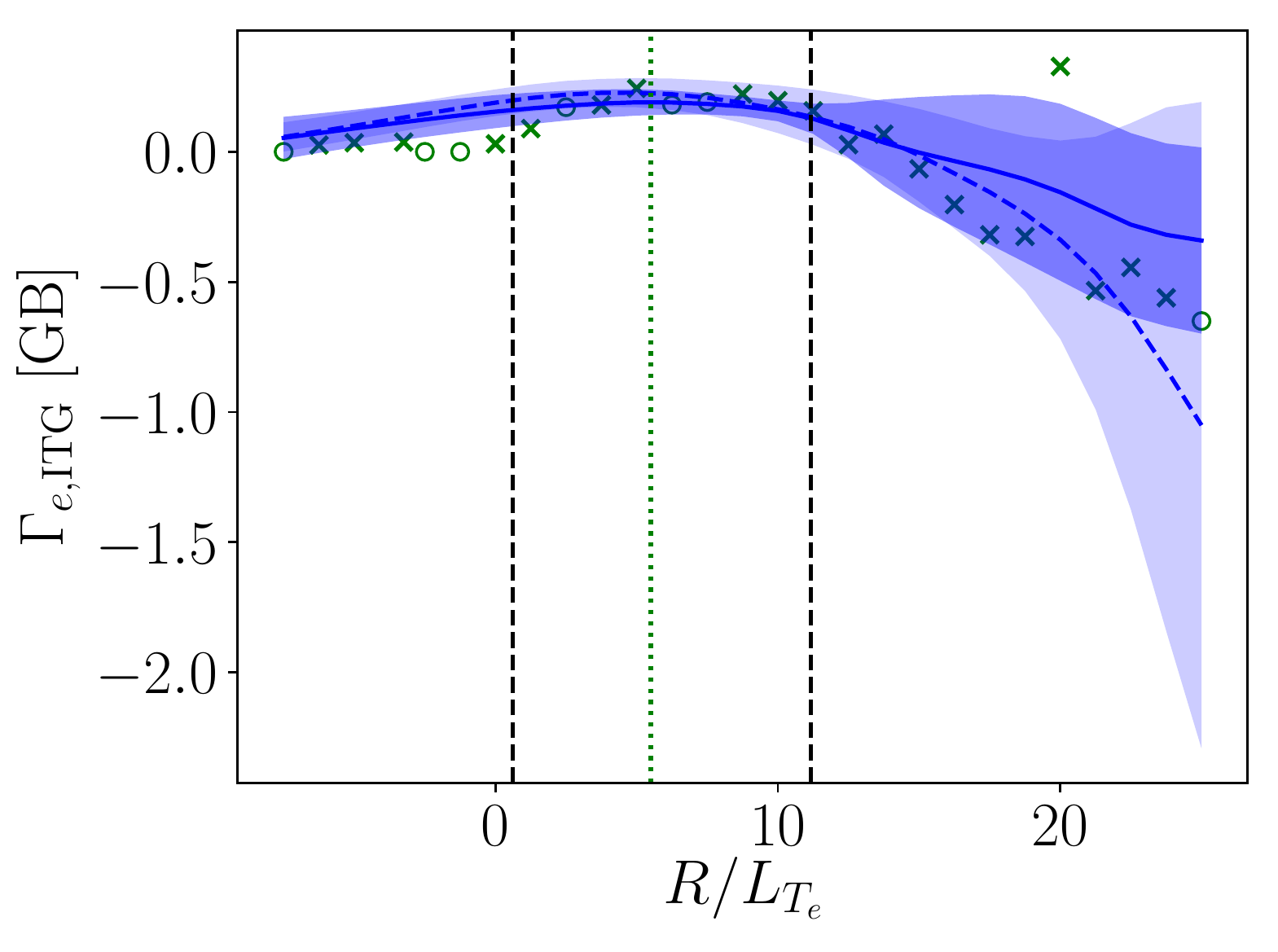}%
	\hspace{2mm}\includegraphics[scale=0.23]{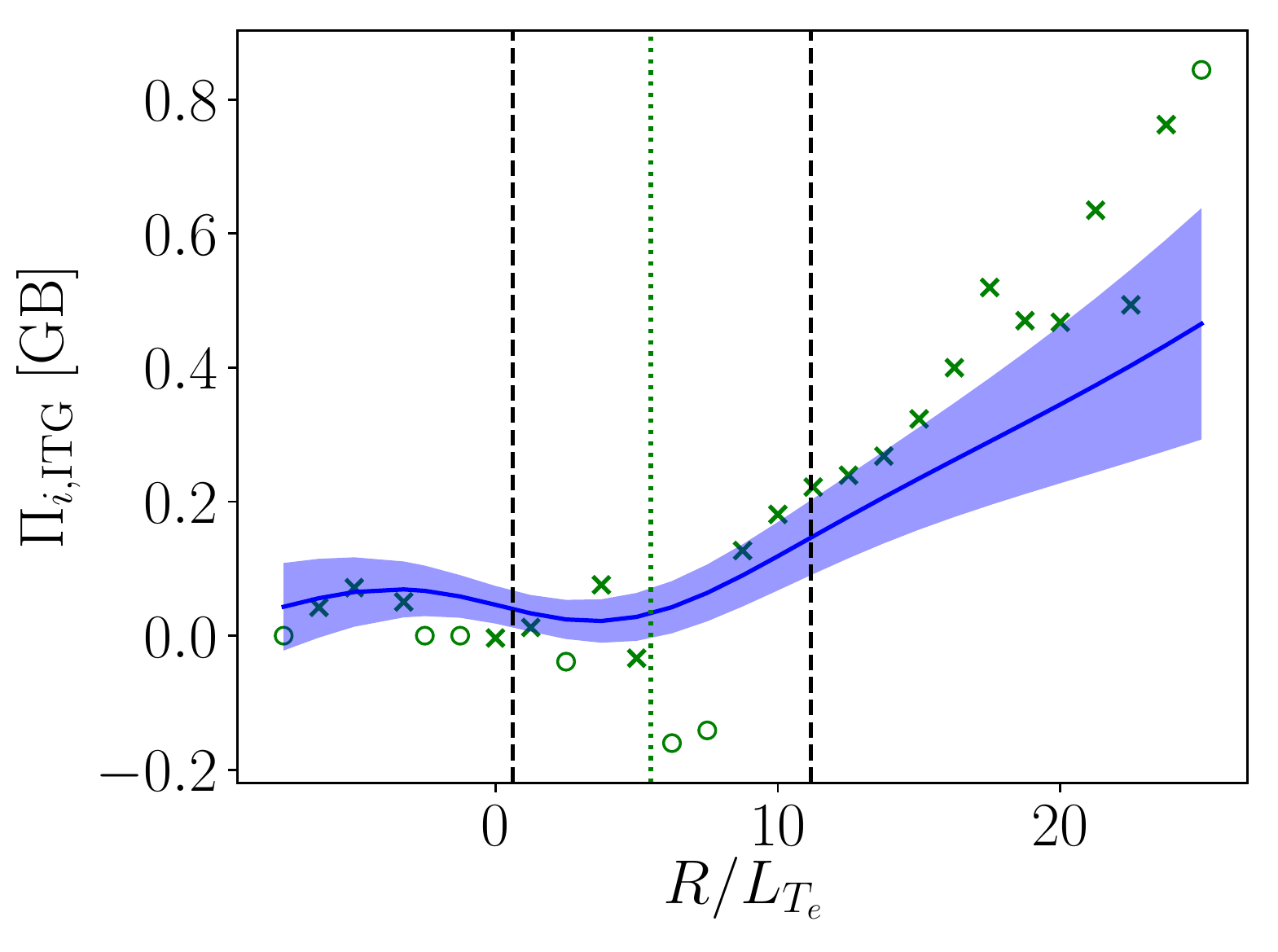}
	\caption{Comparison of main ITG-driven transport fluxes as a function of the logarithmic electron temperature gradient, $R/L_{T_e}$.}
	\label{fig:FluxComparisonsElectronTemperatureGradient}
\end{figure}

\begin{figure}[h]
	\centering
	\includegraphics[scale=0.25]{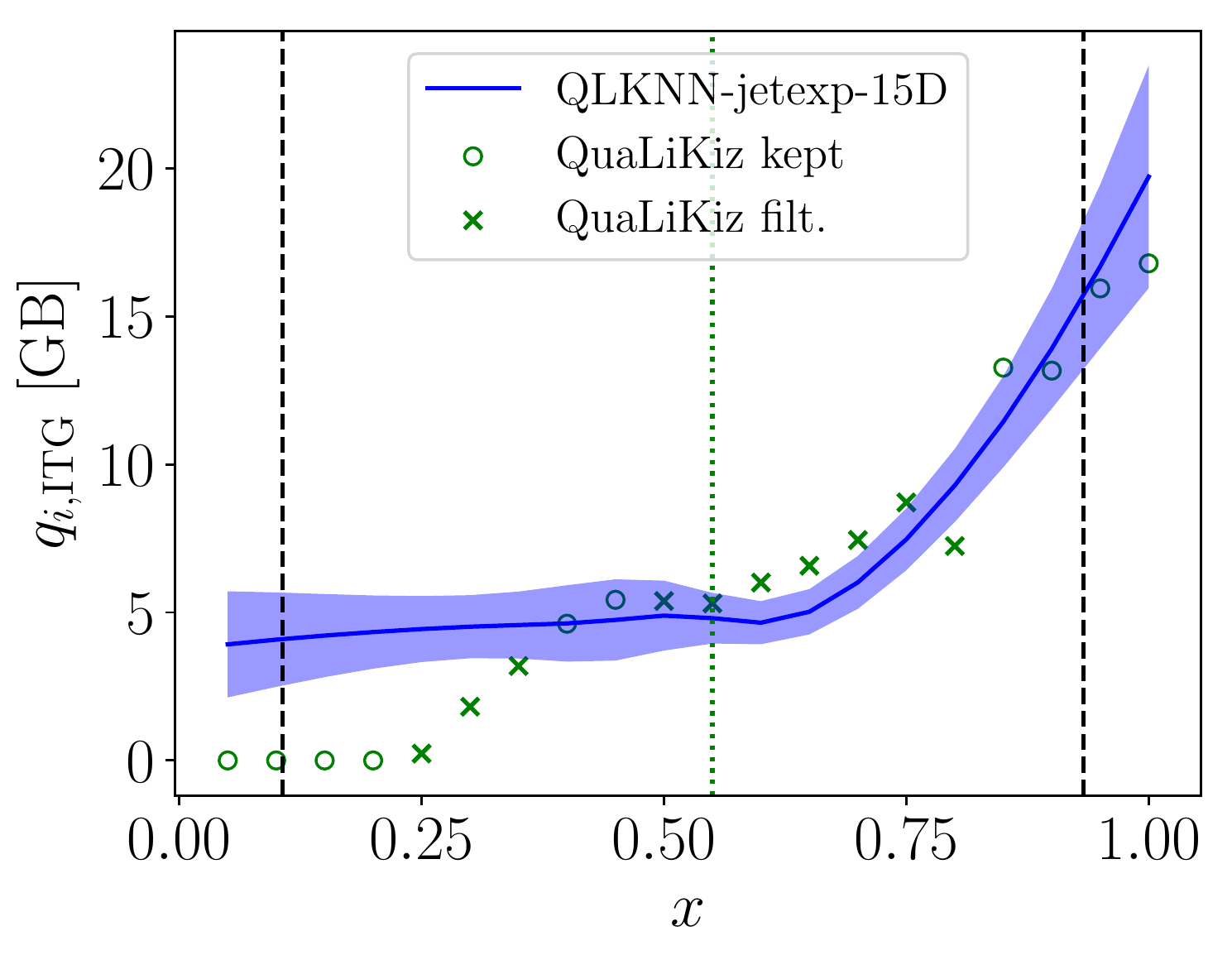}%
	\hspace{2mm}\includegraphics[scale=0.25]{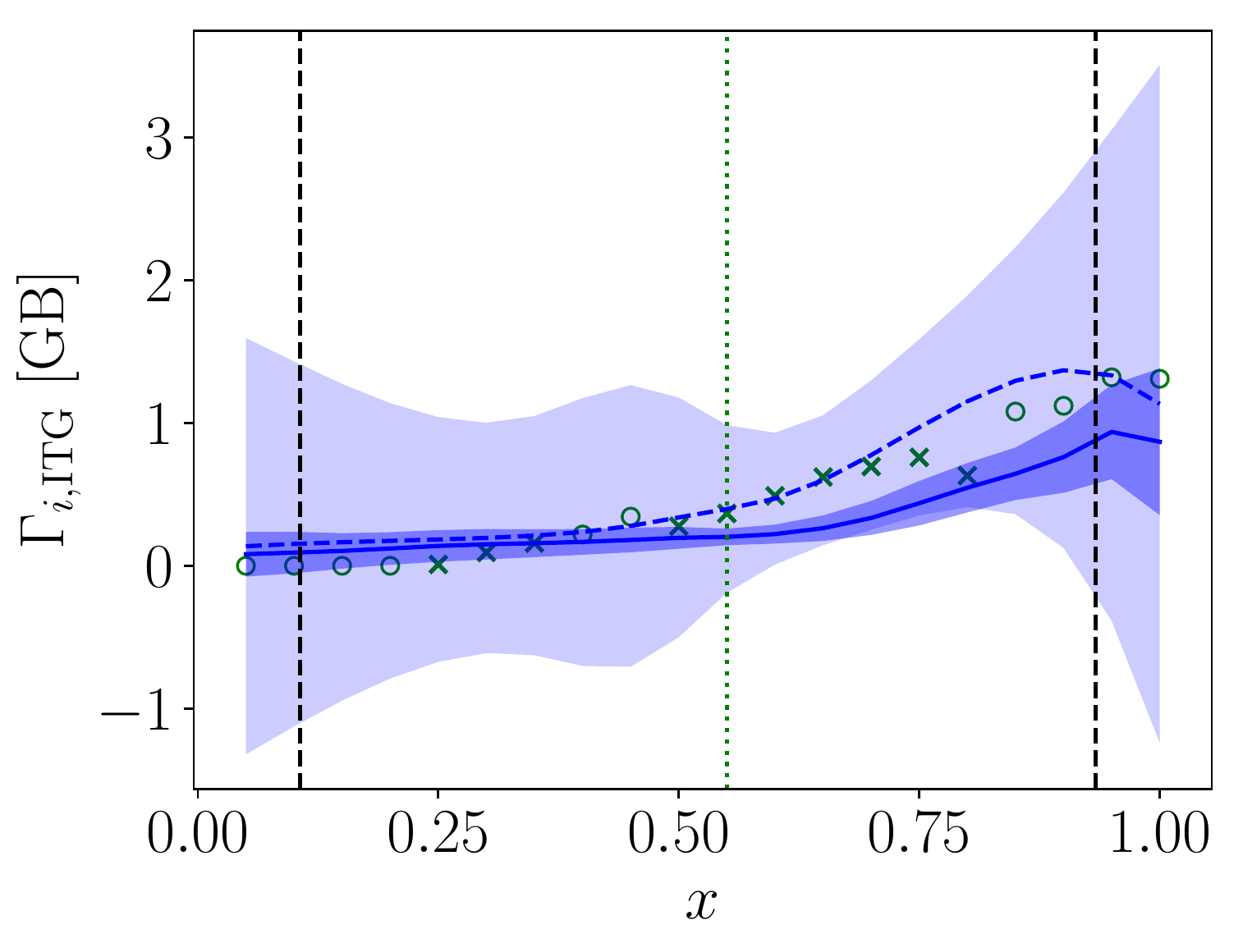} \\
	\hspace{0.25mm}\includegraphics[scale=0.25]{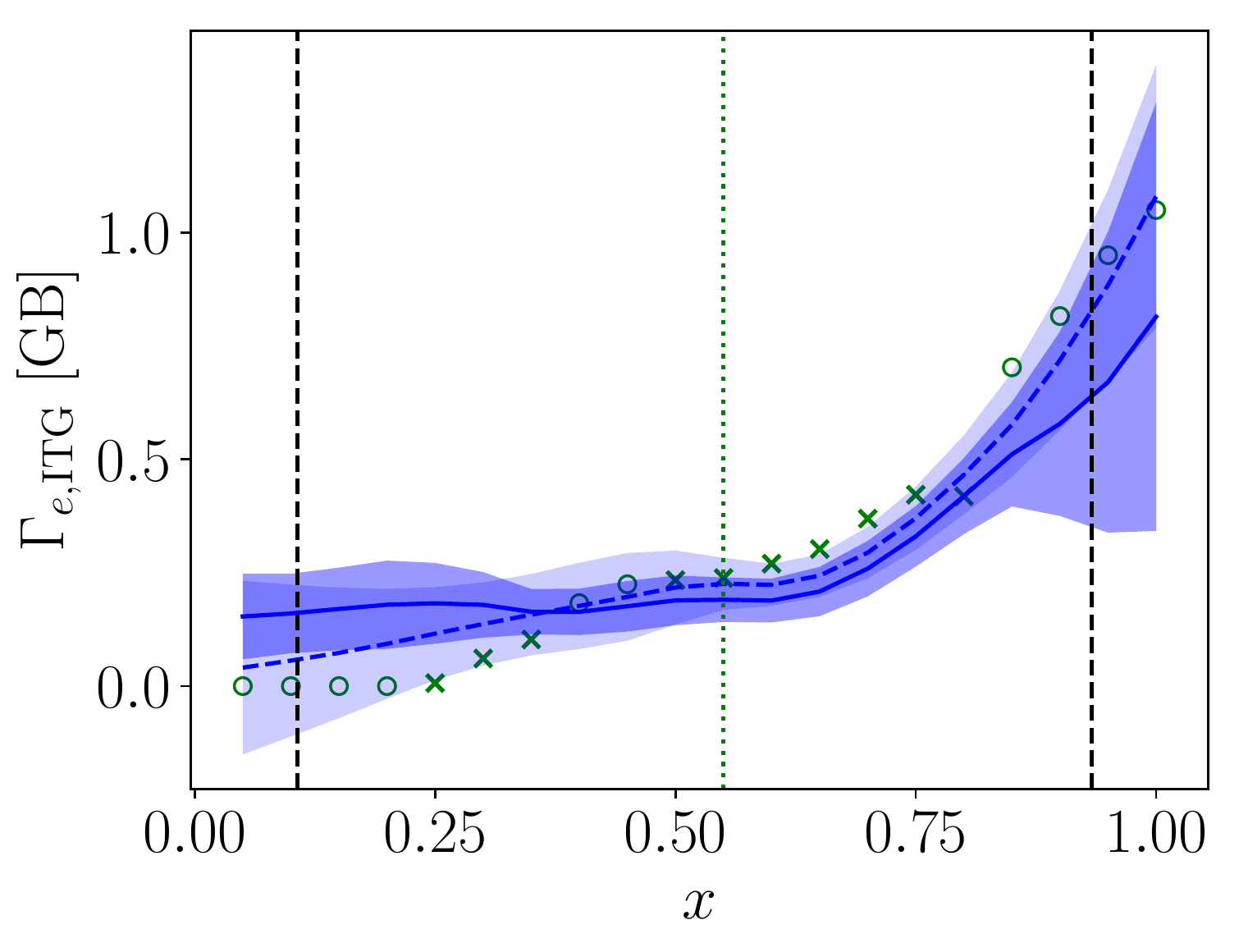}%
	\hspace{2mm}\includegraphics[scale=0.25]{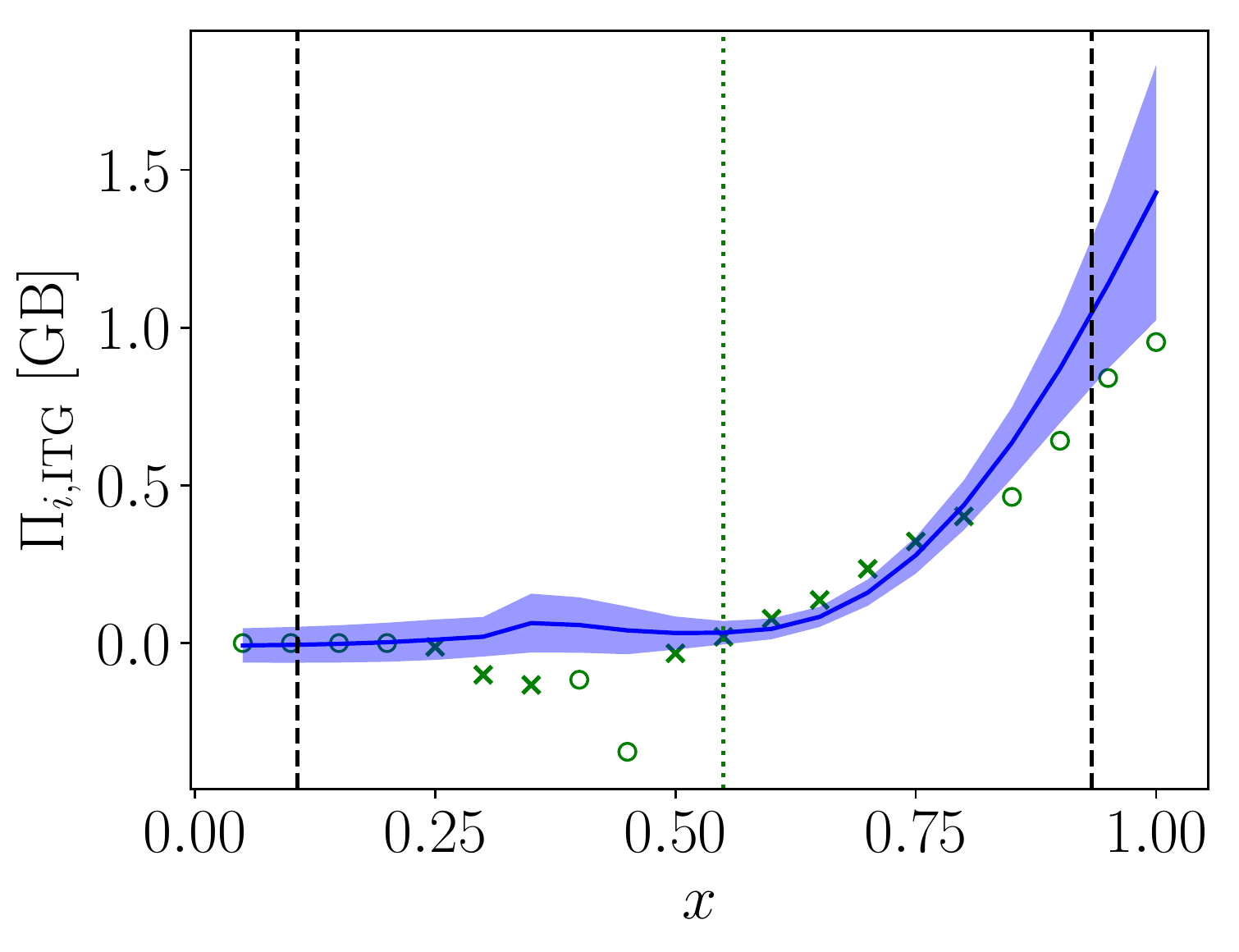}
	\caption{Comparison of main ITG-driven transport fluxes as a function of the normalized midplane-averaged minor radius, $x$.}
	\label{fig:FluxComparisonsMinorRadius}
\end{figure}

\begin{figure}[h]
	\centering
	\includegraphics[scale=0.245]{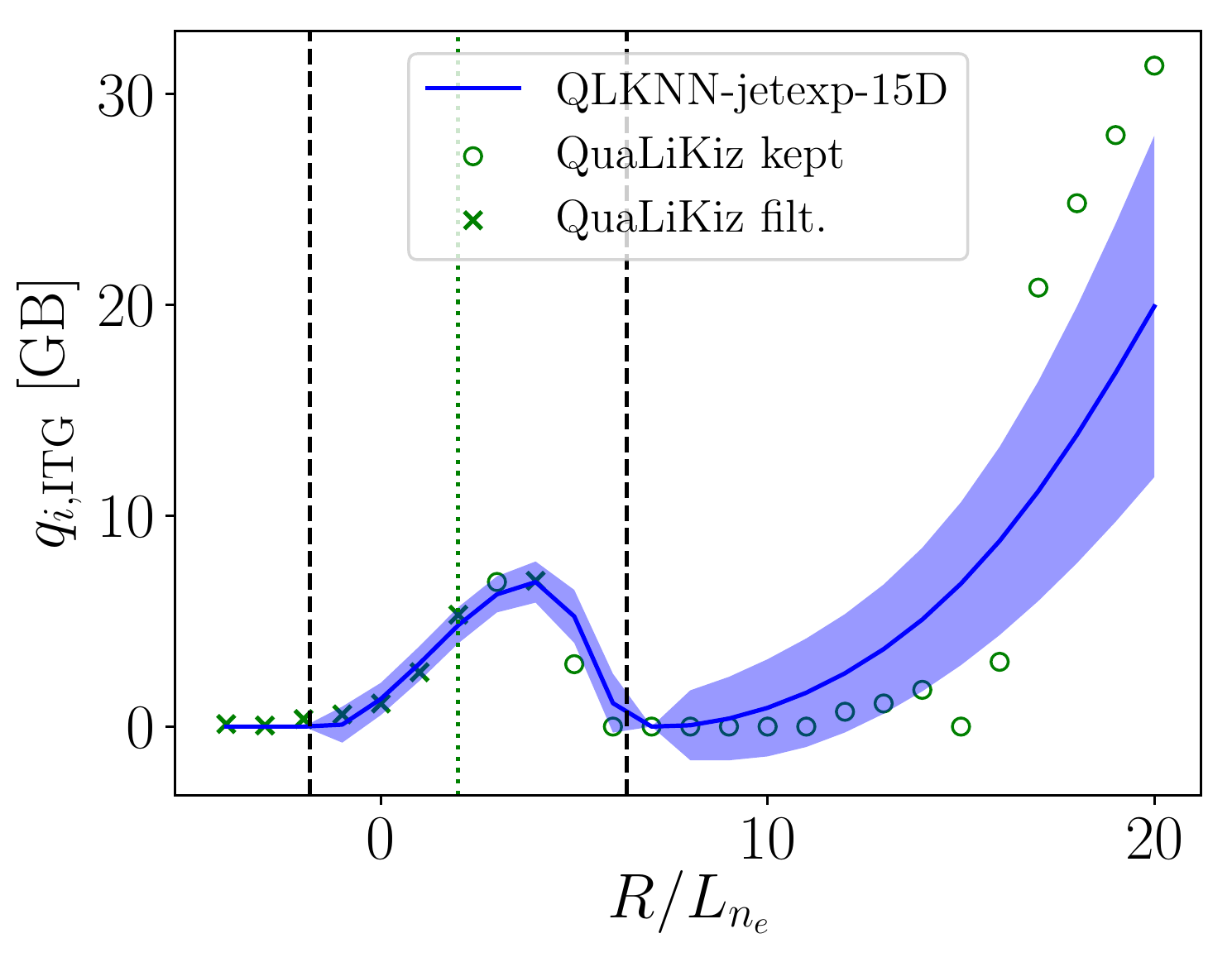}%
	\hspace{2mm}\includegraphics[scale=0.245]{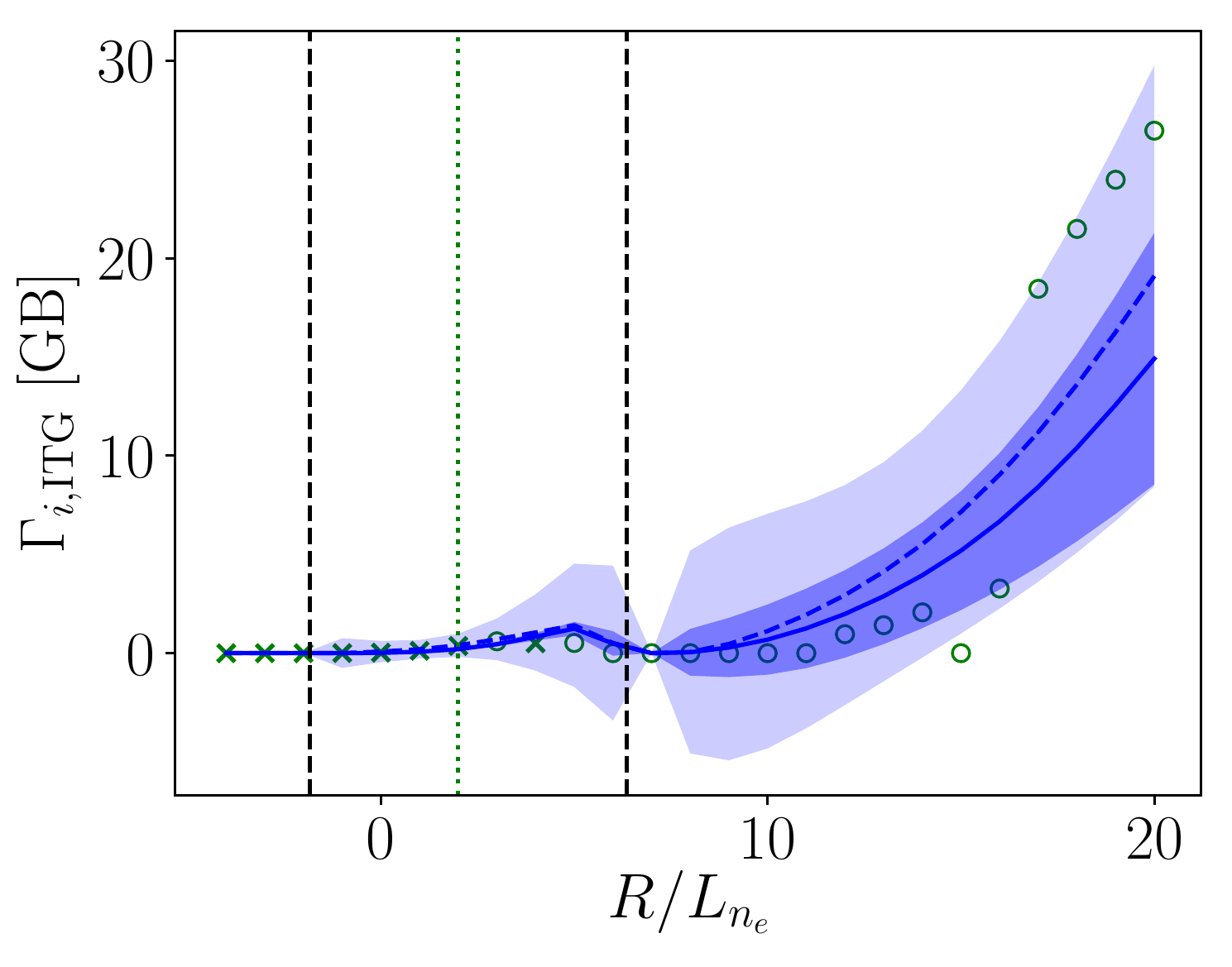} \\
	\hspace{0.25mm}\includegraphics[scale=0.245]{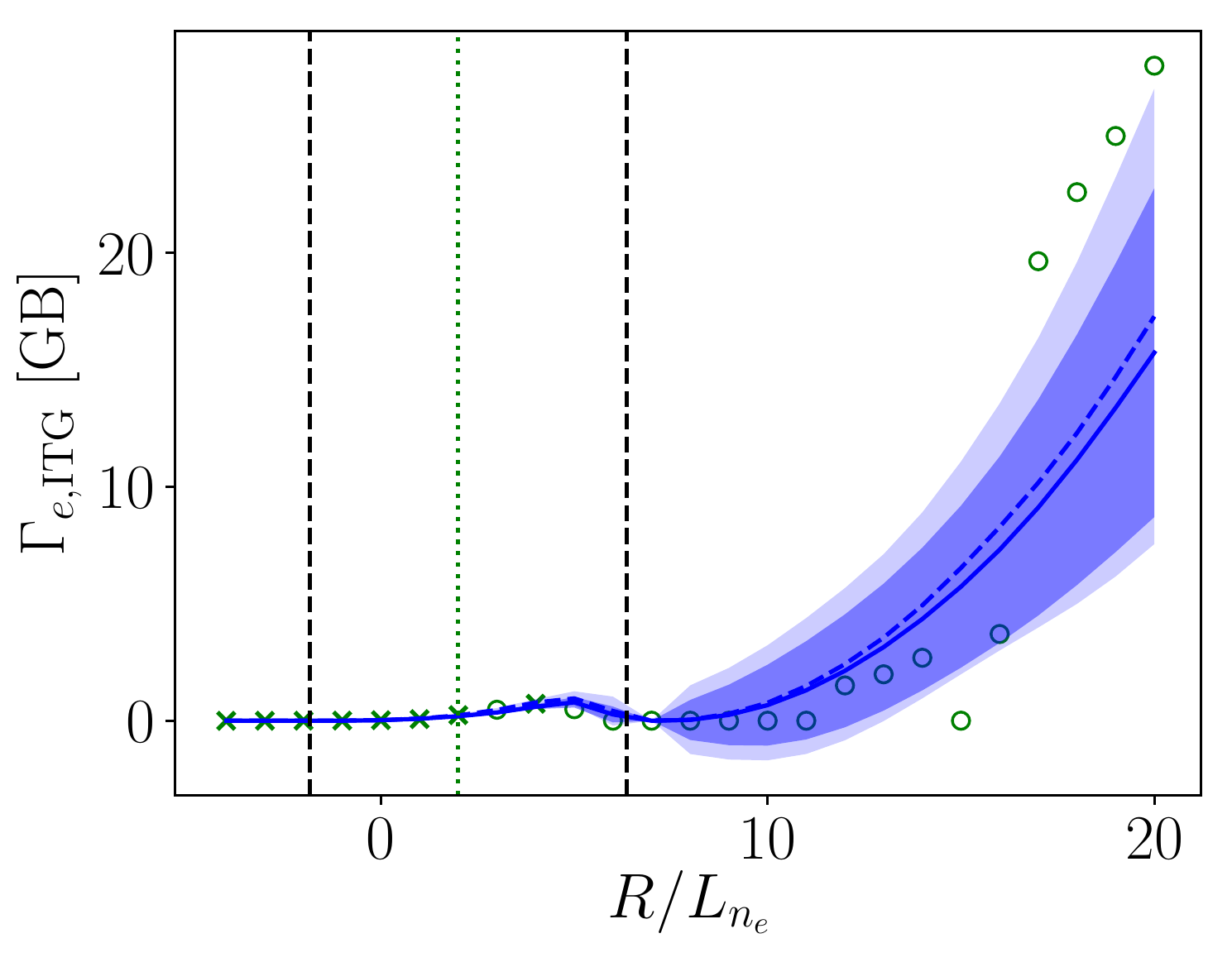}%
	\hspace{2mm}\includegraphics[scale=0.245]{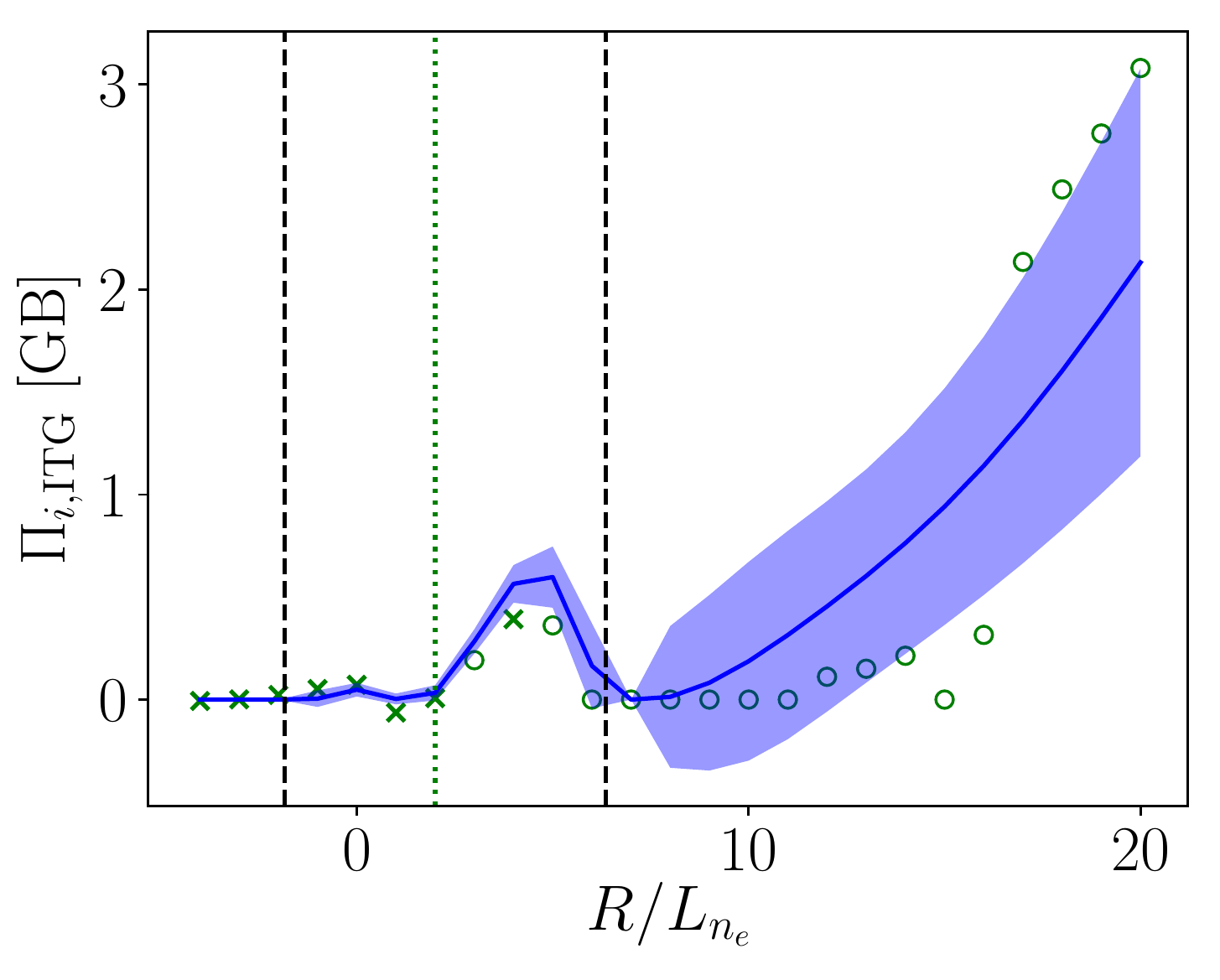}
	\caption{Comparison of main ITG-driven transport fluxes as a function of the logarithmic electron density gradient, $R/L_{n_e}$.}
	\label{fig:FluxComparisonsElectronDensityGradient}
\end{figure}

\begin{figure}[h]
	\centering
	\includegraphics[scale=0.245]{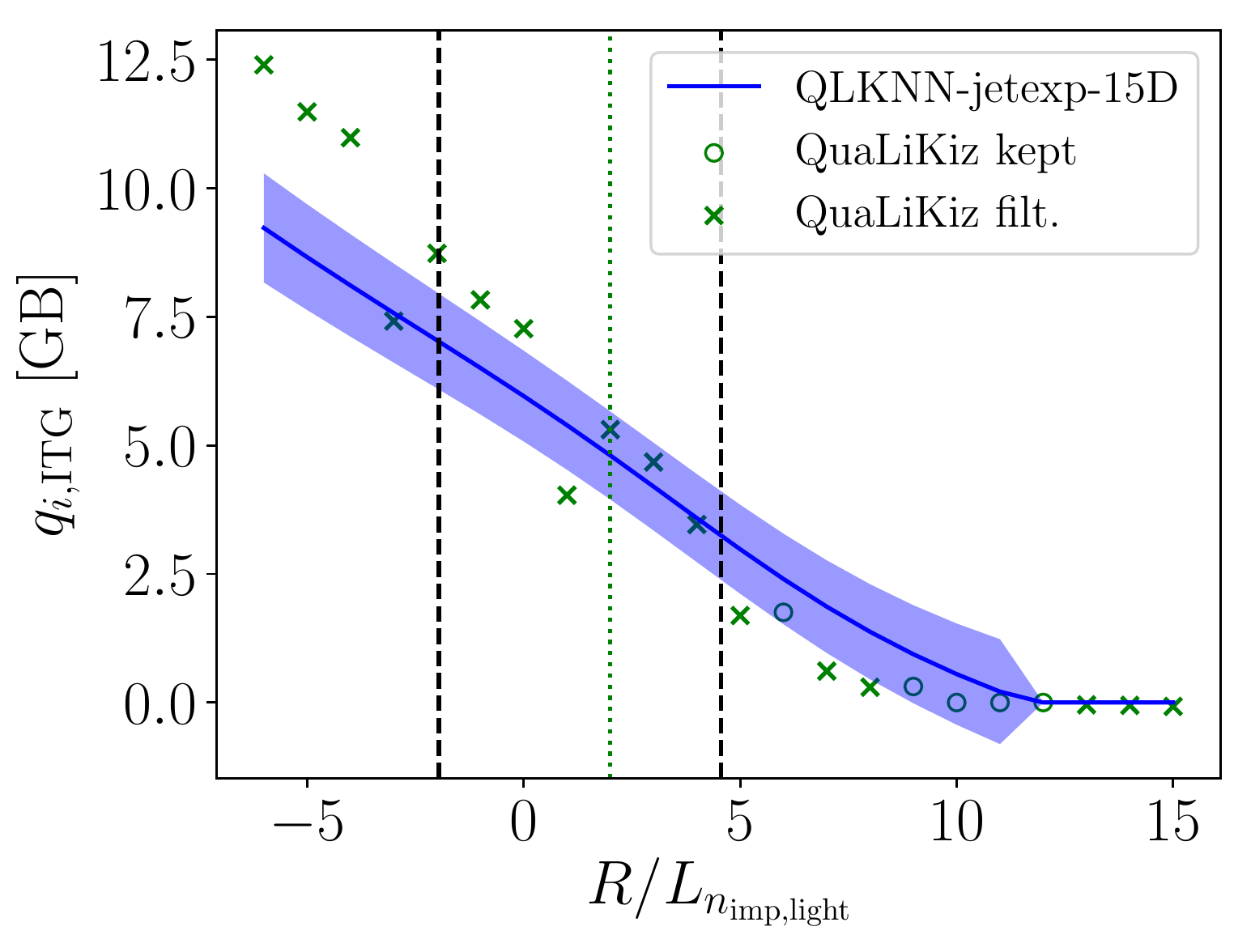}%
	\hspace{2mm}\includegraphics[scale=0.245]{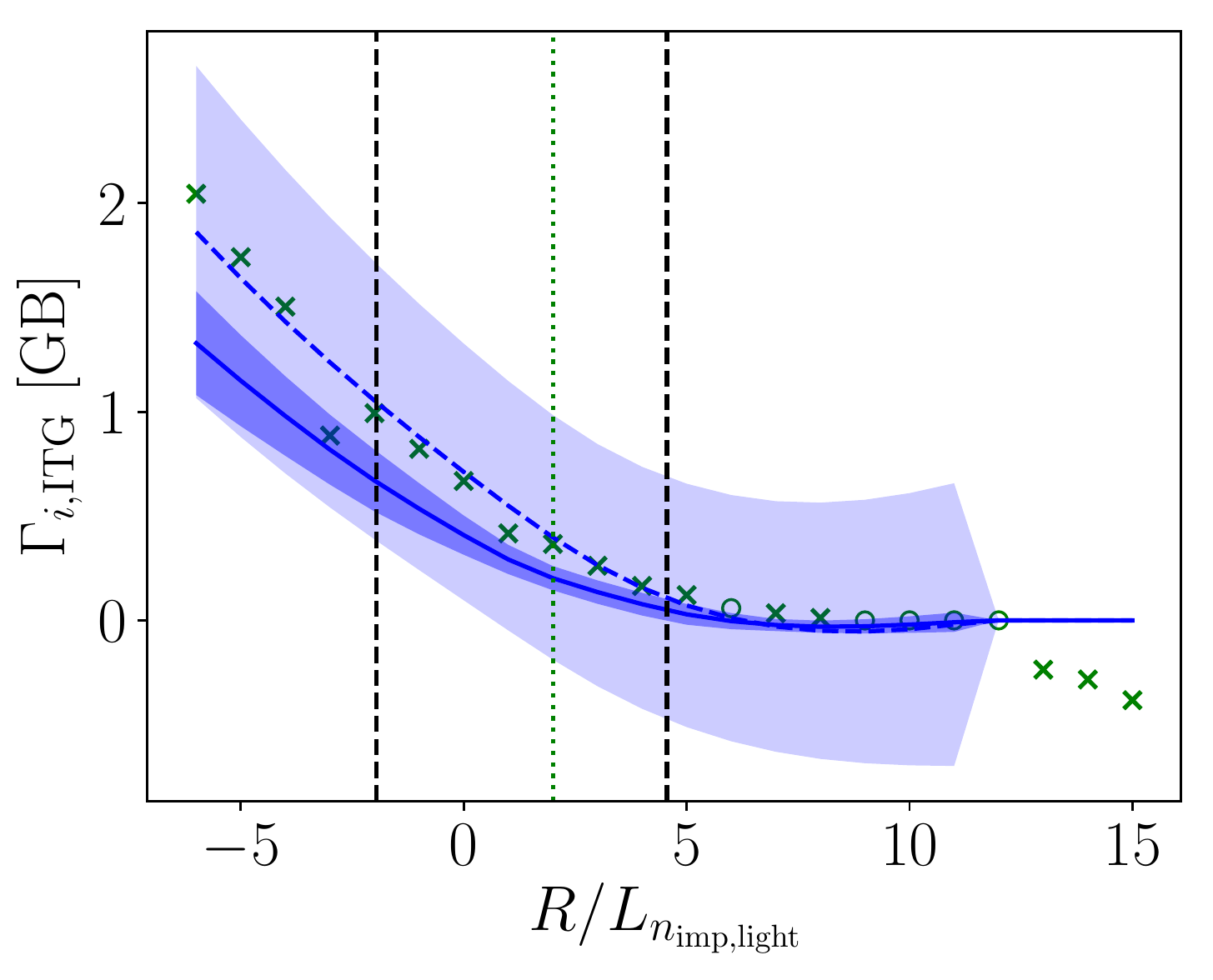} \\
	\hspace{0.25mm}\includegraphics[scale=0.245]{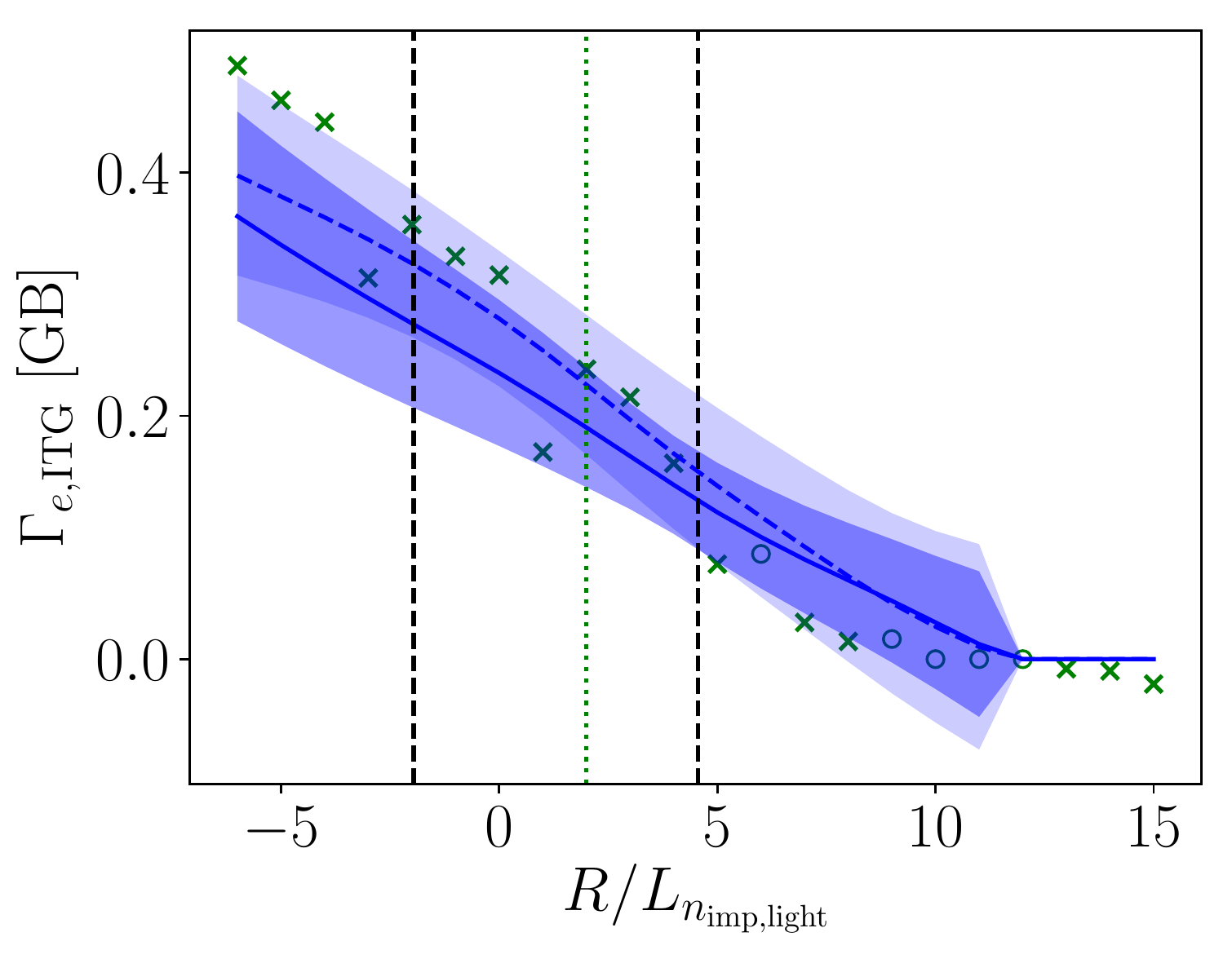}%
	\hspace{2mm}\includegraphics[scale=0.245]{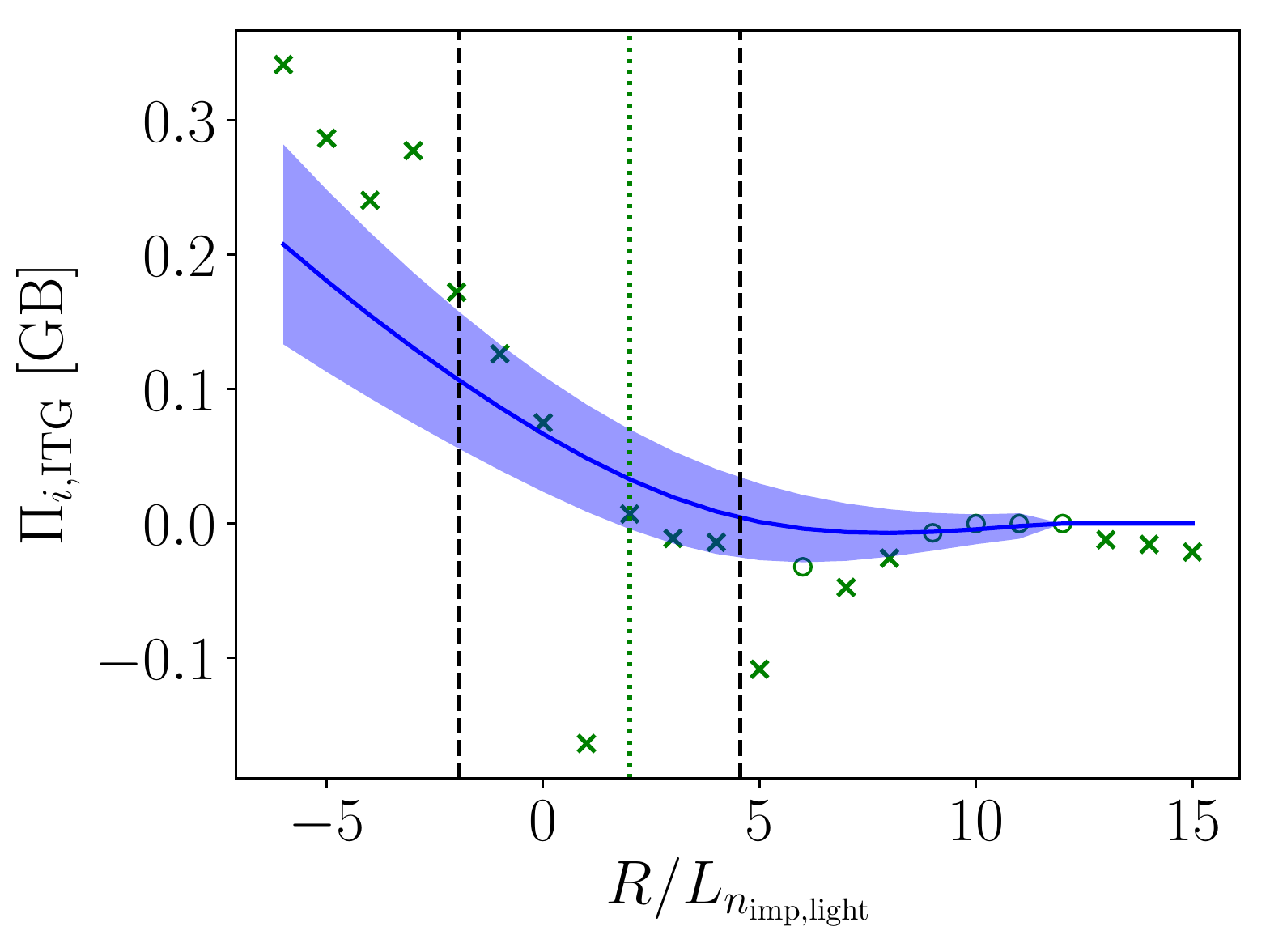}
	\caption{Comparison of main ITG-driven transport fluxes as a function of the logarithmic light impurity ion density gradient, $R/L_{n_{\text{imp,light}}}$.}
	\label{fig:FluxComparisonsIonDensityGradient}
\end{figure}

\begin{figure}[h]
	\centering
	\includegraphics[scale=0.245]{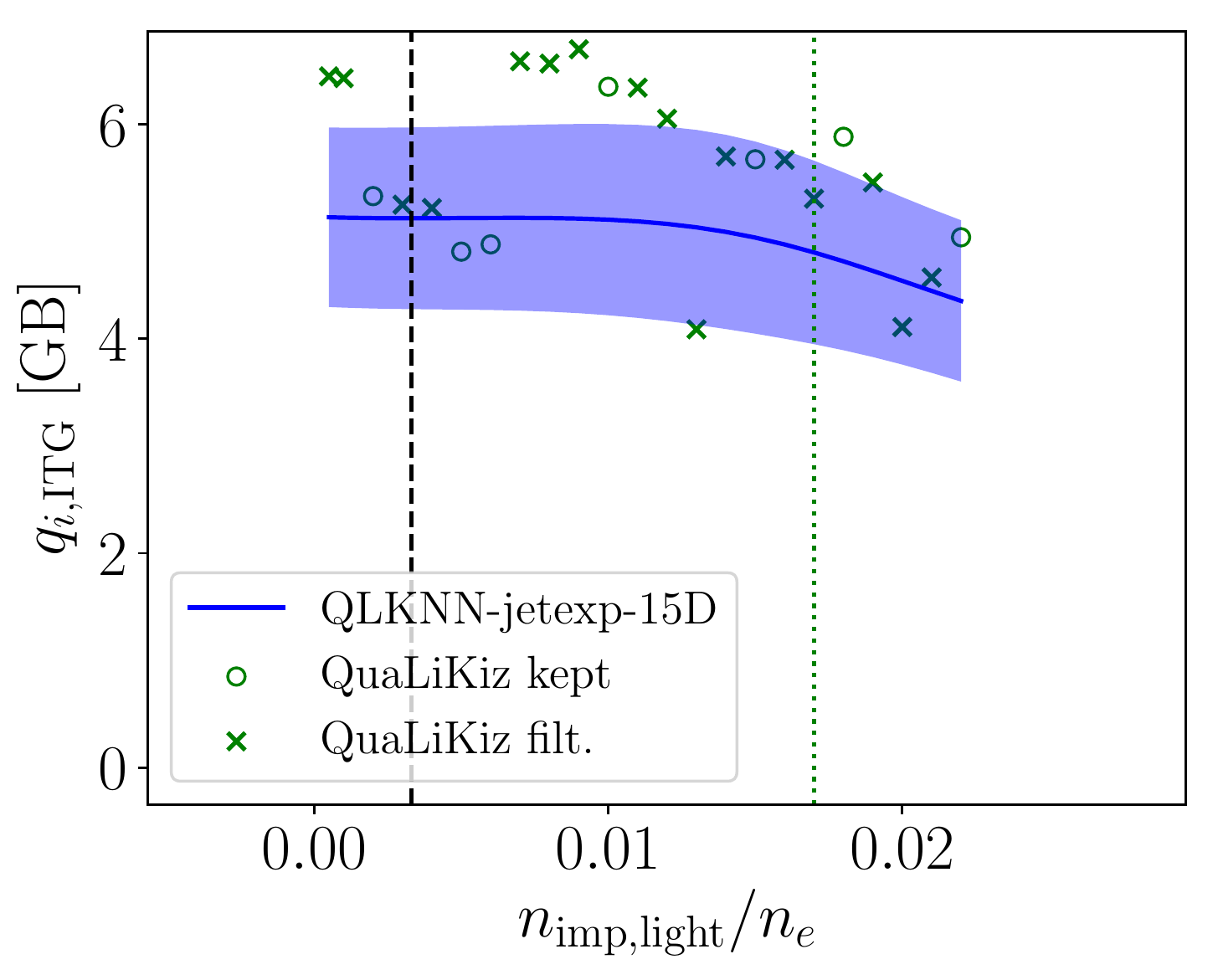}%
	\hspace{2mm}\includegraphics[scale=0.245]{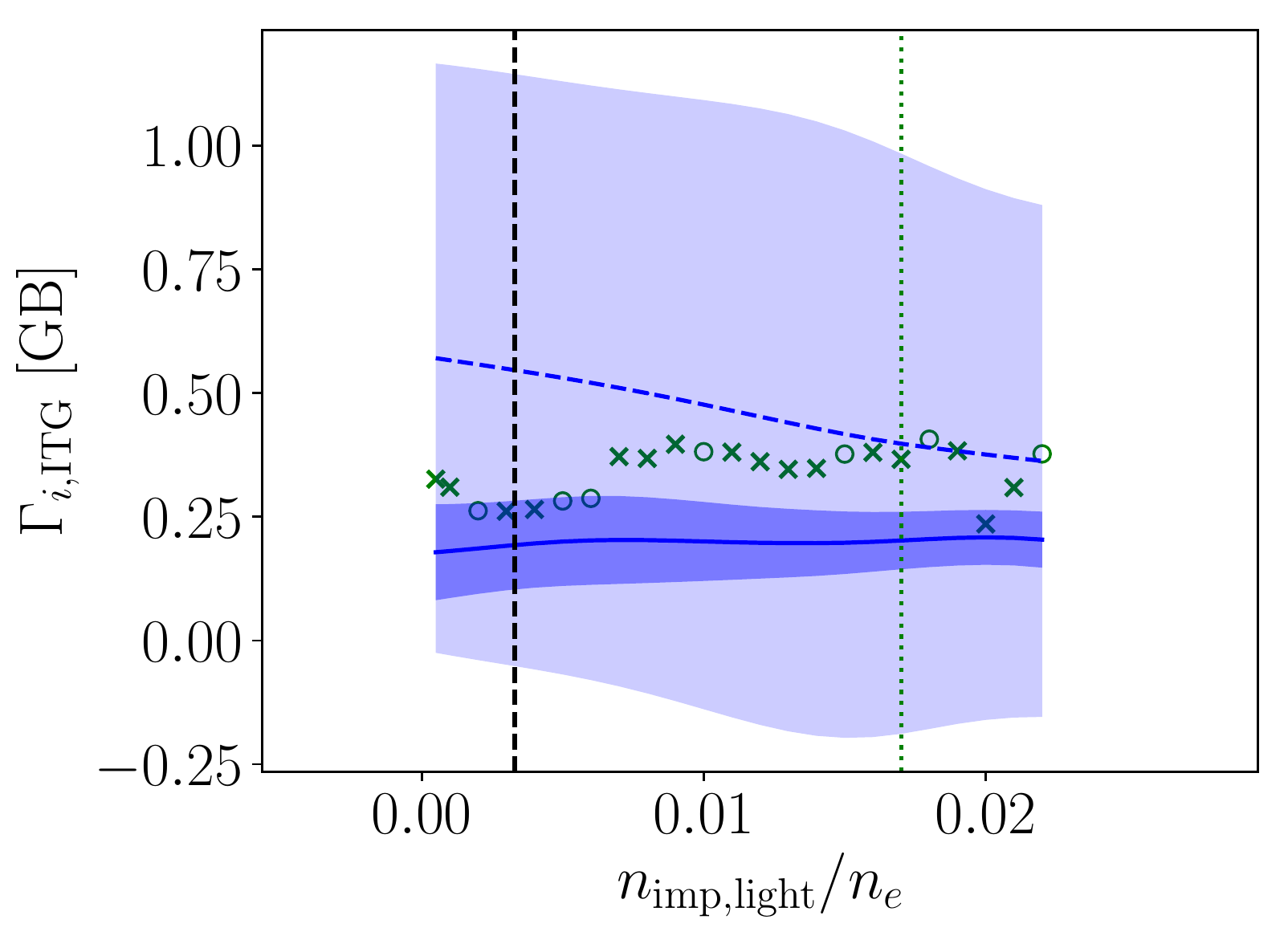} \\
	\hspace{0.25mm}\includegraphics[scale=0.245]{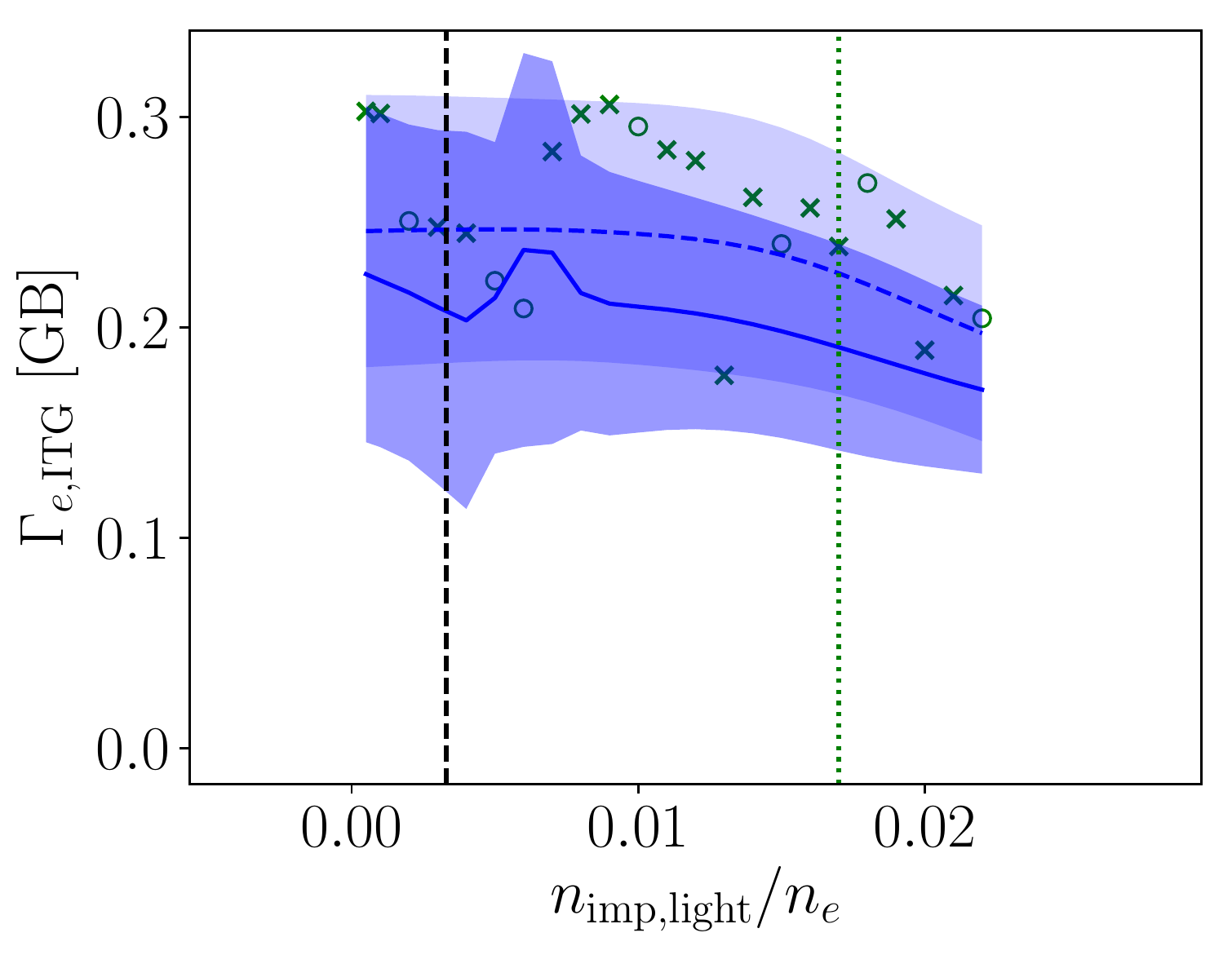}%
	\hspace{2mm}\includegraphics[scale=0.245]{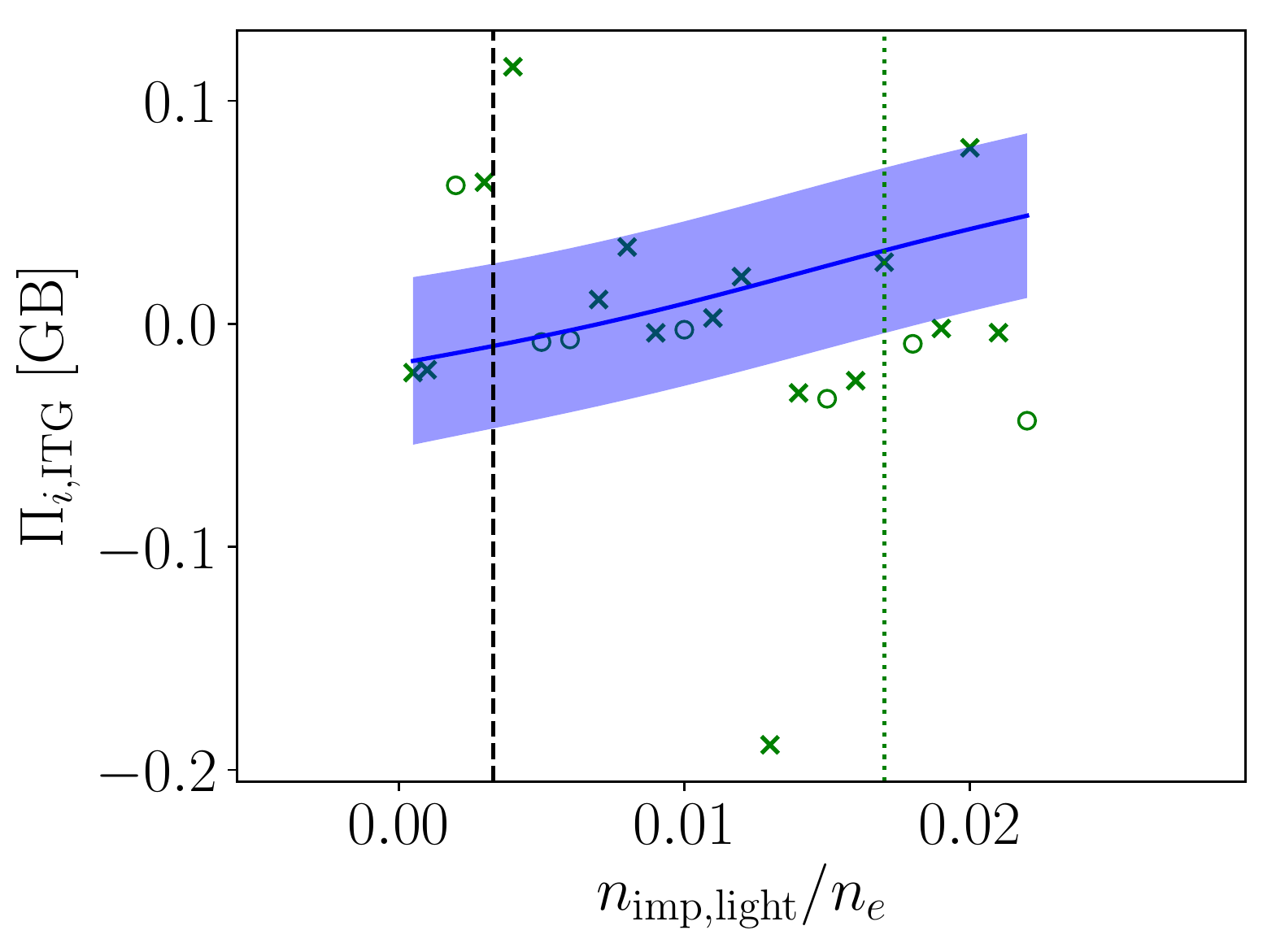}
	\caption{Comparison of main ITG-driven transport fluxes as a function of the normalized light impurity ion density, $N_{\text{imp,light}}$.}
	\label{fig:FluxComparisonsImpurityDensity}
\end{figure}

\begin{figure}[h]
	\centering
	\includegraphics[scale=0.245]{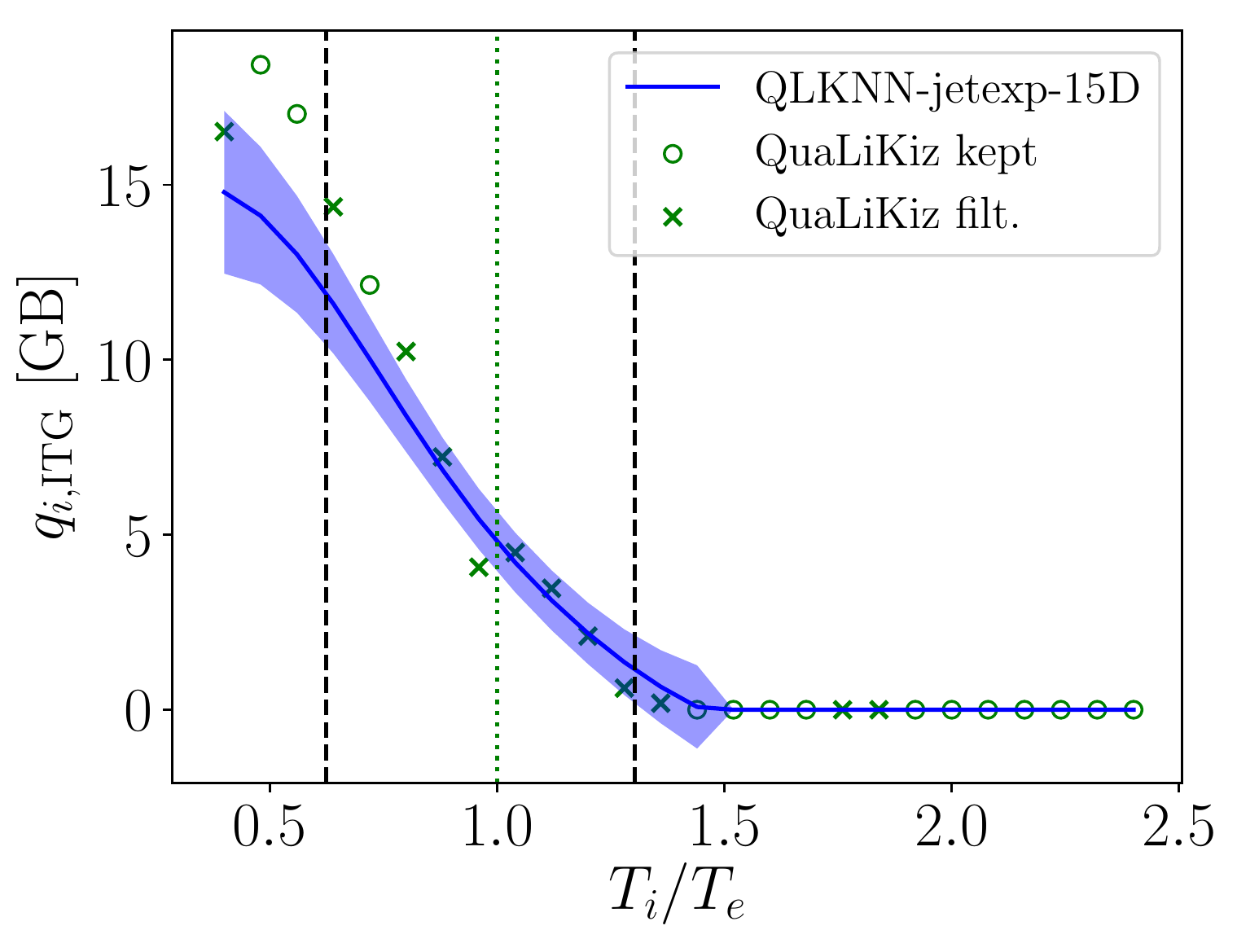}%
	\hspace{2mm}\includegraphics[scale=0.245]{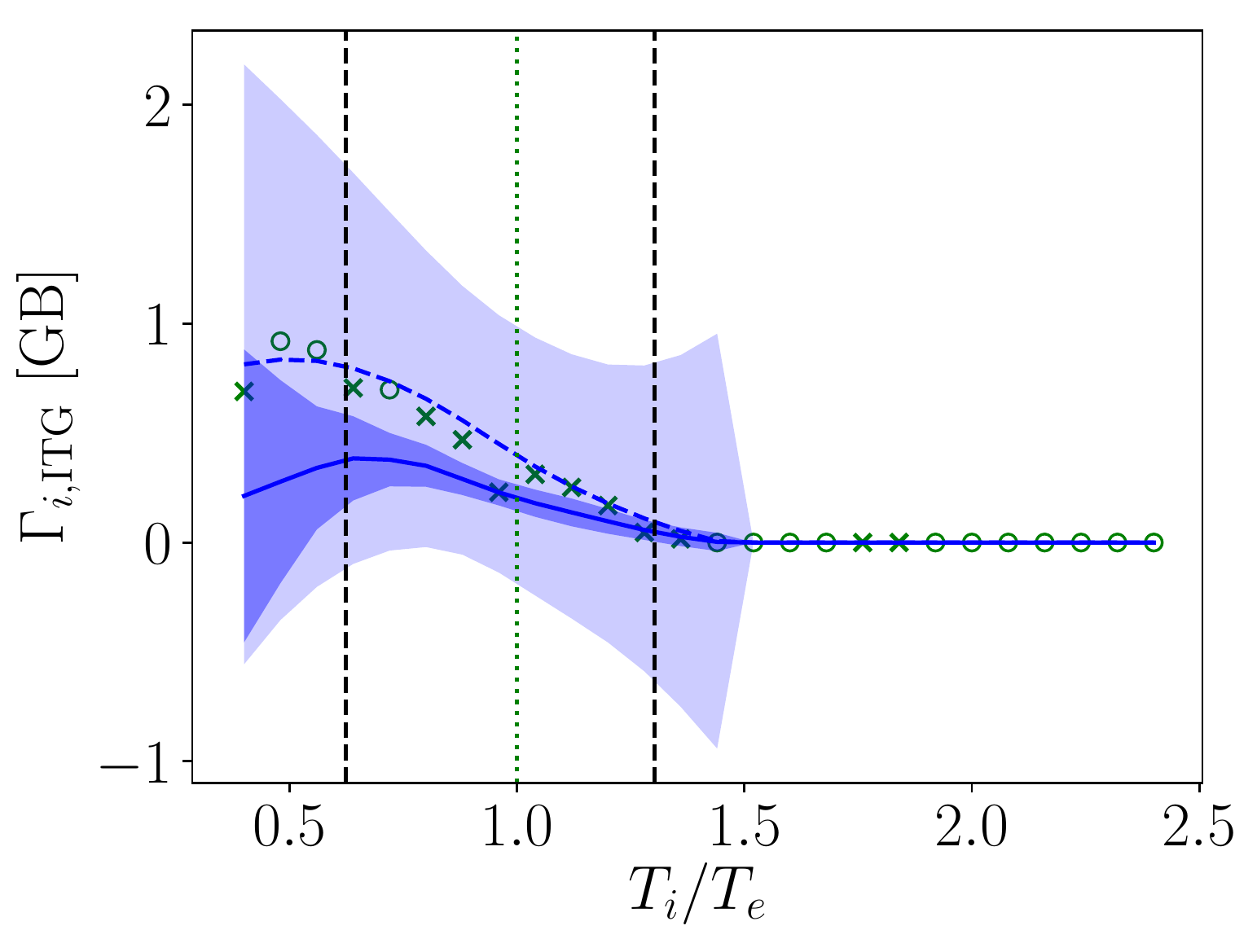} \\
	\hspace{0.25mm}\includegraphics[scale=0.245]{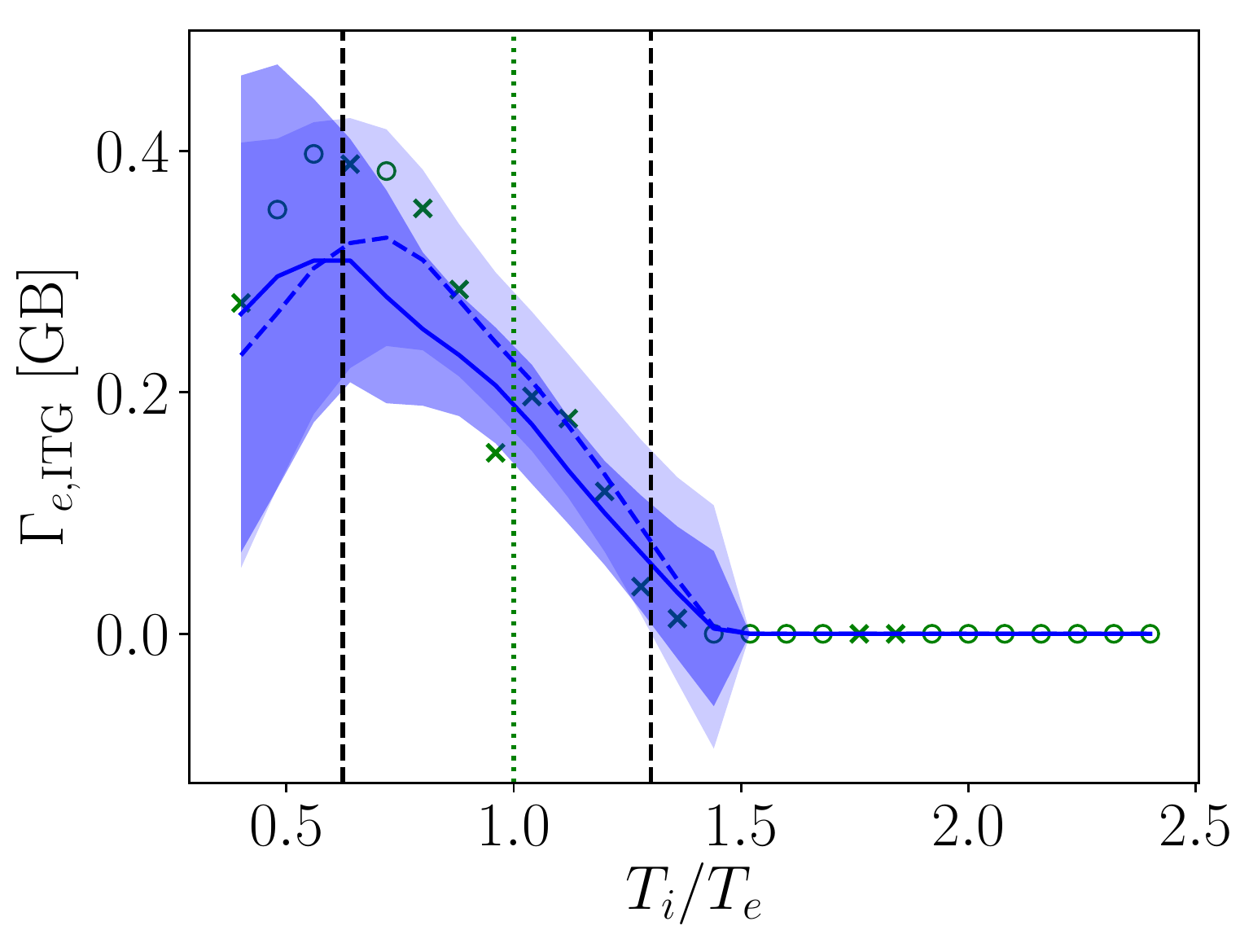}%
	\hspace{2mm}\includegraphics[scale=0.245]{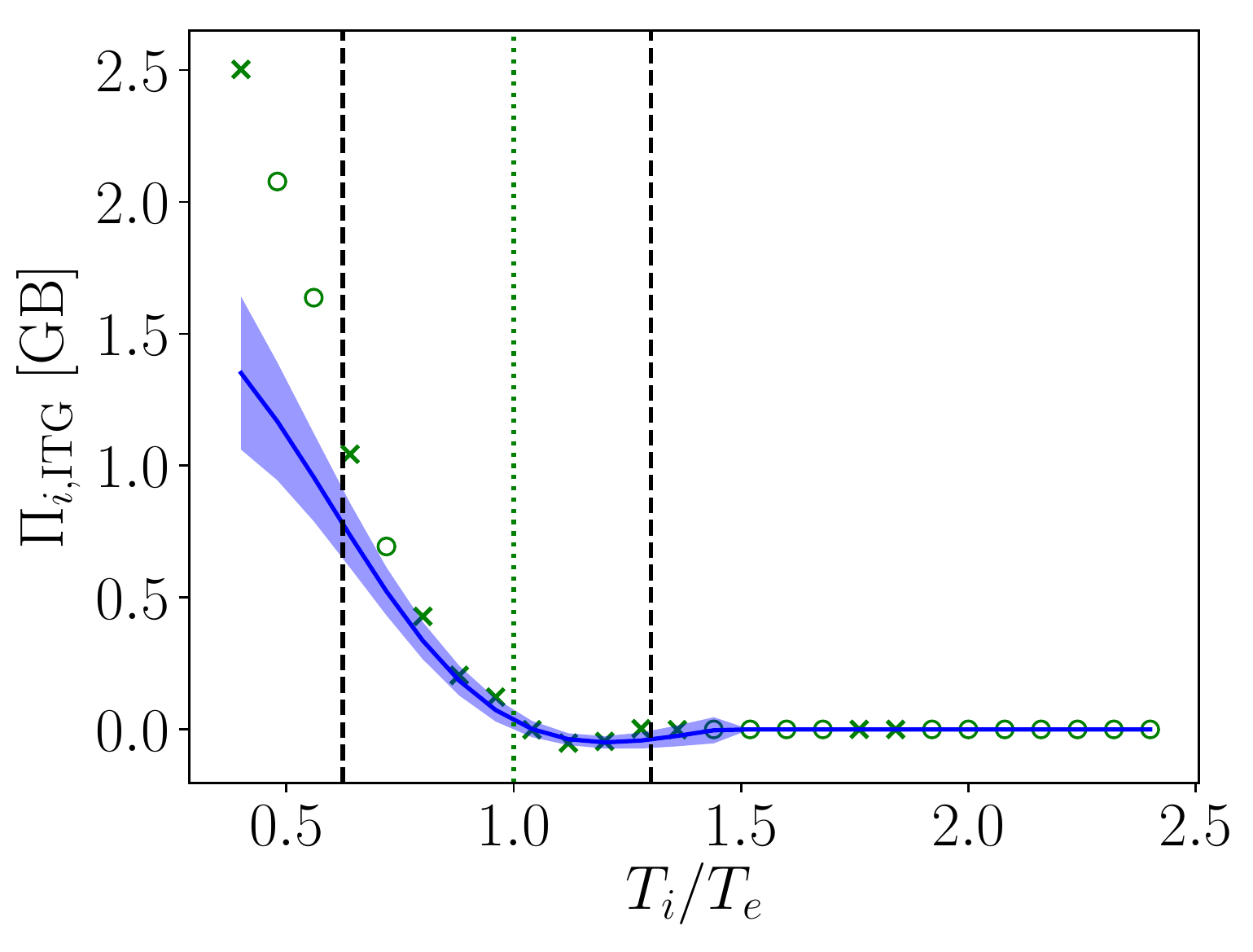}
	\caption{Comparison of main ITG-driven transport fluxes as a function of the ion-to-electron temperature ratio, $T_i/T_e$.}
	\label{fig:FluxComparisonsTemperatureRatio}
\end{figure}

\begin{figure}[h]
	\centering
	\includegraphics[scale=0.245]{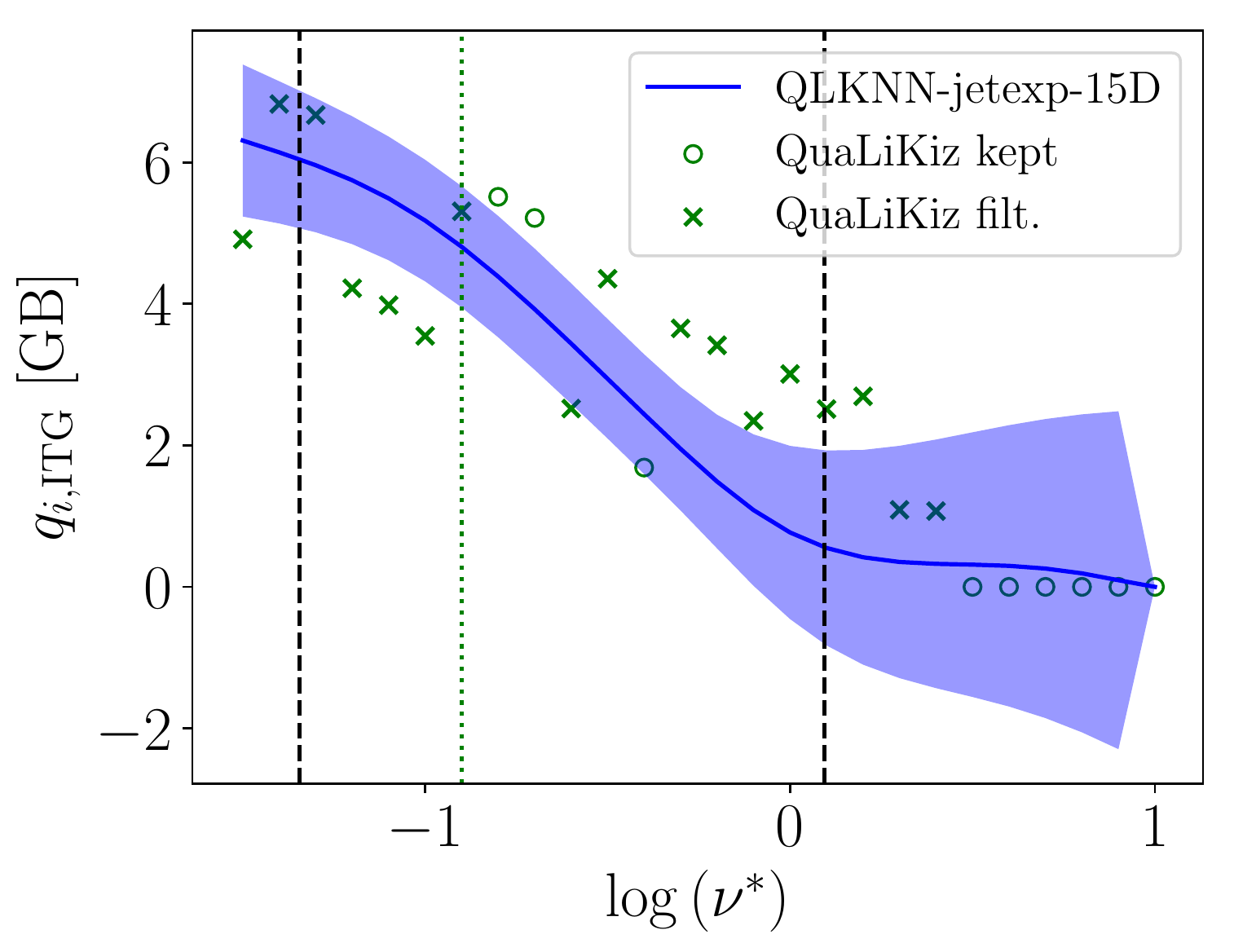}%
	\hspace{2mm}\includegraphics[scale=0.245]{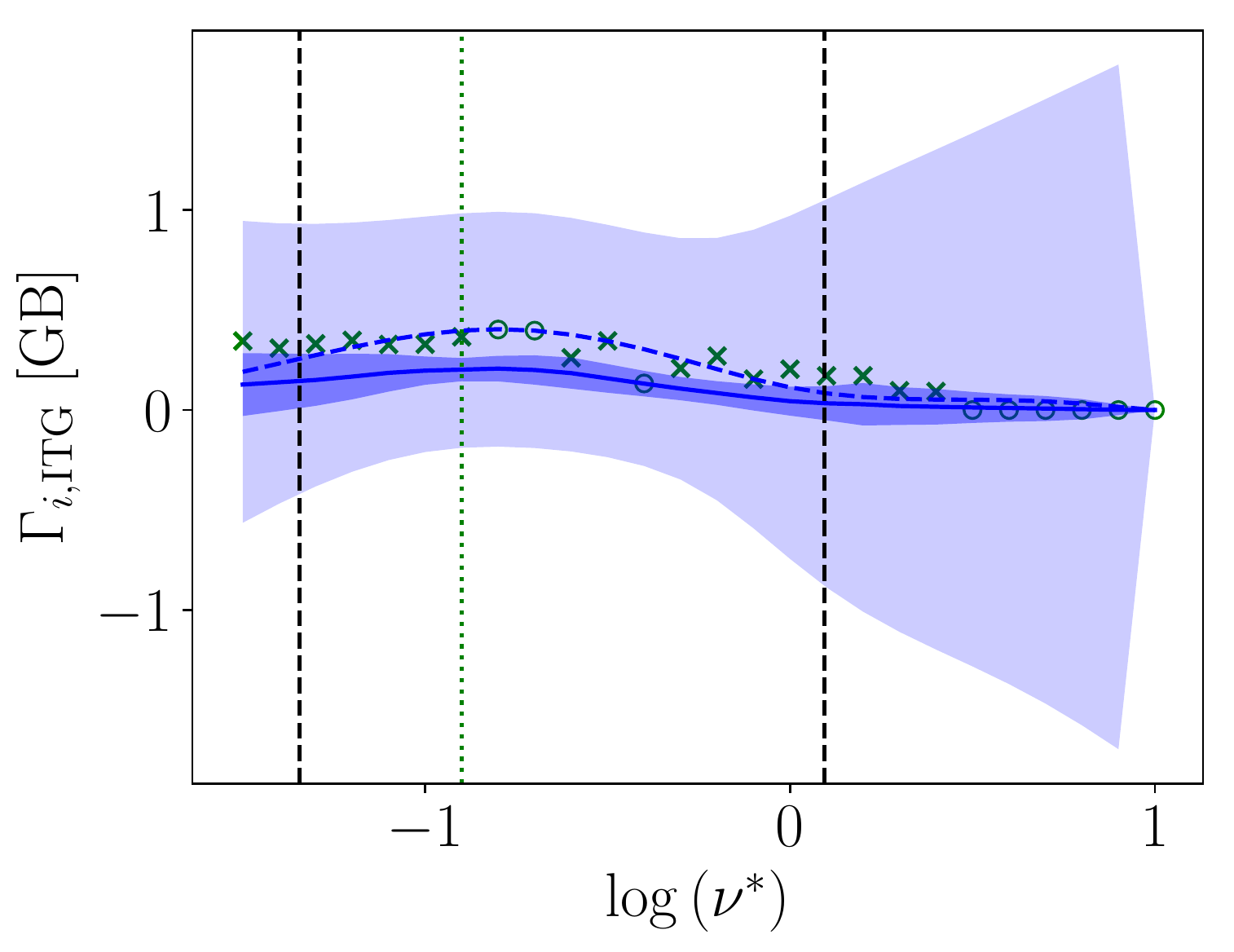} \\
	\hspace{0.25mm}\includegraphics[scale=0.245]{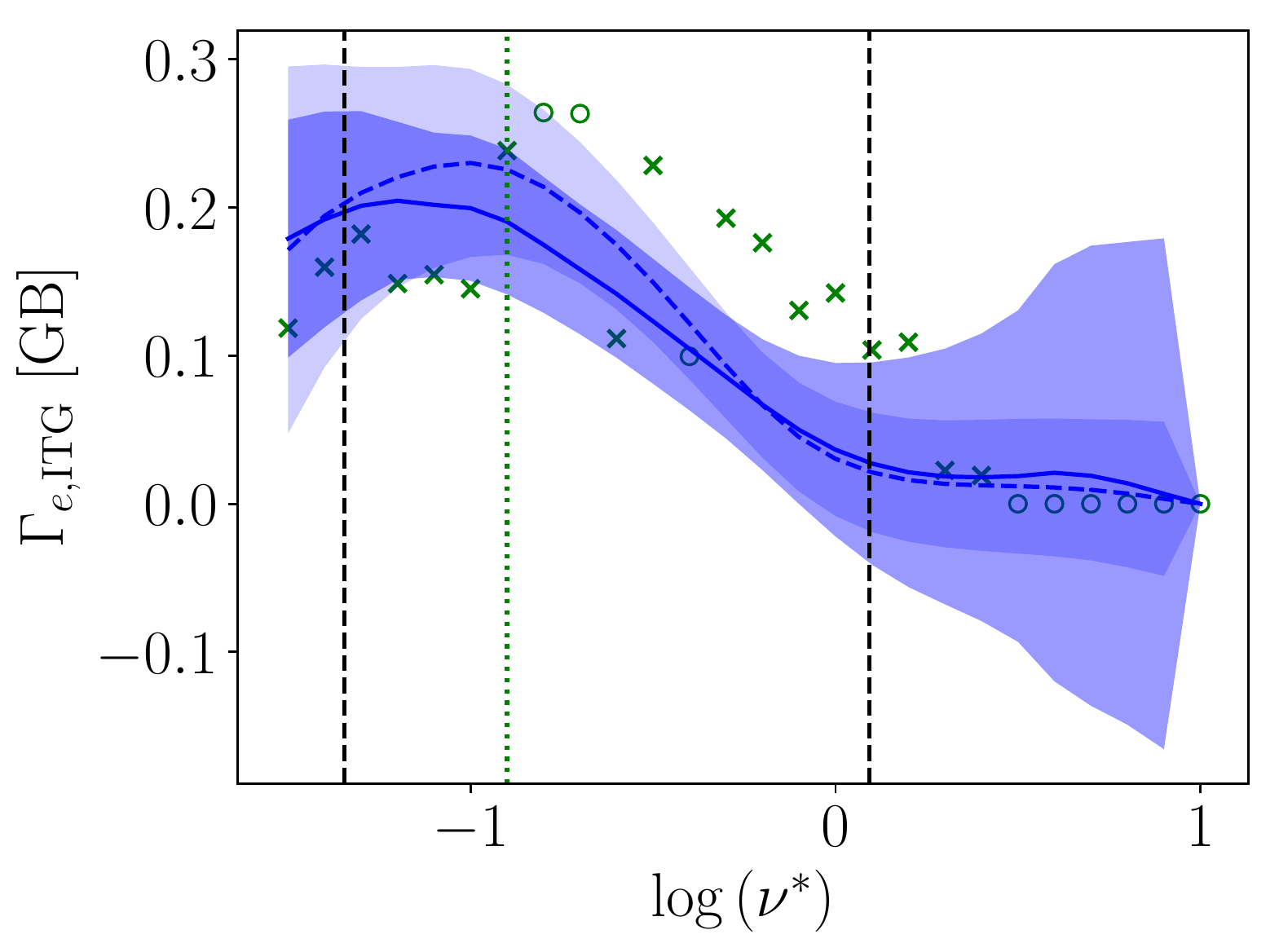}%
	\hspace{2mm}\includegraphics[scale=0.245]{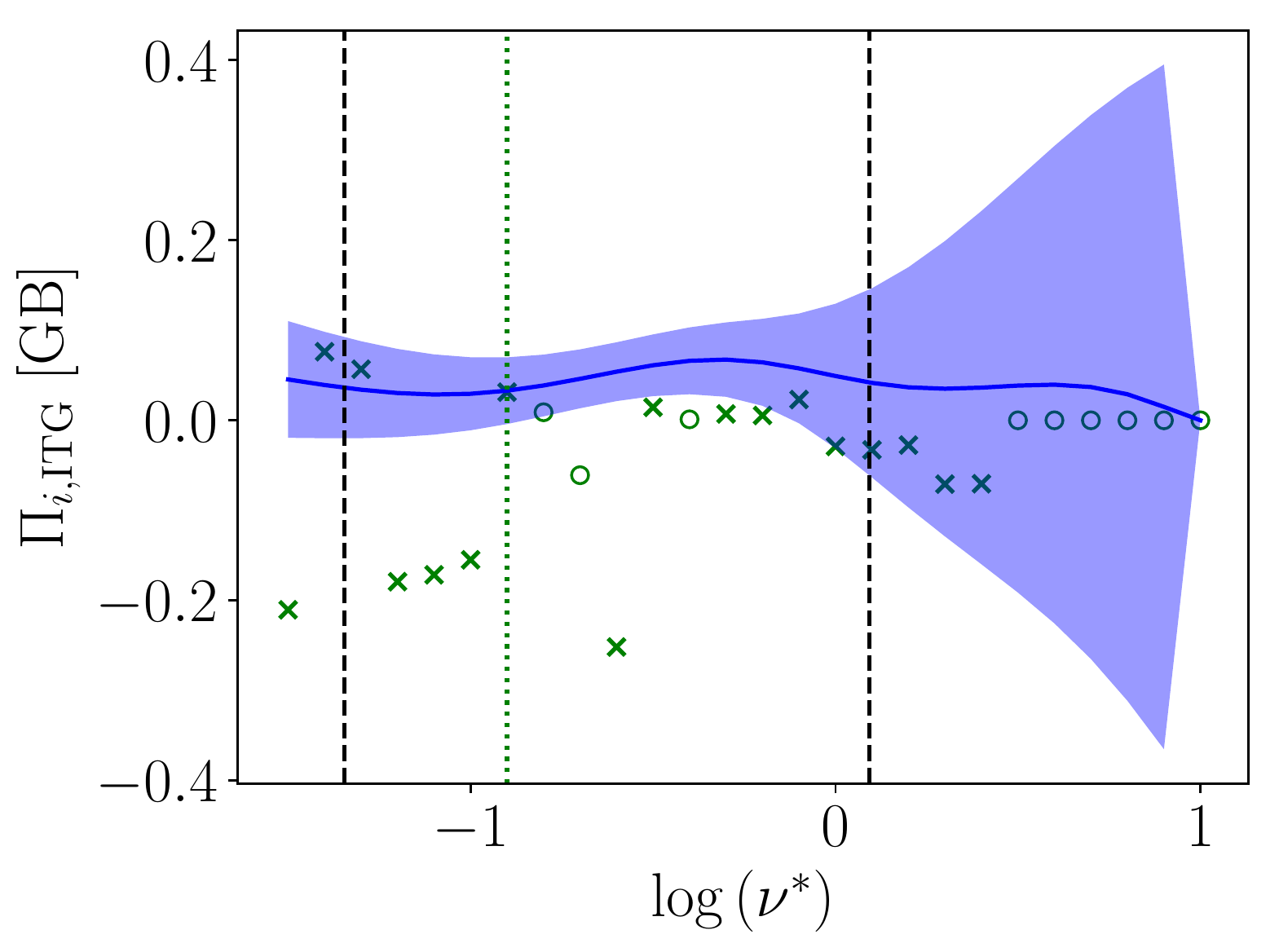}
	\caption{Comparison of main ITG-driven transport fluxes as a function of the logarithm of normalized collisionality, $\log_{10}(\nu^*)$.}
	\label{fig:FluxComparisonsCollisionalityRate}
\end{figure}

\begin{figure}[h]
	\centering
	\includegraphics[scale=0.245]{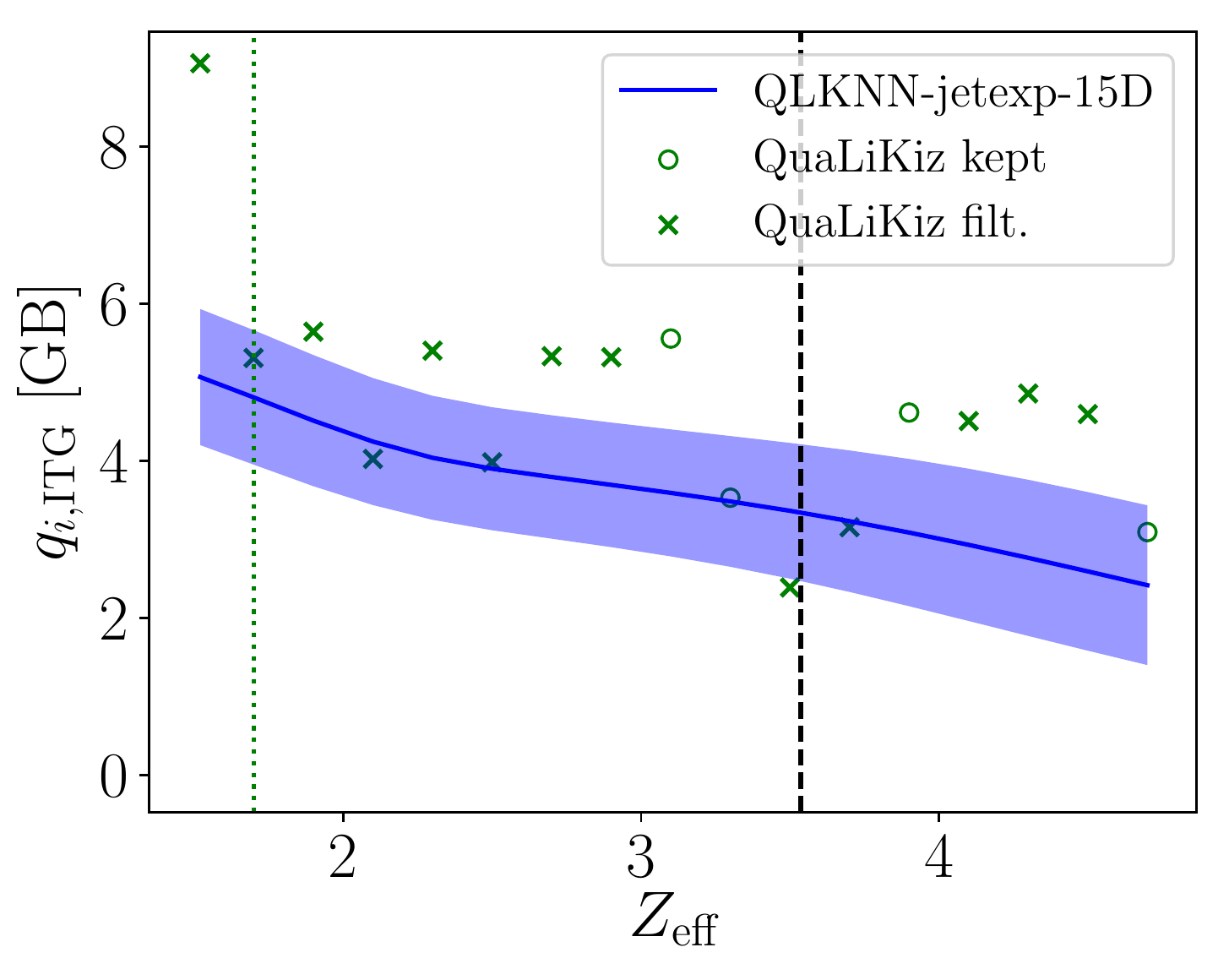}%
	\hspace{2mm}\includegraphics[scale=0.245]{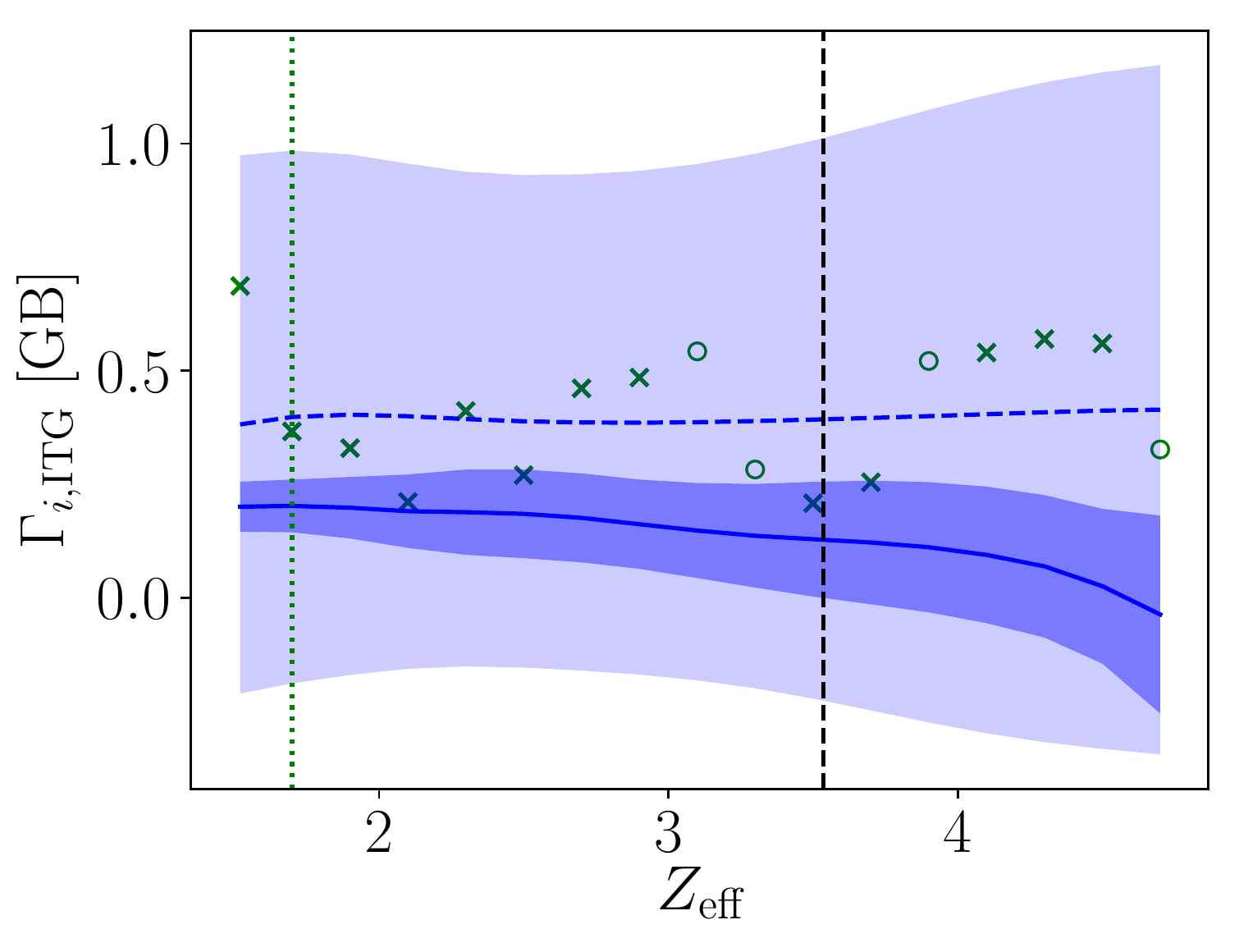} \\
	\hspace{0.25mm}\includegraphics[scale=0.245]{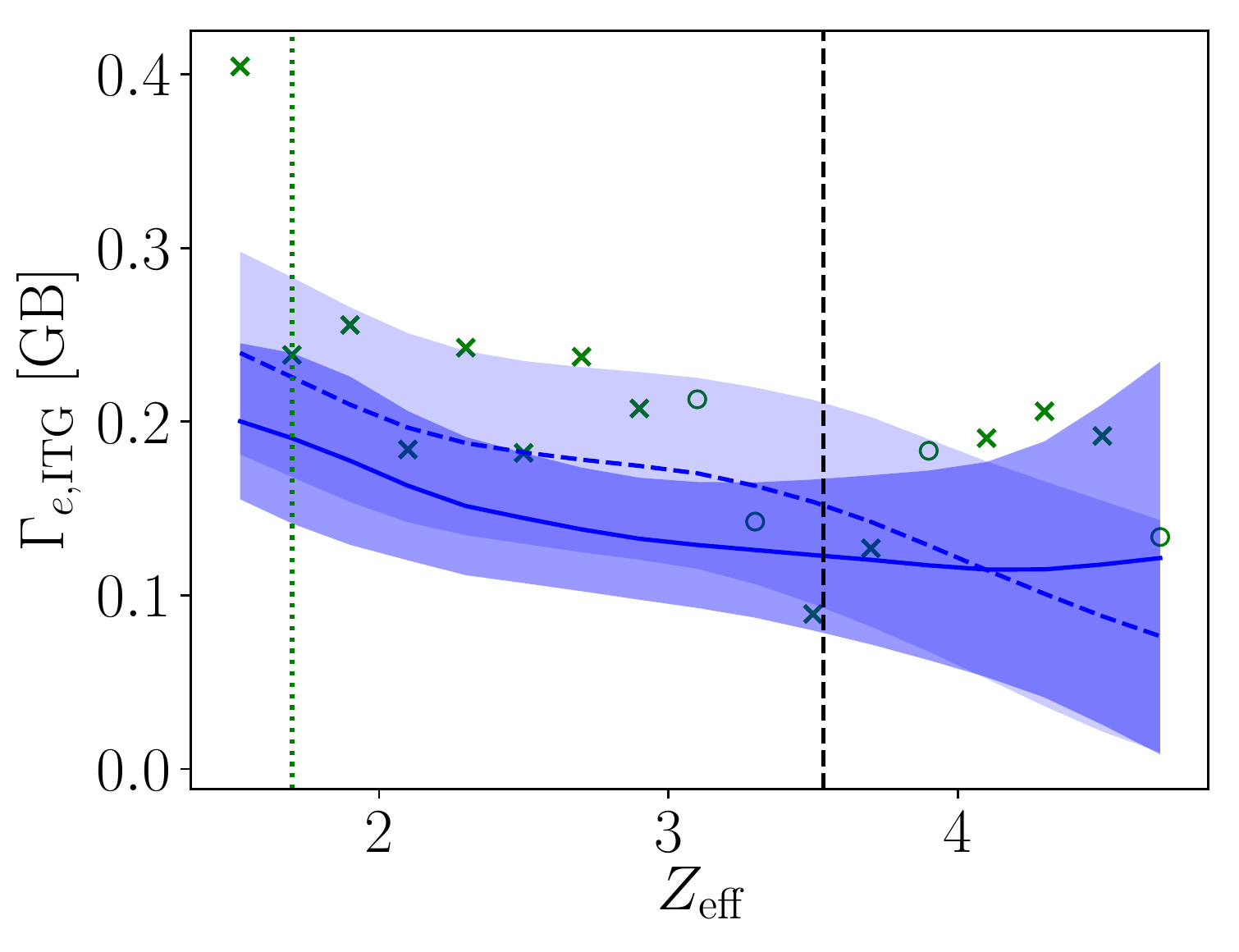}%
	\hspace{2mm}\includegraphics[scale=0.245]{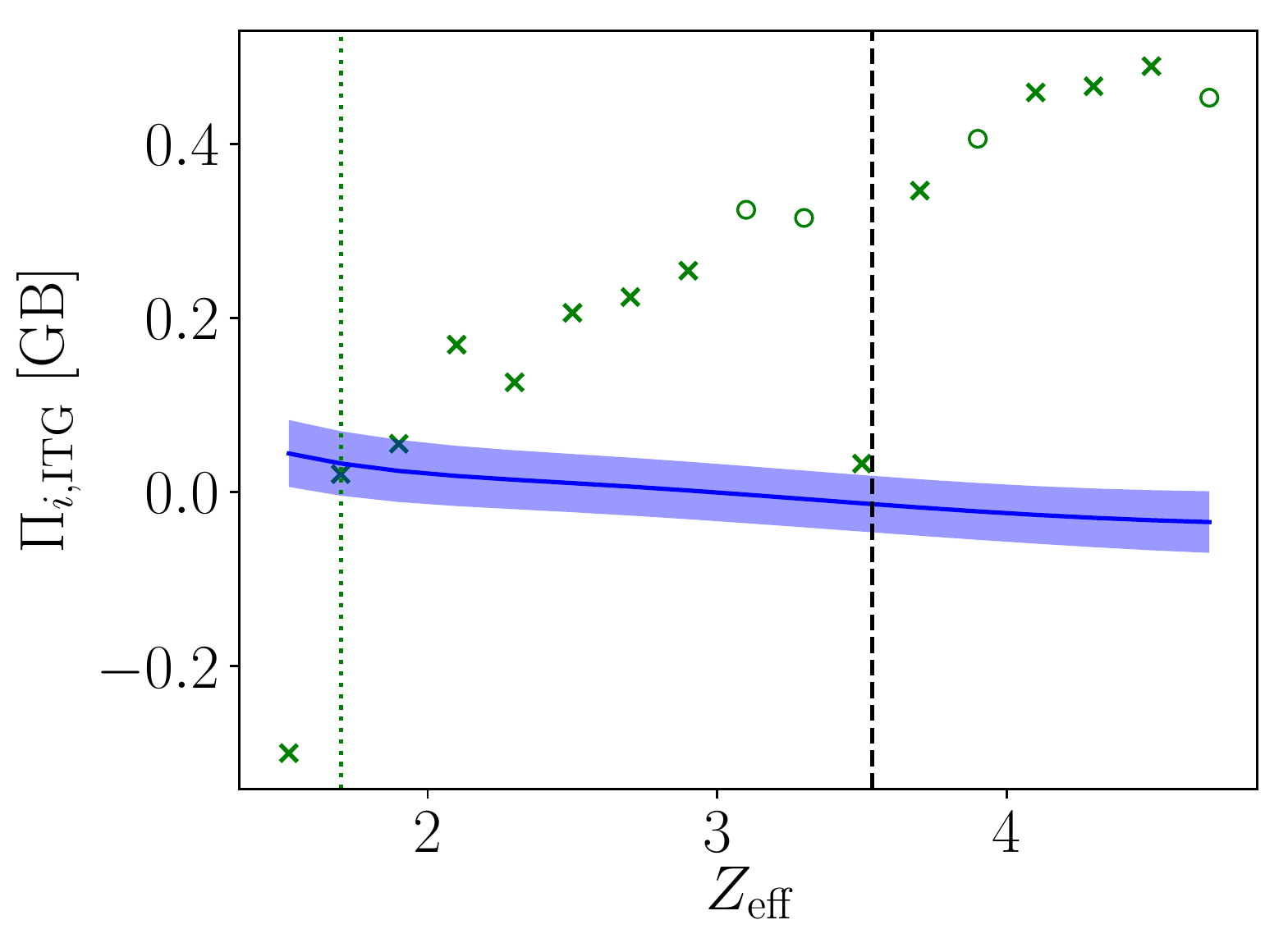}
	\caption{Comparison of main ITG-driven transport fluxes as a function of the effective charge, $Z_{\text{eff}}$.}
	\label{fig:FluxComparisonsEffectiveCharge}
\end{figure}

\begin{figure}[h]
	\centering
	\includegraphics[scale=0.245]{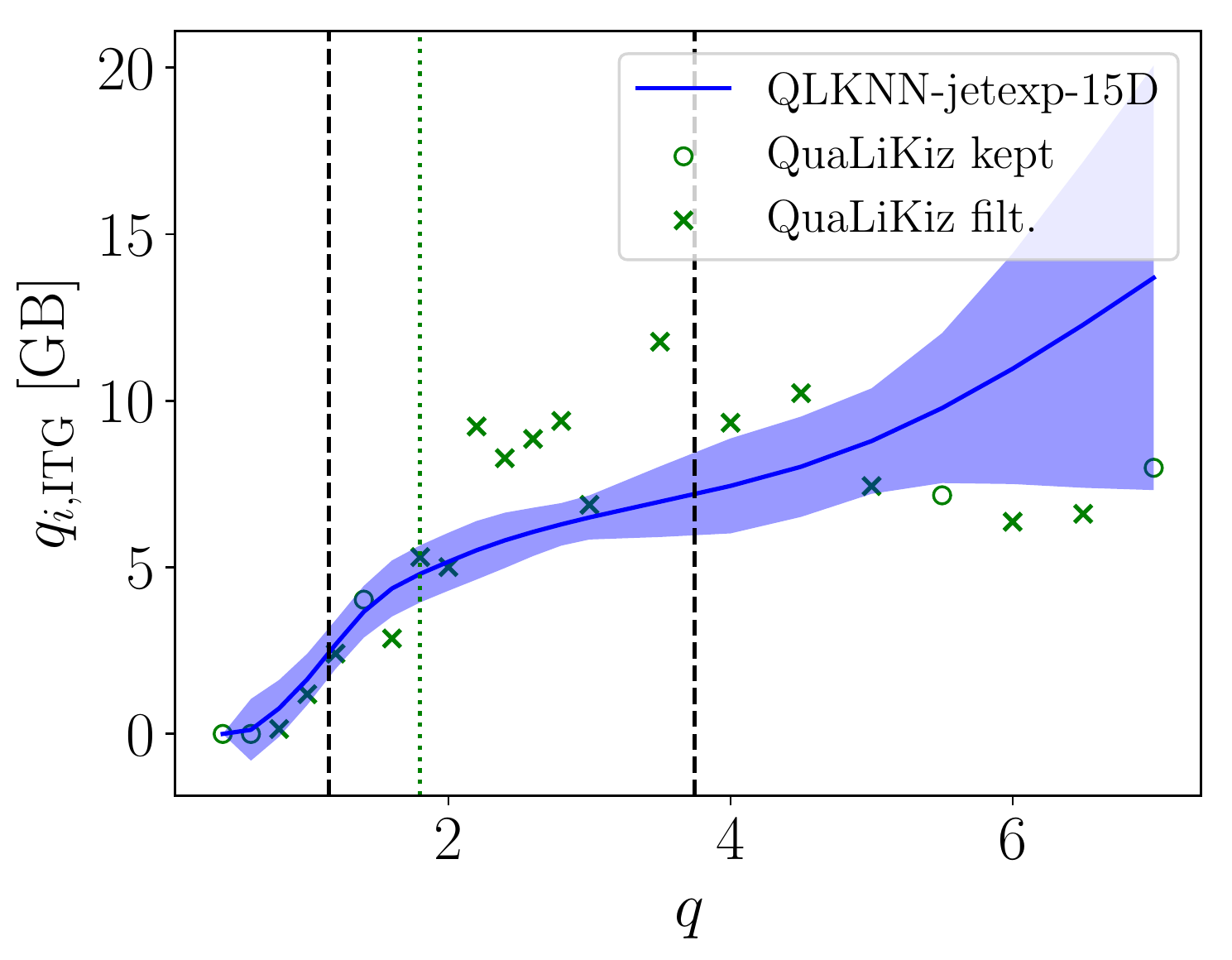}%
	\hspace{2mm}\includegraphics[scale=0.245]{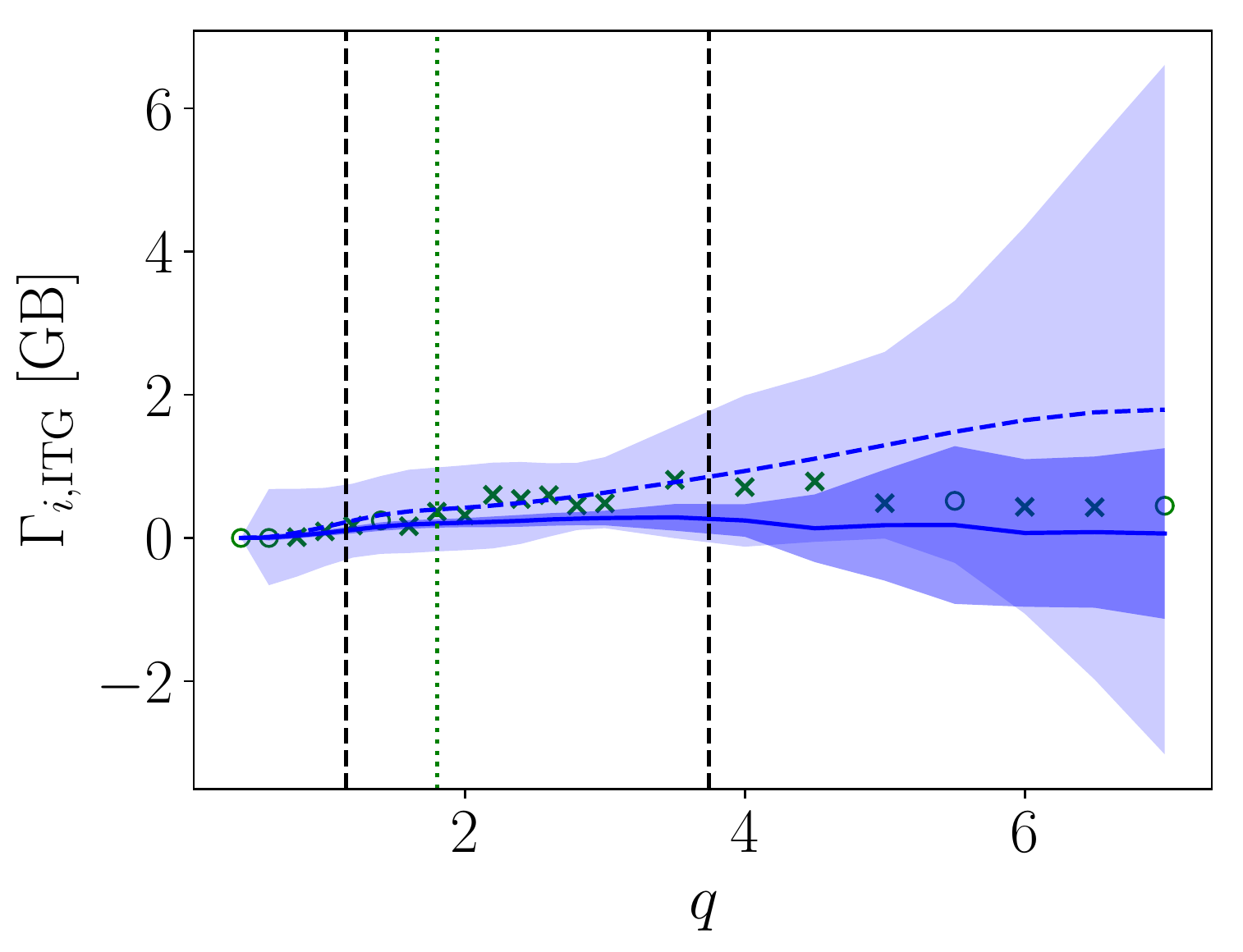} \\
	\hspace{0.25mm}\includegraphics[scale=0.245]{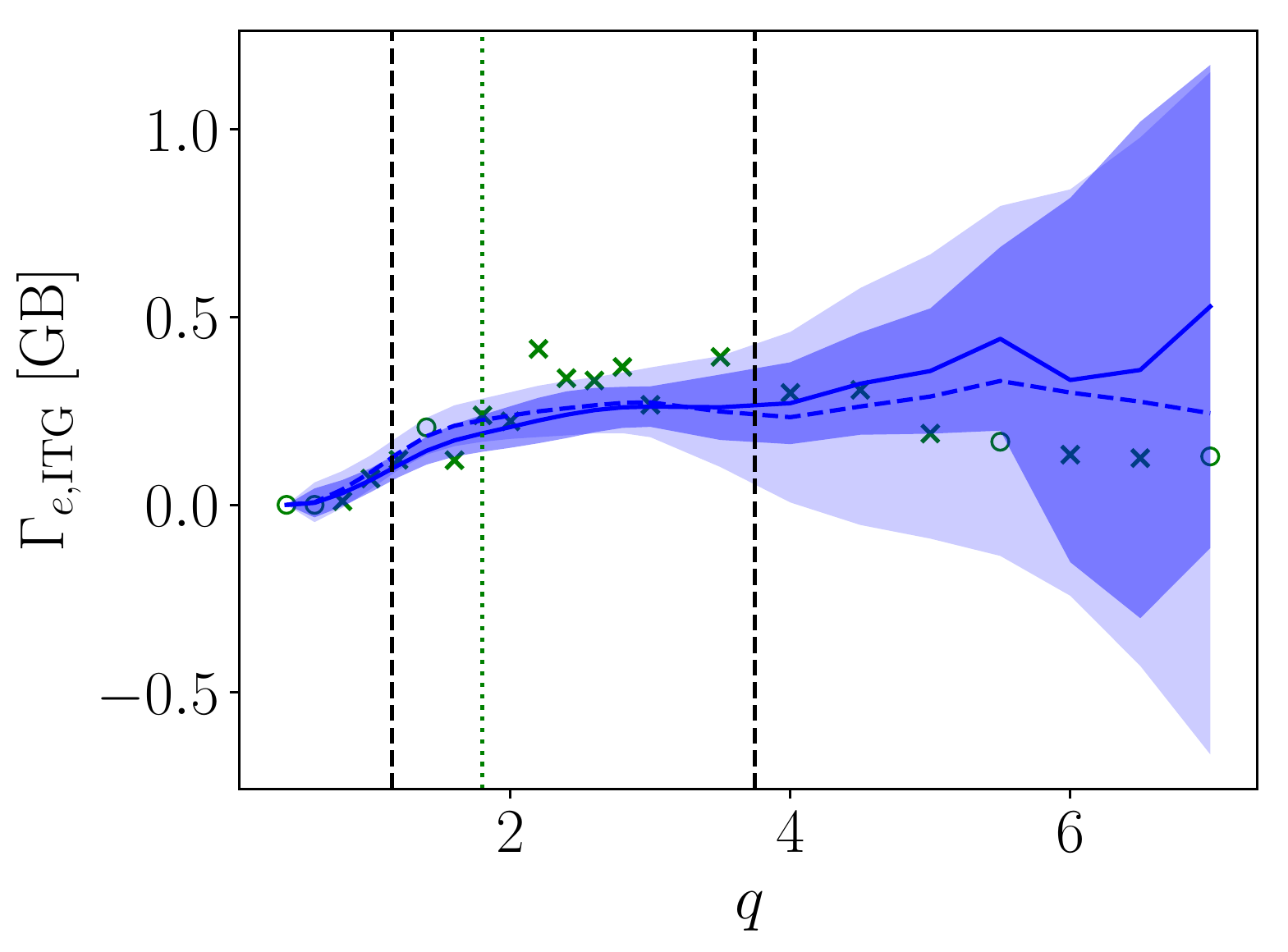}%
	\hspace{2mm}\includegraphics[scale=0.245]{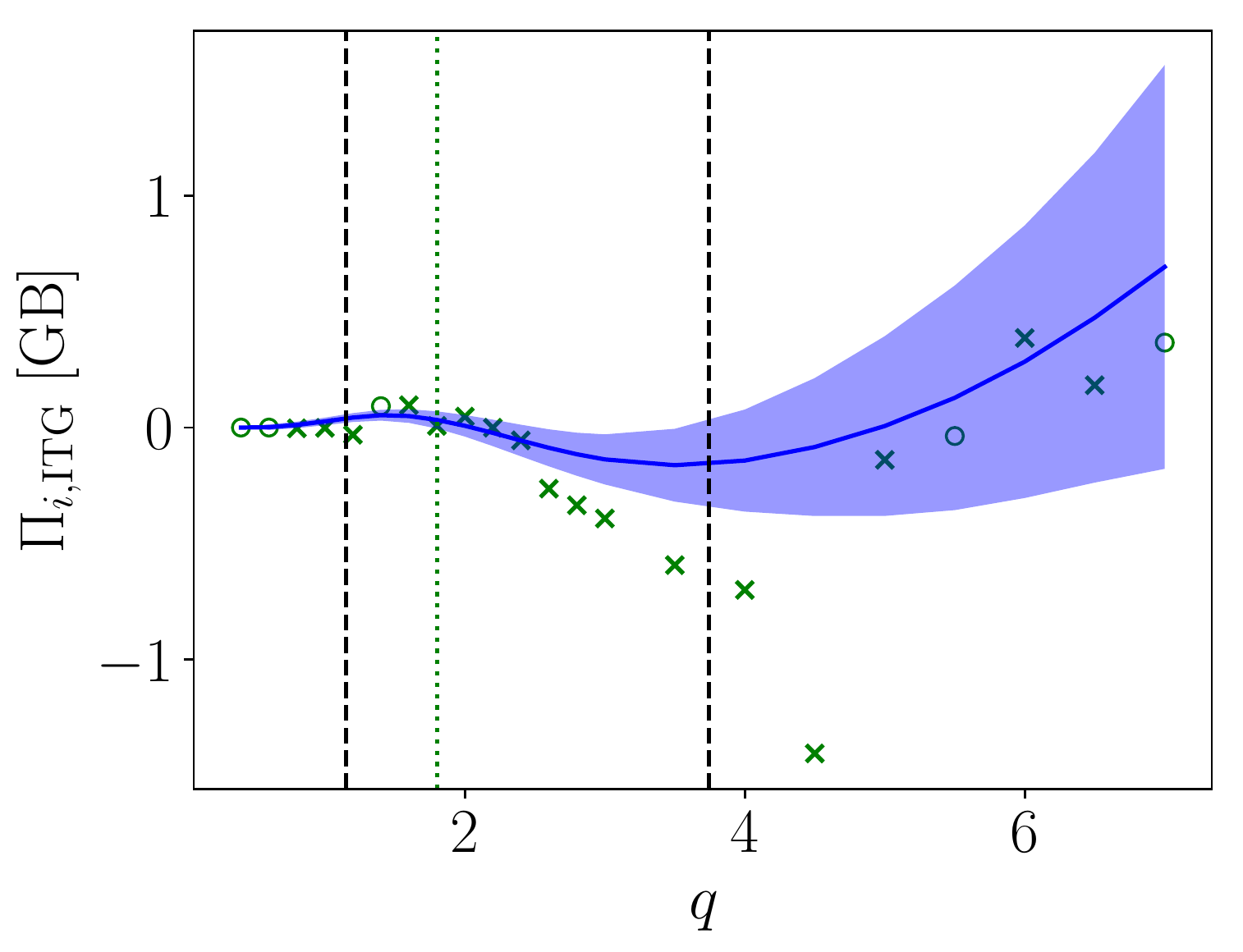}
	\caption{Comparison of main ITG-driven transport fluxes as a function of the safety factor, $q$.}
	\label{fig:FluxComparisonsSafetyFactor}
\end{figure}

\begin{figure}[h]
	\centering
	\includegraphics[scale=0.245]{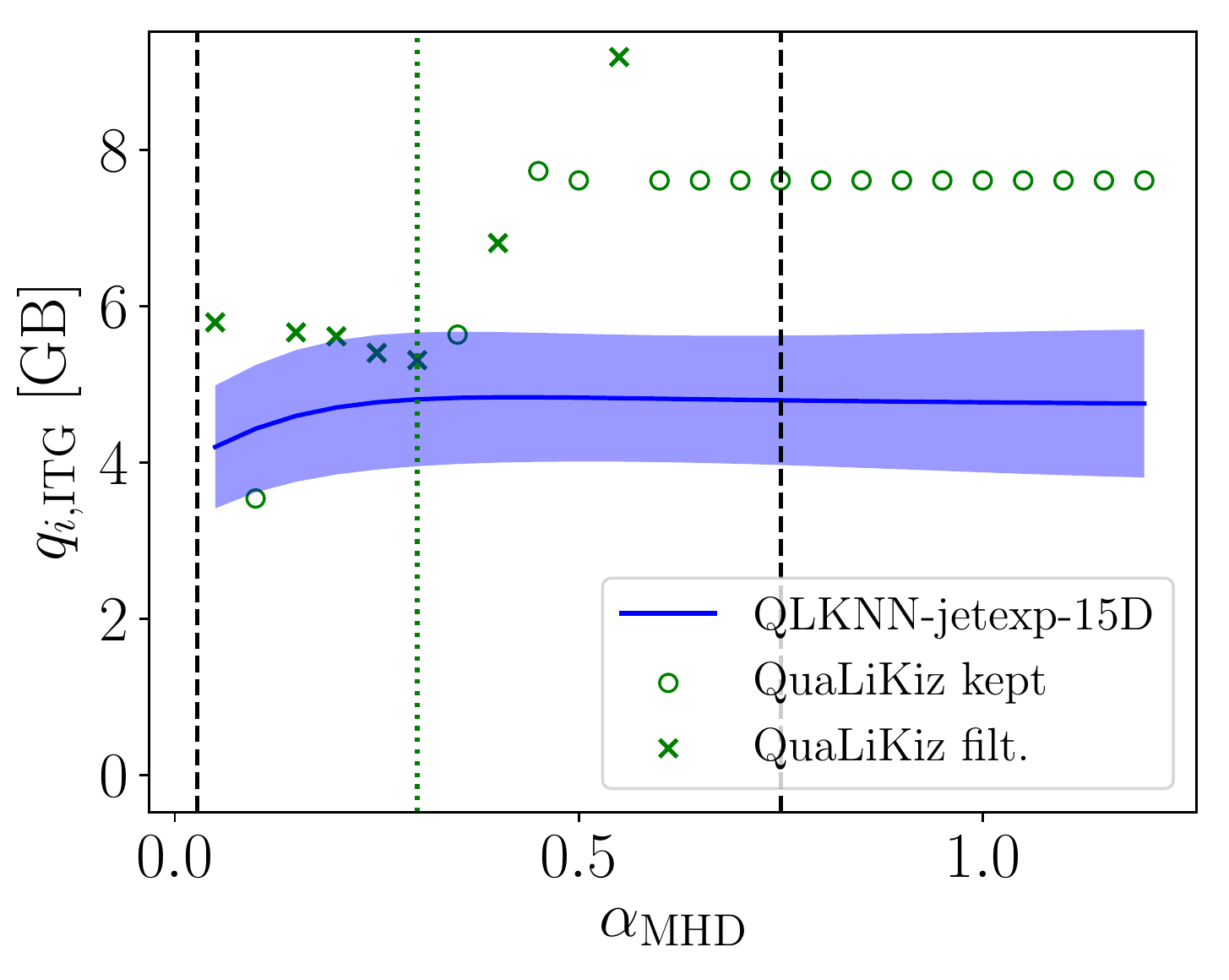}%
	\hspace{2mm}\includegraphics[scale=0.245]{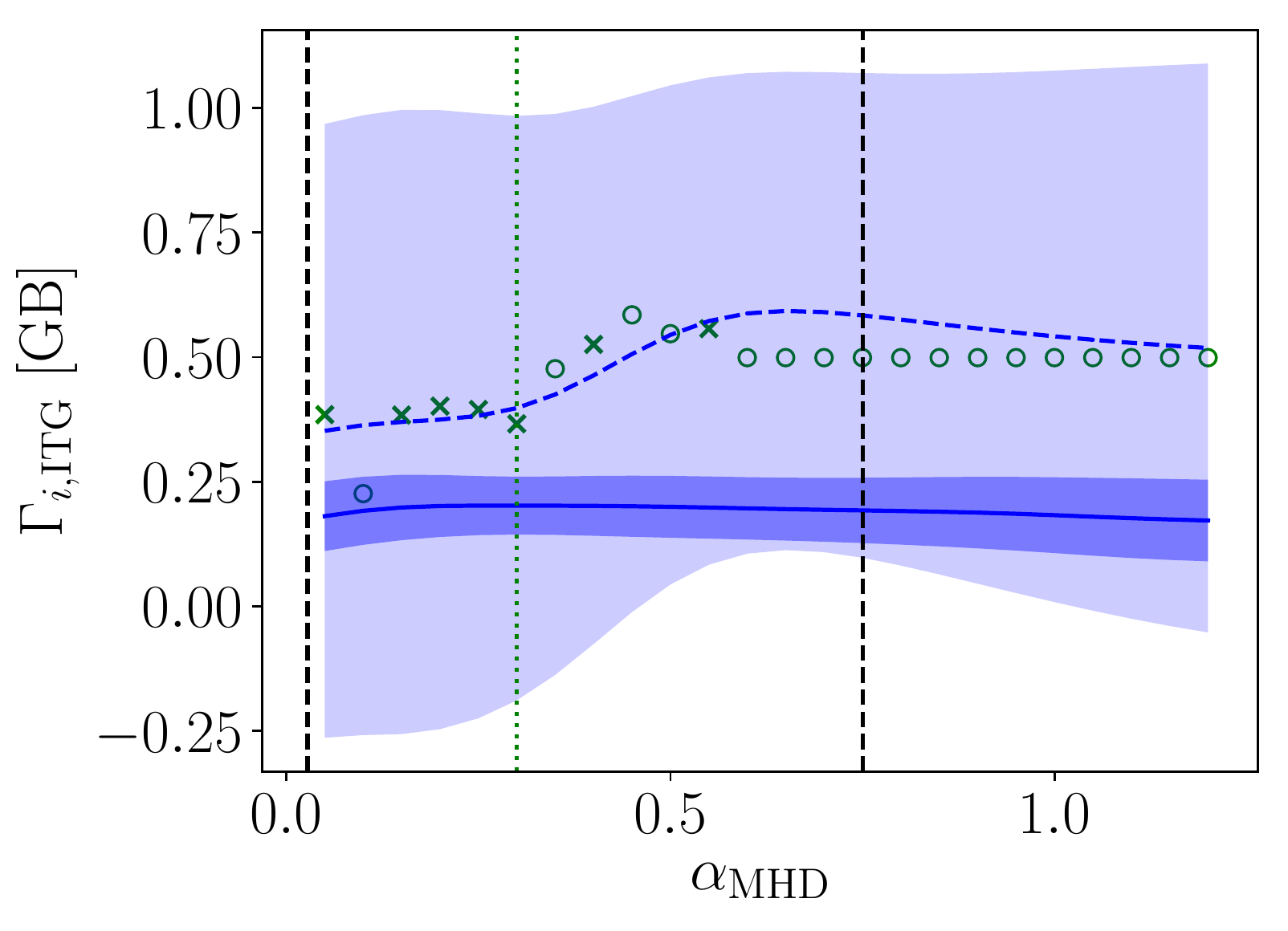} \\
	\hspace{0.25mm}\includegraphics[scale=0.245]{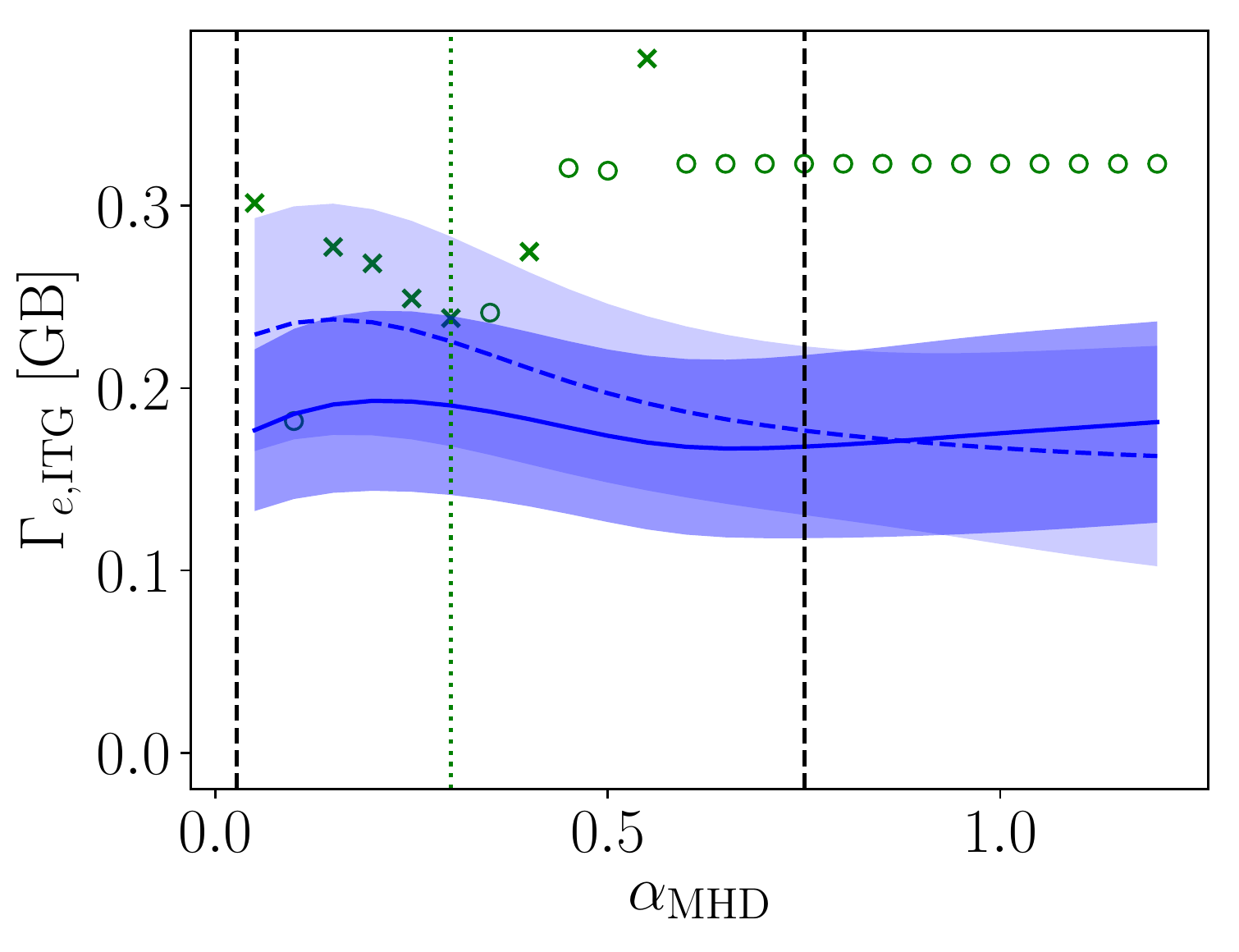}%
	\hspace{2mm}\includegraphics[scale=0.245]{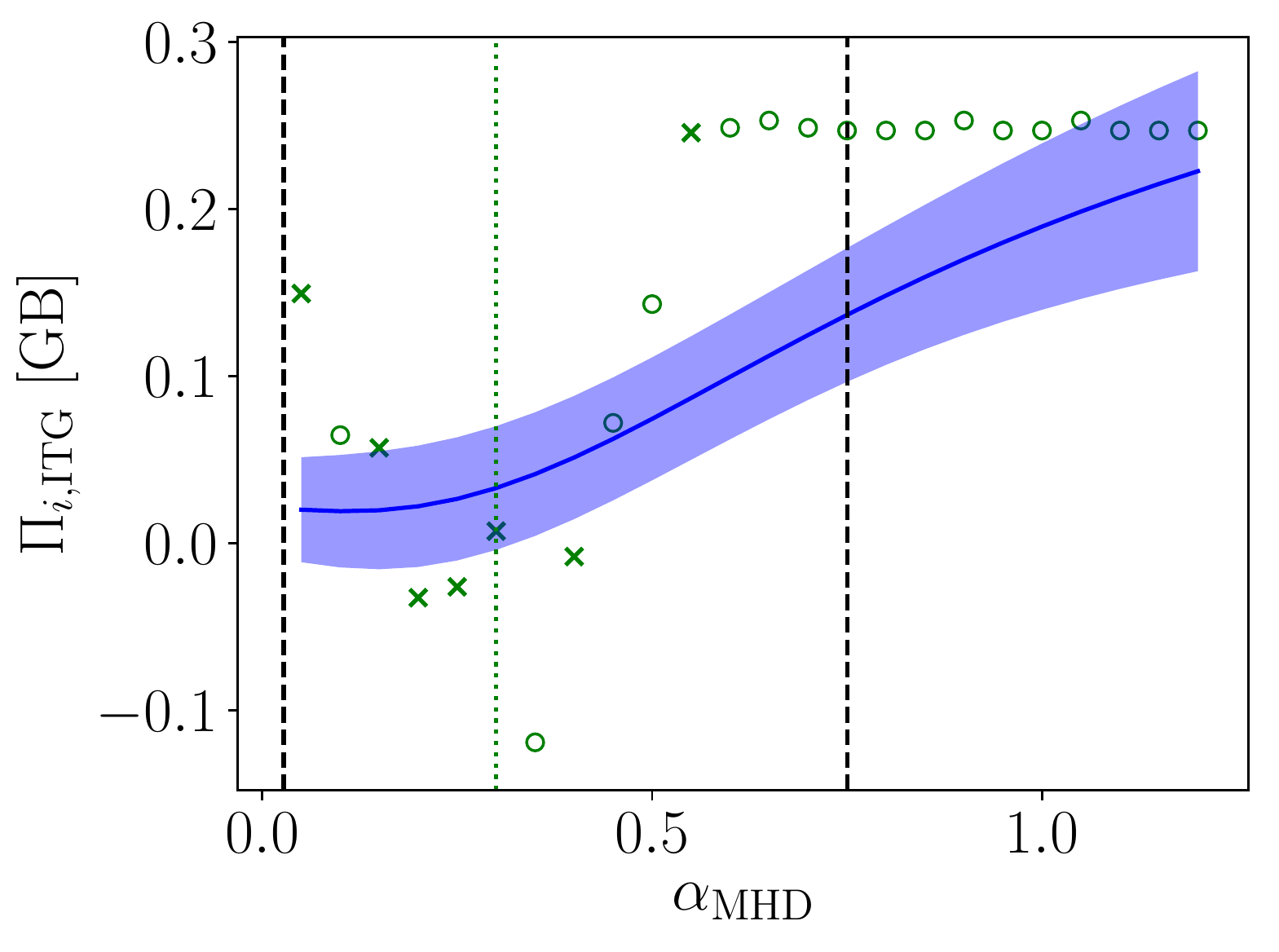}
	\caption{Comparison of main ITG-driven transport fluxes as a function of the normalized pressure gradient, $\alpha_{\text{MHD}}$.}
	\label{fig:FluxComparisonsPressureGradient}
\end{figure}

\begin{figure}[h]
	\centering
	\includegraphics[scale=0.245]{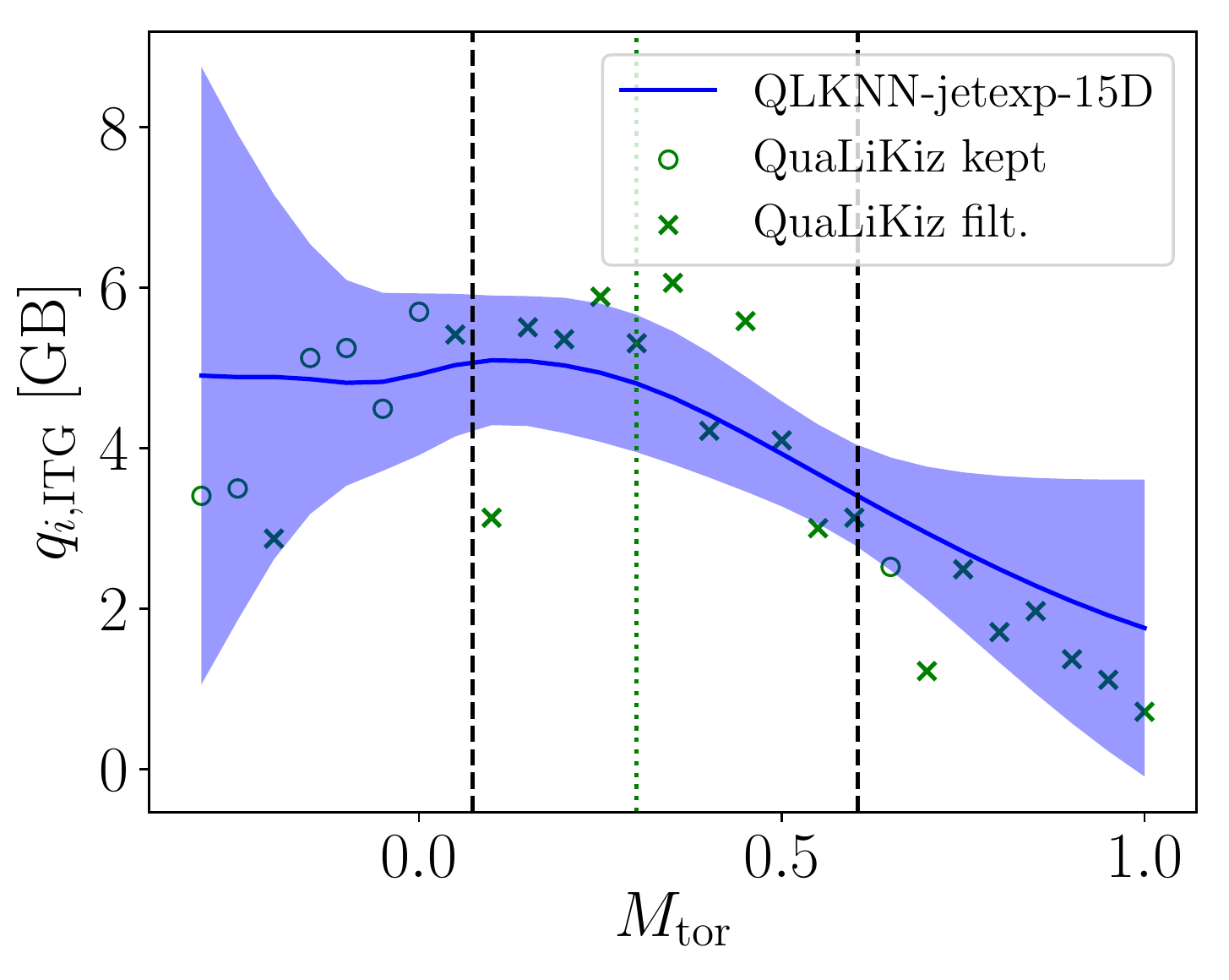}%
	\hspace{2mm}\includegraphics[scale=0.245]{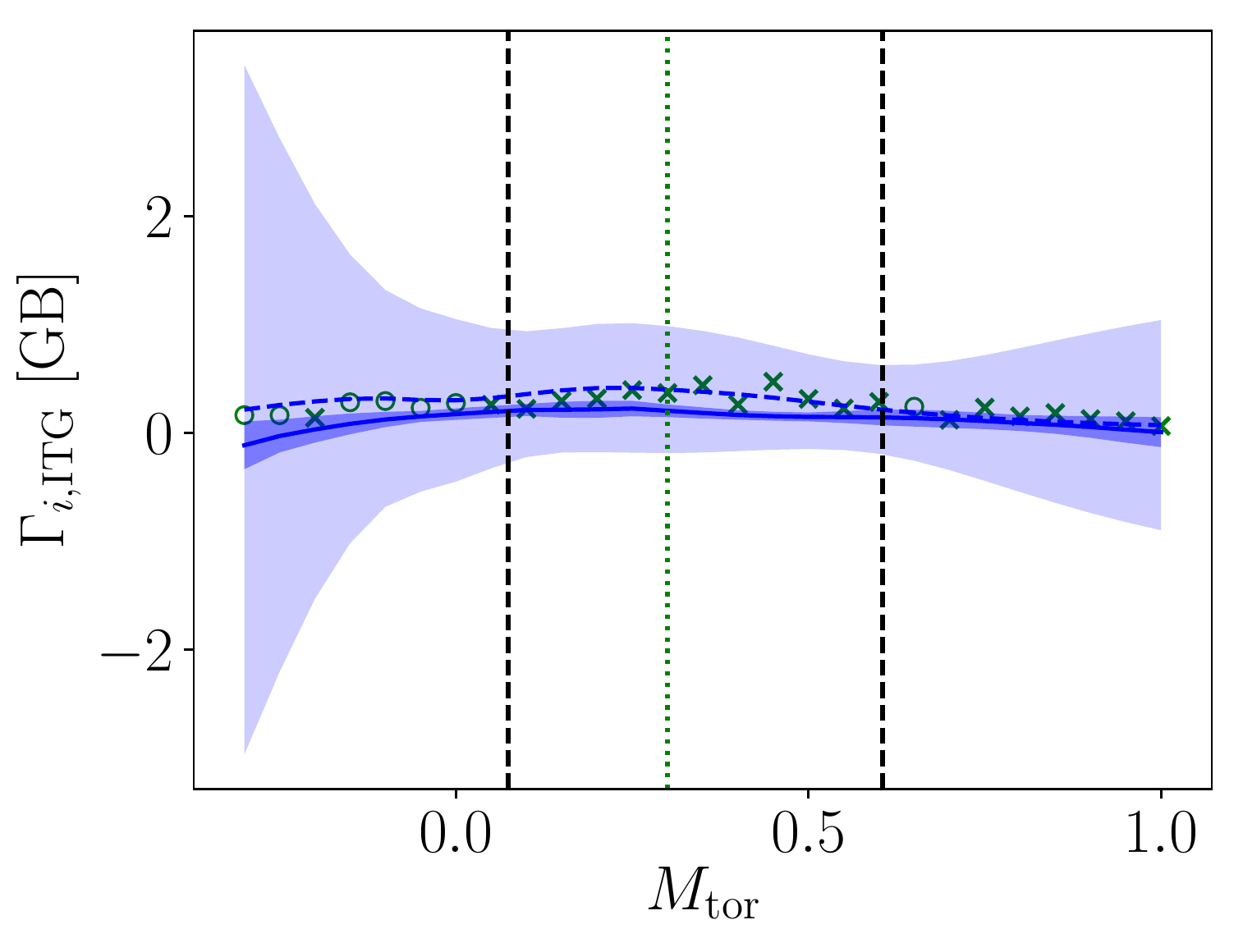} \\
	\hspace{0.25mm}\includegraphics[scale=0.245]{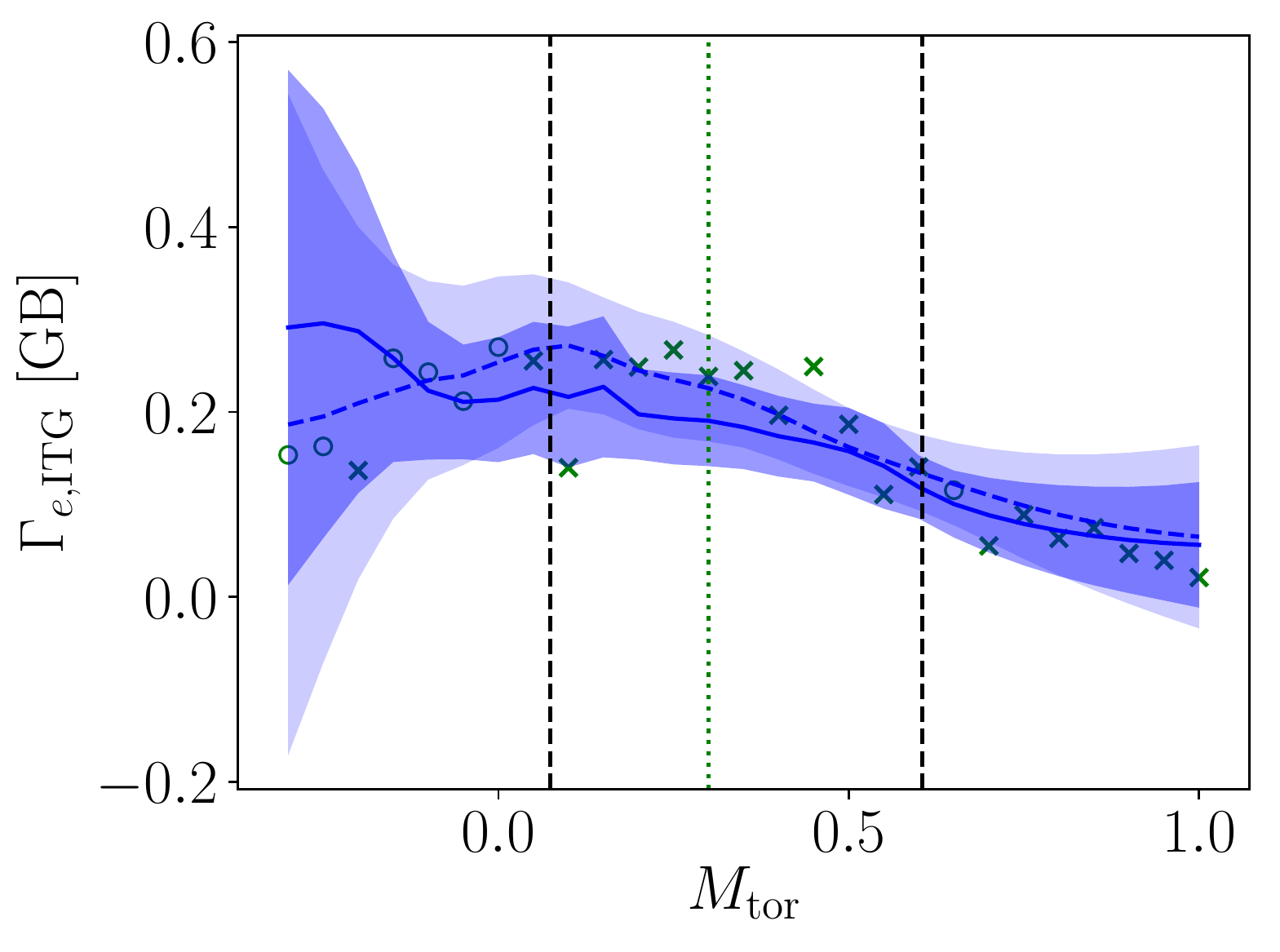}%
	\hspace{2mm}\includegraphics[scale=0.245]{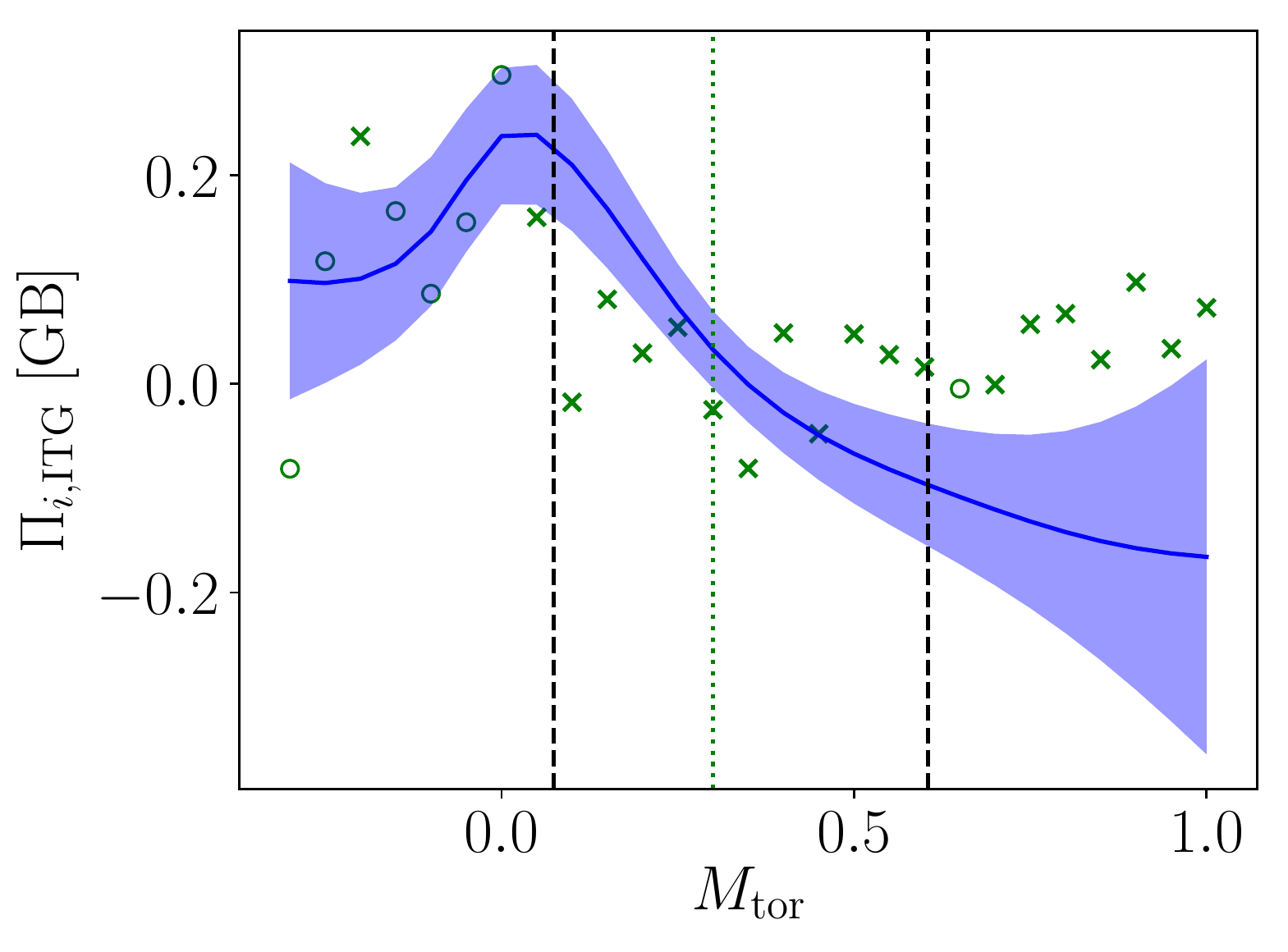}
	\caption{Comparison of main ITG-driven transport fluxes as a function of the toroidal Mach number, $M_{\text{tor}}$.}
	\label{fig:FluxComparisonsRotationVelocity}
\end{figure}

\begin{figure}[h]
	\centering
	\includegraphics[scale=0.245]{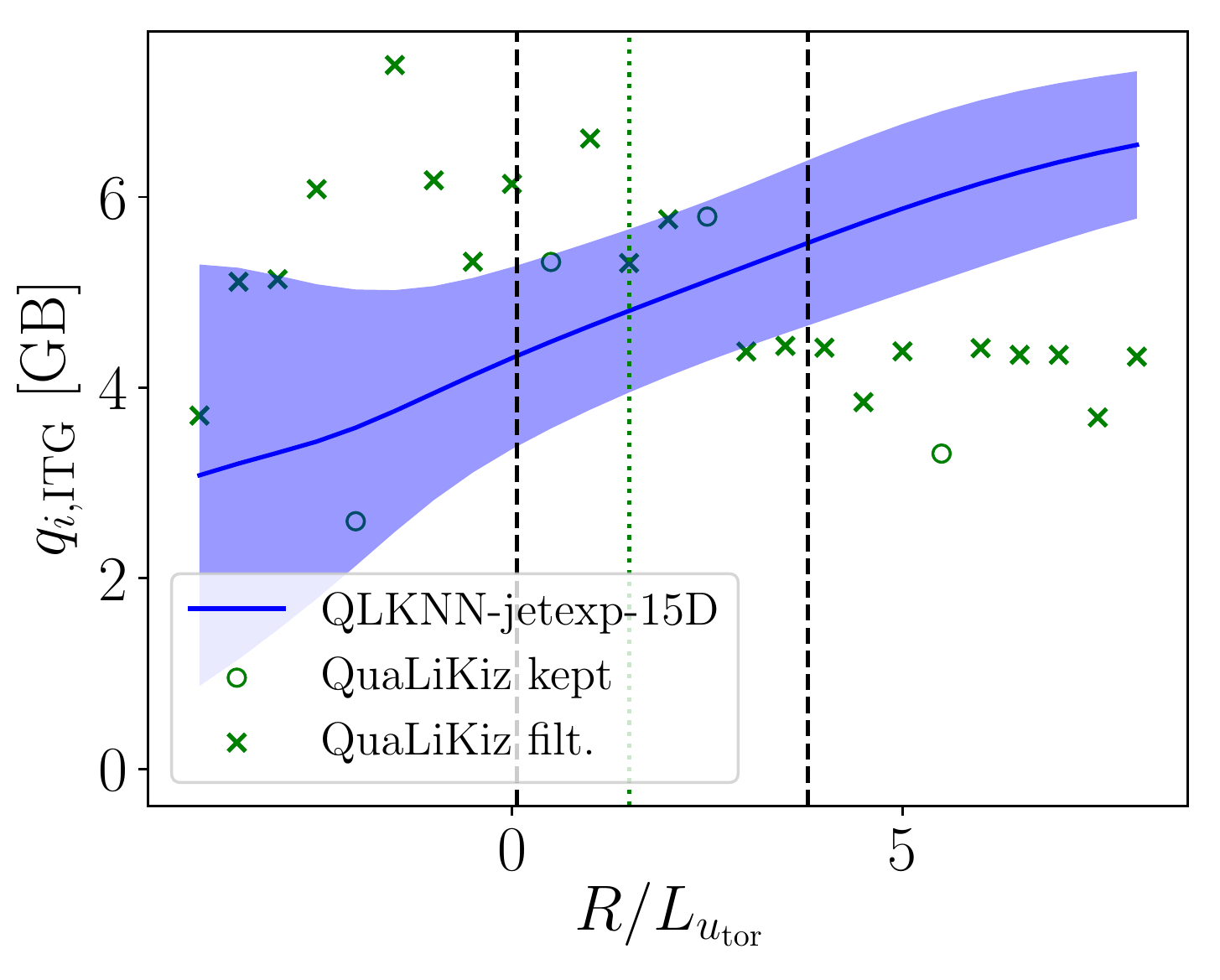}%
	\hspace{2mm}\includegraphics[scale=0.245]{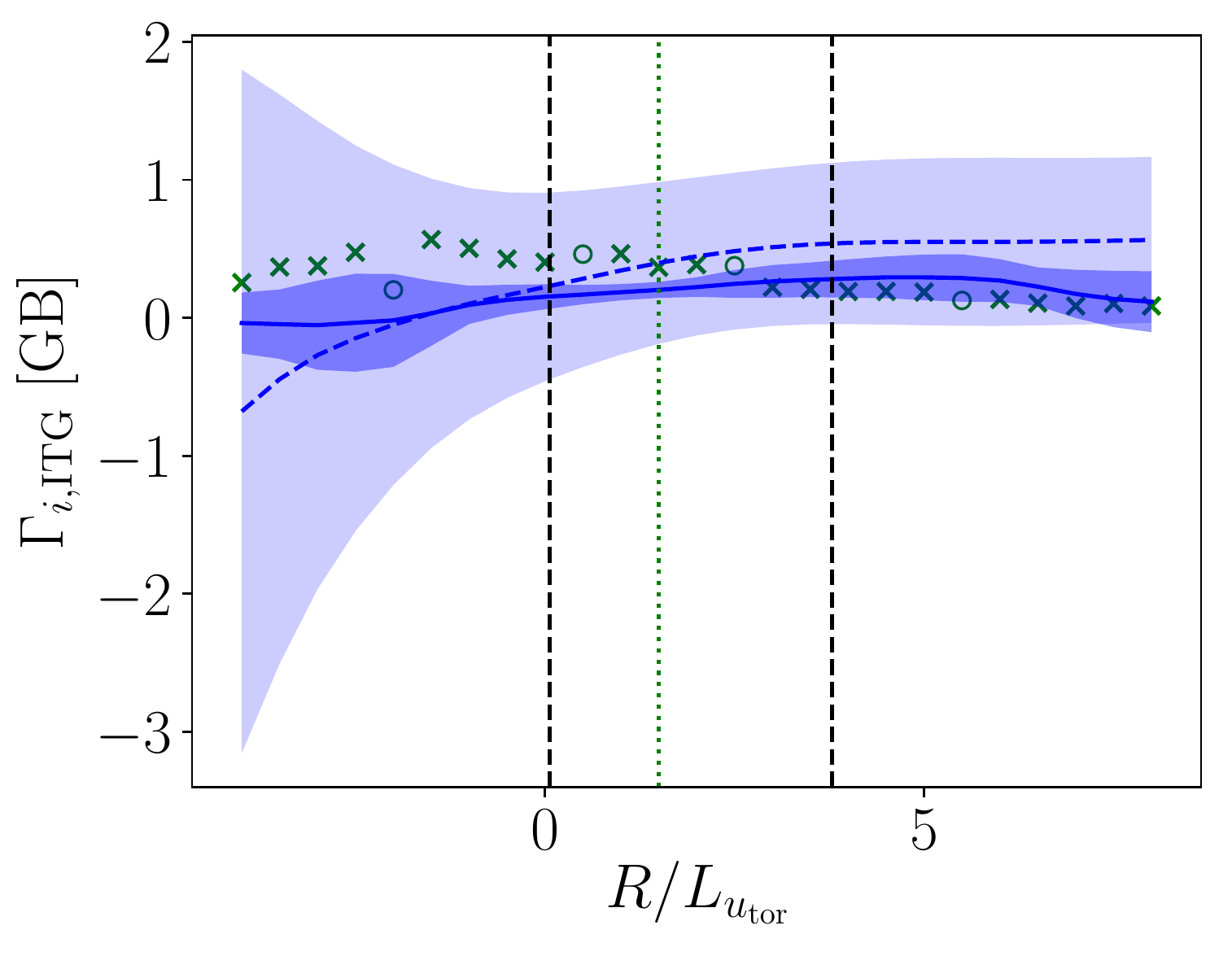} \\
	\hspace{0.25mm}\includegraphics[scale=0.245]{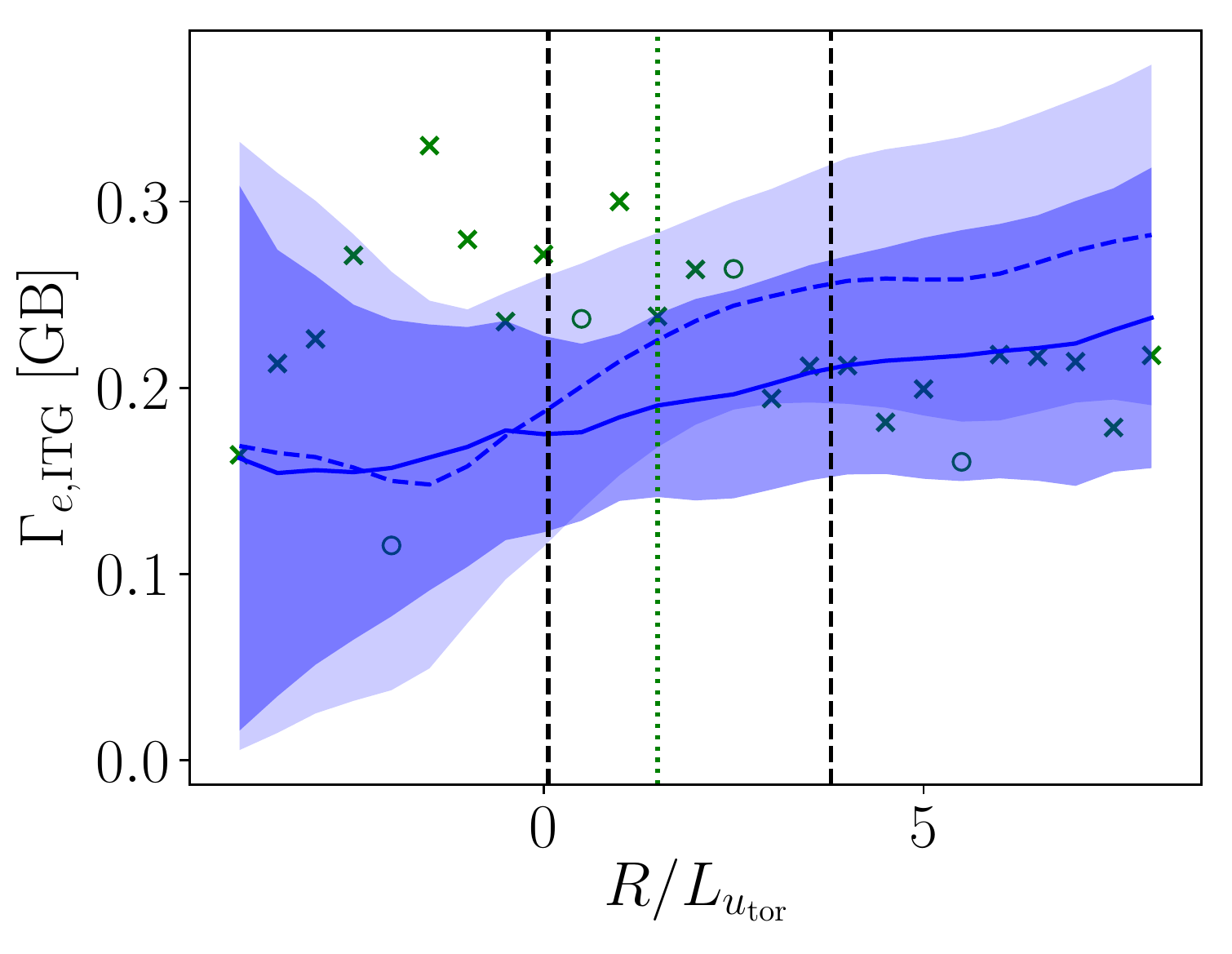}%
	\hspace{2mm}\includegraphics[scale=0.245]{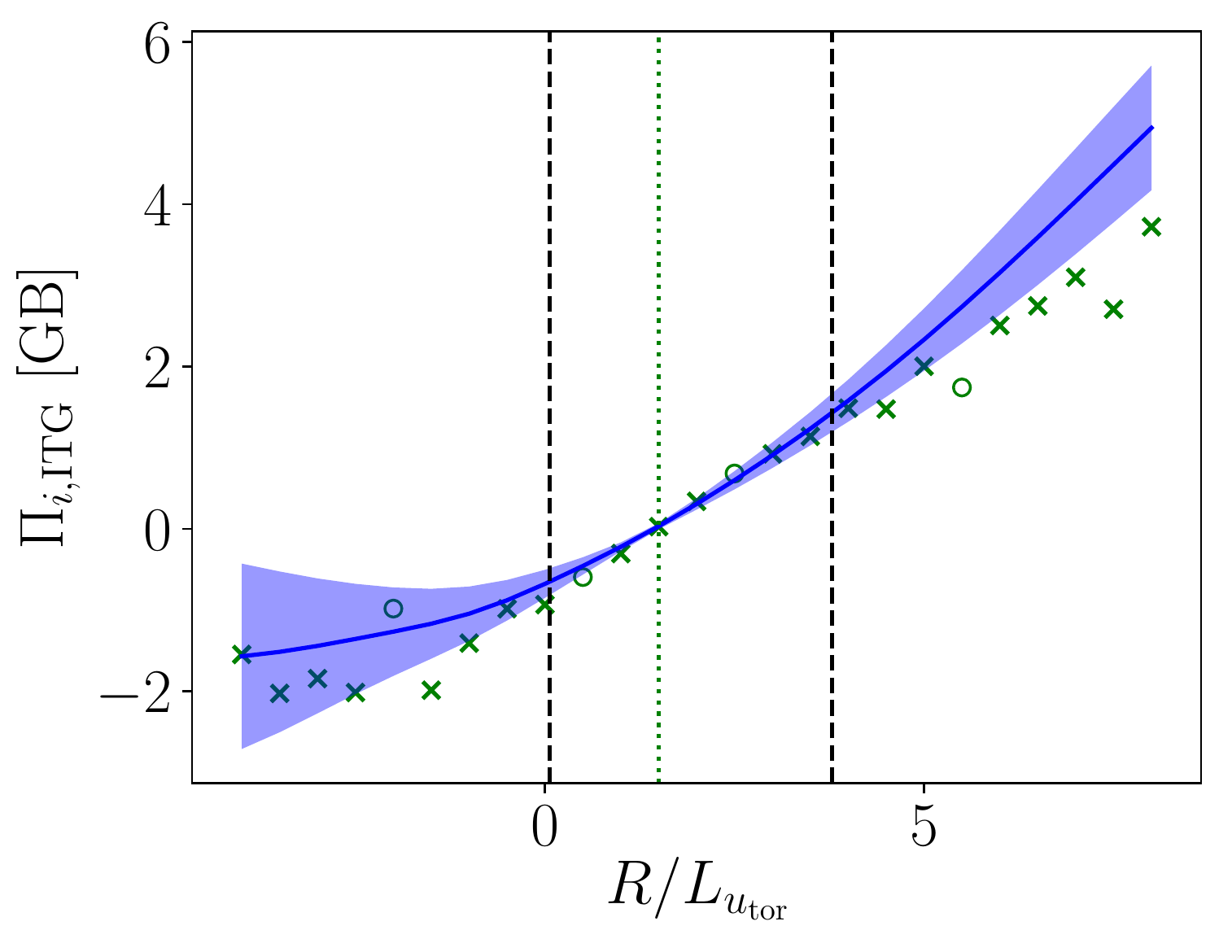}
	\caption{Comparison of main ITG-driven transport fluxes as a function of the normalized toroidal bulk velocity gradient, $R/L_{u_{\text{tor}}}$.}
	\label{fig:FluxComparisonsRotationGradient}
\end{figure}

\end{document}